\documentclass[11pt,oneside,a4paper]{article}
\usepackage{amssymb}
\usepackage{amsmath}
\usepackage[titletoc,title]{appendix}
\usepackage[T1]{fontenc}
\usepackage{graphicx}
\usepackage{float}
\usepackage{subfig}
\usepackage{newtxtext,newtxmath}
\usepackage{url}
\usepackage{color}
\usepackage{hyperref}
\usepackage{miller}
\usepackage{multirow}
\usepackage[gen]{eurosym}
\usepackage{rotating}

\usepackage{multirow}
\usepackage{array}

\usepackage{siunitx}
\usepackage[style=numeric-comp,sorting=none,backend=biber,firstinits=true]{biblatex} % sorting=none : in order references appear in text 
\definecolor{blu}{rgb}{0.,0.,1.}
\definecolor{red}{rgb}{1.,0.,0.}
\definecolor{burgundy}{rgb}{0.5, 0.0, 0.13}
\definecolor{crimsonred}{rgb}{0.6, 0.0, 0.0}
\definecolor{persianblue}{rgb}{0.11, 0.22, 0.73}
\definecolor{forestgreen}{rgb}{0.13,0.35,0.13}

\makeatletter
\g@addto@macro\bfseries{\boldmath}
\makeatother

\hypersetup{colorlinks, citecolor=crimsonred, linkcolor=persianblue, urlcolor=crimsonred}

\newcommand{\Fig}[1]{Figure~\ref{#1}}

\newcommand{\Sec}[1]{Section~\ref{#1}}
\newcommand{\Secs}[1]{Sections~\ref{#1}}

\ifdefined\qtyproduct
\else
  \ifdefined\NewCommandCopy
    \NewCommandCopy\qtyproduct\SI
  \else
    \NewDocumentCommand\qtyproduct{O{}mm}{\SI[#1]{#2}{#3}}
  \fi
\fi

\begin{document}

\begin{center}
\LARGE{\bf High Intensity Kaon Experiments (HIKE) \\ at the CERN SPS}

\vspace{1cm}
\LARGE{\bf Proposal for Phases 1 and 2}

\vspace{1.0cm}
\LARGE{The HIKE Collaboration} 
\end{center}
\vspace{2.0cm}

\begin{figure}[H]
\centering
\includegraphics[width=10cm]{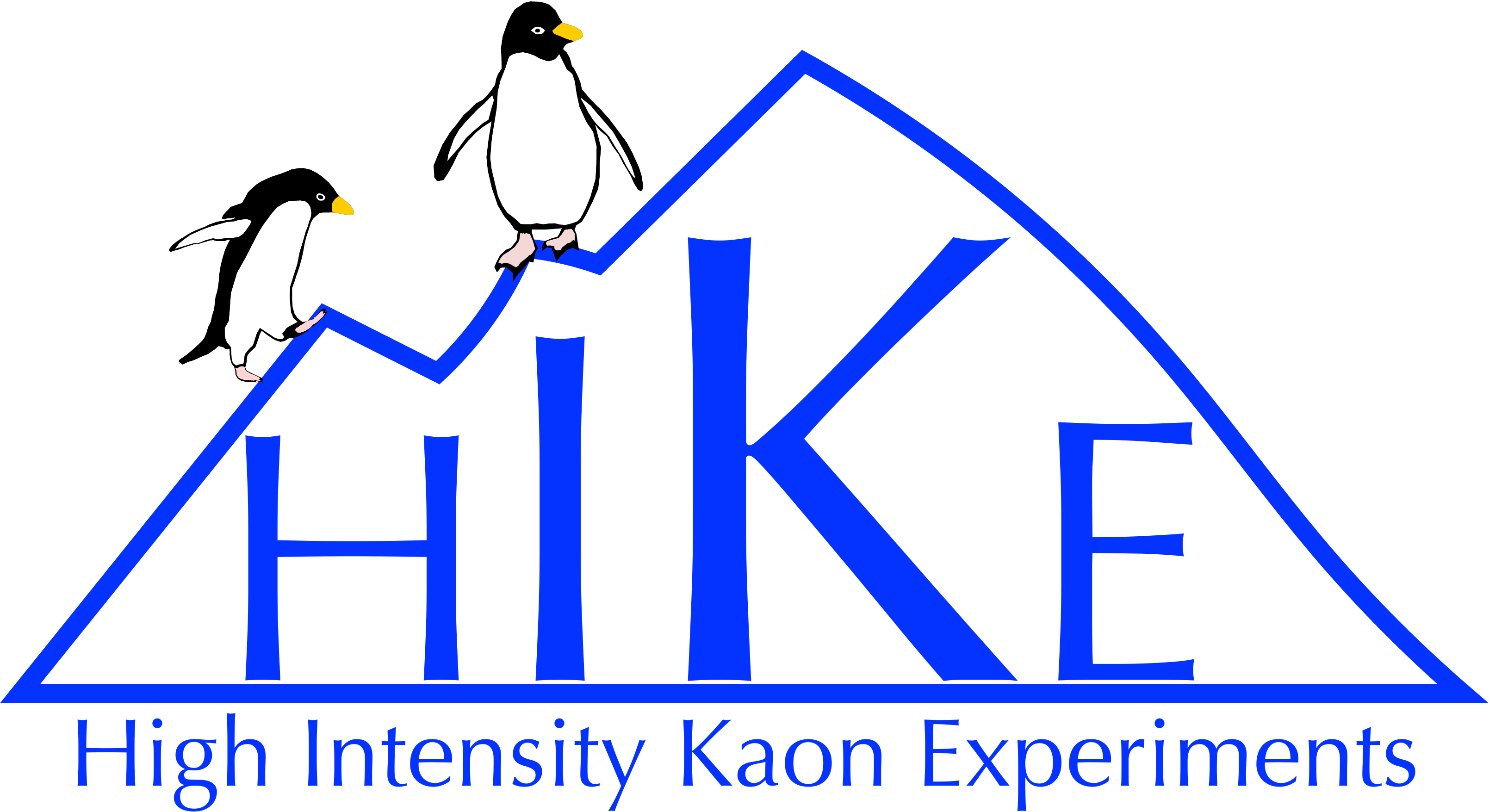}
\label{fig:my_label}
\end{figure}
\vspace{2.0cm}

\begin{abstract}
A timely and long-term programme of kaon decay measurements at an unprecedented level of precision is presented,
leveraging the capabilities of the CERN Super Proton Synchrotron (SPS).
The proposed HIKE programme is firmly anchored on the experience built up studying kaon decays at the SPS over the past four decades, and includes rare processes, CP violation, dark sectors, symmetry tests and other tests of the Standard Model. The programme is based on a staged approach involving experiments with charged and neutral kaon beams, as well as operation in beam-dump mode. The various phases will rely on a common infrastructure and set of detectors.
\end{abstract}

\newpage

\begin{center}
{\Large The HIKE collaboration}
\end{center}
\vspace{-2mm}
\begin{flushleft}
%%%%%%%
%{\bf Universit\'e Catholique de Louvain, Louvain-La-Neuve, Belgium}\\
 M.~U.~Ashraf$^{1}$,  
 E.~Cortina Gil$^{1}$,   
 J.~Jerhot$^{1,8}$,   
 N.~Lurkin$^{1}$,   
%%%%%%%
%{\bf TRIUMF, Vancouver, British Columbia, Canada}\\
 T.~Numao$^{2}$,   
 B.~Velghe$^{2}$,   
 V.~W.~S.~Wong$^{2}$,   
%%%%%%%
%{\bf University of British Columbia, Vancouver, British Columbia, Canada}\\
 D.~Bryman$^{3}$,   
%%%%%%%
%{\bf Charles University, Prague, Czech Republic}\\
 L.~Bician$^{4}$,   
 Z.~Hives$^{4}$
% ...
%
%%%%%%%%%%%%%%%%%%%%%%%%%%%%%%%%%
%
%\newlength{\basefootnotesep}
%\setlength{\basefootnotesep}{\footnotesep}
%
%\begin{flushleft}
%\setcounter{footnote}{0}
%\renewcommand{\thefootnote}{\arabic{footnote}}
\vspace{5mm}

$^{1}$
Universit\'e Catholique de Louvain, B-1348 Louvain-La-Neuve, Belgium \\
$^{2}$
TRIUMF, Vancouver, British Columbia, V6T 2A3, Canada \\
$^{3}$
University of British Columbia, Vancouver, British Columbia, V6T 1Z4, Canada \\
$^{4}$
Charles University, 116 36 Prague 1, Czech Republic \\
% ...
$^{8}$
Max-Planck-Institut f\"ur Physik (Werner-Heisenberg-Institut), D-85748 Garching, Germany \\
% ...
\end{flushleft}
\clearpage

\newpage
\setcounter{page}{1}
\pagestyle{plain}

\section*{Executive summary}

The HIKE programme at the CERN SPS will probe the limits of the precision frontier in the search for physics beyond the Standard Model (SM). 
HIKE's physics reach is complementary to that of the LHC experiments, making its programme synergistic with them in exploring this new territory.
HIKE primarily focuses on flavour physics, aiming to search indirectly for
new phenomena
at mass scales of tens to hundreds of TeV or even higher using high-intensity kaon beams.
HIKE is the only experiment worldwide where the wide-ranging programme described in this proposal can be addressed with comprehensive scope. This programme can only be carried out at CERN, and only in the ECN3 experimental area.

The primary goal of HIKE is to probe the SM with kaon decays at the ${\cal O}(10^{-13})$ level in terms of decay branching fractions; to improve the precision of rare kaon decay measurements to the level needed to match or challenge theory; to measure for the first time channels not yet observed; to search with unprecedented sensitivity for kaon decays forbidden by the SM; and to explore BSM parameter spaces never investigated before. HIKE will also address beam-dump physics in regions of mass--coupling space complementary to those for other existing and planned experiments. The breadth and depth of the HIKE physics programme 
as described in this proposal
is unique, bringing the kaon and FIP programmes to a new level of precision. 

Effects on kaon observables are predicted in several models of BSM physics, ranging from super\-symmetric to non-supersymmetric models, involving heavy $Z'$ bosons, vector-like couplings, leptoquarks, and extra dimensions, in both minimally flavour violating (MFV) scenarios or with new sources of flavour violation (Section~\ref{sec:k-decays}). A common feature is that bounds on the parameter spaces of these models coming from direct searches at LHC, as well as $B$ and $D$ physics, have only a marginal impact on the possible effects on kaon observables, even more so if the models are non-MFV and non-supersymmetric. In contrast, the main constraints on the effects of new physics on kaon observables come from the precisely measured parameters $\varepsilon_K$ and $\Delta M_K$. 
More generally, the comparison between the flavour picture emerging from kaons with that from $B$ mesons is a powerful tool to investigate the indirect effects of new physics
to provide insights into the flavour structure of possible new physics models,
and to push in a model-independent way the sensitivity to mass scales beyond those attainable with $B$ and $D$ mesons only.

A subset of ultra-rare SM processes, called golden modes, in which theoretically clean observables can be measured to reveal new physics contributions that favourably compete with SM processes, play a special role in the indirect search for new physics.
Precision measurements of two of the quintessential golden modes,
$K^+\to\pi^+\nu\bar\nu$ and $K_L\to\pi^0\ell^+\ell^-$,
are at the centre of the HIKE programme.
Measurements of
the branching ratios and spectra of these decays offer model-independent standard candles that can constrain any BSM scenarios, present or future.
The status of BSM models in the future is hard to predict, but measurements made by the unrivalled HIKE experimental programme will be durable standards against which many of those models will be judged. 
Presently, the main limitation to the investigation of BSM models, and more generally to the quest for new physics with kaons, comes from the limited statistical precision in the kaon decay measurements. To a lesser extent, the limitation is due to theoretical uncertainties and the precision of the measurements of the SM parameters. These uncertainties, however, are on track to be reduced in the coming years. Still, the HIKE measurements are a set of firm deliverables even if BSM does not manifest itself.

%%%%%%%%%%%%%%%%%%%%%%
\newpage

HIKE Phase~1 will measure the branching ratio for the ultra-rare decay $K^+\to\pi^+\nu\bar\nu$ with $\mathcal{O}(5\%)$ precision, 
rivalling the precision of theoretical predictions and providing a strong test of the SM, as well as probing many BSM scenarios.
Furthermore, the first comprehensive analysis of the $K^+\to\pi^+\nu\bar\nu$ decay spectrum will probe the nature of the decay in terms of a possible lepton-number-violating scalar contribution, and possible production of hidden-sector mediators (Section~\ref{sec:phase1}).
HIKE Phase~2 will make the first observation
with a significance above $5\sigma$, and measurement with a precision of about 15\% of the ultra-rare golden decay modes $K_L\to\pi^0 e^+e^-$ and $K_L\to\pi^0\mu^+\mu^-$, which will lead to an independent determination of the CKM parameter ${\rm Im}\lambda_t$ to 20\% precision (Section~\ref{sec:phase2}). 
These decays also give unique access to short-distance BSM effects in the photon coupling via the tau loop.
Considering that HIKE has been optimised for the most challenging studies, it 
will carry out a wide range of studies of precision measurements of rare and forbidden kaon decays. These
include (but are not limited to) measurements of $K^+\to\pi^+\ell^+\ell^-$ form factors to sub-percent precision testing lepton flavour universality; $K_L\to\mu^+\mu^-$ rates to percent-level precision; radiative decays to percent or sub-percent precision addressing low-energy QCD parameters; and searches for lepton flavour and lepton number violating decays with ${\cal O}(10^{-13})$ sensitivity. In addition, HIKE measurements will determine $V_{us}$ to a permille precision, elucidating the nature of the first-row CKM unitarity deficit.

From a global fit performed using individual measurement projections as input, it is evident that measurements of rare kaon decays offer powerful constraints on BSM scenarios involving lepton flavour universality violation  (Section~\ref{sec:k-impact}).
When specific BSM models of leptoquarks or $Z^\prime$ are considered, 
it is clear that HIKE will provide a significant step forward: HIKE sensitivity is better than or competitive with, and complementary to, projected bounds on these models from other experiments, even more so at high mass scales.

The long decay volume and detector characteristics needed for kaon physics make HIKE suitable to search for new long-lived feebly interacting particles (FIPs) in beam-dump mode, providing unprecedented sensitivity to forward processes.
The SHADOWS experiment, which is complementary to HIKE in terms of FIPs sensitivity, will be able to run in ECN3 concurrently with HIKE in dump mode. 
Together, HIKE and SHADOWS significantly improve the coverage for FIP masses between $M_K$ and $M_B$ by exploiting on-axis and off-axis production. On the other hand, the HIKE programme includes searches for kaon decays to FIPs, providing sensitivities below the kaon mass where virtually no other experiment dedicated to the search for FIPs has coverage, a line of inquiry already pursued with success by NA62 (Section~\ref{sec:fips}).
With a combination of data taken in kaon and beam-dump modes, HIKE will be able to reach unprecedented sensitivities for all PBC benchmarks (with the exception of BC3 for which the sensitivity is yet to be evaluated). Relying solidly on the data and methods from NA62, HIKE sensitivity curves show marked improvements with respect to present experimental limits for FIP masses below 2~GeV/$c^2$, reaching ${\cal O}(10^{-5})$ to ${\cal O}(10^{-10})$ in the FIP coupling depending on the scenario (Section~\ref{sec:fips_sensitivity}).

The setup and detectors will be optimised for the experimentally most challenging channels, with the key parameters being time resolution, granularity, energy resolution and rate resilience. Some detectors will be upgraded from NA62 while others will be replaced. The prime examples of the latter are the beam tracker, the straw spectrometer and the main calorimeter. The detectors are challenging but generally smaller in size or quantity compared to many other experiments; R\&D and construction of these detectors will be synergetic with on-going developments for High-Luminosity LHC experiments. This both ensures that the HIKE detectors can be developed and constructed on the timescale defined and presents an opportunity to build full-system demonstrators of new technologies that could later be used in other experiments elsewhere (Section~\ref{sec:detectors}). The data-acquisition system will feature state-of-the-art streaming readout (Section~\ref{sec:readout}).

The critical importance of high-intensity kaon experiments at CERN, including future upgrades, to provide a unique probe into BSM physics complementary to the $B$ sector was acknowledged by the European Strategy Group with the inclusion of kaon physics among essential scientific activities in the 2020 strategy update. The HIKE programme in kaon physics is strongly supported in national roadmaps for high-energy physics across Europe. Kaon decays were the central subject of a dedicated working group (RF2) within the Snowmass 2021 Rare Processes and Precision Measurements Frontier, with about 100~authors contributing to the white papers on the theoretical and experimental aspects and the working group report.

HIKE will broaden the CERN flavour physics programme significantly, improving its discovery potential and ensuring long-term leadership in the field. 
Conversely, without HIKE, the currently world-leading CERN kaon physics programme will halt for several decades at least. The only other dedicated high-intensity kaon physics experiment in the world is KOTO at J-PARC.  
Potentially, the measurement of $K_L\to\pi^0\nu\bar\nu$ from KOTO-II could turn up evidence of new physics after 2035; without HIKE, this measurement will exist in isolation. The complementary data from other channels needed for a kaon-based CKM analysis will not exist; the discovery potential and any opportunity to obtain constraints on new physics from other parts of the quark-flavour sector for comparison with results from $B$ physics will have been lost. 

In summary, HIKE provides unique opportunities to constrain new physics, complementing results from other parts of the quark-flavour sector and providing for an independent, kaon-based CKM analysis. HIKE will perform a number of unique tests of fundamental symmetries, and, in conjunction with SHADOWS, will advance the search for hidden sectors. No experimental alternative currently envisioned can come close to completing HIKE's diverse suite of cutting-edge measurements.

\newpage
\setcounter{tocdepth}{2}
\tableofcontents

\section{Introduction}

For 75~years, experimental studies of kaon decays have played a unique role in propelling the development of the Standard Model. As in other branches of flavour physics, the continuing experimental interest in the kaon sector derives from the possibility of conducting precision measurements, particularly of suppressed or rare processes, which may reveal the effects of new physics with mass-scale sensitivity exceeding that which can be explored directly, e.g., at the LHC or a next-generation hadron collider. Because of the relatively small number of kaon decay modes and the relatively simple final states, combined with the relative ease of producing intense kaon beams, kaon decay experiments are in many ways the quintessential intensity-frontier experiments.

Over the past four decades, the CERN North Area has hosted a successful series of precision kaon decay experiments. Among the many results obtained by these experiments is the discovery of direct CP violation, widely quoted to be among the top 10 discoveries made at CERN. Continuation of high-intensity kaon experiments at CERN 
has been identified as an essential scientific activity in the 2020 Update of the European Strategy for particle physics, and is strongly supported in the national roadmaps across Europe.

The High-Intensity-Kaon-Experiments (HIKE) project represents a broad, long-term programme at CERN after Long Shutdown 3~(LS3), based in the North Area ECN3 experimental hall, covering all the main aspects of rare kaon decays and searches accessible via kaon physics, from ultra-rare kaon decays to precision measurements and searches for new phenomena. 
HIKE is intended to continue the very successful tradition of kaon experiments at CERN in ECN3, where NA62, currently operating, is the latest of these experiments. 
NA62 is the first decay-in-flight experiment to measure ${\cal B}(K^+\to\pi^+\nu\bar\nu)$, and the extensive experience learned there, and at its predecessor NA48 with neutral kaon beams, is key towards building improved experiments.

HIKE will profit from a beam intensity up to six times higher than that of NA62 and cutting-edge detector technologies to perform precision measurements and searches for physics beyond the Standard Model (BSM).
This will allow HIKE to play a pivotal role in the quest for BSM physics at the sensitivity required by the present experimental limits and theoretical models, over a wide range of possible masses and interaction couplings. The programme presented in this proposal relates to the first two phases described in the HIKE Letter of Intent submitted in November 2022~\cite{HIKE:2022qra}. Phase~1 will make use of a $K^+$ beam with an intensity of four times that of NA62. The centrepiece of the programme is the measurement of ${\cal B}(K^+\to\pi^+\nu\bar\nu$) to $\mathcal{O}(5\%)$ precision; a broad array of measurements will also be carried out using the intense beam and state-of-the-art detector. Phase~2 will make use of a $K_L$ beam with an intensity of six times that of NA62. The main goals are the $K_L\to\pi^0 e^+e^-$ and $K_L\to\pi^0 \mu^+\mu^-$ decays, sensitive probes for new physics which have never been observed; a broad slate of other $K_L$ decay measurements is similarly planned. HIKE will run periodically in beam-dump mode to perform searches for long-lived, feebly-interacting particles. 
The possibility to measure ${\cal B}(K_L\to\pi^0\nu\bar\nu)$ in a third phase, described in the Letter of Intent, demonstrates that HIKE will continue to be able to produce extremely interesting measurements over an even longer term.
However, because of the time scale and significant additional investment required, Phase~3 will be the subject of a separate proposal at a later time.  

HIKE implementation will follow a general, staged approach, in which new or refurbished detectors are installed as soon as they are needed and ready, while maintaining the principle that changes must serve the remaining phases of the programme once they are applied.
The organisation into phases allow for insertion, repositioning or removal of specific elements depending on the physics requirements, on top of a largely common detector and DAQ suitable for all phases. The changes required between phases mainly concern the beamline.

\newpage

\section{HIKE kaon physics programme}
\label{sec:k-decays}

The Standard Model (SM) of particle physics describes the results obtained so far by a vast array of experiments to exceptional precision. The culmination of its success was the discovery of the Higgs boson by the ATLAS and CMS collaborations in 2012~\cite{ATLAS:2012yve,CMS:2012qbp}. Thus the SM provides a solid foundation for the present understanding of elementary particles and their interactions.
However cosmological observations, experimental tensions, and theoretical motivations strongly suggest that the SM is an approximation of a more fundamental theory.
One paradigm assumes that this theory lies above the electroweak (EW) mass scale and manifests itself in terms of new particles with masses well above the Higgs boson mass having sizeable interactions with the SM fields.
Another paradigm, 
usually referred to as the feebly interacting particle (FIP) scenario, contemplates extensions of the SM that predict particles with masses below the EW scale that interact only very weakly with the SM fields.
Both high-scale BSM and FIP models provide explanations for the principal open questions in modern physics: the nature of dark matter, the baryon asymmetry of the universe, cosmological inflation, the origin of neutrino masses and oscillations, the strong CP problem, the hierarchical structure of the Yukawa couplings, the hierarchy problem, and the cosmological constant.

HIKE intends to study new physics with kaon decays in both high-scale BSM and FIP scenarios.
In this respect, HIKE is perfectly aligned with the recommendation of the 2020 Update of the European Strategy for Particle Physics~\cite{EuropeanStrategyGroup:2020pow}%
, where a chapter is devoted to flavour physics research with the statement: 
``Experimental hints for deviations from SM predictions in flavour processes are one of our best hopes to direct research towards the right energy scale where NP can be found.''

\subsection{Experimental context for high-scale BSM physics}

The experimental techniques for the search for BSM physics at high mass scales are the direct detection of new massive particles or processes forbidden by the SM and precision measurements of observables known precisely within the SM. Experiments of the former type are usually referred to as {\it direct searches}, the latter as {\it indirect searches}.

Direct searches allow unambiguous identification of new physics 
in single processes.
Following the observation and the study of the Higgs boson, direct searches for BSM particles became a primary goal of ATLAS and CMS experiments at the LHC.
However, the LHC centre-of-mass energy limits the sensitivity of this technique to the TeV scale or slightly above. Besides, a direct observation alone gives little information on the BSM structure.
The absence of any significant direct observations of BSM phenomena so far sets strong bounds on several types of BSM models up to the TeV scale.
The bounds are still statistically limited in the TeV region, and the LHC high-luminosity programme will allow them to be extended to the kinematic limits, if no observations emerge in the meantime.
Another type of direct search focuses on the study of processes forbidden by the SM due to accidental SM symmetries such as charged lepton flavour or lepton number conservation.
These searches typically involve the study of hadron or lepton decays forbidden by the SM, e.g. $K_L\to\mu^\pm e^\mp$ or $\tau\to\mu\mu\mu$.
Presently, these searches are statistically limited, and experiments at the high-intensity frontier are planned with increased sensitivity. 

Indirect searches exploit the possibility that new particles affect low-energy observables virtually, via loop-level corrections to the SM prediction.
The golden modes for indirect searches are theoretically clean observables of rare processes in the SM, i.e., those occurring at loop-level at the lowest order, because BSM here may favorably compete with SM.
Indirect searches allow sensitivities to BSM up to mass scales well above the TeV scale, but may require measurements of several observables to provide conclusive evidence of new physics and to determine its nature.
Given the limits on new physics at the TeV scale already set at the LHC through direct searches, indirect searches represent a promising tool to further boost BSM searches.
Experimental flavour physics, measurements of Higgs couplings, and measurements in the gauge sectors of the SM ($W$, $Z$, and top masses and gauge couplings) are typical examples of indirect searches.
The existing measurements are broadly consistent with the SM and already suggest that new physics may occur well beyond the TeV scale.
Nevertheless, tensions between experiments and the SM do exist, mainly in flavour physics observables, and statistical limitations and the lack of measurements of rare processes demonstrate the need for additional efforts and motivate future directions in experimental particle physics.

The sensitivity of indirect searches to BSM physics scales with dataset size, which is particularly important for rare processes.
Moving to higher intensities leads to increased statistics; however, this requires improvements of detector performance to cope with harsher experimental environments.
Upgraded detectors also allow the accuracy of already precise measurements to be improved, overcoming systematic uncertainties whenever they are the main limiting factor.
The indirect-search approach is at the heart of the LHC intensity frontier programme; it is the motivation for the upgrade of the LHCb experiment and for the continuation of the $B$-factory programme with the Belle~II experiment, as well as for the HIKE programme in ECN3. In the case of HIKE, the relatively limited number and the hierarchy of well-measured large branching ratios in 
the kaon system are valuable assets for keeping important categories of background under control.

%%%%%%%%%%%%%%%

\subsection{Flavour-changing neutral currents}
\label{sec:rare-kaon-decays}

Rare kaon decays, defined as those proceeding at the loop level in the SM, typically have branching ratios below $10^{-7}$. They provide a tool independent from $B$ and $D$ physics to test the flavour structure of the SM by constraining the unitary triangle via loop-level observables.

\begin{figure}[htb]
\includegraphics[width=1.0\textwidth]{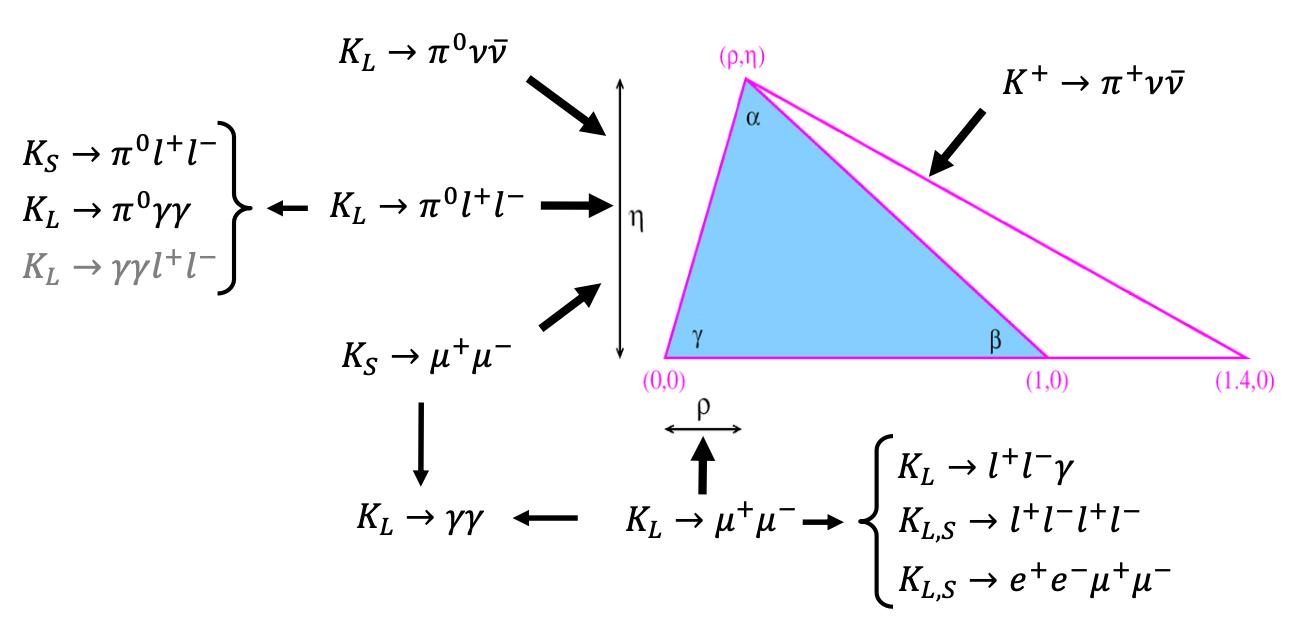}
\vspace{-8mm}
\caption{Relation between kaon rare decay modes and the parameters $\rho$ and $\eta$ of the unitary triangle (UT).
The direct link between decay modes and the UT indicates short distance terms dependent on  $\rho$ or $\eta$ contributing to the corresponding decay amplitudes.
Decays not directly connected to the UT are relevant to interpret the experimental results of the decay modes to which they are related.}
\label{fig:ckmk}
\end{figure}

The link between rare kaon decays and flavour physics in the $\rho-\eta$ plane is schematically shown in Fig.~\ref{fig:ckmk}. 
CP-violating short distance (SD) physics contributes to the $K_L\to\pi^0\nu\bar\nu$, $K_L\to\pi^0e^+e^-$, $K_L\to\pi^0\mu^+\mu^-$ and $K_{S}\to\mu^+\mu^-$ decays.
The amplitudes of these decays thus depend on the height, $\eta$, of the unitary triangle.
The SD physics contributing to the decay $K_L\to\mu^+\mu^-$ is CP-conserving, and its amplitude depends on the base, $\rho$, of the unitary triangle.
The amplitude of the rare decay $K^+\to\pi^+\nu\bar{\nu}$ contains terms depending on both $\eta$ and $\rho$.

The $K_L\to\pi^0e^+e^-$, $K_L\to\pi^0\mu^+\mu^-$ and $K_L\to\mu^+\mu^-$ decays are long-distance (LD) dominated.
The extraction of their SD components from experimental data proceeds through the study of ancillary decay modes listed in Fig.~\ref{fig:ckmk} in association with the rare decays mentioned above.
In contrast, LD contributions are sub-dominant to the amplitudes of the $K^+\to\pi^+\nu\bar\nu$ decay, and negligible for the $K_L\to\pi^0\nu\bar\nu$ decay. As a matter of fact, these two decay modes belong to the theoretically cleanest probes of the SM flavour structure among all the kaon and $B$-meson decays.

%%%%%%%%%%%%%%%%%%%

\subsubsection{$K\to\pi\nu\bar\nu$ decays}
\label{sec:PhysicsKpinunu}

The $K\to\pi\nu\bar\nu$ decays involve $s\to d$ quark transitions. They proceed through box and penguin diagrams, which are SD dominated even in the SM due to the leading contribution of the virtual $t$-quark exchange.
The quadratic GIM mechanism and the Cabibbo suppression of the $t\to d$ transition place these processes among the rarest meson decays in the SM.
Their decay amplitudes can be parameterised in terms of the precisely measured $K^+\to\pi^0e^+\nu$ decay, allowing theoretical computation free from hadronic uncertainties.

It is convenient to express the formulas for the branching fractions in terms of contributions from the different loop functions. Following Ref.~\cite{Buras:2015qea}, the SM predictions can be written as
\begin{displaymath}
{\cal B}(K^+ \to \pi^+ \nu \bar{\nu})= \kappa_{+}
(1+\Delta_{\rm EM})\left[ \left(\frac{{\rm Im} \lambda_t}{\lambda^5} X(x_t)\right)^2  + \left(\frac{{\rm Re} \lambda_c}{\lambda}P_c(X)+\frac{{\rm Re} \lambda_t}{\lambda^5}X(x_t)\right)^2  \right],
\end{displaymath}
with $\Delta_{EM}=-0.003$ the electromagnetic radiative corrections, $x_t= m_t^2/M_W^2$, $\lambda = |V_{us}|$, $\lambda_i=V^*_{is}V_{id}$ the relevant combinations of CKM matrix elements, $X$ and  $P_c(X)$ the loop functions for the top and charm quark respectively, and 
$$ \kappa_+ = (5.173 \pm 0.025)\times 10^{-11}\left[\frac{\lambda}{0.225}\right]^8 $$the parameter encoding the relevant hadronic matrix elements extracted from semileptonic decay rates. As the formula shows, $\mathcal{B}(K^+ \to \pi^+\nu\bar\nu)$ depends on the sum of the square of the imaginary part of the top loop (CP violating)  and the square of the sum of the charm contribution and the real part of the top loop.
The corresponding formulas for $K_L$ are:
\begin{displaymath}
{\cal B}(K_L\to\pi^0 
\nu\bar\nu)= \kappa_L \left(\frac{{\rm Im}~\lambda_t}{\lambda^5} X(x_t)\right)^2,
\end{displaymath}
and
\begin{displaymath}
\kappa_L = (2.231 \pm 0.013)\times 10^{-10}\left[\frac{\lambda}{0.225}\right]^8.
\end{displaymath}
The quantity $\mathcal{B}(K_L\to \pi^0\nu\bar\nu)$ depends only on the square of the imaginary part of the top loop which is CP violating. The charm contributions drop out because $K_L$ is mostly an odd linear combination of $K^0$ and $\bar{K}^0$. 

The SM predictions of the  $K^+\to\pi^+\nu\bar\nu$ and $K_L\to\pi^0\nu\bar\nu$ branching ratios can also be written more explicitly in terms of directly measurable CKM parameters as~\cite{Buras:2015qea}
\begin{align}
{\cal B}(K^+\to\pi^+\nu\bar{\nu})&=(8.39\pm0.30)\times10^{-11}\left(\frac{|V_{cb}|}{0.0407}\right)^{2.8}\left(\frac{\gamma}{73.2^\circ}\right)^{0.74},
\label{eq:pnnsm_Kp}
\\
{\cal B}(K_L\to\pi^0\nu\bar{\nu})&=(3.36\pm0.05)\times10^{-11}\left(\frac{|V_{ub}|}{3.88\times10^{-3}}\right)^2\left(\frac{|V_{cb}|}{0.0407}\right)^2\left(\frac{\sin{\gamma}}{\sin{73.2^\circ}}\right)^2.
\label{eq:pnnsm_KL}
\end{align}
Here $V_{cb}$ and $V_{ub}$ are elements of the CKM matrix, and $\gamma$ is the angle of the unitary triangle defined as $\arg[(-V_{ud}V_{ub}^*)/(V_{cd}V_{cb}^*)]$.
The theoretical uncertainty depends on the NNLO approximation of the computation of the Feynman diagrams, on the radiative corrections, and on the estimates of the long-distance contributions due to the exchange of the $u$, $d$ and $c$ quarks.
The prediction for the neutral mode is more precise than that for the charged mode because the neutral amplitude is purely imaginary, allowing cancellation of the long-distance corrections.
In either case, the main contribution to the  uncertainties for the branching ratio calculations comes from the precision of the measurements of the CKM matrix elements, i.e., the parametric uncertainty, which is as large as 9\% if the current measurements of $V_{cb}$, $V_{ub}$ and $\gamma$ from the CKM fit are considered.
Eqs.~(\ref{eq:pnnsm_Kp}) and~(\ref{eq:pnnsm_KL}) also indicate that precision measurements of $K\to\pi\nu\bar\nu$ decays would shed light on CKM parameters historically measured with $B$ mesons that suffer from long-standing experimental tensions, such as $|V_{cb}|$.
A global fit to the CKM unitary triangle allows the parametric uncertainty to be factored out and offers a powerful tool to compare the SM flavour structure arising from kaon and $B$-meson physics.

Recently, a suitable combination of external parameters nearly independent of new physics contributions has been identified, which allows the parametric uncertainty to be decreased, reducing the overall relative uncertainties in the branching ratios to 5\%~\cite{Buras:2021nns}.
Writing the meson branching ratios as functions of $\epsilon_K$ and the CKM angles $\beta$ and $\gamma$ only, the following $|V_{cb}|$-independent quantities are obtained~\cite{buras2023kaon}:

\begin{align}
\frac{{\cal B}(K^+\to\pi^+\nu\bar{\nu})}{|\varepsilon_K |^{0.82}} &=(1.31\pm0.05)\times10^{-8}\left(\frac{\sin\gamma}{\sin 64.6^\circ}\right)^{0.015}\left(\frac{\sin 22.2^\circ}{\sin \beta}\right)^{0.71},
\label{eq:pnnsm_Kp1}
\\
\frac{{\cal B}(K_L\to\pi^0\nu\bar{\nu})}{|\varepsilon_K |^{1.18}} &=(3.87\pm0.06)\times10^{-8}\left(\frac{\sin\gamma}{\sin 64.6^\circ}\right)^{0.03}\left(\frac{\sin\beta}{\sin 22.2^\circ}\right)^{0.98}.
\label{eq:pnnsm_KL1}
\end{align}
which, using the experimental values of $\Delta M_s$, $\Delta M_d$, and $|\varepsilon_K|$, leads to
\begin{align}
{\cal B}(K^+\to\pi^+\nu\bar{\nu})&=(8.60\pm0.42)\times10^{-11},
\label{eq:pnnsm_Kp2}
\\
{\cal B}(K_L\to\pi^0\nu\bar{\nu})&=(2.94\pm0.15)\times10^{-11}.
\label{eq:pnnsm_KL2}
\end{align}
The precision of the SM estimate in Eq.~(\ref{eq:pnnsm_Kp2}) is well matched to the target precision of HIKE Phase~1. Other recent calculations reach a similar quantitative conclusion~\cite{Brod:2021hsj}.

The extreme SM suppression makes these decays particularly sensitive to new physics. From a model-independent point of view, the $K\to\pi\nu\bar\nu$ decays probe BSM at very large mass scales, on the order of hundreds of TeV~\cite{Buras:2015qea}.
Existing experimental constraints on NP affect the $K\to\pi\nu\bar\nu$ branching ratio weakly.
Model-dependent scenarios predict sizeable deviations of the branching ratios from the SM, as well as correlations between the branching ratios for the  charged and neutral modes, depending on the model (see for example Refs.~\cite{Chen:2018ytc,Bobeth:2017ecx,Bobeth:2016llm,Endo:2016tnu,Endo:2017ums,Crivellin:2017gks,Blanke:2015wba,Bordone:2017lsy}).

The NA62 experiment at CERN has so far observed 20 candidate events for the decay $K^+\to\pi^+\nu\bar\nu$ with 7~expected background events and 10~expected SM signal events~\cite{NA62:2021zjw}, leading to the measurement  $\mathcal{B}(K^{+}\to\pi^{+}\nu\bar{\nu})=(10.6^{+4.0}_{-3.4}|_{\text{stat}}\pm0.9_{\text{syst}})\times10^{-11}$ at 68\% confidence level. 
This represents the most precise measurement to date of this process, providing first evidence for its existence and falsifying the background-only hypothesis with $3.4\sigma$ significance.
The experiment is currently taking data in Run~2 (2021--LS3) with the aim of reaching an accuracy between 15\% and 20\% on the branching ratio measurement, and has demonstrated the ability to sustain nominal beam intensity. The NA62 experiment has therefore shown that the decay-in-flight technique works well and is scalable to larger data samples.

The principal goal of HIKE Phase~1 is an ${\cal O}(5\%)$  measurement of the $K^+\to\pi^+ \nu\bar\nu$ decay rate. Beyond the rate measurement, it is important to establish if the decay has a purely vector nature as expected within the SM, considering that an additional scalar contribution to the decay is predicted in certain BSM scenarios~\cite{Deppisch:2020oyx,Crosas:2022quq,Aebischer:2022vky}.
The HIKE sensitivity to the $K^+\to\pi^+\nu\bar\nu$ decay is discussed is detail in Section~\ref{sec:phase1:kpinn}.

The current upper limit on the branching ratio of the $K_L\to\pi^0\nu\bar\nu$ decay is $3\times10^{-9}$ at 90\%~CL, set by the KOTO experiment at J-PARC~\cite{KOTO:2018dsc}. The KOTO experiment is currently taking data with the goal of reaching ${\cal O}(10^{-11})$ sensitivity in the next five years. In the longer term, an ambitious upgrade, KOTO Step-2, is planned to begin construction after 2025 in a proposed extension of the Hadron Experimental Facility, with the goal of measuring the branching ratio to a 20\% precision in about three years of data taking~\cite{NA62KLEVER:2022nea}.
A possible third HIKE phase focusing on the measurement of $K_L\to\pi^0\nu\bar\nu$, outside the scope of this document but under discussion for several years within the collaboration and the kaon community at large, is summarised in the HIKE Letter of Intent~\cite{HIKE:2022qra}.

%%%%%%%%%%%%%%%%

\subsubsection{$K_L\to\pi^0\ell^+\ell^-$ decays}
\label{sec:physics:fcnc:klpi0ll}

The ultra-rare $K_L\to\pi^0\ell^+\ell^-$ decays ($\ell=e,\mu$) represent a set of theoretically clean golden modes in kaon physics, 
allowing for the direct exploration of new physics contributions in the $s\to d\ell\ell$ short-distance interaction (to be compared to the $b\to s\ell\ell$ transition).

The SM description of the $K_L\to\pi^0\ell^+\ell^-$ decays is provided in Refs.~\cite{DAmbrosio:1998gur,Isidori:2004rb,Mescia:2006jd}. The branching ratios depend on the CKM parameter $\lambda_t=V_{ts}^*V_{td}$, and can be written~\cite{Isidori:2004rb}
\begin{eqnarray}
{\cal B}_{\rm SM}(K_L\to\pi^0e^+e^-) &=&
\left(15.7|a_S|^2
\pm 6.2|a_S|
\left(\frac{{\rm Im}~\lambda_t}{10^{-4}}\right)
+ 2.4
\left(\frac{{\rm Im}~\lambda_t}{10^{-4}}\right)^2
\right)
\times 10^{-12},\nonumber \\
{\cal B}_{\rm SM}(K_L\to\pi^0\mu^+\mu^-) &=& 
\left(3.7|a_S|^2
\pm 1.6|a_S|
\left(\frac{{\rm Im}~\lambda_t}{10^{-4}}\right)
+ 1.0
\left(\frac{{\rm Im}~\lambda_t}{10^{-4}}\right)^2
+5.2
\right)
\times 10^{-12}.
\nonumber
\end{eqnarray}
In the above expressions, the first three terms represent the indirect CPV contribution due to $K_S$--$K_L$ mixing, the interference of the indirect and direct CPV contributions (of unknown sign), and the direct CPV contribution determined by short-distance dynamics, respectively. The fourth term in the $K_L\to\pi^0\mu^+\mu^-$ case accounts for the long-distance CP-conserving component due to two-photon intermediate states, which is negligible in the helicity-suppressed $K_L\to\pi^0 e^+e^-$ case. The parameter $a_S$ describes the decay form factor, and has been measured with $K_S\to\pi^0\ell^+\ell^-$ decays to be $|a_S|=1.2\pm0.2$ by the NA48/1 experiment at CERN~\cite{NA481:2003cfm,NA481:2004nbc}.
While the value of $|a_S|$ can be determined from the $K_S\to\pi^0\ell^+\ell^-$ branching ratios, measuring the differential decay distributions as a function of the di-lepton invariant mass would give the form factor parameters, including, in principle, the sign of $a_S$.
LHCb has plans to measure $a_S$ with $K_S\to\pi^0\mu^+\mu^-$ decays. Prospects for this measurement are discussed in \cite{AlvesJunior:2018ldo}.
The sign and value can in principle also be determined theoretically, since the calculation is stable, with no large cancellations. 

The SM expectation for the branching ratios is~\cite{Mescia:2006jd}
\begin{eqnarray}
{\cal B}_{\rm SM}(K_L\to\pi^0 e^+e^-) & = & 3.54^{+0.98}_{-0.85}~ \left(1.56^{+0.62}_{-0.49}\right)\times 10^{-11}, \nonumber\\
{\cal B}_{\rm SM}(K_L\to\pi^0\mu^+\mu^-) & = & 1.41^{+0.28}_{-0.26}~ \left(0.95^{+0.22}_{-0.21}\right)\times 10^{-11}, \nonumber
\end{eqnarray}
where the two sets of values correspond to constructive (destructive) interference between the direct and indirect CP-violating contributions. Beyond the SM, the decay rates can be enhanced significantly in the presence of large new CP-violating phases, in a manner correlated with the effects in $K_L\to\pi^0\nu\bar\nu$ and $\varepsilon^\prime/\varepsilon$~\cite{Aebischer:2022vky}.
These decays also give unique access to short-distance BSM effects in the photon coupling via the tau loop~\cite{isidori-seminar}.

Numerically in the SM framework, in the assumption of constructive interference (which is preferred theoretically), using ${\rm Im}~\lambda_t=1.4\times 10^{-4}$ and $|a_S|=1.2$, the sensitivities to the input parameters are 
\begin{displaymath}
\frac{\delta{\cal B}(K_L\to\pi^0e^+e^-)}{{\cal B}(K_L\to\pi^0e^+e^-)}=
0.53 \cdot \frac{\delta\,{\rm Im}~\lambda_t}{{\rm Im}~\lambda_t} ,
\quad
\frac{\delta{\cal B}(K_L\to\pi^0\mu^+\mu^-)}{{\cal B}(K_L\to\pi^0\mu^+\mu^-)} =
0.44 \cdot \frac{\delta\,{\rm Im}~\lambda_t}{{\rm Im}~\lambda_t} 
\end{displaymath}
and
\begin{displaymath}
\frac{\delta{\cal B}(K_L\to\pi^0e^+e^-)}{{\cal B}(K_L\to\pi^0e^+e^-)} =
1.48 \cdot \frac{\delta|a_S|}{|a_S|},
\quad
\frac{\delta{\cal B}(K_L\to\pi^0\mu^+\mu^-)}{{\cal B}(K_L\to\pi^0\mu^+\mu^-)} =
0.88\cdot \frac{\delta |a_S|}{|a_S|}.
\end{displaymath}

Experimentally, the most stringent upper limits (at 90\% CL) on the branching ratios have been obtained by the KTeV experiment~\cite{KTeV:2003sls,KTEV:2000ngj}:
\begin{displaymath}
{\cal B}(K_L\to\pi^0 e^+e^-) < 28\times 10^{-11}, \quad {\cal B}(K_L\to\pi^0 \mu^+\mu^-) < 38\times 10^{-11}.
\end{displaymath}

The principal goal of HIKE Phase~2 is the first observation and measurement of both $K_L\to\pi^0 \ell^+\ell^-$ decay modes. The HIKE sensitivity is presented in Section~\ref{sec:phase2:Klpi0ll}.

%%%%%%%%%%%%%%%%%%

\subsubsection{$K\to\ell^+\ell^-$ decays}
\label{sec:KLmumu}

The branching ratio of the helicity-suppressed $K_L\to\mu^+\mu^-$ decay is dominated by LD contributions, with a smaller, CP-conserving SD component~\cite{Buras:1997fb}.
The SM prediction exhibits large uncertainties due to the sign ambiguity of the interference between the LD and SD terms contributing to the decay amplitude.
The sign ambiguity leads to two SM predictions:
\begin{equation}
\text{[LD$+$]:}~~\left(6.82^{+0.77}_{-0.24}\pm0.04\right)\times10^{-9},~~~\text{[LD$-$]:}~~\left(8.04^{+1.66}_{-0.97}\pm0.04\right)\times10^{-9}.
\end{equation}
The first uncertainty comes from the calculation of the LD terms, dominates the overall precision, and creates the asymmetry in the uncertainty which is reflected in the computation of the $K_L\to\mu^+\mu^-$ rate with the inclusion of new physics.
The decay has been studied experimentally and the measured branching ratio is $(6.84\pm0.11)\times10^{-9}$~\cite{E871:2000wvm}.
The sign ambiguity and the LD uncertainties prevent a clean theoretical interpretation of this result at present.

\begin{figure}[t!]
\begin{center}
\includegraphics[width=0.65\textwidth]{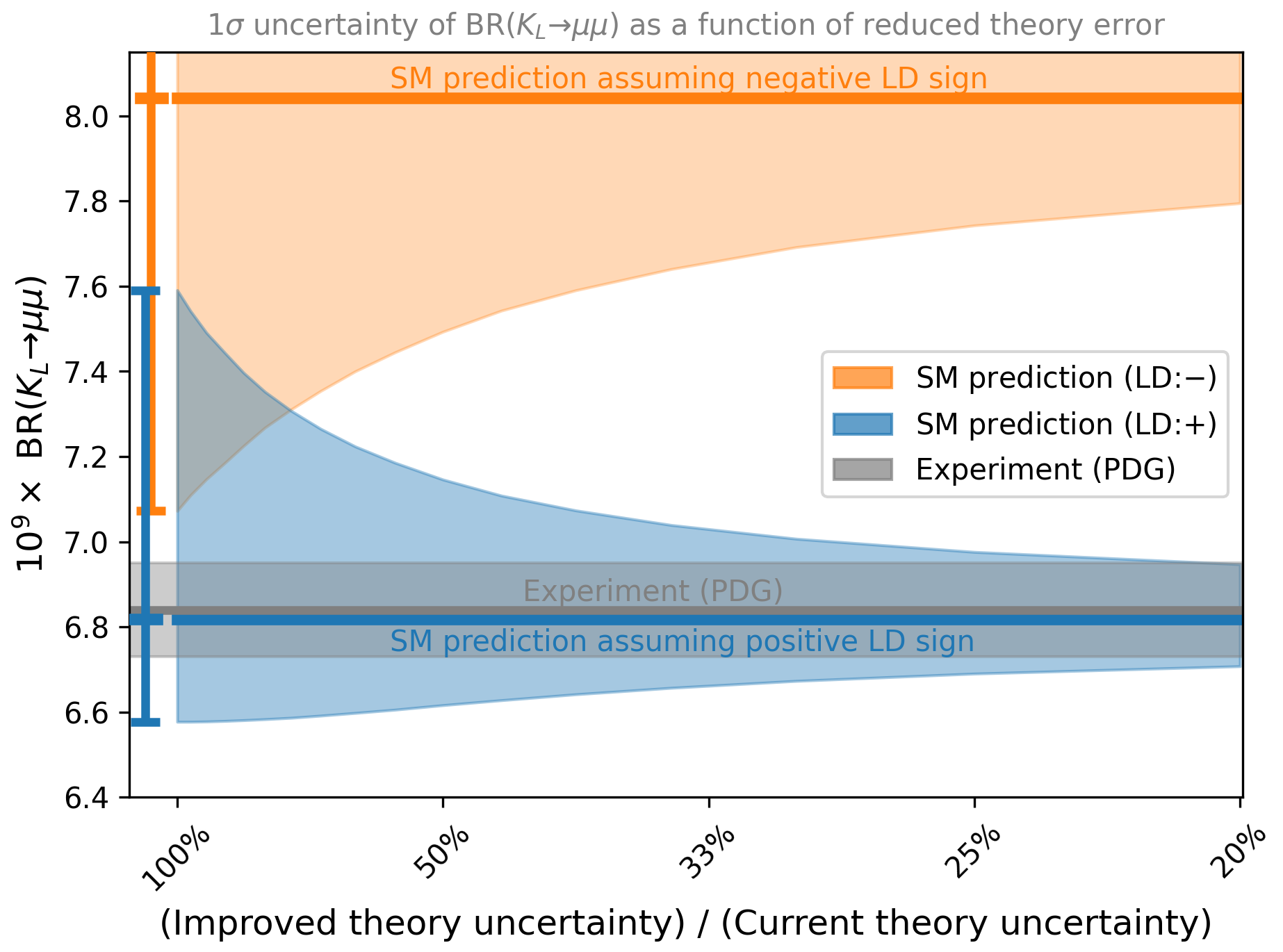}
\vspace{-2mm}
\caption{The impact of reducing theory error for the two signs of the LD contribution to the $K_L\to\mu^+\mu^-$ decay.
\label{fig:LDsign}}
\end{center}
\vspace{-8mm}
\end{figure}

The sign of LD contributions to $K_L\to \mu^+ \mu^-$ has a clear impact on the precision with which BSM models can be probed. If the precision of the theoretical predictions is improved in the future, it will become important to decrease the experimental uncertainty in order to extracting information on SD physics as well as to identify the correct sign of the interference term. (Fig.~\ref{fig:LDsign}).

In any case, the $K_L\to\mu^+\mu^-$ channel provides additional information on top of a potential $K_S\to\mu^+\mu^-$ observation. Theoretical efforts are actively ongoing to clarify the sign ambiguity of the interference term by an appropriate matching between LD and SD contributions, to reduce the LD uncertainty by at least a factor of two. This is expected to be resolved well before the start of HIKE~\cite{https://doi.org/10.48550/arxiv.2203.10998}, and
could allow the uncertainty on the SM prediction to be reduced sizeably, opening up the possibility to exploit the high sensitivity of this decay to BSM physics models.
Once the LD sign is established, even considering the most adverse new physics scenario---minimal flavor violation, in which new operators have SM structures and the SD relation between $K_S\to\mu^+\mu^-$ and $K_L \to\mu^+\mu^-$ is dictated by the CKM electroweak phase $V_{td} V^*_{ts}$---a ${\cal B}(K_L\to\mu^+\mu^-)$ measurement to 1\% precision will have better sensitivity to BSM than the expected boundary on ${\cal B}(K_S\to\mu^+\mu^-)$ from LHCb by the end of Run~6 (i.e.~mid-2040s)~\cite{DAmbrosio:2022kvb,Crivellin_2016}. 
Due to the unprecedented $K_L$ flux, HIKE will measure the $K_L\to\mu^+\mu^-$ decay rate to 1\% precision.

The study of the $K_S\to\mu^+\mu^-$ decay is the main goal of the kaon physics programme of the LHCb experiment. The high kaon production rate in proton-proton collisions at 13~TeV partly compensates the small acceptance for kaon decays due to the long lifetime. LHCb has reported an upper limit on the $K_S\to\mu^+\mu^-$ branching ratio using the combined Run~1+2 dataset, ${\cal B}(K_S\to\mu^+\mu^-) < 2.1\times 10^{-10}$ at 90\%~CL~\cite{LHCb:2020ycd}, to be compared to the SM prediction, ${\cal B}_{\rm SM}(K_S\to\mu^+\mu^-) = (5.18\pm1.50)\times 10^{-12}$~\cite{Ecker:1991ru, DAmbrosio:2017klp}. The ultimate expected LHCb sensitivity is close to the SM branching ratio, and is expected to be limited by the statistical uncertainties on background subtraction and signal yield~\cite{AlvesJunior:2018ldo}.
However, even considering the most favourable BSM model, in which the $K_S \to\mu^+\mu^-$ branching ratio is enhanced up to $35\times 10^{-12}$~\cite{Chobanova:2017rkj}, a $5\sigma$ observation of the $K_S\to\mu^+\mu^-$ decay at LHCb by the end of Run~6 is unlikely~\cite{LHCb:2018roe}.

From a theoretical point of view, a high-precision $K_L\to e^+ e^-$ measurement would be interesting.
The $K\to\ell^+\ell^-$ decays are helicity suppressed by a factor proportional to $m_\ell^2$. The LD contribution to the amplitude from photon-photon exchange depends on $\ln[(1-\beta)/(1+\beta)]$ and is therefore partially enhanced by the small electron mass. Since the SD component does not have this term, the helicity suppression in $K_L\to e^+ e^-$ relative to $K_L\to \mu^+ \mu^-$ increases the LD dominance in the former case. However, BSM scalar or pseudoscalar contributions would not be helicity suppressed, so a measurement at the percent level would be sensitive to these contributions.

%%%%%%%%%%%%%%%%%%%

\subsection{Lepton flavour universality tests}
\label{sec:lfu}
Lepton flavour universality (LFU) is a cornerstone of the SM postulating that the lepton coupling to gauge bosons is independent of lepton type (``flavour''), in contrast with the flavour-dependence of quark interactions. The origin of the observed LFU, which is not associated with any known symmetry and is thus not a fundamental conservation law, is a major question in modern physics.
Tensions between SM predictions and experimental results, such as the anomalous magnetic moment of the muon~\cite{Muong-2:2021ojo}, the Cabibbo angle anomaly~\cite{Coutinho:2019aiy}, and the LFU tensions in the $B$ sector~\cite{LFUReview},
suggest possible LFU violation and further motivate the quest for improved measurements.

The kaon sector offers opportunities for precision lepton flavour universality tests by measuring the ratios ${\cal B}(K^+\to\pi^+e^+e^-)/{\cal B}(K^+\to\pi^+\mu^+\mu^-)$,
${\cal B}(K^+\to e^+\nu)/{\cal B}(K^+\to\mu^+\nu)$ and
${\cal B}(K\to\pi e\nu)/{\cal B}(K\to\pi\mu\nu)$,
exploiting cancellation of hadronic effects~\cite{Bryman:2021teu,Crivellin:2016vjc}.

There is no general relation between 
possible LFU-violation effects in the $B$ and kaon sectors. 
The tests of the LFU principle in the kaon and $B$-meson sectors are thus complementary. 
Relations between the two sectors are expected only in certain model-dependent scenarios. Considering, for example, a very specific case of a MFV scenario with $U(3)$ symmetry, the LFU observables from kaon FCNC processes ($sd\ell\ell$) must be measured with a precision $\sim |V_{tb}|/|V_{td}|={\cal O}(100)$ times better than the corresponding ones from the $B$-sector ($bs\ell\ell$), to reach the same NP sensitivity. 
On the other hand, the LFU observables from kaon charged-current processes ($su\ell\nu$, e.g. ${\cal B}(K^+\to e^+\nu)/{\cal B}(K^+\to \mu^+\nu)$) are $\sim |V_{us}|/|V_{cb}|={\cal O}(10)$ times more sensitive to new physics than the corresponding ones from the $B$-sector ($bc\ell\nu$, e.g. $R(D), R(D^*)$) for the same experimental precision.
Even in MFV, if its realisation is via $U(2)$ symmetry, which is a favoured scenario with respect to $U(3)$ given the strong constraints from $K^0-\bar{K}^0$ mixing, the $U(2)$ operator for $bs\ell\ell$ is unrelated to that for $sd\ell\ell$ (independent and different coefficients), requiring dedicated measurements from both the kaon and $B$ sectors~\cite{Bordone_2017}.

\subsubsection{$K^+\to\pi^+\ell^+\ell^-$ }

The amplitudes for the $K^+\to\pi^+\ell^+\ell^-$ decays ($\ell=e,\mu$) are LD dominated and the branching ratios for these decays are $\mathcal{O}(10^{-7})$.
The differential decay amplitudes with respect to the dilepton invariant mass depend on two form factor parameters, denoted $a_+$ and $b_+$, which lepton universality predicts to be independent from the flavour of the leptons.
Differences between $a_+$ and $b_+$ between the electron and muon channels can be correlated to LFU-violating effects that can explain the
tensions observed in $B$ physics~\cite{Crivellin:2016vjc,DAmbrosio:2018ytt}.

The NA62 experiment has recently reported a new measurement of the $K^+\to\pi^+\mu^+\mu^-$ decay form factor from a background-free sample of $3\times10^4$ decays collected in 2017--2018~\cite{NA62:2022qes}.
Form-factor parameter values $a_+ = -0.575 \pm 0.013$ and $b_+ = -0.722 \pm 0.043$ were extracted from a fit to the dilepton mass spectrum.
This improves significantly on the previous measurements in the muon channel while remaining consistent within stated uncertainties with measurements of the same parameters performed both in the muon and electron channels~\cite{E865:1999ker,NA482:2009pfe,e865:1999kah,NA482:2010zrc}.
The NA62 experiment is expected to improve the precision of the LFU test in the next years, addressing also the measurement of the $K^+\to\pi^+e^+e^-$ form factors.
However, a significant increase in precision is required to perform an LFU test with comparable sensitivity to BSM models as in current results from $B$ physics~\cite{Crivellin:2016vjc,DAmbrosio:2022kvb}. HIKE will offer a significantly larger data sample, further reducing statistical uncertainties, which remain a large fraction of the total uncertainty in the most recent NA62 measurement~\cite{NA62:2022qes}. 
HIKE will perform new model-independent measurements of LFU observables that may provide a direct signal of LFU violation or, otherwise, lead to stringent constraints on LFU-violating NP models.
The sensitivity is discussed further in Section~\ref{sec:phase1-other-measurements}.

\subsubsection{Leptonic and semileptonic kaon decays}

The ratio of purely leptonic decay rates of the charged kaon $R_K = \Gamma(K^+ \to e^+ \nu)/\Gamma(K^+ \to \mu^+ \nu)$ is strongly suppressed in the SM by conservation of angular momentum and is an extremely sensitive probe of~LFU.
Its SM expectation, $R^{\rm SM}_{K} = (2.477 \pm 0.001) \times 10^{-5}$~\cite{Cirigliano:2007xi}, is known to excellent (0.4\textperthousand) precision, while the currently most precise (4\textperthousand) experimental result is $R_K = (2.488 \pm 0.007_{\rm stat} \pm 0.007_{\rm syst}) \times 10^{-5}$~\cite{NA62:2012lny}. The ratio $R_K$ is highly sensitive to the possible violation of LFU naturally arising in new physics scenarios involving sterile neutrinos~\cite{Abada:2012mc, Bryman:2019bjg}, leptoquarks~\cite{Crivellin:2021egp}, massive gauge bosons~\cite{Capdevila:2020rrl}, or an extended Higgs sector~\cite{Girrbach:2012km, Fonseca:2012kr}.
Variations of $R_K$ up to a few per mille from its SM expectation are predicted by these models, without contradicting any present experimental constraints.

An analogous LFU test is based on the ratio of semileptonic kaon decay rates $R_K^{(\pi)} = \Gamma(K \to \pi e \nu)/\Gamma(K \to \pi \mu \nu)$, where both neutral and charged kaon decays ($K_L \to \pi^\pm \ell^\mp\nu$ and $K^+ \to \pi^0\ell^+\nu$) can be used.
For a given neutral or charged initial state kaon, the Fermi constant,
$V_{us}$, short-distance radiative corrections, and the hadronic form factor at zero momentum transfer cancel out when taking the ratio $R_K^{(\pi)}$~\cite{Bryman:2021teu}. Therefore, in the SM, this ratio is entirely determined by phase space factors and long-distance radiative corrections~\cite{Cirigliano:2001mk, Cirigliano:2004pv, Cirigliano:2008wn, Seng:2021boy,Seng:2022wcw}.

%%%%%%%%%%%%%%%%%%

\subsection{Lepton flavour and number violation}
\label{sec:lfv}

Individual lepton flavours---electron, muon and tau numbers---are conserved in the SM but known to be violated in nature, which is evident from neutrino oscillations. No LFV has yet been observed in the charged-lepton sector; however, it is generically expected in many extensions of the SM, notably those that aim to generate neutrino masses~\cite{Calibbi:2017uvl,Davidson:2022jai}. An observation would provide groundbreaking indirect evidence for new elementary particles, e.g.~heavy neutrinos~\cite{Atre:2009rg}, additional Higgs bosons, leptoquarks~\cite{Pati:1974yy,Shanker:1981mj,,Bordone:2018nbg,Mandal:2019gff}, light pseudoscalar bosons (ALPs)~\cite{Cornella:2019uxs}, or a $Z^\prime$ boson~\cite{Landsberg:2004sq,Langacker:2008yv}. Furthermore, models which predict LFU violation naturally predict also LFV processes~\cite{Glashow:2014iga,Calibbi:2015kma,Greljo:2015mma}. In the absence of model-independent predictions, it is essential to explore all possible LFV signatures~\cite{Davidson:2022jai}.

While there is no doubt that the individual lepton flavours are not conserved in nature, the same is not known for total lepton number: no lepton number violating process has ever been observed, in agreement with the SM prediction~\cite{FileviezPerez:2022ypk}. An observation would again provide evidence for additional particles beyond the SM and have wide-ranging consequences for our understanding of fundamental physics and even cosmology, since lepton number violation could be the reason for the observed dominance of matter over antimatter~\cite{Davidson:2008bu}. Neutrino masses can serve to motivate the violation of lepton number as well: if neutrinos are Majorana particles, then lepton number is broken and we expect corresponding signatures, the most sensitive of which is arguably neutrinoless double beta decay $(A,Z)\to (A,Z+2)+2 e^-$~\cite{Rodejohann:2011mu}. Meson decays provide a complementary probe that is sensitive to different flavour structures~\cite{Littenberg:1991ek}. Below the kaon mass, the constraints on active-sterile mixing angles between Majorana neutrinos obtained from searches in the kaon sector are already competitive with those coming from neutrinoless double beta decay~\cite{Littenberg:2000fg,Atre:2005eb,Atre:2009rg,Abada:2017jjx}.

HIKE will make significant improvements on the world data in terms of probing the possible LF and LN violating decays of the $K^+$, $K_L$ and $\pi^0$ mesons. In Phase~1, HIKE will address the LFV decays $K^+\to\pi^+(\pi^0)e^\pm\mu^\mp$ and $\pi^0\to e^\pm\mu^\mp$, LNV decays $K^+\to\pi^-(\pi^0)e^+e^+$, $K^+\to\pi^-\mu^+\mu^+$ and $K^+\to \pi^-e^+\mu^+$, as well as decays $K^+\to e^-\nu\mu^+\mu^+$ and
$K^+\to\mu^-\nu e^+e^+$ violating either LF or LN conservation depending on the flavour of the emitted neutrino (Section~\ref{sec:phase1-other-measurements}). In Phase~2, HIKE will address a range of LFV decays of the $K_L$ meson including  $K_L\to\pi^0(\pi^0) e^\pm \mu^\mp$ and $K_L\to e^\pm e^\pm \mu^\mp\mu^\mp$.

%%%%%%%%%%%%%%%%%%%%%%%

\subsection{Tests of low-energy QCD}
\label{sec:chpt}

Kaon decays governed by long-distance physics are described by chiral perturbation theory (ChPT), the low-energy effective field theory of QCD.
Kaon decay amplitudes are evaluated in the ChPT framework using the low-energy constants determined from experimental data. Comprehensive measurements of kaon decay rates and form factors provide both essential tests of the ChPT predictions and inputs to the theory~\cite{Cirigliano:2011ny}. The HIKE dataset will provide a unique opportunity to perform a wide range of precision measurements of rare and radiative decays of $K^+$ and $K_L$ mesons. The highlights of this programme are described below.

\begin{itemize}
\item $K^+\to\pi^+\ell^+\ell^-$: Precise experimental knowledge of the form factors for these decays would have a profound impact. First, simultaneous accurate measurements of the form factors for both $K^+\to\pi^+e^+e^-$ and $K^+\to\pi^+\mu^+\mu^-$ would provide a powerful LFU test~\cite{DAmbrosio:2022kvb} (see also Section~\ref{sec:lfu}); the recent fit to the $K\to3\pi$ decays~\cite{DAmbrosio:2022jmd} is important for this purpose. Second, these modes are related to the $K_S\to\pi^0e^+e^-$ decay, allowing determination of the sign of the form-factor $a_S$ (see also Section~\ref{sec:physics:fcnc:klpi0ll}), since different combinations of ChPT parameters enter the ${\cal O}(p^4)$  
chiral Lagrangian~\cite{DAmbrosio:1998gur}. Third, these modes contribute to the global fit for new physics in the kaon sector~\cite{DAmbrosio:2022kvb}. Finally, these modes may provide short-distance tests through CP or forward-backward asymmetries~\cite{DAmbrosio:1998gur}.
\item $K^+\to\pi^+\gamma\gamma$, $K^+\to\pi^+\gamma\ell^+\ell^-$: 
The leading contribution to the former decay is at ${\cal O}(p^4)$ in the chiral expansion, while a large next-to-leading ${\cal O}(p^6)$ contribution is determined by unitarity through the $K^+\to\pi^+\pi^+\pi^-$ loop~\cite{DAmbrosio:1996cak}. The latter decay also receives leading contributions at ${\cal O}(p^4)$, 
with contributions from $K^+\to\pi^+\gamma\gamma^*$ and from $K^+\to\pi^+\ell^+\ell^-$ with bremsstrahlung, providing interesting chiral tests when compared with $K^+\to\pi^+\gamma\gamma^*$, including determination of the ${\cal O}(p^4)$ weak chiral Lagrangian and relations among low-energy observables~\cite{Gabbiani:1998tj}. It may also provide P- and CP-violating short-distance tests.
\item $K^+\to\pi^+\pi^0\gamma$,
$K^+\to\pi^+\pi^0\ell^+\ell^-$:
These decays have electric and magnetic contributions according to the Lorentz structure of the amplitude (angular state decomposition).
The electric part of $K^+\to\pi^+\pi^0\gamma$ is interesting for the determination of the weak chiral Lagrangian~\cite{Cappiello:2017ilv} and to study CP asymmetries. The magnetic contribution is dominant and important for the interplay of chiral dynamics, vector dominance, and vector anomalies.
In addition, the $K^+\to\pi^+\pi^0\ell^+\ell^-$ decay may exhibit interference between the electric and magnetic contributions~\cite{DAmbrosio:2018ytt}.
\item $K^+\to e^+\nu\gamma$: The inner bremsstrahlung contribution to this decay is determined by the kaon decay constant $F_K$ only. The structure-dependent contribution is described by the vector and axial form factors. These can be calculated to next-to-leading order in ChPT, ${\cal O}(p^6)$, while non-trivial contributions arise at ${\cal O}(p^4)$~\cite{Bijnens:1992en}. 
A measurement aiming at ${\cal O}(p^6)$ would be very interesting, because the ChPT Lagrangian terms here are not known from other data, and a recent measurement from J-PARC~\cite{J-PARCE36:2022wfk} departs from the $O(p^4)$ theory result.
\item $K^+\to\pi^0 e^+\nu\gamma$: This radiative decay has been accurately studied theoretically, and can limit to 1\% the novel structure-dependent contributions of new physics~\cite{Bijnens:1992en,NA62:2023lnp}. Experiments have not yet reached this precision, which is an important target.
A T-odd correlation can be constructed as a triple product, $\xi = \vec p_\gamma \cdot (\vec p_\ell \times \vec p_\pi)/m^3_K$, of the three independent momenta in $K_{e3\gamma}$ decays. The asymmetry in the $\xi$ distribution, $A_\xi$, is important in the search for new physics with vectors or pseudovectors. Values of $A_\xi$ of the order of $-10^{-5}$ to $-10^{-4}$ are expected in the SM and its extensions~\cite{Braguta:2003wf,Muller:2006gu}.
\item $K\to 2\pi$ and $K\to 3\pi $: Measurements of the branching ratios of these principal decay modes provide overall information on all isospin amplitudes, $\pi\pi$ phase shifts, the $\Delta I =1/2$ rule, and a test of the weak chiral Lagrangian~\cite{Cirigliano:2011ny,DAmbrosio:2022jmd}, as well as inputs for theoretical and experimental studies of the form factors of the $K^+\to\pi^+\gamma\gamma$, $K^+\to\pi^+\ell^+\ell^-$ and $K^+\to\pi^+\gamma\ell^+\ell^-$ decays whose amplitudes receive substantial loop contributions that depend on the $K\to 3\pi$ amplitude. Moreover, accurate new measurements of $K\to 2\pi/3\pi$ decays (and  kaon lifetimes) are needed since large-scale factors appear in the PDG for the $K^\pm$ lifetime and the $K_L\to 3\pi^0$ branching ratio. These measurements will also allow for a constrained fit leading to a stringent limit on the $K_L$ decaying into a fully invisible final state.

\item $K_L\to\gamma\gamma$: This decay provides an interesting test of chiral perturbation theory. The leading-order $\mathcal{O}(p^4)$ term vanishes in the SU(3) limit. However, large $\mathcal{O}(p^6)$ contributions mediated by pseudoscalar mesons are expected with values depending on the amount of singlet–octet mixing. Precise measurements of the $K_L\to\gamma\gamma$ decay rate are also of interest in connection with the $K_L\to\mu^+\mu^-$ decay. In fact the absorptive part of the decay rate, $\Gamma_{\rm abs}(K_L\to\mu^+\mu^-)$, is proportional to $\Gamma(K_L\to\gamma\gamma)$. This constrains the dispersive part, $\Gamma_{\rm disp}(K_L\to\mu^+\mu^-)$, and eventually the possibility of determining the $\eta$ parameter of the CKM matrix and more generally short-distance physics.

\end{itemize}

%%%%%%%%%%%%%%%%%%

\subsection{Tests of first-row CKM unitarity}

Measurements of semileptonic kaon decays $K\to\pi\ell\nu$ provide the principal input for the extraction of the CKM parameter $V_{us}$, while the ratios of (semi)leptonic $K^+$ and $\pi^+$ decay rates are used to extract the ratio $V_{us}/V_{ud}$, with inputs provided from lattice QCD~\cite{Aoki:2021kgd}. Determination of $V_{us}$ from kaon, pion, and $\tau$ decays, combined with that of $V_{ud}$ from super-allowed beta decays~\cite{Hardy:2020qwl} and neutron decays~\cite{Czarnecki:2018okw,Seng:2020wjq}, gives rise to a $3\sigma$ deficit in first-row CKM unitarity relation, known as the Cabibbo angle anomaly. A tension of similar significance is observed between $K\to\ell\nu$ and $K\to\pi\ell\nu$ rates~\cite{Cirigliano:2022yyo,Bryman:2021teu}.

\begin{figure}[t!]
\begin{center}
\includegraphics[width=\textwidth]{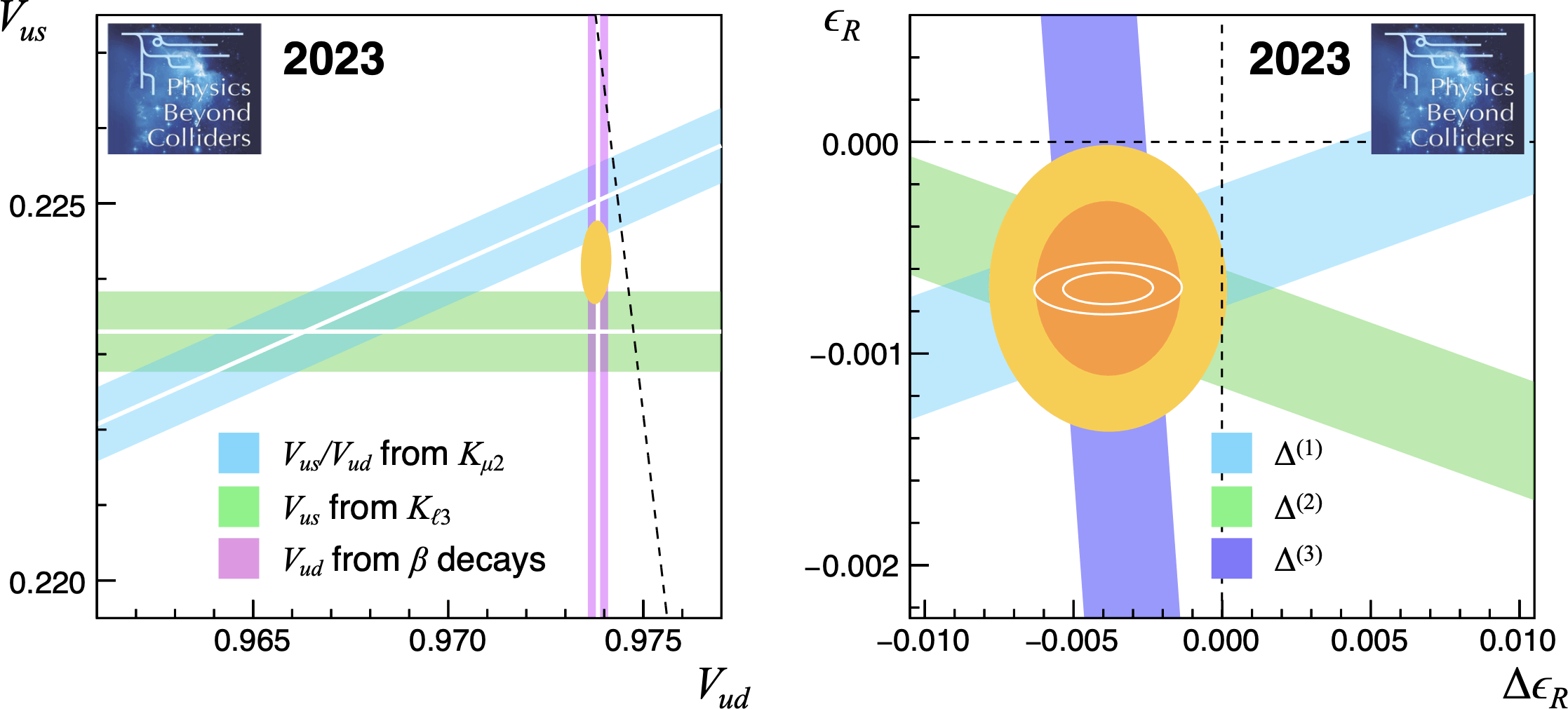}
\vspace{-6mm}
\caption{Status of first-row CKM unitarity in 2023. Left: measurements of $V_{us}$, $V_{us}/V_{ud}$, and $V_{ud}$ and relation to CKM unitarity.
Right: constraints on right-handed currents from observed unitarity deficits.} 
\label{fig:ckm_2023}
\end{center}
\end{figure}

Fig.~\ref{fig:ckm_2023}~(left) shows the current experimental constraints in the $V_{us}$--$V_{ud}$ plane from $K_{\ell3}$ decays ($V_{us}$, green band), $K_{\mu2}$ decays ($V_{us}/V_{ud}$, light blue band), and nuclear and neutron beta decays ($V_{ud}$, magenta band). The tension between the values of $V_{us}$ from $K_{\mu2}$ and $K_{\ell3}$ decays is seen in the fact that these bands do not intersect at a common point. The yellow ellipse represents the 68.27\% CL confidence interval from a fit for the best values of $V_{us}$ and $V_{ud}$. The unitarity curve is illustrated by the dashed line. The significance of the unitarity deficit
from $K_{\ell3}$ and beta decays is $-3.1\sigma$, that from $K_{\mu2}$ and beta decays is $-1.7\sigma$, and that
from the fit result is $-2.7\sigma$. 

Fig.~\ref{fig:ckm_2023}~(right) illustrates the constraints from CKM unitarity on the contributions to the leptonic and semileptonic kaon decay amplitudes from right-handed quark currents, following the analysis of Ref.~\cite{Cirigliano:2022yyo}. Specifically, denoting by $\epsilon_R$ the contributions of right-handed currents to the decays of non-strange quarks and by $\epsilon_R^{(s)}$ those to the decays of strange quarks, and noting that 
the CKM elements as extracted from the vector-current mediated semileptonic decays are contaminated by $(1 + \epsilon)$ and those from the axial-current mediated leptonic decays by $(1 - \epsilon)$, the following relations to the unitarity deficits are obtained:
\begin{align}
\Delta_{\rm CKM}^{(1)} &\equiv |V_{ud}^{\beta}|^2 + |V_{us}^{K_{\ell3}}|^2 - 1 = 2\epsilon_R + 2\Delta\epsilon_R V_{us}^2,\\
\Delta_{\rm CKM}^{(2)} &\equiv |V_{ud}^\beta |^2\, \left[1 + (|V_{us}/V_{ud}|^{K_{\mu2}})^2\right] - 1 = 2\epsilon_R - 2\Delta\epsilon_R V_{us}^2,\\
\Delta_{\rm CKM}^{(3)} &\equiv |V_{us}^{K_{\ell3}}|^2\, \left[(|V_{us}/V_{ud}|^{K_{\mu2}})^{-2} + 1\right] - 1 
= 2\epsilon_R - 2\Delta\epsilon_R (2 - V_{us}^2),
\end{align}
with $\Delta\epsilon_R \equiv \epsilon_R - \epsilon_R^{(s)}$. 
The coloured bands in the plot show the $\pm1\sigma$ constraints from $\Delta^{(1)}$ ($K_{\ell3}$ and beta decays, green),
$\Delta^{(2)}$ ($K_{\mu2}$ and beta decays, light blue), and $\Delta^{(3)}$ ($K_{\ell3}$ and $K_{\mu2}$ decays, dark blue) in the plane of $\epsilon_R$  vs.\ $\Delta\epsilon_R$; note that the bands intersect by construction. The yellow and orange ellipses illustrate the 95.45\% CL and 68.27\% CL confidence intervals from a fit. The point $\epsilon_R = \Delta\epsilon_R = 0$ is excluded with $3.1\sigma$ confidence. At face value, this is evidence for the existence of right-handed currents, which is as good an illustration as any of the first-row unitarity crisis.  

The uncertainty in $V_{us}$ comes in equal parts from the experimental errors and theoretical uncertainties in the ratio of decay constants, $f_K/f_\pi$, and the $K\to\pi\ell\nu$ form factor, $f_+(0)$. 
The ellipses traced in white in Fig.~\ref{fig:ckm_2023}~(right) indicate the much smaller confidence intervals that would be obtained if the only source of uncertainty were from the kaon decay measurements. If the uncertainties on the lattice determinations of $f_+(0)$, the $K_{\ell3}$ form factor at zero momentum transfer, and $f_K/f_\pi$, the ratio of kaon to pion decay constants, and, above all, on the determination of $V_{ud}$ from beta decays were to be reduced, the evidence for right-handed currents would be even stronger.
Substantial improvements in the lattice QCD calculations of the hadronic factors are expected in the next five years, thanks to decreased lattice spacing and accurate evaluation of
electromagnetic effects~\cite{Carrasco:2015xwa, Boyle:2019rdx,Feng:2021zek,Seng:2020jtz}. Significant progress on the calculation of radiative corrections has been achieved recently, reducing the uncertainties on the long-distance electromagnetic corrections to a negligible level~\cite{Seng:2019lxf,Seng:2021boy,Seng:2021wcf,Seng:2022wcw}.
In order to shed light on the Cabibbo angle anomaly, improved measurements of the principal $K^+$ and $K_{L,S}$ branching ratios are essential (note that no $K_L$ decay measurements have been made in the past decade). In particular a precision measurement of the ratio of $K^+\to\pi^0\mu^+\nu$ and $K^+\to\mu^+\nu$ rates is well-motivated~\cite{Cirigliano:2022yyo}. 
It should be noted that the Belle~II experiment is planning to extract $V_{us}$ at an improved precision from a suite of inclusive and exclusive measurements of $\tau$ decays, including ${\cal B}(\tau^-\to K^-\nu)/{\cal B}(\tau^-\to\pi^-\nu)$, combined with theory improvements~\cite{Belle-II:2022cgf}. This would provide a possibility to cross-check the kaon results.

\newpage

If the Cabibbo angle anomaly persists, HIKE would be able to make precision measurements of (semi)leptonic $K^+$ and $K_L$ decays based on special datasets collected with minimum-bias triggers at low beam intensity. 
Given that the dominant contribution to the uncertainty on the measurement of the first-row unitarity deficit is from the determination of $V_{ud}$ from nuclear beta decays, the fact that experimental and theoretical sources contribute approximately equally to the current overall uncertainty on $V_{us}$, and the substantial set of kaon decay measurements in world data,
HIKE can contribute to the understanding of the anomaly mainly by providing experimental confirmation of the leptonic and semileptonic branching ratio values, to help to exclude an experimental origin. Indeed, as discussed in \Secs{sec:ckm_phase1} and~\ref{sec:ckm_phase2}, the experimental situation is complex, with a few measurements of the branching ratios playing an outsize role in the overall determination of $V_{us}$. 

\subsection{Benchmark channels for the HIKE kaon programme}

The ultra-rare decay $K^+\to\pi^+\nu\bar\nu$ represents one of the main golden modes of kaon flavour physics because it is short-distance dominated, well predicted theoretically and sensitive to a large variety of BSM scenarios. Similarly, the ultra-rare $K_L\to\pi^0\ell^+\ell^-$ decays ($\ell=e,\mu$) also belong to the theoretically clean golden modes. These decay modes share the characteristic of being the most experimentally challenging, because of the smallness of their branching ratios and the suppression with respect to the potential experimental backgrounds. For these reasons, they are taken as benchmark channels towards which the HIKE design is optimised. Once this is achieved, HIKE will allow for an unprecedented exploration of the kaon sector with rare decays, precision measurements and searches for exotic processes.

\section{Phase 1: a multi-purpose $K^+$ decay experiment}
\label{sec:phase1}

%%%%%%%%%%%%%%%

\subsection{Experimental layout}

\begin{figure}[ht]
\begin{center}
\resizebox{\textwidth}{!}{\includegraphics{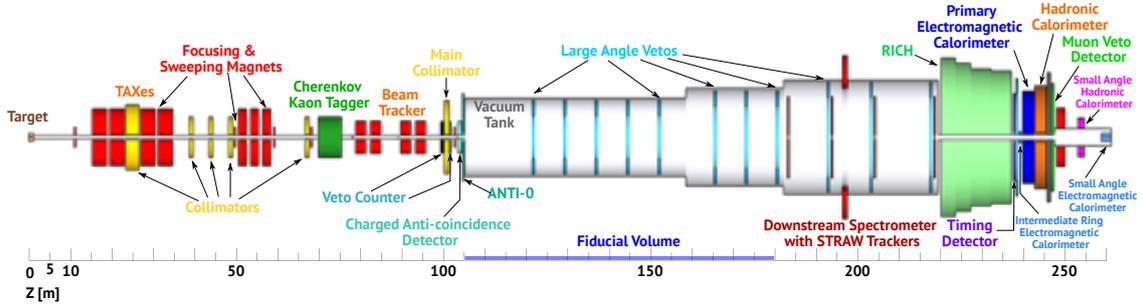}}
\end{center}
\vspace{-7mm}
\caption{HIKE Phase~1 layout, with an aspect ratio of 1:10.}
\label{fig:phase1-layout}
\end{figure}

The success of the NA62 experiment has proven that its layout is suitable for a precision ${\cal B}(K^+\to \pi^+\nu\bar\nu)$ measurement. 
For HIKE, new or upgraded detectors will replace those used for NA62, with the goal of sustaining secondary-beam rates four times higher to boost the statistical sensitivity. 
To this end, new technologies will be used, in synergy with the upgrades of the LHC experiments.
While some detectors will be rebuilt for the start of the HIKE $K^+$ phase (Phase~1), others already use intrinsically fast detector technologies but will need readout upgrades.

In the baseline configuration, the $K^+$ beam is produced by $1.3\times10^{13}$ protons/spill (ppp) at $400~{\rm GeV}/c$ and zero angle on target. The above figure of protons/spill assumes four times the nominal intensity of the NA62 experiment, which is $33\times10^{12}$~ppp.  The mean $K^+$ momentum at the entrance to the decay volume is 75~GeV/$c$. The layout (Fig.~\ref{fig:phase1-layout}) includes a beam tracker, a kaon-identification detector, and a veto counter upstream of the decay volume; a main tracking system with the MNP33 magnet in the decay volume; 
calorimeters, veto, and particle-identification detectors downstream of the decay volume; and a large-angle veto system surrounding the decay volume and part of the region downstream.

The HIKE setup will coexist with the SHADOWS detector. 
A dedicated simulation of the setup when including the passive SHADOWS apparatus has been performed, to inspect whether the flux and distributions of ``beam-halo" muons are affected. The simulation has an equivalent statistics of $5.5 \times 10^9$ POT, and is compared with a simulation of the setup in absence of SHADOWS.
The total muon flux reaching HIKE detectors is the same within a few per cent, and there is no indication of any issues for HIKE operation in presence of SHADOWS.

The first phase of HIKE will be optimised for a precision measurement of the $K^+\to\pi^+\nu\bar\nu$ decay, with a target of reaching a branching ratio measurement with $\mathcal{O}(5\%)$ precision, competitive with the theoretical uncertainty. The approach towards this goal will continue, and build upon, the successful strategy of the NA62 experiment.
The optimisation of the setup for this challenging objective will allow a broad range of other excellent $K^+$ measurements to be carried out.

%%%%%%%%%%%%%%%%%%%%

\subsection{The benchmark channel $K^+\to\pi^+\nu\bar\nu$ }
\label{sec:phase1:kpinn}

The keystones of the ${\cal B}(K^+\to\pi^+\nu\bar\nu)$ measurement, in which signal events must be efficiently selected while suppressing backgrounds at the $\mathcal{O}(10^{11})$ level, are the following:
\begin{itemize}
\item High-efficiency and high-precision tracking of both the $K^+$ upstream and the $\pi^+$ downstream. This, coupled with a careful choice of signal regions, will allow kinematic suppression of backgrounds by a factor of $\mathcal{O}(10^{3})$.
\item High-precision time measurements, allowing time-matching between upstream and downstream detectors with $\mathcal{O}(20~\text{ps})$ precision. Simulations show that with this time-matching performance, the effect of the higher intensity can be totally compensated in the matching between upstream and downstream tracks, without losing efficiency for signal selection and without increasing the relative contamination of background coming from upstream decays and interactions with respect to NA62.
\item Comprehensive and hermetic veto systems:
\begin{itemize}
\item Photon veto detectors with hermetic coverage of the 0--50~mrad range for photons from $\pi^0\to\gamma\gamma$ decays, to suppress these decays by a factor of $\mathcal{O}(10^{8})$.
\item Veto detectors to cover downstream regions not covered by the principal detectors. 
\item Rejection of upstream decays and interactions, suppressing possible upstream backgrounds by an additional factor of up to six, achieved via beamline modification and installation of new systems dedicated to the detection of upstream background events.
\end{itemize}
\item High-performance particle identification systems for $\pi$/$\mu$ discrimination, to suppress backgrounds with muons by a factor of $\mathcal{O}(10^{7})$.
\end{itemize}

%%%%%%%%%%%%%%%%%%%
\subsubsection{From NA62 to HIKE} 
The NA62 experiment was designed and constructed to study the $K^+\to\pi^+\nu\bar\nu$ decay and has solidly demonstrated the success of the decay-in-flight experimental strategy, providing the most precise measurement to date of the branching ratio ${\cal B}(K^+\to\pi^+\nu\bar\nu)$ using data collected in 2016--2018. This period of data taking is referred in the following as Run~1.~\cite{NA62:2021zjw}.
NA62 is still taking data and since 2021 has been operating at a significantly higher intensity than during Run~1. The period of data taking of NA62 started in 2021 and planned up to Long Shutdown 3 (LS3) is referred in the following as Run~2.
The NA62 experiment is collecting about $2\times 10^{18}$~POT per full year equivalent (standard year or FYE = 200 days and 3000 spills per day), corresponding to $5\times 10^{12}$ kaon decays in the fiducial volume per FYE. So far, in four years of operation (2017, 2018, 2021, and 2022), NA62 has collected 3 FYE in terms of POT delivered to the experiment. 
Despite the success of the NA62 programme, it is clear that the experiment, with restricted data collection time and hardware limitations at higher intensities, will not be able to provide a branching ratio measurement to better than 15\% precision.
This leaves a critical gap between the precision of the measurement and that of the theoretical prediction ($\mathcal{O}(5\%)$), a gap that can only be bridged with a new experiment, i.e. HIKE.

HIKE Phase~1 will fully exploit the extensive experience already gained and continuing to be gained by the NA62 experiment, building upon its successful strategy and implementing improvements in all required areas.
The geometric design of the HIKE experiment is very close to that of NA62, meaning that the overall geometric acceptance and coverage will be similar. 
However, increasing the collection rate for signal $K^+\to\pi^+\nu\bar\nu$ decays requires use of a high-intensity beam. This generally necessitates the detectors to be faster by the same factor ($\times4$) as the intensity increase, as well as to have finer granularity, to sustain the higher particle rate expected while maintaining performance that is the same or better than that achieved at NA62. 

%%%%%
The precision on the measurement of ${\cal B}(K^+\to\pi^+\nu\bar\nu)$ is driven by three key factors, outlined below, with details provided in the following subsections.
\begin{itemize}
\item \textbf{The statistical sample size of signal candidates.} 
The objective of HIKE is to collect between 400 and 500 signal candidates in 4--5 standard years of Phase~1 operation. This leads to a 5\% statistical uncertainty.
\item \textbf{The systematic uncertainty on the estimated sensitivity to the (SM) signal.} 
This uncertainty will be improved relative to NA62 by a factor about three, with a target of reaching below 3\%. 
This will, at least in part, be achieved due to the significant reduction of inefficiency by migrating from a hardware to a software trigger.
\item \textbf{The level of background contamination.} 
There is a trade-off between increasing acceptance for signal and minimising backgrounds. However, simple statistical considerations imply that the fractional uncertainty on the branching ratio is given by 
\begin{equation}
\frac{\sigma_{\mathcal{B}}}{\mathcal{B}} = \frac{\sqrt{S+B}}{S},
\label{eq:pnnstaterror}
\end{equation}
where $S$ and $B$ are the numbers of signal and background events. Under this assumption, a level of background similar to NA62, i.e., $B/S$ in the range 0.5--0.7, allows excellent precision to be obtained by maintaining a high signal yield. Therefore, 
the HIKE strategy does not place a premium on reducing the background below this level. However, it should be noted that with significantly increased statistics the knowledge of the background will also be improved allowing better precision to be achieved for the background estimation.
\end{itemize}

%%%%%%%%%%%%%%%%%%%

\subsubsection{$K^+\to\pi^+\nu\bar\nu$ decay measurement}

The $K^+\to\pi^+\nu\bar\nu$ analysis at HIKE follows the strategy of NA62.
The steps of the analysis (referred to as the PNN analysis in the following) are
\begin{itemize}
\item reconstruction and identification of the charged pion;
\item reconstruction of the parent kaon;
\item photon and multiplicity rejection;
\item kinematic selection of the signal.
\end{itemize}

\paragraph{Pion reconstruction and identification} \hspace{5pt} \newline
The analysis starts with the selection of candidate charged particle tracks reconstructed in the straw spectrometer.
The track reconstruction efficiency of NA62 is, on average, 94--95\%, with a weak intensity dependence.
The same performance is expected for HIKE at $\times4$ the NA62 intensity, thanks to the smaller straw radius and the improved trailing-time resolution, which mitigates the effects of the larger pileup on pattern recognition.

Each track is associated to signals in the fast timing array (CHOD for NA62 and the timing plane for HIKE), the signals from which provide track-time measurements.
In NA62, the association relies only on spatial information, with negligible impact on the PNN analysis.
The situation is expected to be similar in HIKE, where the association will benefit from the 1~ns time resolution of the spectrometer track and a finer granularity of the timing plane compared to the CHOD.
The HIKE timing plane provides track-time measurements with 100~ps resolution. 
This information allows precise matching of the track with the activity in the RICH and calorimeters.
The signals from these detectors are then used for particle identification (PID) and kaon--pion association.

\begin{figure}
\centering
\includegraphics[scale=0.5]{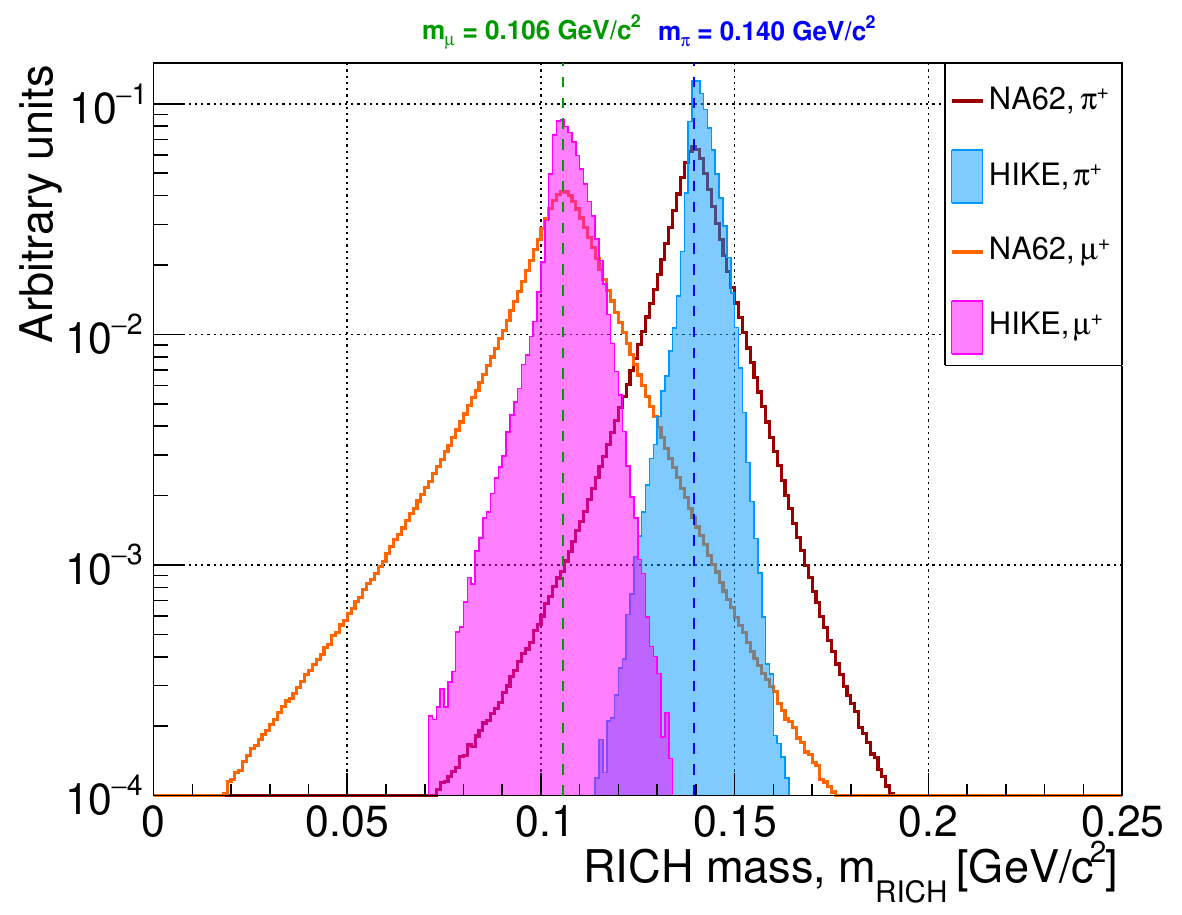}
\vspace{-3mm}
\caption{Expected HIKE Phase~1 RICH performance (filled histograms) compared to NA62 RICH performance (histogram outlines), for the reconstructed particle mass, for $\pi^+$ from $K^{+}\to\pi^+\pi^0$ decays and $\mu^+$ from $K^{+}\to\mu^{+}\nu$ decays.}
\vspace{-3mm}
\label{fig:mRICH}
\end{figure}

The NA62 PNN analysis selects tracks in the 15--45~GeV$/c$ momentum range for optimum RICH particle identification performance. A likelihood technique is used to distinguish pions from muons and electrons using the RICH information. A similar procedure is envisaged for the HIKE analysis.
The RICH reconstruction efficiency depends on Cherenkov photon yield and photodetector time resolution to resolve rings overlapping in space.
The photo\-detector size is the dominant factor limiting the RICH PID performance in NA62.
Compared to the NA62 RICH, the HIKE RICH photo\-detectors are about half the size and have a quantum efficiency twice as high, as well as a time resolution of 100~ps, compared to 250~ps in NA62.
Fig.~\ref{fig:mRICH} shows the expected improvement for HIKE compared to NA62 of the particle mass reconstructed by the RICH, or RICH mass, $m_{\rm RICH}$, with the particle momentum provided by the spectrometer.
The ring time provides the most precise pion time measurement, when associated to a candidate track. The ring-time resolution expected for the HIKE RICH is about 20~ps, compared to 80~ps for NA62. 

% --------------------------------------------------

The NA62 muon detector (MUV3) provides a muon rejection factor of $10^{3}$ with a loss of 10\% of pions due to accidental activity.
In order to veto muons, NA62 relies only on the timing of the MUV3 signals because the size of the MUV3 tiles does not allow efficient spatial matching with the spectrometer track.
The HIKE muon veto plane profits from a $\times4$ better time resolution than the NA62 MUV3 and a finer granularity.
The smaller tile size will allow HIKE to use spatial information to provide a more efficient muon veto.
As a consequence, the muon veto performance for HIKE is expected to match that of NA62, despite the significantly higher intensity.

Given the performance of the RICH and muon detectors, the HIKE calorimeters must suppress muons by an additional factor of 50 with respect to pions to maintain the $K^+\to\mu^+\nu$ background at the NA62 level.
NA62 reaches a muon suppression factor of 100 using a boosted decision tree (BDT) algorithm trained on data to exploit information from the electromagnetic and hadronic calorimeters.
The transverse slab size of the NA62 hadronic calorimeters is the main limitation for the calorimetric PID performance at higher intensity.
The design of the HIKE hadronic calorimeters therefore specifically aims to overcome this issue and to keep calorimeteric PID performance similar to that measured in NA62.
A new PID algorithm based on a convolutional neural network, under study at NA62, could be ported to HIKE to further improve the PID.

Based on the performance described above, simulations indicate that HIKE identifies pions with at least 10\% higher efficiency than NA62 when keeping the same muon--pion misidentification probability.
The improvement mostly comes from the improved RICH performance, while the muon veto and calorimetric performance are comparable to those of NA62.

\paragraph{Parent kaon reconstruction} \hspace{5pt}\newline
Parent kaon reconstruction involves combination of the reconstructed pion with the signals detected in the differential Cherenkov counter (KTAG) and in the beam tracker.

HIKE envisages the same parent kaon identification procedure in the KTAG detector (Section~\ref{sec:ktag}) as NA62: a five-fold coincidence of the eight octants defines a kaon with 99\% efficiency.
The parent kaon is the KTAG kaon closest in time to the pion\footnote{Here, and in the following, the RICH time used to define the candidate pion time}.
The identification performance depends on the kaon rate, and on the time resolution of the RICH and KTAG.
As a consequence, HIKE is expected to maintain the same performance as NA62, because the fraction of kaons in the secondary beam for HIKE is the same as for NA62 and the time resolutions are four times better.

\begin{figure}
\centering
\includegraphics[scale=0.5]{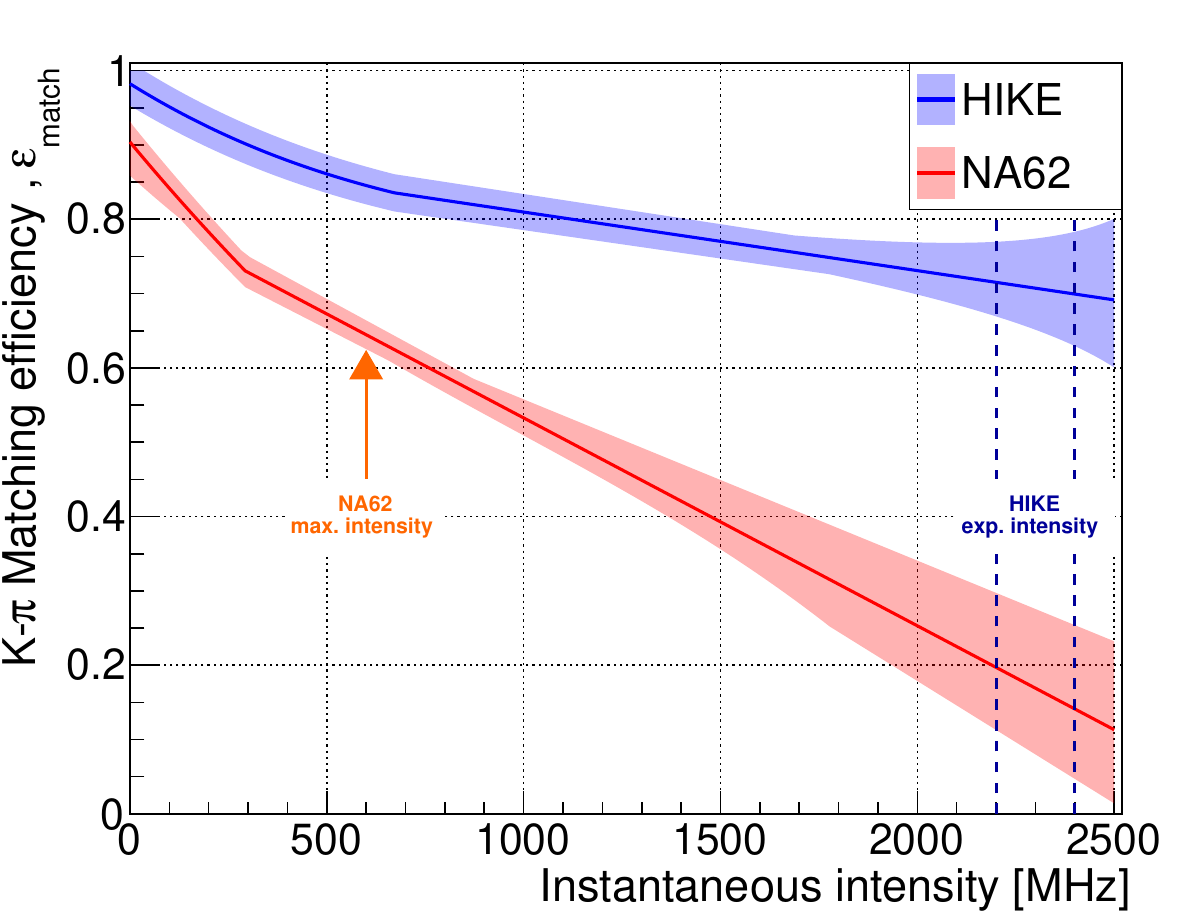}
\vspace{-2mm}
\caption{Expected HIKE Phase~1 kaon--pion matching efficiency (blue) compared to NA62 (red) as a function of the instantaneous beam intensity. The lines and shaded areas represent the central value and uncertainties, respectively.
The maximum NA62 intensity is indicated by the arrow and the expected range of intensities for HIKE, which depends on beam settings and quality, is shown as the region between vertical dashed lines.}
\vspace{-2mm}
\label{fig:KpiMatchVsI}
\end{figure}

%%%%%%%%%

Following the NA62 analysis procedure, the parent kaon track is defined as the GTK track best matching the KTAG parent kaon in time, and the pion in time and space. The matching variables are the differences between the KTAG, RICH and GTK times, and the minimum distance (closest distance of approach, CDA) between the GTK and pion tracks extrapolated to the decay region. NA62 combines these inputs into a likelihood that defines the matching quality. NA62 reconstructs the parent kaon with the intensity-dependent efficiency shown in Fig.~\ref{fig:KpiMatchVsI}.
The corresponding misidentification probability is around 2\%, with a weak intensity dependence.
The time resolution and pixel size of the NA62 GTK and the resolution on the measurement of the slope of the pion track are the main limiting factors.
The first factor is responsible for the intensity dependence of kaon--pion matching performance: pixel size limits the GTK pattern recognition leading to reconstruction of fake tracks, with a probability increasing with intensity.
The resolution on the slope of the pion track is the primary limiting factor for the CDA resolution.
The kaon--pion matching for HIKE benefits from the $\times4$ improvement in time resolution compared to NA62, the $\times3$ smaller pixel size in the beam tracker, and the 40\% lower material budget in the straw spectrometer.
All these improvements compensate for the deterioration of the performance with intensity.
Fig.~\ref{fig:KpiMatchVsI} shows the HIKE kaon--pion matching efficiency. 
The corresponding misidentification probability is about 2\%, similar to in NA62.
Overall, the HIKE kaon--pion efficiency is expected to be a factor of $\times1.1$ higher than that for NA62, which runs at $\times4$ lower intensity. 
Complex matching algorithms based on supervised neural networks are under study to improve the matching performance further. 

%%%%%%%%%%%%

\paragraph{Photon and multiplicity rejection\label{sec:pnnPhotonRejection}}
\hspace{5pt} \newline
The suppression of kaon decays with photons or multiple charged particles in the final state relies on vetoing events with signals in the calorimeters and fast timing detectors. 
These signals must be not spatially associated to, but must be in time with, the pion.
The algorithm developed by NA62 leads to a $10^8$ suppression factor of the $K^+\to\pi^+\pi^0$ decays, which is a benchmark channel to evaluate the photon rejection performance.
The NA62 algorithm also suppresses the $K^+\to\pi^+\pi^+\pi^-$ and $K^+\to\pi^+\pi^-e^+\nu$ decays to a level sufficient to make these backgrounds sub-leading.
This performance is possible due to the 45~GeV/$c$ maximum $\pi^+$ momentum, which guarantees that at least 30~GeV/$c$ of energy is available to other particles produced, in addition to the daughter pion, in the $K^{+}$ decay.

% Random Veto -----
\begin{figure}[t]
\centering
%\vspace{-2mm}
\includegraphics[scale=0.47]{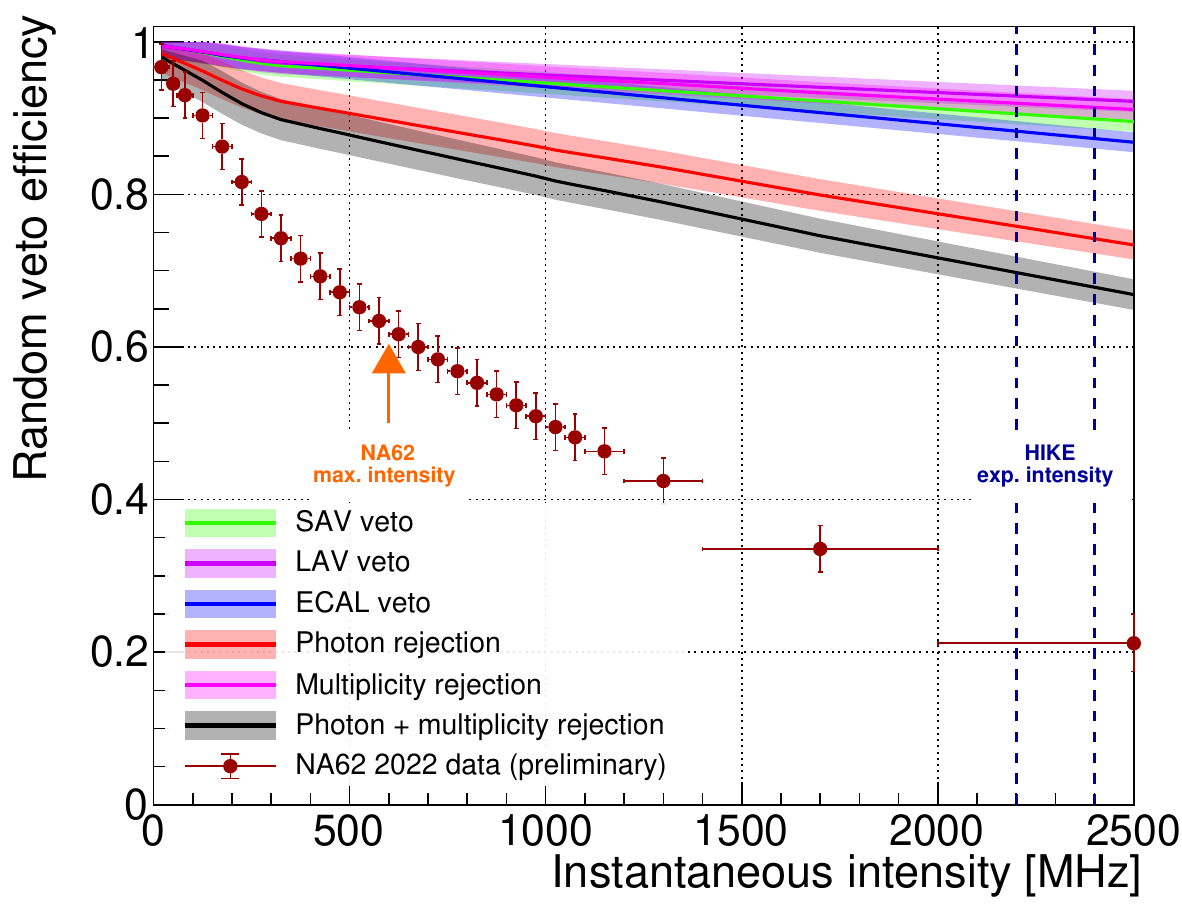}
\vspace{-3mm}
\caption{Expected HIKE Phase~1 random veto efficiency $\varepsilon_{\rm RV}$ for the $K^+\to\pi^+\nu\bar\nu$ analysis as a function of instantaneous beam intensity: the total effect (grey band) and contributions from the individual photon/multiplicity veto conditions. The NA62 graph is also shown for comparison: the lower values of $\varepsilon_{\rm RV}$ with respect to HIKE Phase~1 are due to worse time resolution.
The maximum intensity for NA62 is indicated by the arrow and the expected range of intensities for HIKE, which depends on beam settings and quality, is shown as the region between vertical dashed lines.}
\vspace{-3mm}
\label{fig:RVprojections}
\end{figure}

HIKE envisages using the same selection criteria as NA62 to suppress decays into photons and multiple charged particles.
The photon rejection algorithm is sensitive to the presence of accidental events in time with $K^+\to\pi^+\nu\bar\nu$ decays, because it employs sharp cuts on the times of the signals recorded by the subdetectors.
The consequence is a loss of $K^+\to\pi^+\nu\bar\nu$ events due to random activity, leading to a signal selection efficiency at NA62 of about 65\% at the maximum intensity delivered by the SPS in Run~2\footnote{NA62 measured an instantaneous beam rate of 600~MHz on average in 2022. However, the rate strongly depends on the smoothness of the spill shape, and the quoted value refers to the quality of the beam delivered on average by the SPS to NA62 from 2022 onward. In the following we will refer to a beam rate of 600\,MHz as the maximum NA62 intensity.}.
This efficiency, referred to here as the random-veto efficiency, $\varepsilon_{\rm RV}$, is measured from data with $K^+\to\mu^+\nu$ decays, and exhibits a quasi-linear dependence on the instantaneous beam intensity.
The limiting factor is the timing precision of the detectors.
HIKE plans to maintain or improve the random-veto efficiency, which requires an improvement in the time resolution for the veto systems at least by the same factor as the intensity increase. 
A projection for the random-veto efficiency as a function of intensity at HIKE is shown in Fig.~\ref{fig:RVprojections}. 
Individual contributions from photon veto subsystems (SAV, LAV, ECAL) are indicated, which in total give the curve labelled ``photon rejection''. 
The curve labelled ``multiplicity rejection'' relates to the selection criteria required to reject additional activity in an event, apart from the signals from the photon veto detectors. 
Finally, the total random veto efficiency, combining photon and multiplicity rejection, is shown. 
At a secondary beam intensity $\times4$ higher than NA62, with the detector updates and corresponding changes to the selection shrinking veto windows by a factor of four, the random veto efficiency is maintained at the same level as currently achievable at NA62.

%%%%%%%%%%%%%%%%%%%%%%%%

\begin{figure}[t]
\centering
\includegraphics[width=\linewidth]{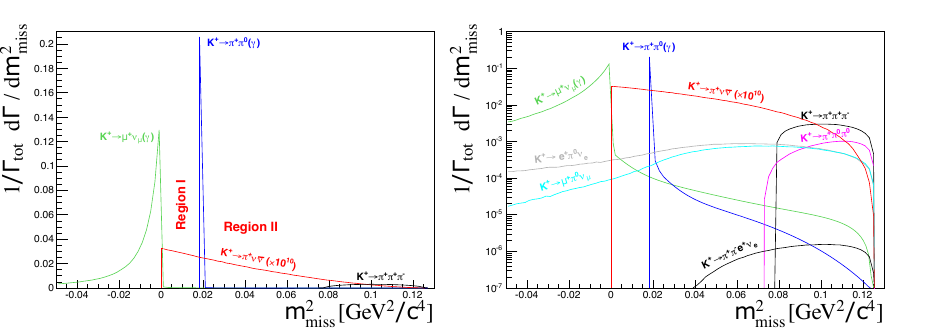}
\vspace{-5mm}
\caption{Distribution of the $m_{\rm miss}^{2}$ variable for kaon decay modes with the largest contribution (left, linear scale); all decay modes (right, logarithmic scale).}
\label{fig:m2miss}
\end{figure}
% --------------------------

%%%%%%%%%%%%%%%%%%

\begin{figure}[p]
\centering
\vspace{-3mm}
\includegraphics[width=0.49\columnwidth]{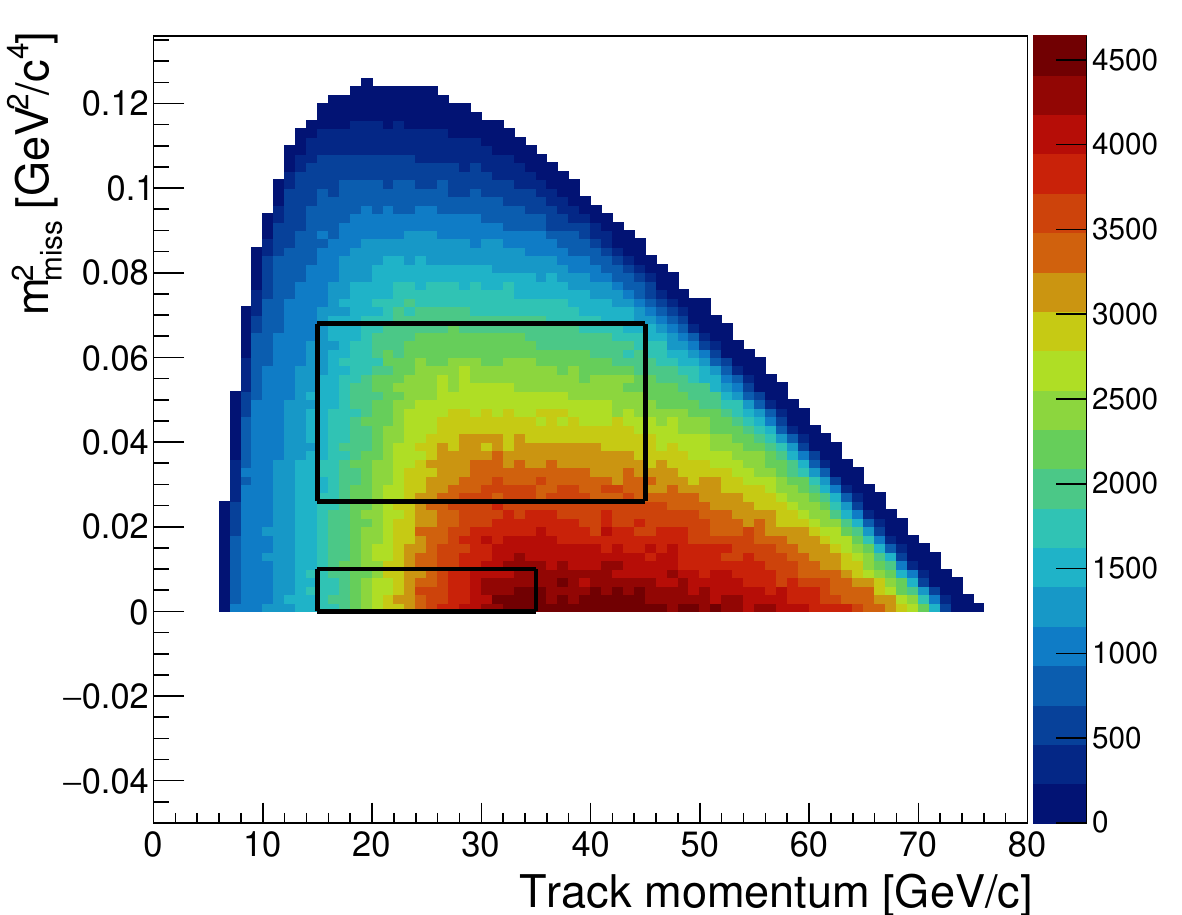}
\includegraphics[width=0.49\columnwidth]{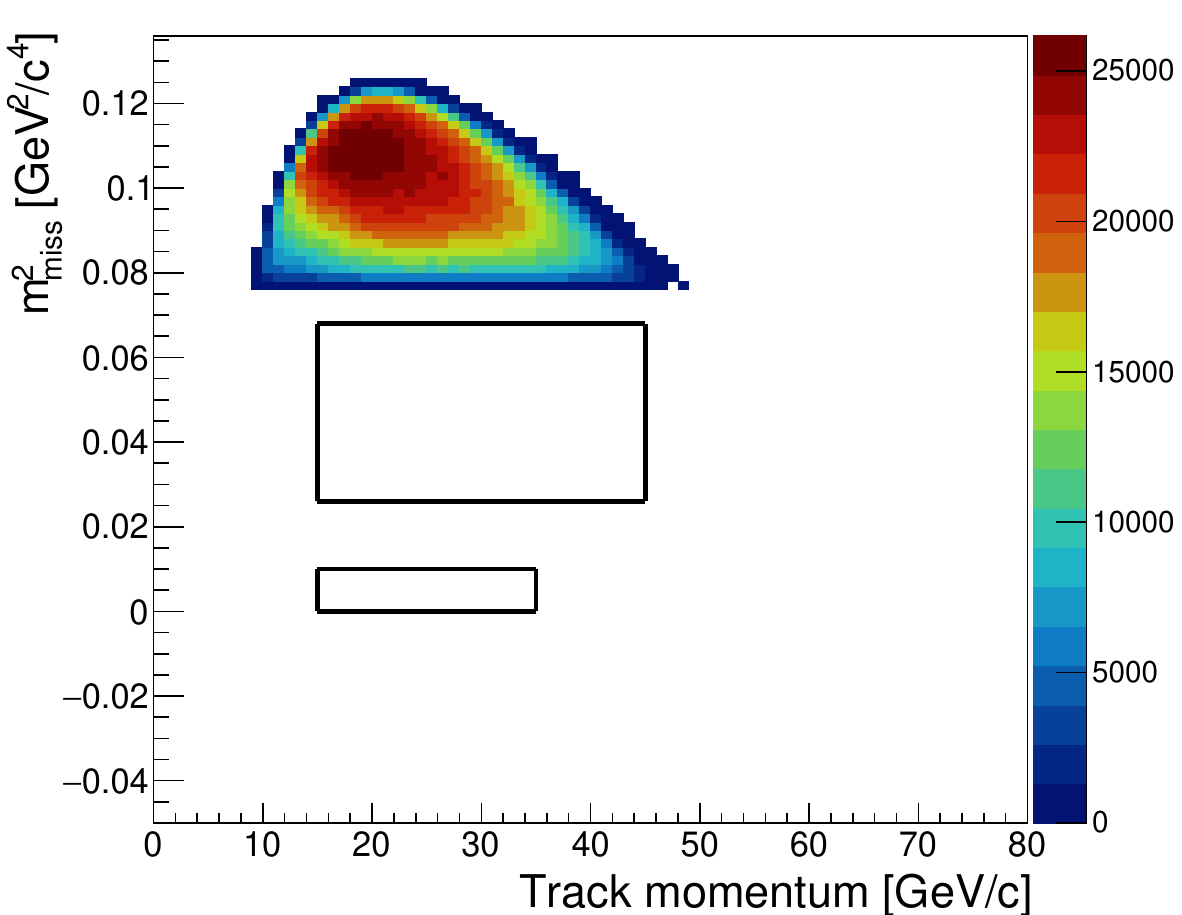}
\put(-330,142){\color{red}{\bf\LARGE $K^{+}\rightarrow\pi^{+}\nu\bar{\nu}$}}
\put(-110,142){\color{black}{\bf\large $K^{+}\rightarrow\pi^{+}\pi^{+}\pi^{-}$}}
\\
\includegraphics[width=0.49\columnwidth]{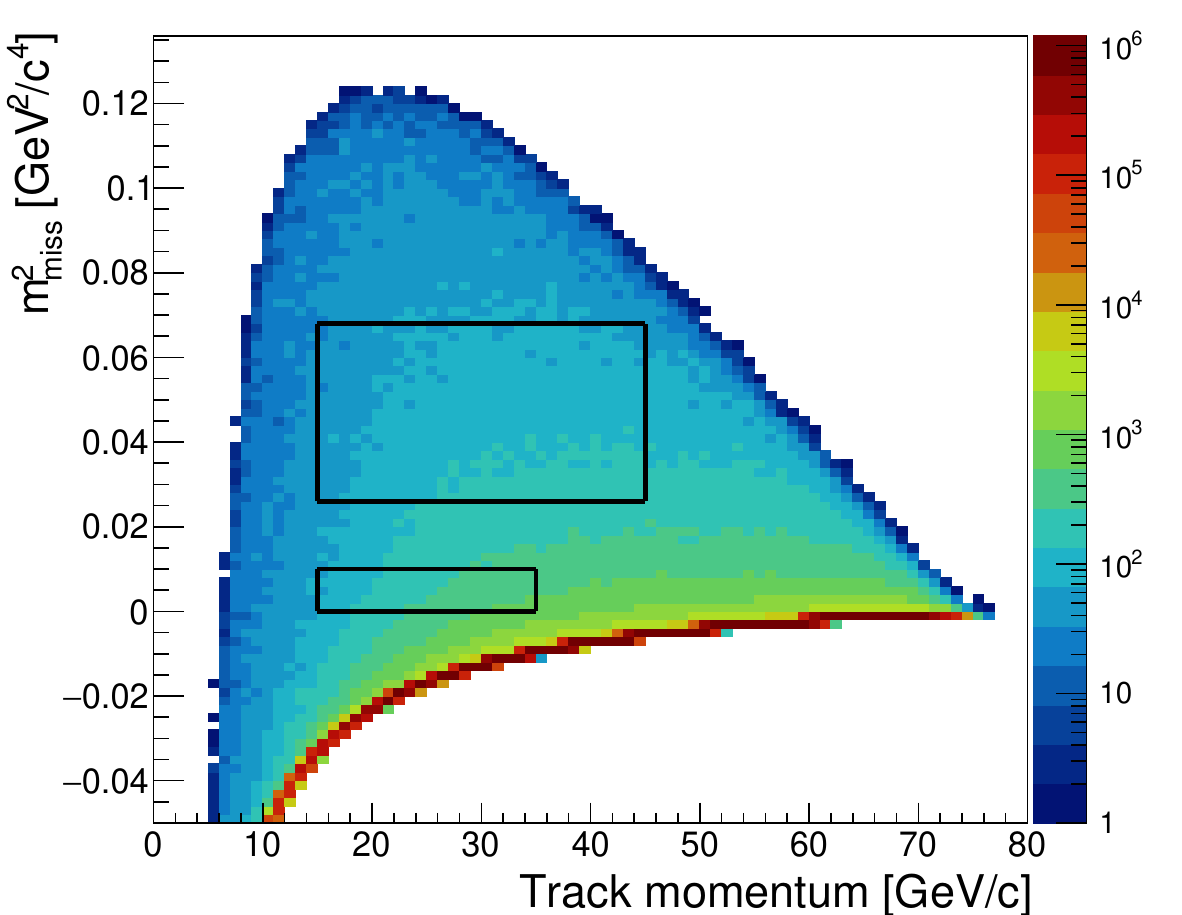}
\includegraphics[width=0.49\columnwidth]{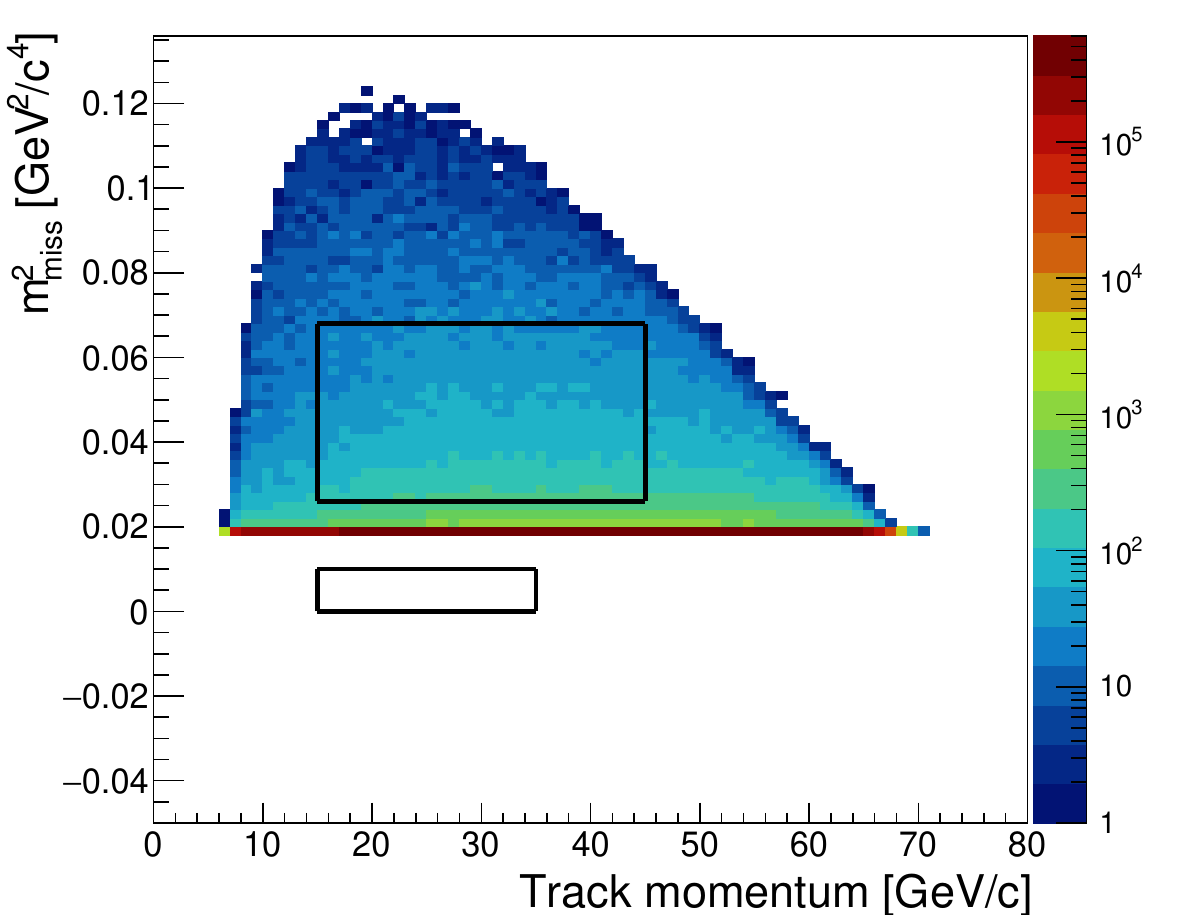}
\vspace{-1mm}
\put(-320,140){\color{green}{\bf\large $K^{+}\rightarrow\mu^{+}\nu(\gamma)$}}
\put(-110,140){\color{blue}{\bf\large $K^{+}\rightarrow\pi^{+}\pi^{0}(\gamma)$}}
\caption{
Simulated kinematic distributions (before reconstruction) of the squared missing mass $m_{\rm miss}^{2}$ vs.\ $K^+$ daughter particle momentum, for signal $K^{+}\to\pi^{+}\nu\bar{\nu}$ (top left) and backgrounds decays
$K^{+}\to\pi^{+}\pi^{+}\pi^{-}$ (top right), $K^{+}\to\mu^{+}\nu(\gamma)$ (bottom left) and 
$K^{+}\to\pi^{+}\pi^{0}(\gamma)$ (bottom right), after requiring that the $K^+$ daughter particle remains inside the geometric region covered by detectors downstream of the fiducial volume. The signal regions are indicated by the black rectangles. The colour scales are arbitrary.
}
\label{fig:mm2vspMCT}
\end{figure}
% -----------------------------------------------------

% KinematicTails -----
\begin{figure}[p]
\centering
\vspace{-7mm}
\includegraphics[width=0.5\columnwidth]{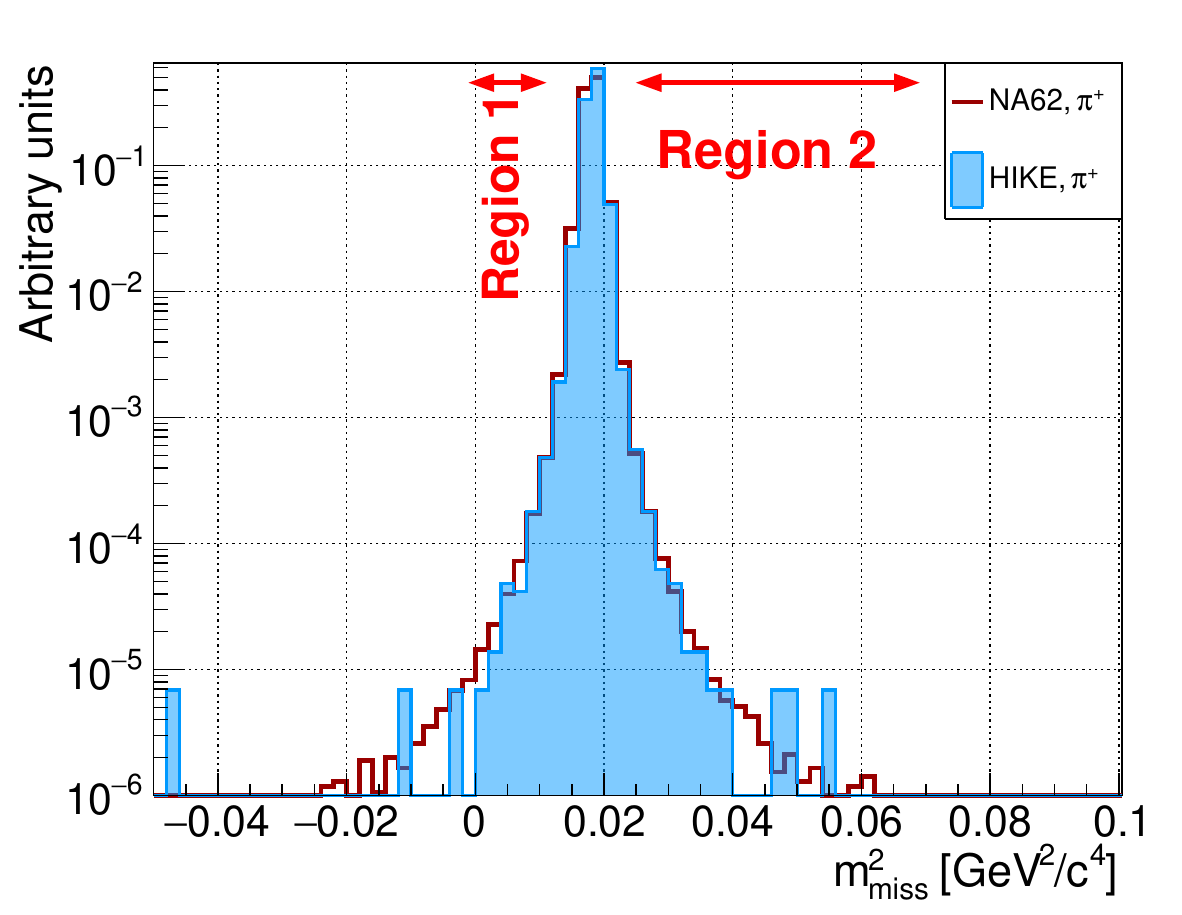}%
\includegraphics[width=0.5\columnwidth]{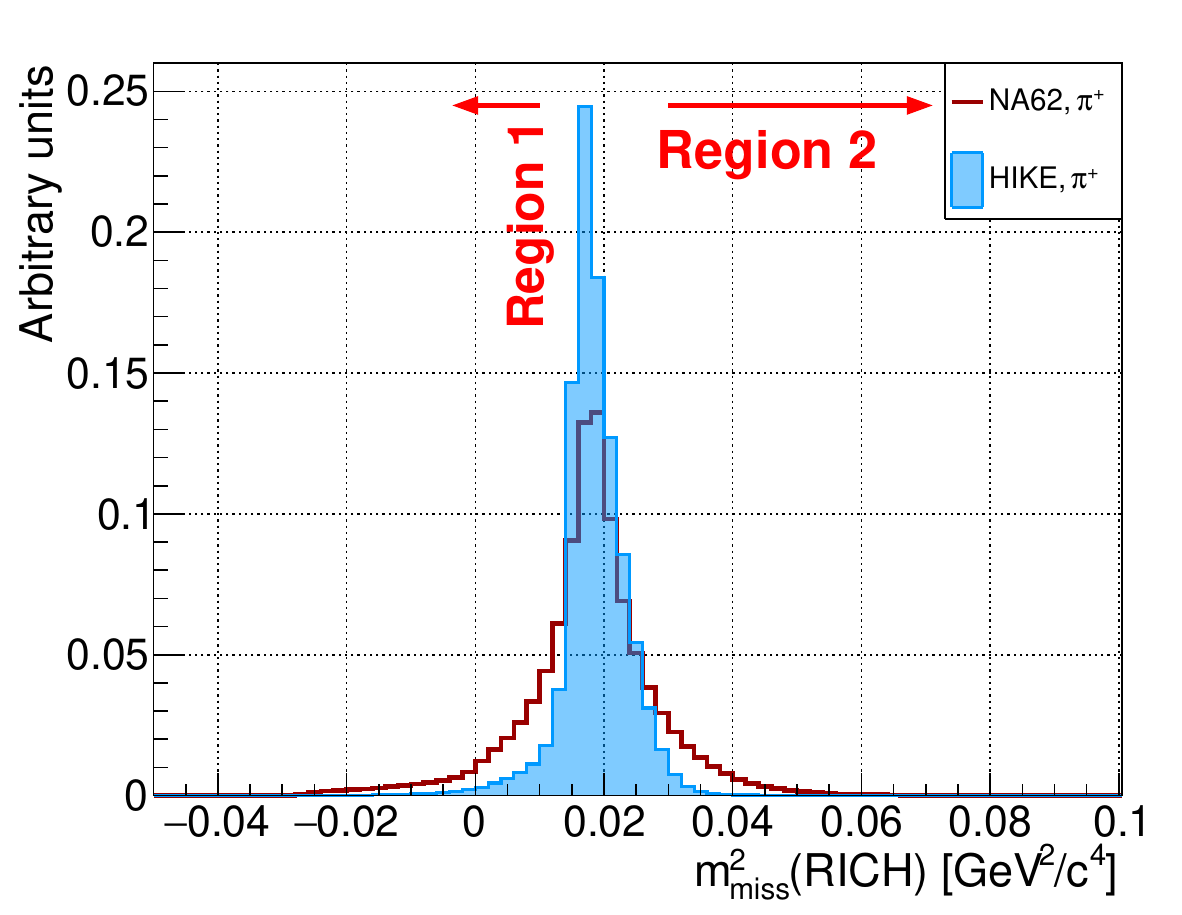}%
\vspace{-3mm}
\caption{HIKE Phase~1 (filled histograms) and NA62 (histogram outlines) performance for the kinematic resolution and tails in $m_{\rm miss}^{2}$~(left) and $m_{\rm miss}^{2}(\text{RICH})$~(right) variables for $K^{+}\to\pi^{+}\pi^{0}$ decays
in the full 15--45~GeV/$c$ momentum range.}
\label{fig:KineTailsMm2}
\end{figure}
% -----------------------------------------

%%%%%%%%%%%%%%%%%%%

\paragraph{Kinematic selection} \hspace{5pt}\newline
The squared mass of the two neutrinos, i.e., the squared missing mass $m^2_{\rm miss}$, defines the kinematics of the $K^+\to\pi^+\nu\bar\nu$ decay. This variable is reconstructed as $m^2_{\rm miss}=(P_K-P_\pi)^2$, where $P_K$ and $P_\pi$ are the four-momenta of the reconstructed $K^+$ and $\pi^+$, respectively, and is used to separate the signal from the main kaon decay modes as shown in Fig.~\ref{fig:m2miss}. 
The signal regions are two intervals of $m^2_{\rm miss}$ that are forbidden by kinematics to $K^+\to\pi^+\pi^0$, $K^+\to\mu^+\nu$, and $K^+\to\pi^+\pi^+\pi^-$ decays, located on either side of the peak from the $K^+\to\pi^+\pi^0$ decays.
These signal regions are referred to as regions 1 and 2, for $m^2_{\rm miss}$ below and above the $m^2_{\pi^0}$ peak, respectively.
For both regions, the lower momentum bound is at 15~GeV/$c$, while the upper bound is at 35~(45)~GeV/$c$ for signal region 1~(2). This is shown in Fig.~\ref{fig:mm2vspMCT}, overlaid onto the kinematic distributions of signal and background decays, where the candidate final-state particle remains inside the geometric region covered by HIKE detectors downstream of the fiducial volume.  

The main kaon decay modes enter the signal regions via the resolution tails in the reconstructed $m^2_{\rm miss}$ value.
Semileptonic kaon decay modes, instead, span over the signal regions.
The choice of the signal regions is determined by the $m^2_{\rm miss}$ resolution. NA62 defines the signal regions to keep the fraction of $K^+\to\pi^+\pi^0$ and $K^+\to\mu^+\nu$ decays that enter the signal regions below $10^{-3}$.
NA62 measures this fraction in data, and finds it to be well reproduced by simulations.
The tails mostly depend on the material budget of the straw spectrometer, while the intensity dependence is negligible.
Fig.~\ref{fig:KineTailsMm2}~(left) shows the $m^2_{\rm miss}$ distribution expected at HIKE compared to that of NA62.
The HIKE resolution is better than NA62 due to the 40\% lower material budget of the HIKE straw spectrometer.
Similarly to NA62, HIKE can define the signal regions combining $m^2_{\rm miss}$ with $m^2_{\rm miss} ({\rm RICH})$, the latter reconstructed using the magnitude of the $\pi^+$ momentum measured with the RICH under the $\pi^+$ hypothesis and the direction from the spectrometer, as shown in Fig.~\ref{fig:KineTailsMm2}~(right).
Simulation indicates that HIKE can optimise the signal regions to increase the acceptance by 10\% compared to NA62, while maintaining the resolution tails at the same level.

%%%%%%%%%%%%%%%%%%%%%%%

\paragraph{Additional selection criteria\label{sec:AdditionalCriteria}}
\hspace{5pt}\newline
The maximum pion momentum is limited to 35~GeV$/c$ for events entering region 1 to keep the $K^+\to\mu^+\nu$ background at the level of 5\% of the signal. Note that the $m^2_{\rm miss}$ spectrum of $K^+\to\mu^+\nu$ events approaches region 1 at high momentum because the $\pi^+$ mass hypothesis is used to calculate $m^2_{\rm miss}$ (Fig.~\ref{fig:mm2vspMCT}, bottom left).

The reconstructed decay vertex is required to be in a 55--65~m long region in the decay volume starting from at least 7~meters downstream from the final GTK station (GTK3).
The size of the decay region depends on the pion momentum, and is chosen to minimise the background from $K^+$ decays.
The initial cut is meant to suppress a source of background from single pions produced by hadronic interactions of beam kaons with the material of the GTK3.
The requirement of no activity in time with the pion in a guard-ring detector around the GTK3 (CHANTI) further suppresses this potential background.

Kaon decays occurring upstream of the final collimator defining the start of the decay region are a potential source of background if the daughter $\pi^+$ enters the decay region by passing through the collimator hole.
To suppress this background NA62 makes use of a BDT algorithm combining the slope of the pion track, the position of the reconstructed decay vertex, and the transverse position of the pion at the final collimator.
In addition, a scintillator array (veto counter, Section~\ref{sssec:vetocounter}) covering a region above and below the entrance hole of the collimator is used to tag kaon decays through the detection of additional daughter particles produced together with the $\pi^+$.

The above criteria will be applied to the HIKE PNN analysis. In NA62, the CHANTI and veto counter conditions cause a 3\% signal inefficiency at the maximum intensity, due to the random activity within the veto time windows.
Because the fraction of lost signal events is proportional to intensity for a fixed time window, the improved time resolution of the HIKE CHANTI and veto counter fully compensates for the $\times4$ intensity increase of HIKE compared to NA62. 
%%%%% >>>> 
In addition, beamline and detector optimisation will improve coverage for the HIKE veto counter, assisting with further suppression of backgrounds arising from decays upstream of the final collimator.
%%%%% <<<<<<

\newpage

\paragraph{Expected yield of $K^+\to\pi^+\nu\bar{\nu}$ signal events} \hspace{5pt}\newline
The yield of SM $K^+\to\pi^+\nu\bar{\nu}$ decays selected by NA62 per spill is the starting point for the evaluation of the number of SM signal events at HIKE. NA62 measures a yield of $2.5\times10^{-5}$ selected $K^+\to\pi^+\nu\bar\nu$ per spill for the 2022 data.
This value includes the effects of the trigger, data acquisition and offline selection, and accounts for the signal loss due to the beam intensity-dependent effects.

\begin{figure}[t]
\centering
\includegraphics[width=0.8\textwidth]{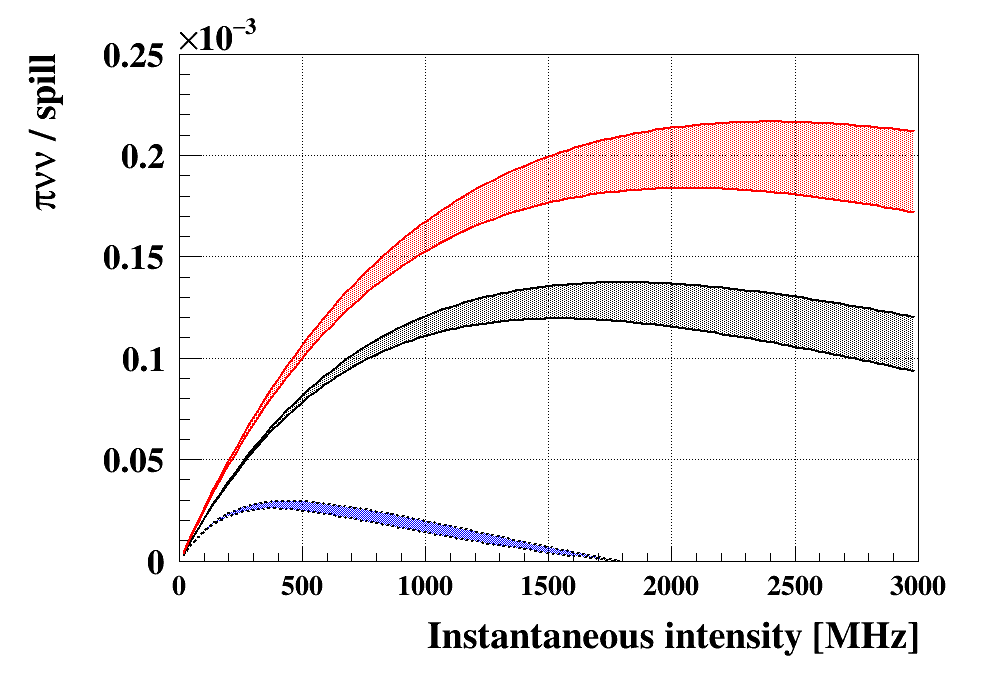}
\vspace{-4mm}
\caption{Numbers of selected $K^+\to\pi^+\nu\bar\nu$ events per spill as a functions of the instantaneous beam intensity. 
The blue shaded area shows the number of events from a data-driven model of the NA62 signal yield.
The black shaded area represents the same model but with detector time resolutions improved by a factor of 4 with respect to NA62, assuming also a software trigger.
The red shaded area represents the final HIKE signal yield model with all improvements included.
The width of the shaded areas illustrates the uncertainty in the intensity dependence model.
}
\label{fig:intdep}
\end{figure}

Using data taken at different intensities, NA62 describes the intensity dependence of the signal yield by the model shown in blue in Fig.~\ref{fig:intdep}, which has two main components.
The first component is a dead-time-equivalent paralyzable model that accounts for the intensity dependence of the trigger, DAQ, and all selection criteria except for the photon and multiplicity rejection.
The second component is a polynomial description of the random veto efficiency dependence on the intensity, due to the photon and multiplicity rejection.
The model shows that NA62 has a maximal signal yield at a beam intensity of $(400\pm25)$~MHz, corresponding to 65--70\% of the maximum NA62 intensity. 
Nevertheless, the yield has a broad maximum and differs by only a few percent between the peak and the maximum intensity.
The saturation of the $K^+\to\pi^+\nu\bar\nu$ event yield occurs both online and offline, and the two contributions are comparable.
The online part depends on the trigger, limited memory size and bandwidth of the L0 trigger processor, and the local trigger unit (LTU) system used to dispatch the triggers to the subdetectors. The online time resolution and limited flexibility of the hardware-based L0 trigger to veto muons and photons are the main contributing factors from the trigger. 
The LTU alone introduces an irreducible 75~ns dead time, which accounts for a loss of more than 10\% of signal events at the maximum intensity. The offline contribution to signal yield saturation is the result of the intensity-dependent effects described in the previous sections, dominated by the kaon--pion association and random veto efficiency.

HIKE is designed to overcome the intensity limitations of NA62: the time resolution of most detectors is $\times4$ better than in NA62; a software trigger and a new DAQ system are envisaged.
The software trigger has two main advantages: it allows recovery of the 15--20\% signal loss measured by NA62 due to the hardware trigger veto conditions, which do not fully overlap with the software veto criteria, and it weakens the intensity dependence of the trigger efficiency.
With these improvements alone, assuming the same behaviour of the signal yield as at NA62 (shown in blue in Fig.~\ref{fig:intdep}) occurring in HIKE at a 4 times higher intensity, the corresponding signal yield is shown in black in Fig.~\ref{fig:intdep}. The black shaded area accounts for the uncertainty in the extrapolation of the NA62 signal yield model.
This uncertainty stems from the errors of the parameters of the NA62 model, estimated comparing the behaviour with intensity of different data-sets collected during the 2023 data taking of NA62.

%%%%%%%%%%%%%%%%%%%%%

In addition to the detector time resolution improvements and use of a software trigger, 
HIKE expects the following improvements in signal efficiency (Section~\ref{sec:detectors}), 
which leads to the final expectation for HIKE shown by the red curve of Fig.~\ref{fig:intdep}:
\begin{itemize}
\item The recovery of the dead time of the LTU at DAQ level ($\times 1.1$ increase of signal efficiency between 1.6 and 1.8~GHz).
\item Improved kaon--pion association ($\times 1.1$).
\item Improved performance of the RICH ($\times 1.1$).
\item Better kinematic resolution ($\times 1.1$).
\end{itemize}
The first two factors weaken the intensity dependence of the signal yield.
The last two factors increase the signal acceptance at zero intensity.
A relative uncertainty of 10\% is assigned to the fractional improvement of each of the above factors. This spread reflects the uncertainties of the Monte Carlo simulations, which are estimated as an educated guess from the NA62 analysis experience.
The uncertainty is included in the red shaded area of figure~\ref{fig:intdep} and is subdominant to the uncertainty of the intensity model.
The improvements lead to a signal yield increase by a factor of about 1.5 between 1.6 and 1.8~GHz, and further weaken the intensity dependence.
This allows HIKE to move the peak yield running point to around 2.3~GHz leading to $(2.00\pm0.15)\times10^{-4}$ selected SM $K^+\to\pi^+\nu\bar\nu$ per spill.
This number weakly depends on the intensity in a range between 2.1 and 2.5~GHz.

Assuming $6\times10^5$ spills per standard year of HIKE data taking, 
we expect HIKE to select $(480\pm50)$ $K^+\to\pi^+\nu\bar\nu$ decays in four years (FYE) of data taking.
The uncertainty quoted accounts for the intensity dependence model, and the accuracy on the SM decay branching ratio.

\paragraph{Backgrounds} \hspace{5pt}\newline 
The NA62 analysis shows that the background can be divided in two categories:
\begin{enumerate}
\item background from $K^+$ decays in the fiducial volume;
\item background from $K^+$ decays or interactions occurring upstream to the decay volume, or ``upstream background''.
\end{enumerate}
NA62 measures the main sources of background from control data samples.
The background depends on the particle energy range, detector geometry and performance.
Therefore, the background expected in HIKE can be evaluated starting from the background observed by NA62. 

\paragraph{Kaon decay background} \hspace{5pt}\newline
\label{sec:KaonBackgrounds}
The main sources of background from kaon decays in the fiducial region are the $K^+\to\pi^+\pi^0$, $K^+\to\mu^+\nu$, $K^+\to\pi^+\pi^-e^+\nu$, and $K^+\to\pi^+\pi^+\pi^-$ decays.

The main factors suppressing the $K^+\to\pi^+\pi^0$ decays are the $\pi^0$ rejection and the kinematic definition of the signal regions.
The $\pi^0$ rejection depends on the energy of the selected particles, and the geometry and efficiency of the electromagnetic calorimeters.
The HIKE calorimeters are designed to have at least the same efficiencies as those of NA62. Studies performed at NA62 have shown that the $\pi^0$ rejection does not depend on beam intensity, which, instead, affects the signal efficiency (Section~\ref{sec:pnnPhotonRejection}).
The efficiency of the main electromagnetic calorimeter (LKr) for single photons above 10~GeV, measured to be $10^{-5}$, is the limiting factor for the $\pi^0$ rejection power at NA62. The HIKE MEC is designed to satisfy the above efficiency requirement.
Another source of photon detection inefficiency at HIKE is particle pileup in the seed cell of the MEC because unresolved overlapping clusters may shift the photon time from a $K^+$ decay out of the veto time window.
HIKE expects the photon rate to be above 10~MHz, so the MEC signals must have a FWHM below 20~ns to resolve overlapping photon clusters.
With this precaution, HIKE is expected to have the same $\pi^0\to\gamma\gamma$ rejection factor as NA62, which is of the order of $10^8$. 
The suppression of the $K^+\to\pi^+\pi^0$ decays is proportional to the fraction of events entering signal regions due to the resolution tails of the $m^2_{\rm miss}$ distribution.
As stated above, HIKE exploits the improved kinematic resolution by increasing the signal efficiency while keeping this fraction the same as that for NA62.
Therefore, HIKE expects the same background from $K^+\to\pi^+\pi^0$ decays as NA62, i.e. a fraction $0.10\pm0.02$ of the signal.

Evaluation of the $K^+\to\mu^+\nu$ background follows the same procedure as for the $K^+\to\pi^+\pi^0$ background.
The suppression depends on the kinematic definition of the signal regions and PID performance.
The fraction of events entering the signal regions due to the $m^2_{\rm miss}$ resolution tails and pion--muon misidentification are expected to be the same as those of NA62.
As a consequence, the $K^+\to\mu^+\nu$ background at HIKE is similar to NA62, i.e. a fraction $0.08\pm0.02$ of the signal.

NA62 evaluates the $K^+\to\pi^+\pi^-e^+\nu$ background using a sample of more than $10^9$ simulated events. The simulation results are validated using control data samples. The background level depends on the geometrical acceptance, particle energies, and multi-track rejection and positron identification performance.
These factors are comparable between HIKE and NA62, leading to the same level of $K^+\to\pi^+\pi^-e^+\nu$ background at HIKE and NA62.
Based on the NA62 results, HIKE expects the $K^+\to\pi^+\pi^-e^+\nu$ background to be a fraction $0.04\pm0.02$ of the signal.

The NA62 studies show that the $K^+\to\pi^+\pi^+\pi^-$ background depends on the
multiplicity rejection performance.
As a consequence, this background for HIKE is also expected similar to NA62, i.e., a fraction $0.02\pm0.01$ of the signal.

Following the NA62 analysis, the background from other $K^+$ decays can be considered negligible.
The uncertainties in the background fractions quoted above quantify the uncertainty from the knowledge of the future HIKE detector performance.
The total background from $K^+$ decays is estimated to be $0.24\pm0.04$ of the signal, i.e. $115\pm20$ events assuming 480~signal events. 

\paragraph{Upstream backgrounds\label{sec:UpsrtreamBackgrounds}} \hspace{5pt}\newline
Analysis of the NA62 data has established the existence of an accidental background originating from the beamline.
Upstream events are defined as interactions or decays of beam particles upstream of the fiducial volume.
In most cases, the $\pi^+$ detected downstream comes from a $K^+$ decaying in the region upstream of GTK3. The parent $K^+$ track is not reconstructed in the GTK due to the missing signals in the GTK stations, but a pileup GTK track is accidentally matched to the $\pi^+$.
The KTAG signal produced by the parent $K^+$ tags the matched GTK track as a kaon.

The PNN analysis employs specific selection criteria against this background (Section~\ref{sec:AdditionalCriteria}).
Upstream events can be background if the association between a pileup GTK track and the $\pi^+$ occurs accidentally; 
the $\pi^+$ track is misreconstructed because it undergoes a large-angle scattering at the first plane of the straw spectrometer;
the longitudinal coordinate of the decay vertex is accidentally reconstructed inside of the fiducial volume;
the parent $K^+$ does not leave additional hits in the GTK stations before decaying; and
the additional particles produced by the $K^+$ decay escape the veto counter acceptance.
 
Upstream interactions and decays are the dominant background source in the analysis of the NA62 Run~1 data, amounting to a fraction of the SM signal of $0.50\pm0.15$ at 60--70\% of the full intensity, as measured from data. The veto counter designed to reduce the upstream background was not present in Run~1; it was constructed during LS2 and has been fully operational since 2022.

The background-to-signal ratio is expected to increase linearly with the intensity for a fixed kaon--pion association time window, as a consequence of the increased rate of pileup particles in the time window.
In addition, a greater-than-linear effect could be present due to the increase in the fake-track reconstruction probability with intensity.
NA62 is currently studying the intensity dependence of the upstream background using data collected in Run~2.
The preliminary results are consistent with expectations, suggesting a nearly linear dependence.

A preliminary analysis of NA62 Run~2 data shows that the veto counter reduces the upstream background by a factor of 2 to 2.5. The restricted geometrical coverage of the beam area is the main limitation to the veto counter rejection power.
Full coverage was not possible in NA62 due to the beamline design just upstream of the final collimator, which was originally developed without fully taking into account the upstream background. Simulations show that a reduction of the upstream background by a factor of up to~6 is possible with an appropriate design of the upstream beam region.

The $\times4$ improved time resolution at HIKE with respect to NA62 allows the time window to be reduced proportionally to the increase of the intensity compared to NA62.
This fully compensates for the linear increase of the upstream background with intensity with respect to the signal.
Benefits are also expected from the smaller pixel size of the HIKE GTK and the corresponding decrease in the fake-track reconstruction probability.
Using the analysis of the NA62 data for guidance, the upstream background at HIKE at $\times4$ the maximum NA62 intensity can be estimated to be a fraction of $1.0\pm0.4$ of the signal without using the veto counter.
The uncertainty takes into account the 30\% precision of the upstream background estimate for NA62 Run~1 data, and an additional 30\% due to the uncertainty in the linear-dependence hypothesis.

Including an NA62-like veto counter without considering any beamline optimisation, the background fraction is expected to be reduced to $0.5\pm0.3$ of the signal.
The error quoted includes a 20\% uncertainty of the knowledge of the reduction factor of the NA62-like veto counter.
Optimisation of the 
beamline and veto counter geometry can further reduce the upstream background by a factor between 2 and 3.
This leads to an upstream background of the order of, or even lower than, the $K^+$ decay background.
Additional reduction might be obtained from the reduced amount of material of the HIKE straw spectrometer; this effect has not been quantified so far.
Studies are underway to reduce the uncertainty on the expected upstream background.

\paragraph{Precision on the branching ratio measurement}
\hspace{5pt}\newline
The projected numbers of collected events from the HIKE PNN analysis in four years (FYE) of data-taking are summarised in Table~\ref{tab:pnnexp}.
According to Eq.~(\ref{eq:pnnstaterror}), the relative statistical uncertainty for the measurement of ${\cal B}(K^+\to\pi^+\nu\bar\nu)$ is between 5.4\% and 6.1\%.

Systematic uncertainties arise from the precision of the background and signal sensitivity estimates.
Based on NA62 experience, these can be minimised by splitting the analysis into categories fitting the shapes of kinematic distributions.
Part of the systematic uncertainty for the branching ratio measurement performed using NA62 Run~1 data arises from limited statistics, which will not be the case at HIKE due to the increase in the size of the data sample by a factor of 40. 
Studies are ongoing with NA62 Run~2 data to reduce the uncertainties. The projected total relative systematic uncertainty for HIKE is below 3\%.

A measurement with a precision of 3.5\% would allow matching the purely theoretical error of the SM $K^+\to\pi^+\nu\bar\nu$ prediction. Statistically, HIKE would need about 7 years to reach such an uncertainty, but an excellent control of the systematic uncertainty becomes crucial. 
Just as an example, a systematic error at the level of 2.5\% would require a statistical error also at the level of 2.5\%, meaning about 12 years of data taking, to reach an overall 3.5\% uncertainty on the branching ratio measurement. 
While most of the systematic uncertainty is of statistical nature and can therefore be improved, more detailed studies are on-going to assess the possibility to further reduce the systematic uncertainty.

\begin{table}[tb]
\caption{Summary of expectations for the $K^+\to\pi^+\nu\bar\nu$ analysis in four years (FYE) of HIKE Phase~1 operation. The values quoted have a 10\% relative uncertainty.}
\begin{center}
\vspace{-6mm}
\begin{tabular}{l|l}
\hline
Number of spills & $2.4\times10^6$\\  
Protons on target & $3.2\times10^{19}$ \\ 
$K^+$ decays in FV & $8.0\times10^{13}$ \\ 
Expected SM $K^+\to\pi^+\nu\bar\nu$ events & 480 \\
Background from $K^+$ decays & 115 \\
Upstream/accidental background & 80--240 \\ 
Expected statistical precision $\sigma(\mathcal{B})/\mathcal{B}$ & $5.4\%$--$6.1\%$ \\
\hline
\end{tabular}
\end{center}
\vspace{-7mm}
\label{tab:pnnexp}
\end{table}

%%%%%%%%%%%

\subsubsection{Test for scalar amplitudes in $K^+\to\pi^+\nu\bar\nu$ decay}

\begin{figure}[tb]
\centering
\vspace{-7mm}
\includegraphics[width=0.5\columnwidth]{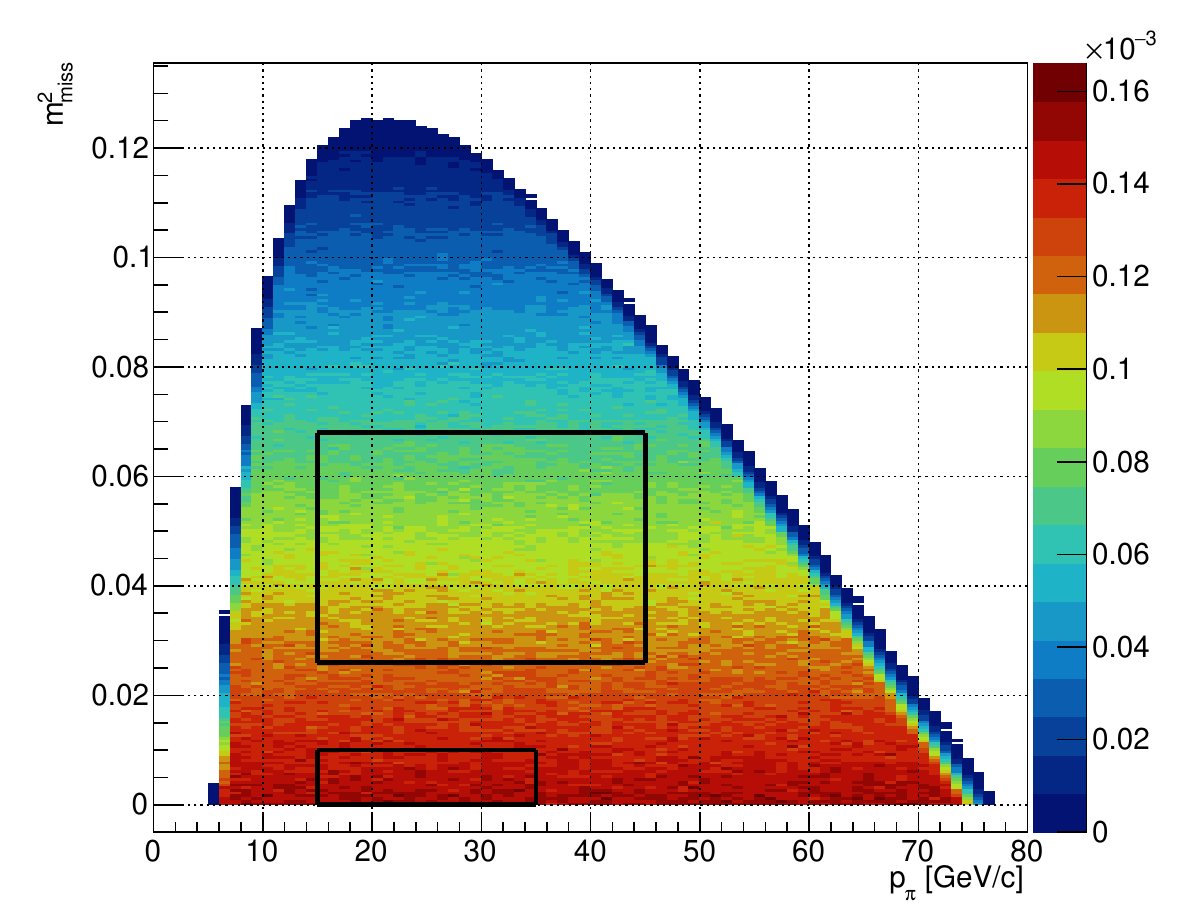}%
\includegraphics[width=0.5\columnwidth]{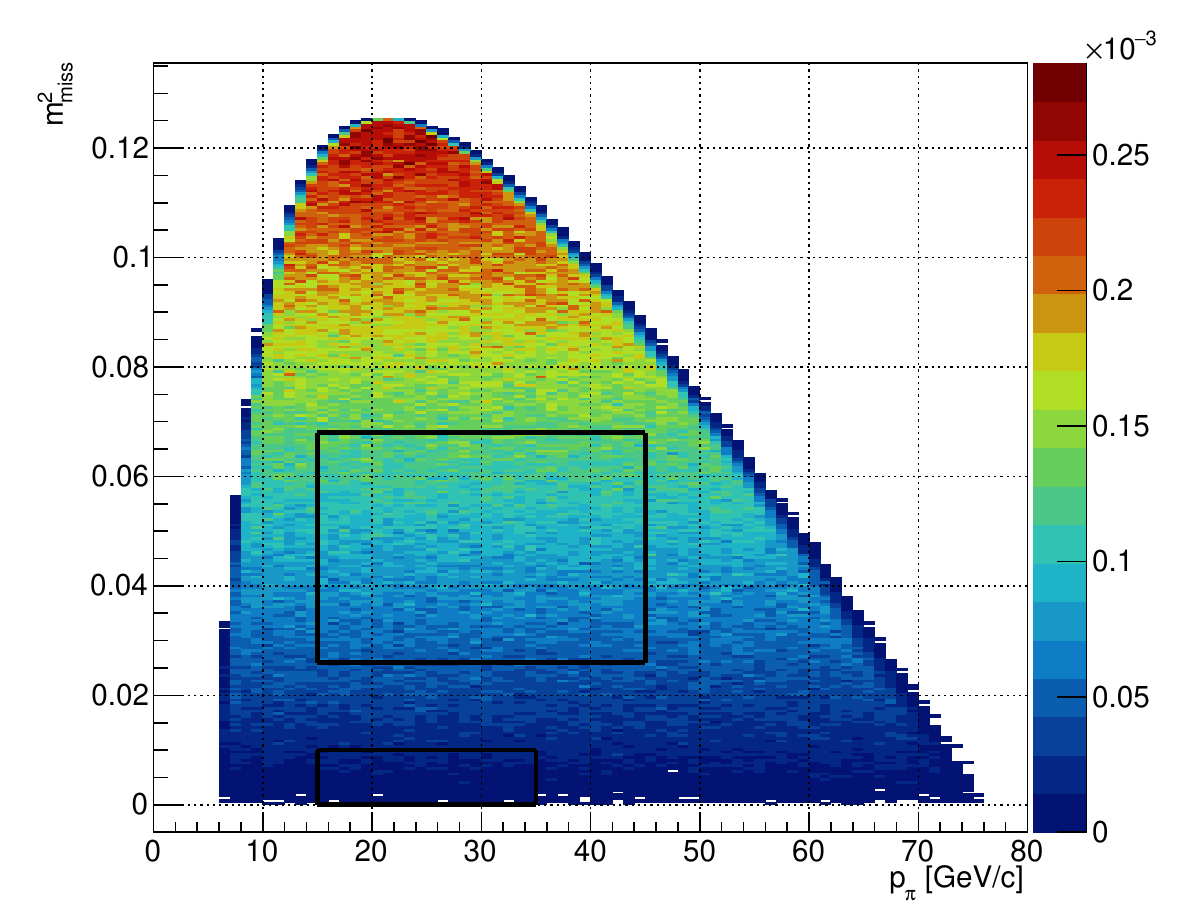}
\vspace{-8mm}
\caption{Simulated distributions of the squared missing mass $m_{\rm miss}^{2}=(P_{K^+}-P_{\pi^+})^2$ vs pion momentum in the laboratory frame for the SM $K^+\to\pi^+\nu\bar\nu$ decay with a vector nature (left) and the LNV $K^+\to\pi^+\nu\nu$ decay with a scalar nature (right). Black boxes indicate the signal regions for which the experimental sensitivity is highest.}
\label{fig:SMvsLNVdistributions}
\end{figure}

In BSM scenarios, the $K^+\to\pi^+\nu\bar\nu$ branching ratio may receive contributions from both the SM process $K^+\to\pi^+\nu\bar\nu$ with a vector nature, and a possible
BSM process $K^+\to\pi^+\nu\nu$ with a scalar nature~\cite{Deppisch:2020oyx,Crosas:2022quq}.
Therefore the experimentally measured branching ratio is given by~\cite{Aebischer:2022vky,Deppisch:2020oyx,Buras:2023qaf}
\begin{displaymath}
\mathcal{B}(K^+\to\pi^+\nu\bar\nu) 
= \mathcal{B}_{\rm SM}(K^+\to\pi^+\nu\bar\nu) + \sum_{i \leq j}^{3} \mathcal{B}_{\rm LNV}(K^+\to\pi^+\nu_i\nu_j) 
= \mathcal{B}^{\text{SM}}_{\text{v}} + \mathcal{B}^{\text{LNV}}_{\text{s}},
\end{displaymath}
with the two terms being due to the SM (vector) and BSM (scalar) processes, respectively.
The SM and BSM contributions are expected to have different kinematic distributions (Fig.~\ref{fig:SMvsLNVdistributions}). To identify the nature of the decay, an investigation of the shape of the distribution of selected signal candidates as a function of kinematic variables will be performed.
In absolute terms, within the $K^+\to\pi^+\nu\bar\nu$ signal regions, the ratio of acceptances for the two contributions is $A_{\text{LNV}}/A_{\text{SM}} = 0.76$.

The main background to a search for a scalar (LNV) $K^+\to\pi^+\nu\nu$ decay is the SM (vector) $K^{+}\to\pi^{+}\nu\bar\nu$ decay itself, which, however, has a different kinematic distribution.
The simplest way to detect a BSM (scalar) component would be to observe an excess~\cite{Buras:2023qaf} of candidate events with respect to the SM expectation. The expected upper limit of $\mathcal{B}^{\text{LNV}}_{\text{s}}/\mathcal{B}^{\text{SM}}_{\text{v}}$ for the complete HIKE dataset using this approach is 0.11 at 90\% CL. Exploitation of the kinematic variables will allow a higher sensitivity and a measurement independent of the absolute SM branching ratio prediction.

Other BSM scenarios can also allow tensor contributions to this decay~\cite{Aebischer:2022vky,Li:2019fhz}. This hypothesis will also be investigated
through the decay spectrum measurement.

%%%%%%%%%%%%%%%%%

\subsection{Other decay modes}
\label{sec:phase1-other-measurements}

\subsubsection{Rare and forbidden $K^+$ decays}

HIKE Phase~1 is expected to increase the world samples of rare $K^+$ decays by an order of magnitude, building upon the strategies developed for the NA62 experiment. This will bring many rare $K^+$ and $\pi^0$ decay measurements to a new level of precision. 

In terms of precision lepton universality tests (Section~\ref{sec:GlobalNPSensitivity}), we expect to collect background-free samples of several times $10^5$ events of both $K^+\to\pi^+e^+e^-$ and $K^+\to\pi^+\mu^+\mu^-$ decays to compare the form-factor parameters ($a_+$, $b_+$) of the two decay modes. Considering that the precision of the recent NA62 measurement in the muon channel~\cite{NA62:2022qes} is limited by the size of the dataset, HIKE Phase~1 is expected to measure the difference $\Delta a_+^{e\mu}$ of the form-factor parameter $a_+$ between the two decay modes to a precision of $\pm0.007$, and the corresponding difference $\Delta b_+^{e\mu}$ to a precision of $\pm0.015$. Exploitation of the correlations between the measured parameters $a_+$ and $b_+$ for each mode will provide additional power for tests of LFU.
Furthermore, HIKE aims to achieve a precision at the level of per mille on the ratio $R_K={\cal B}(K^+\to e^+\nu)/{\cal B}(K^+\to\mu^+\nu)$, providing another powerful LFU test.

In terms of tests of lepton number and flavour conservation in $K^+$ and $\pi^0$ decays (Section~\ref{sec:lfv}), which involve searches for the forbidden decays $K^+\to\pi^-(\pi^0)\ell_1^+\ell_2^+$, $K^+\to\pi^+\mu^\pm e^\mp$, $K^+\to\ell_1^-\nu\ell_2^+\ell_2^+$ and $\pi^0\to\mu^\pm e^\mp$, the experimental technique based on dedicated dilepton trigger chains has been established by the NA62 experiment, leading to world-leading upper limits of ${\cal O}(10^{-11})$ on the branching ratios of a number of processes with the NA62 Run~1 dataset~\cite{NA62:2019eax,NA62:2021zxl,NA62:2022tte,NA62:2022exp}. These searches are not limited by backgrounds, and HIKE Phase~1 sensitivity is expected to improve in the future essentially linearly with the size of the dataset to the ${\cal O}(10^{-12})$ level or below. Comprehensive searches for LNV processes with displaced vertices involving emission and decay of a heavy Majorana neutrino, $K^+\to\ell_1^+N$, $N\to\pi^-\ell_2^+$, will also be performed at unprecedented sensitivity.

In terms of low-energy QCD tests (Section~\ref{sec:chpt}), major improvements are foreseen at HIKE Phase~1 in the measurements of rare and radiative kaon decays. These include the rates of the $K^+\to\pi^+\ell^+\ell^-$, $K^+\to e^{+}\nu\gamma$, $K^+\to\pi^0e^+\nu\gamma$, $K^+\to\pi^+\pi^0\gamma$ and $K^+\to\pi^+\gamma\gamma$ decays which will be measured to relative precision of a few per mille, and the rates for the $K^+\to\pi^+\gamma \ell^+\ell^-$, $K^+\to\pi^+\pi^0e^+e^-$ decays, which will be measured to a relative precision of a few percent. In all cases, precision measurements of the decay spectra will also be performed. These results will provide unique inputs for tests of leading-order and next-to-leading order ChPT calculations, and for the determination of the low-energy constants.

A broad, well-established HIKE programme of searches for production of hidden-sector mediators in $K^+$ decays, which addresses many of the PBC benchmarks and a range of non-minimal scenarios, is discussed in Section~\ref{sec:fips:kaons}. The physics reach for the PBC benchmarks is summarised in Section~\ref{sec:fips_sensitivity}.

Finally, thanks to the high beam flux and the unique beam tracker, HIKE Phase~1 is expected to provide the first observation of tagged neutrinos~\cite{Perrin-Terrin:2021jtl} produced in the $K^+\to\mu^+\nu$ decay and detected via charged-current scattering in the calorimeters, thereby providing a proof of principle for this technique. A preliminary estimate based on extrapolation of the recent NA62 result~\cite{na62-tagged-nu} suggests a signal yield of about 30~events in the total HIKE Phase~1 dataset with a signal-to-background ratio $S/B>1$, possibly reaching a $5\sigma$ significance.

%%%%%%%%%%%%%%%%%

\subsubsection{Main $K^+$ branching ratios and CKM unitarity}

\label{sec:ckm_phase1}

HIKE Phase 1 is poised to make a significant impact on the knowledge of the main $K^\pm$ branching ratios, including not only those for the semileptonic decays used to obtain $V_{us}$, but also that for the dominant $K_{\mu2}$ decay. 
Not only does the value of $V_{us}/V_{ud}$ used in the unitarity analysis derive from a single measurement of ${\cal B}(K_{\mu2})$ with a 0.27\% total uncertainty~\cite{KLOE:2005xes}; this measurement also impacts the normalisation of all other branching ratio measurements in the $K^+$ decay rate fit to world data, e.g., by the PDG or the analysis of \cite{FlaviaNetWorkingGrouponKaonDecays:2010lot}. The importance of the measurement of the ratio ${\cal B}(K_{\mu3})/{\cal B}(K_{\mu2})$ to settle this question is discussed in \cite{Cirigliano:2022yyo}. HIKE could also make a very precise measurement of ${\cal B}(K_{\mu3})/{\cal B}(K_{e3})$, an important test of lepton universality, as well as of other important ratios amenable to measurement with good precision, such as ${\cal B}(K_{e3})/{\cal B}(K_{\pi2})$, ${\cal B}(K_{\mu3})/{\cal B}(K_{\pi2})$, and ${\cal B}(K_{\pi2})/{\cal B}(K_{\mu2})$, possibly with a unified analysis.
With the ratios between the widths for four of the six main $K^+$ decay modes thus determined, current world data on the branching ratios for $K_{\mu2}$, $K_{\pi2}$, $K_{e3}$, and $K_{\mu3}$ can be omitted from the $K^+$ rate fit, allowing HIKE to make a nearly independent determination of the $K_{\mu2}$ and $K_{\ell3}$ branching ratios.

In order to perform high-precision measurements of the five ratios mentioned above, HIKE could collect the needed data in a short data-taking period with a minimum-bias trigger, at low intensity and with special emphasis on maintaining conditions as stable as possible. With a run of less than two weeks in duration, the uncertainties on the measurement of these ratios would be dominated by systematics. While the limiting systematic uncertainties are difficult to predict, the HIKE sensitivity can be estimated on the basis of past experience. NA48/2 measured the ratios
${\cal B}(K_{e3})/{\cal B}(K_{\pi2})$, 
${\cal B}(K_{\mu3})/{\cal B}(K_{\pi2})$, and 
${\cal B}(K_{e3})/{\cal B}(K_{\mu3})$ at the level of 0.4\%~\cite{NA482:2006vnw}. 
With HIKE's high-precision tracking for both beam and secondary particles, redundant particle identification systems, and excellent calorimetry and photon detection, it should be easy to match or exceed this precision, especially for the ratios 
${\cal B}(K_{\mu3})/{\cal B}(K_{\mu2})$ and ${\cal B}(K_{\mu3})/{\cal B}(K_{e3})$, for which significant cancellations of systematic uncertainties are expected. 
The HIKE Phase-1 sensitivity estimate
assumes 0.2\% total uncertainty for the measurements of these latter two ratios, and 0.4\% total uncertainty for the measurements of ${\cal B}(K_{e3})/{\cal B}(K_{\pi2})$, ${\cal B}(K_{\mu3})/{\cal B}(K_{\pi2})$, and ${\cal B}(K_{\pi2})/{\cal B}(K_{\mu2})$.
When the HIKE Phase~1 measurements of these ratios at values near those expected from the current fit replace all of the older measurements involving the branching ratios for the  $K_{\mu2}$, $K_{\pi2}$, $K_{e3}$, and $K_{\mu3}$ decays, the uncertainty on ${\cal B}(K_{\mu2})$ can be reduced by 40\%, while those on ${\cal B}(K_{e3})$ and ${\cal B}(K_{\mu3})$ can be reduced by well over 50\%.

%%%%%%%%%%%%%%%

\subsection{Summary of physics sensitivity}

HIKE will provide a measurement of the branching ratio of the ultra-rare decay $K^+\to\pi^+\nu\bar\nu$ to $\mathcal{O}(5\%)$ accuracy. This precision rivals the theoretical precision, and therefore provides a strong test of the SM~\cite{buras2023kaon}.
Building on the successful NA62 strategy and using the knowledge gained to inform projections for HIKE, it is expected that about 480 SM $K^+\to\pi^+\nu\bar\nu$ events will be observed in 4--5 standard years of HIKE Phase~1 operation.
Furthermore, the first comprehensive analysis of the $K^+\to\pi^+\nu\bar\nu$ decay spectrum will allow investigation of the nature of the decay in terms of a possible lepton number violating contribution and possible production of hidden-sector mediators.
Considering that HIKE Phase~1 is being developed to meet the demands of the extremely challenging measurement of $K^+\to\pi^+\nu\bar\nu$, the experiment has a wide reach for precision measurements of both dominant and rare $K^+$ and $\pi^0$ decays, as well as searches for decays forbidden in the SM.

\section{Phase 2: a multi-purpose $K_L$ decay experiment}
\label{sec:phase2}

\subsection{Experimental layout}

\begin{figure}[ht]
\begin{center}
\resizebox{\textwidth}{!}{\includegraphics{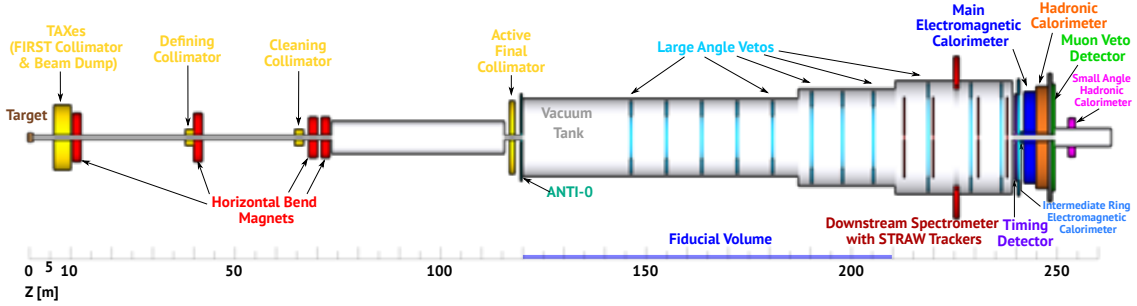}}
\end{center}
\vspace{-7mm}
\caption{HIKE Phase~2 layout, with an aspect ratio of 1:10.}
\label{fig:phase2-layout}
\end{figure}

The layout of the proposed multi-purpose $K_L$ decay experiment is shown schematically in Fig.~\ref{fig:phase2-layout}. The 120~m long beamline studied extensively with simulations~\cite{Gatignon:2018xxx} involves four stages of collimation with an active final collimator. This provides a secondary neutral beam with a 0.4~mrad opening angle, and allows for adequate suppression of the short-lived $K_S$ and $\Lambda$ components (Section~\ref{sec:neutral_beam}). The beamline is followed by a 90~m long fiducial decay volume (FV) located in the vacuum tank. The detector located downstream of the FV is similar to the HIKE Phase~1 downstream setup, with minimal adjustments introduced. Specifically, the beam tracker, KTAG and RICH 
detectors are removed, the STRAW spectrometer is shortened to a total length of 25~m and the central holes of the STRAW chambers are realigned on the neutral beam axis; the SAC is moved; and the LAV detectors will be moved and possibly reduced in number. The magnetic field strength in the spectrometer dipole magnet is reduced by about 20\% with respect to Phase~1, providing a momentum kick of 210~MeV/$c$. This does not lead to a significant degradation of the mass resolution.

\begin{figure}[tb]
\begin{center}
\resizebox{0.5\textwidth}{!}{\includegraphics{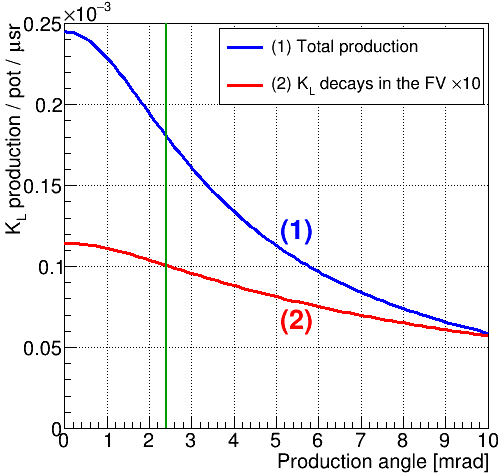}}%
\resizebox{0.5\textwidth}{!}
{\includegraphics{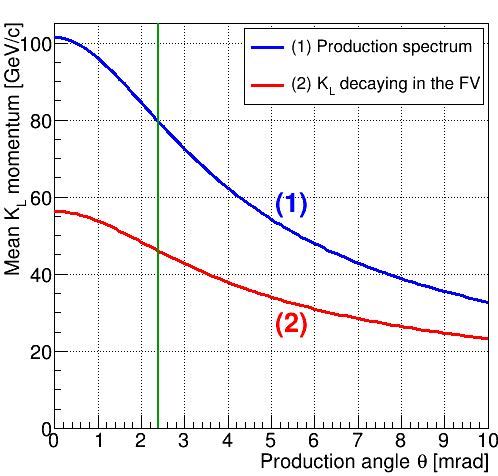}}%
\end{center}
\vspace{-6mm}
\caption{Left: $K_L$ yield in the beam and number of $K_L$ decays in the FV (scaled by 10) per proton on target and per unit solid angle as functions of the production angle. Right: mean $K_L$ momentum at production and for kaons decaying in the FV as functions of the production angle. The production model used is based on a FLUKA simulation of $K_L$ production on a 400~mm beryllium target~\cite{vanDijk:2018aaa}. The production angle of 2.4~mrad considered for HIKE Phase~2 is shown with vertical lines.}
\label{fig:KL_spectrum}
\end{figure}

A production angle of 2.4~mrad, used earlier by the NA48 experiment, is considered for the HIKE neutral beam. The $K_L$ flux per proton on target increases and the $K_L$ spectrum hardens towards small production angles (Fig.~\ref{fig:KL_spectrum}). A harder $K_L$ spectrum increases the signal acceptances (which peak at a $K_L$ momentum of 100~GeV/$c$ for the $K_L\to\pi^0e^+e^-$ decay and 70~GeV/$c$ for the $K_L\to\pi^0\mu^+\mu^-$ decay), and improves the suppression of backgrounds with extra photons, which tend to be intercepted by the main EM calorimeter and have larger energies in the laboratory frame. Though the $K_L$ decay probability in the FV decreases for a harder $K_L$ spectrum, the number of $K_L$ decays in the FV per proton on target still increases towards small production angles due to the increased $K_L$ flux (Fig.~\ref{fig:KL_spectrum}, left). However the neutron and photon fluxes in the beam also increase. The simulations used in this section include the neutral beam line as described in Section~\ref{sec:neutral_beam} and the detector description seen in Fig.~\ref{fig:phase2-layout}.

Assuming $0.6\times 10^6$ useful SPS spills per year of operation at an intensity of $2\times 10^{13}$ POT/spill, we expect an exposure of $1.2\times 10^{19}$~POT/year. For a production angle of 2.4~mrad and a beam opening angle of 0.4~mrad, the expected $K_L$ yield in the beam evaluated with simulations is $5.4\times 10^{-5}$ per proton on target (which accounts for losses due to the beamline and collimation). The total integrated $K_L$ flux in the beam is 
$6.5\times 10^{14}$/year, which corresponds to $3.8\times 10^{13}$ decays/year in the FV. The mean momentum of $K_L$ mesons entering the decay volume is 80~GeV/$c$, while the mean momentum of $K_L$ mesons decaying in the FV is 45~GeV/$c$ (Fig.~\ref{fig:KL_spectrum}, right).

%%%%%%%%%%%%%%%%%%

%%%%%%%%%%%%%%%%%%%%

\subsection{The benchmark channels $K_L\to\pi^0\ell^+\ell^-$}
\label{sec:phase2:Klpi0ll}

The principal goal of HIKE Phase~2 is the first observation and measurement of the ultra-rare decays $K_L\to\pi^0\ell^+\ell^-$  ($\ell=e,\mu$). The motivations for the measurement of these golden modes and the current experimental status are summarised in Section~\ref{sec:physics:fcnc:klpi0ll}. Sensitivity to the $K_L\to\pi^0\ell^+\ell^-$ decays has been evaluated using a full Geant4-based simulation, reconstruction and analysis chain using the HIKE 
software platform developed on the basis of the NA62 software.

The $K_L\to\pi^0\ell^+\ell^-$ decays are modelled according to their expected SM spectra~\cite{Isidori:2004rb}, assuming constructive interference between the direct and indirect CP-violating contributions, which is preferred from the theory point of view. The principal and irreducible background comes from the Greenlee decays $K_L\to\gamma\gamma\ell^+\ell^-$~\cite{Greenlee:1990qy}. These decays are modelled according to their expected SM kinematic distributions~\cite{Husek:2015sma}. The kinematic regions used for the simulation of the Greenlee decays and the corresponding SM branching ratios (which exceed those of the signal modes by orders of magnitude) are listed in Table~\ref{tab:greenlee}.

The final state for both the signal and Greenlee background modes consists of a lepton pair ($\ell^+\ell^-$) and a photon pair ($\gamma\gamma$), with the latter produced in a prompt $\pi^0\to\gamma\gamma$ decay in the case of signal. In the $K_L\to\pi^0e^+e^-$ case, a selection condition $m_{ee}>140~{\rm MeV}/c^2$ is applied on the reconstructed mass of the lepton pair to suppress backgrounds from major $K_L$ decays into neutral pions followed by the Dalitz decay $\pi^0\to\gamma e^+e^-$. Suppression of the Greenlee background in both cases is based on the reconstructed masses of the $\gamma\gamma\ell^+\ell^-$ and $\gamma\gamma$ systems ($m_{\gamma\gamma\ell\ell}$, $m_{\gamma\gamma}$), and on the kinematic variables. 

\begin{table}[tb]
\caption{SM branching ratios of the Greenlee decays in the phase space regions used for simulations.}
\begin{center}
\vspace{-6mm}
\begin{tabular}{llc}
\hline
Mode & Phase space region & Branching ratio \\
\hline
$K_L\to\gamma\gamma e^+e^-$ &
$x=(m_{ee}/m_K)^2>0.05$,
& 
$(1.55\pm0.05)\times 10^{-7}$ \\
& $x_\gamma=(m_{\gamma\gamma}/m_K)^2>0.01$ \\
\hline
$K_L\to\gamma\gamma\mu^+\mu^-$ &
$x_\gamma=(m_{\gamma\gamma}/m_K)^2>0.01$ &
$(1.49\pm0.28)\times 10^{-9}$ \\
\hline
\end{tabular}
\end{center}
\label{tab:greenlee}
\end{table}

\begin{figure}[p]
\begin{center}
\resizebox{0.46\textwidth}{!}{\includegraphics{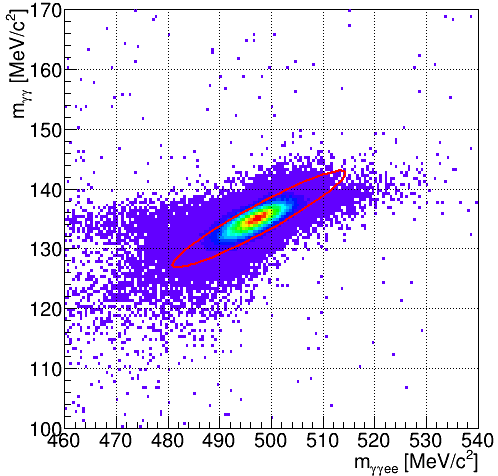}}%
\resizebox{0.46\textwidth}{!}{\includegraphics{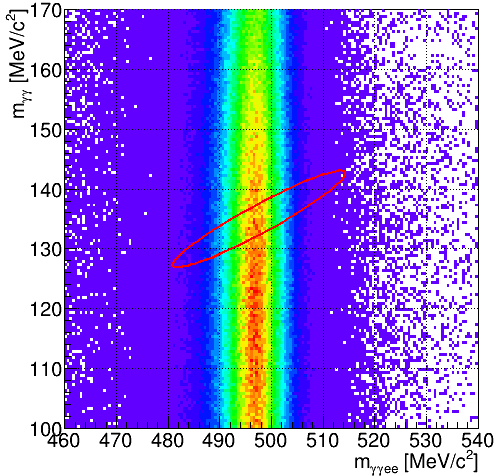}}%
\put(-362,170){\colorbox{white}{\bf $K_L\to\pi^0e^+e^-$}}%
\put(-164,170){\colorbox{white}{\color{red}\bf $K_L\to\gamma\gamma e^+e^-$}}
\end{center}
\vspace{-7mm}
\caption{Reconstructed masses $m_{\gamma\gamma ee}$ and $m_{\gamma\gamma}$ for $K_L\to\pi^0 e^+e^-$ signal (left) and $K_L\to\gamma\gamma e^+e^-$ background (right), with the elliptic selection condition shown.}
\label{fig:kpi0ee-mass}
\end{figure}

\begin{figure}[p]
\begin{center}
\resizebox{0.46\textwidth}{!}{\includegraphics{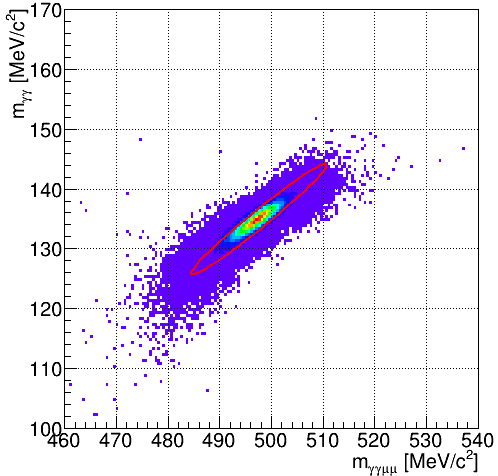}}%
\resizebox{0.46\textwidth}{!}{\includegraphics{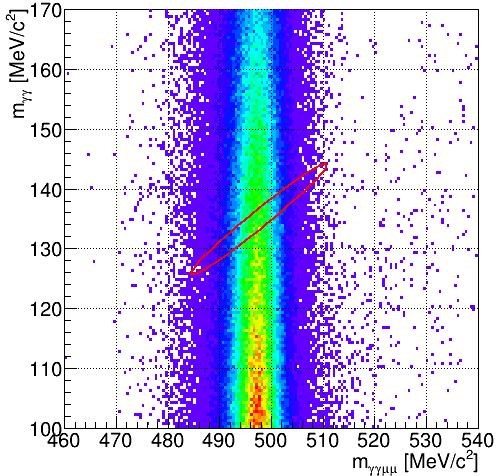}}%
\put(-362,170){\colorbox{white}{\bf $K_L\to\pi^0\mu^+\mu^-$}}%
\put(-164,170){\colorbox{white}{\color{red}\bf $K_L\to\gamma\gamma\mu^+\mu^-$}}
\\
\resizebox{0.46\textwidth}{!}
{\includegraphics{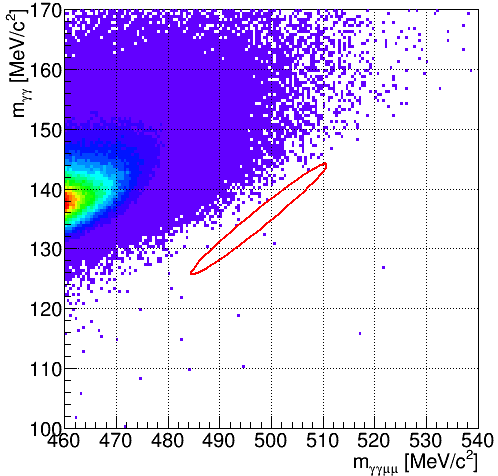}}%
\put(-166,28){\colorbox{white}{\color{red}\bf $K_L\to\pi^+\pi^-\pi^0$}}%
\end{center}
\vspace{-7mm}
\caption{Reconstructed masses $m_{\gamma\gamma\mu\mu}$ and $m_{\gamma\gamma}$ for $K_L\to\pi^0\mu^+\mu^-$ signal (left), $K_L\to\gamma\gamma\mu^+\mu^-$ (right) and $K_L\to\pi^+\pi^-\pi^0$ (bottom) backgrounds, with the elliptic selection condition shown.}
\label{fig:kpi0mm-mass}
\end{figure}

The signal and background distributions in the ($m_{\gamma\gamma\ell\ell}$, $m_{\gamma\gamma}$) plane are shown in Fig.~\ref{fig:kpi0ee-mass} for the $K_L\to\pi^0e^+e^-$ case, and in Fig.~\ref{fig:kpi0mm-mass} for the $K_L\to\pi^0\mu^+\mu^-$ case. Elliptic selection conditions (also shown in Figs.~\ref{fig:kpi0ee-mass},~\ref{fig:kpi0mm-mass}) are applied to exploit the fact that the signal peaks at the $\pi^0$ mass in the $m_{\gamma\gamma}$ projection, taking into account the correlation between the reconstructed $m_{\gamma\gamma\ell\ell}$ and $m_{\gamma\gamma}$ variables. The background suppression is determined by the photon energy resolution provided by the EM calorimeter. The energy resolution used in the present study is identical to that of the present NA62 LKr calorimeter~\cite{NA62:2017rwk}, which leads a 2.2~MeV resolution on the diphoton mass $m_{\gamma\gamma}$ for $K_L\to\pi^0\ell^+\ell^-$ decays for the experimental setup considered.

The kinematic selection exploited in the present study is similar to that used for the search for $K_L\to\pi^0\ell^+\ell^-$ decays at the KTeV experiment~\cite{KTeV:2003sls,KTEV:2000ngj}, and is based on two reconstructed variables. Firstly, the photon energy asymmetry is defined as
\begin{displaymath}
y_\gamma = \frac{2P\cdot(k_1-k_2)}{m_K^2\cdot\lambda^{1/2}(1,x,x_\gamma)},
\end{displaymath}
where $P$ is the kaon four-momentum, $k_{1,2}$ are the photon four-momenta, $x=(m_{ee}/m_K)^2$, $x_\gamma=(m_{\gamma\gamma}/m_K)^2$, and $\lambda(a,b,c)=a^2+b^2+c^2-2(ab+bc+ac)$. The Greenlee background peaks at $|y_\gamma|=1$, especially strongly at low values of $x=(m_{\ell\ell}/m_K)^2$, while the signal distribution is uniform as a consequence of the isotropic nature of the $\pi^0\to\gamma\gamma$ decay. Secondly, the smallest angle between any photon and any lepton in the kaon frame, $\theta_{\ell\gamma}^{\rm min}$, is considered. The Greenlee background peaks at $\theta_{e\gamma}^{\rm min}=0$ in the $K_L\to\pi^0e^+e^-$ case, which allows for a significant background reduction. The discrimination of signal and background provided by the $\theta_{\mu\gamma}^{\rm min}$ variable in the $K_L\to\pi^0\mu^+\mu^-$ case is marginal, however. The selection conditions applied in the ($\theta_{\ell\gamma}^{\rm min}$, $|y_\gamma|$) plane are optimised to maximise the quantity $N_S/\sqrt{N_S+N_B}$, where $N_S$ and $N_B$ are the expected numbers of signal and background events. The optimisation does not depend on the assumptions regarding the total integrated kaon flux, and leads to optimal statistical precision for the measurement of the signal branching ratio.

\begin{figure}[p]
\begin{center}
\resizebox{0.5\textwidth}{!}{\includegraphics{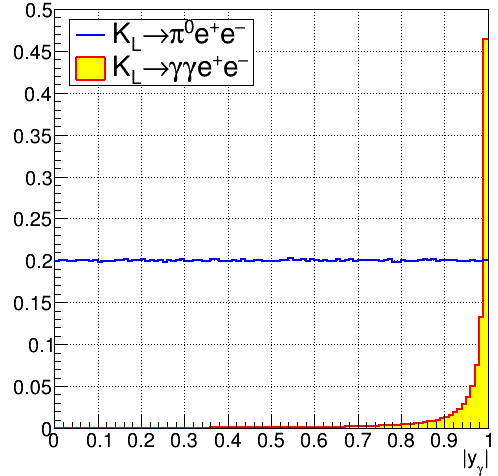}}%
\resizebox{0.5\textwidth}{!}{\includegraphics{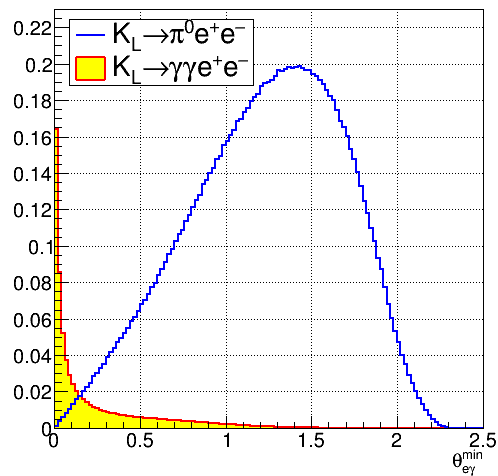}}%
\end{center}
\vspace{-6mm}
\caption{Distributions of the true values of the kinematic variables $|y_\gamma|$ (left) and $\theta_{e\gamma}^{\rm min}$ (right) for simulated samples of $K_L\to\pi^0 e^+e^-$ and $K_L\to\gamma\gamma e^+e^-$ decays.}
\label{fig:kpi0ee-true}
\end{figure}

\begin{figure}[p]
\begin{center}
\resizebox{0.5\textwidth}{!}{\includegraphics{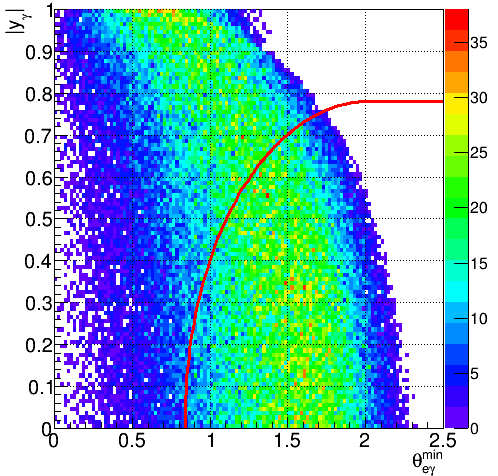}}%
\resizebox{0.5\textwidth}{!}{\includegraphics{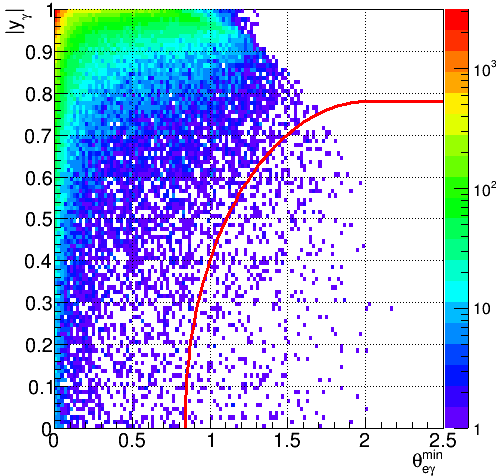}}%
\put(-305,187){\bf $K_L\to\pi^0e^+e^-$}%
\put(-94,187)
{\color{red}\bf $K_L\to\gamma\gamma e^+e^-$}
\end{center}
\vspace{-6mm}
\caption{Distributions of reconstructed kinematic variables $|y_\gamma|$ and $\theta_{e\gamma}^{\rm min}$ for simulated samples of $K_L\to\pi^0 e^+e^-$ (left) and $K_L\to\gamma\gamma e^+e^-$ (right) decays. The selection condition applied in the ($\theta_{e\gamma}^{\rm min}$, $|y_\gamma|$) plane is shown with solid lines. The signal region is below the line.}
\label{fig:kpi0ee-selection}
\end{figure}

\begin{figure}[p]
\begin{center}
\resizebox{0.5\textwidth}{!}{\includegraphics{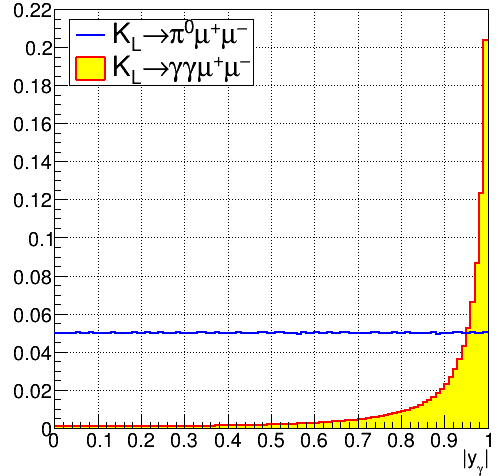}}%
\resizebox{0.5\textwidth}{!}{\includegraphics{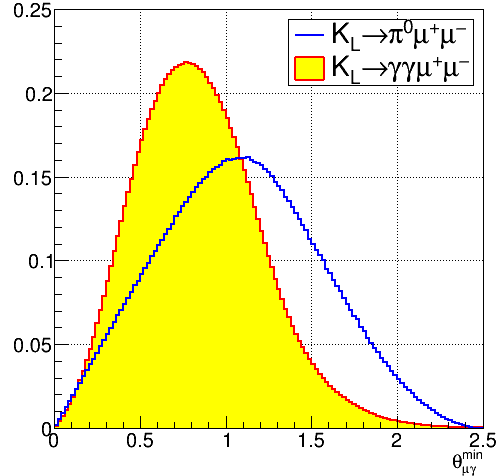}}%
\end{center}
\vspace{-6mm}
\caption{Distributions of the true values of the kinematic variables $|y_\gamma|$ (left) and $\theta_{\mu\gamma}^{\rm min}$ (right) for simulated samples of $K_L\to\pi^0\mu^+\mu^-$ and $K_L\to\gamma\gamma\mu^+\mu^-$ decays.}
\label{fig:kpi0mm-true}
\end{figure}

\begin{figure}[p]
\begin{center}
\resizebox{0.5\textwidth}{!}{\includegraphics{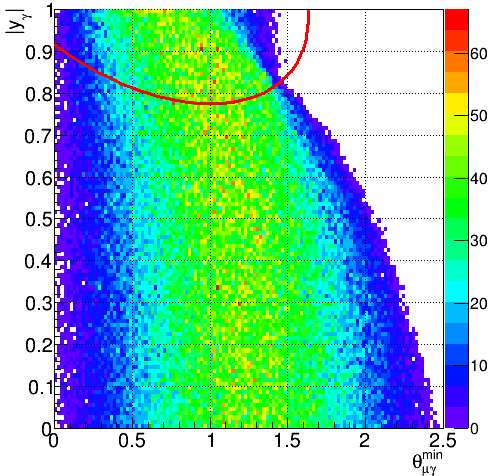}}%
\resizebox{0.5\textwidth}{!}{\includegraphics{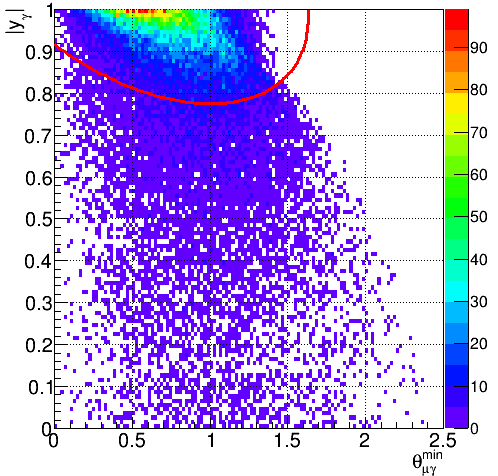}}%
\put(-306,187){\bf $K_L\to\pi^0\mu^+\mu^-$}%
\put(-95,187){\color{red}\bf $K_L\to\gamma\gamma\mu^+\mu^-$}
\end{center}
\vspace{-6mm}
\caption{Distributions of reconstructed kinematic variables $|y_\gamma|$ and $\theta_{\mu\gamma}^{\rm min}$ for simulated samples of $K_L\to\pi^0\mu^+\mu^-$ (left) and $K_L\to\gamma\gamma\mu^+\mu^-$ (right) decays. The selection condition applied in the ($\theta_{\mu\gamma}^{\rm min}$, $|y_\gamma|$) plane is shown with solid lines. The signal region is below the line.}
\label{fig:kpi0mm-selection}
\end{figure}

The distributions of the true values of the kinematic variables $|y_\gamma|$ and $\theta_{e\gamma}^{\rm min}$ for the $K_L\to\pi^0e^+e^-$ and $K_L\to\gamma\gamma e^+e^-$ decays are shown in Fig.~\ref{fig:kpi0ee-true}. The reconstructed $K_L\to\pi^0e^+e^-$ and $K_L\to\gamma\gamma e^+e^-$ distributions in the ($\theta_{e\gamma}^{\rm min}$, $|y_\gamma|$) plane and the selection condition applied in that plane are shown in Fig.~\ref{fig:kpi0ee-selection}. The corresponding distributions for the $K_L\to\pi^0\mu^+\mu^-$ case are shown in Figs.~\ref{fig:kpi0mm-true},~\ref{fig:kpi0mm-selection}.

The $K_L\to\pi^+\pi^-\pi^0$ decay, with a branching ratio of 12.5\%, followed by  $\pi^\pm\to\mu^\pm\nu$ decays in flight, represents another background source to the $K_L\to\pi^0\mu^+\mu^-$ decay. Pion decays upstream of the spectrometer typically lead to reconstructed $m_{\gamma\gamma\mu\mu}$ values significantly below the kaon mass. In case of misreconstruction of the $K_L$ decay vertex position due to the $\pi^\pm$ decays in flight, the $m_{\gamma\gamma\mu\mu}$ value may become compatible with the kaon mass; however in this case $m_{\gamma\gamma}$ becomes higher than the $\pi^0$ mass. In all cases with $\pi^\pm$ decays upstream of the spectrometer, the reconstructed ($m_{\gamma\gamma\mu\mu}$, $m_{\gamma\gamma}$) values are incompatible with the signal definition (Fig.~\ref{fig:kpi0mm-mass}, bottom). However $\pi^\pm\to\mu^\pm\nu$ decays in the vicinity of the spectrometer magnet lead to incorrect momentum reconstruction. This, often coupled with photon conversions in the STRAW chambers, results in reconstructed events scattered in the ($m_{\gamma\gamma\mu\mu}$, $m_{\gamma\gamma}$) plane, as seen in Fig.~\ref{fig:kpi0mm-mass} (bottom).

It has been established with dedicated simulations that the background from $K_L\to\pi^+\pi^-\pi^0$ decays followed by two $\pi^\pm$ decays in flight does not exceed 20\% of the number of reconstructed SM $K^+\to\pi^0\mu^+\mu^-$ events at 90\% CL. The estimate is limited by the size of the simulated background sample. The background is therefore subdominant; moreover it can be suppressed further by improving the spectrometer reconstruction algorithm in order to identify tracks with a kink in the vertical plane, by rejecting events with photon conversions, and by applying dedicated kinematic conditions similar to those exploited in the KTeV analysis~\cite{KTEV:2000ngj}. Background $K_L\to\pi^+\pi^-\pi^0$ decays with $\pi^\pm$ misidentification (due to punch-through or accidental activity) is expected to be smaller than that due to $\pi^\pm$ decays in flight according to the NA62 experience~\cite{NA62:2019eax}.

\begin{table}[t]
\caption{Expected integrated beam flux, numbers of SM $K_L\to\pi^0\ell^+\ell^-$ events ($N_S$) and Greenlee background events ($N_B$) to be collected in five years of HIKE Phase~2 operation. The signal significance and the precision on the signal branching ratio measurement are also shown.}
\begin{center}
\vspace{-6mm}
\begin{tabular}{lcccc}
\hline
Number of spills & \multicolumn{3}{c}{$3\times10^6$} \\
Protons on target & \multicolumn{3}{c}{$6\times10^{19}$} \\
$K_L$ decays in FV &
\multicolumn{3}{c}{$1.9\times10^{14}$} \\
\hline
Mode & $N_S$ & $N_B$ & $N_S/\sqrt{N_S+N_B}$ & $\delta{\cal B}/{\cal B}$ \\
\hline
$K_L\to\pi^0 e^+e^-$ & 70 & 83 & 5.7 & 18\% \\
$K_L\to\pi^0\mu^+\mu^-$ & 100 & 53 & 8.1 & 12\% \\
\hline
\end{tabular}
\end{center}
\label{tab:KLpi0ll}
\end{table}

The expected numbers of SM signal events and Greenlee background events (which are the dominant background) to be collected in five years of HIKE Phase-2 operation for each of the $K_L\to\pi^0\ell^+\ell^-$ decay modes, and the corresponding signal significance and precision on the signal branching ratio measurement, are summarised in the Table~\ref{tab:KLpi0ll}. The signal significance is higher for $K_L\to\pi^0\mu^+\mu^-$ decay than for the $K_L\to\pi^0e^+e^-$ decay, despite the weaker background suppression, thanks to the smaller branching ratio of the corresponding Greenlee decay (Table~\ref{tab:greenlee}). We conclude that the experiment will make the first observation, with a significance above $5\sigma$, and measurement of both ultra-rare decay modes.

The primary interest in the $K_L\to\pi^0\ell^+\ell^-$ measurements lies in the exploration of new physics contributions in the $s\to d\ell\ell$ short-distance interaction. On the other hand within the SM framework, considering the sensitivities to the CKM parameters given in Section~\ref{sec:physics:fcnc:klpi0ll}, and that the LHCb Phase-I upgrade is expected to measure the form-factor parameter $|a_S|$ to 5\% relative precision from the $K_S\to\pi^0\mu^+\mu^-$ decay~\cite{AlvesJunior:2018ldo}, making the corresponding uncertainties in ${\cal B}(K_L\to\pi^0\ell^+\ell^-)$ negligible, the HIKE measurements of ${\cal B}(K_L\to\pi^0\ell^+\ell^-)$ will lead to the following accuracy in the determination of the CKM parameter $\lambda_t=V_{ts}^*V_{td}$:
\begin{displaymath}
\left.\frac{\delta({\rm Im}~\lambda_t)}{{\rm Im}~\lambda_t}\right|_{K_L\to\pi^0e^+e^-} = 0.33, \quad
\left.\frac{\delta({\rm Im}~\lambda_t)}{{\rm Im}~\lambda_t}\right|_{K_L\to\pi^0\mu^+\mu^-} = 0.28.
\end{displaymath}
In combination, the $K_L\to\pi^0e^+e^-$  and $K_L\to\pi^0\mu^+\mu^-$  measurements at HIKE will determine the parameter ${\rm Im}~\lambda_t$ to a  20\% relative precision, which is equivalent to a ${\cal B}(K_L\to \pi^0\nu\bar\nu)$ measurement to a 40\% precision (Section~\ref{sec:PhysicsKpinunu}).

Beyond the SM paradigm, the unique $K_L\to\pi^0\ell^+\ell^-$ measurements will be exploited to discover or constrain new physics at the ${\cal O}(100~{\rm TeV})$ scale inducing corrections to the SM rate via loop diagrams in a correlated way with other observables in the kaon sector~\cite{Aebischer:2022vky}, in the framework of a global LFU test (Section~\ref{sec:GlobalNPSensitivity}), and to search for production of feebly-interacting particles at the ${\cal O}(100~{\rm MeV})$ scale~\cite{Beacham:2019nyx,Goudzovski:2022vbt}.

%%%%%%%%%%%%%%%%%

%%%%%%%%%%%%%%%%%%

\clearpage
\newpage

\subsection{Other decay modes}
\label{sec:phase2:other}

\subsubsection{Rare and forbidden $K_L$ decays}

Considering that HIKE Phase~2 is being developed for the challenging measurements of the ultra-rare $K_L\to\pi^0\ell^+\ell^-$ decays, the experiment will be well suited to carry out a wide range of studies, from precision measurements of rare $K_L$ decays to searches for $K_L$ decays forbidden in the SM at a record level of sensitivity. The development of a broad rare and forbidden decay programme has started; a few specific examples are provided below.

For the rare FCNC decay $K_L\to\mu^+\mu^-$, with an indicative SM branching ratio of $7\times 10^{-9}$ (Section~\ref{sec:KLmumu}), the estimated acceptance of 17\% leads to a sample of $2\times 10^5$ decays, allowing statistical precision of 0.2\% on the  branching ratio to be reached in five years of operation. We expect that an overall precision better than 1\% is attainable when normalising to the $K_L\to\pi^+\pi^-$ decay, improving significantly on the previous measurement by the BNL-E871 experiment~\cite{E871:2000wvm}.

HIKE Phase~2 may have the potential to make a first measurement of $K_L\to e^+e^-$. The acceptance for this decay is estimated to be 5\%, leading to a single-event sensitivity of ${\cal B}_{\rm SES} \approx 10^{-13}$. Compared to the Standard Model expectation of ${\cal B}_{\rm SM} \approx 10^{-11}$, the expected signal yield is about 100~events. However, the backgrounds, including those from the $K_L\to e^+e^-e^+e^-$ and $K_L\to e^+e^-\gamma$ decays, have yet to be evaluated.

The LFV decay $K_L\to\mu^\pm e^\mp$ was studied by the dedicated BNL-E871 experiment, which has established a stringent upper limit of $4.7\times 10^{-12}$ at 90\% CL on its branching ratio~\cite{BNL:1998apv}. The sensitivity to this channel is limited by the background from the $K_L\to\pi^\pm e^\mp\nu$ decay (with a branching ratio of 41\%) followed by $\pi^\pm\to\mu^\pm\nu$ decay in the spectrometer and subsequent misreconstruction of the pion/muon track. It has been established with dedicated simulations that the acceptance of the signal selection for the $K_L\to\pi^\pm e^\mp\nu$ background is suppressed in the HIKE Phase~2 conditions to the level below $10^{-10}$ (this estimate is limited by the size of the simulated sample), while the signal acceptance is kept at the 3\% level, by a requirement on the asymmetry of the reconstructed electron and muon track momenta: $(p_e-p_\mu)/(p_e+p_\mu)>0.2$. This requirement exploits the fact that the misrecontructed muon momentum in a background event must be necessarily overestimated in order for the two-track mass, $m_{\mu e}$, to be consistent with the kaon mass. For a competitive search at HIKE, suppression of this background to the level of at least $10^{-11}$ must be demonstrated. Possible avenues for improvement in the background rejection include the development of dedicated algorithms for identification of tracks with a kink, and increasing the magnetic field in the spectrometer.

The $K_L\to\gamma\gamma$ branching ratio is currently known to within about 1\%, from the average of measurements from KLOE and NA48. The NA48 measurement is systematics-dominated. While dedicated studies are required in order to evaluate the HIKE potential, from considerations similar to those for the HIKE measurements of the principal $K_L$ branching ratios (Section~\ref{sec:ckm_phase2}), we expect to substantially improve on the precision for $\mathcal{B}(K_L\to\gamma\gamma)$, perhaps reaching the 0.5\% level.  

Searches for $K_L\to\pi^0\pi^0\nu\bar{\nu}$ and $K_L\to\pi^0X$ are very similar in concept to those for $K_L\to\pi^0\nu\bar{\nu}$, and while HIKE Phase~2 may be able to do some pilot studies involving these channels for a possible dedicated third phase, given the extra material in the decay volume from the tracking detectors and the limited coverage ($\theta < 50$~mrad) of the large-angle photon vetoes, it is unlikely that HIKE Phase~2 results from these channels would be competitive with results from KOTO Step-2, if in fact it is built and achieves its intended sensitivity.

These and other ongoing sensitivity studies for rare $K_L$ decays will be developed into a fully fledged programme for the Technical Design Report.

%%%%%%%%%%%%%%%%%%%%%

\subsubsection{Main $K_L$ branching ratios and CKM unitarity}
\label{sec:ckm_phase2}

The potential improvements from HIKE Phase~2 to the knowledge of the semileptonic branching ratios for $K_L$ decays are more challenging to evaluate than those for Phase~1.
The $K_L\to3\pi^0$ and $K_L\to\pi^+\pi^-\pi^0$ channels account for 32\% of the $K_L$ decay width, but are topologically quite different from $K_{e3}$ and $K_{\mu3}$, making high-precision measurements of ratios of semileptonic to $3\pi$ widths difficult. The $K_L\to\pi^+\pi^-$ decay is topologically suitable for normalising the $K_{\ell3}$ decays. Its branching ratio is only $1.967(7)\times10^{-3}$, but, even taking data at very low intensity for maximum systematic control, HIKE could accumulate several million $K_L\to\pi^+\pi^-$ decays within about two weeks, so that systematic uncertainties would dominate.

The HIKE measurements would still depend on the current values of ${\cal B}(K_L\to3\pi^0)$
and ${\cal B}(K_L\to\pi^+\pi^-\pi^0)$; it may be possible for HIKE to measure ratios involving one or both channels, but it is difficult to extrapolate with what precision. On the other hand, the world data on $K_L$ decays feature greater redundancy and higher consistency than that for $K^+$ decays.
One possible set of HIKE Phase~2 measurements that could be added to the current $K_L$ world dataset to improve the precision of the $K_{\ell3}$ branching ratios consists of high-precision measurements of ${\cal B}(K_{e3})/{\cal B}(K_{\mu3})$ and ${\cal B}(K_L\to\pi^+\pi^-)/{\cal B}(K_{e3})$, as well as a good measurement of ${\cal B}(K_L\to\pi^+\pi^-)/{\cal B}(K_L\to\pi^+\pi^-\pi^0)$ with less stringent precision requirements, to assist in normalisation via the global fit. The Phase~2 sensitivity estimate assumes total uncertainties of 0.3\%, 0.4\%, and 0.6\%, respectively for these measurements. These are consistent with or slightly more conservative than the assumptions for HIKE Phase~1. In particular, NA48 made a statistically dominated measurement of ${\cal B}(K_L\to\pi^+\pi^-)/{\cal B}(K_{e3})$ with a systematic uncertainty of 0.3\%~\cite{NA48:2006jeq}. When added to the existing fit to the world dataset, HIKE Phase~2 measurements of these ratios at values near those expected from the current fit would reduce the uncertainties on ${\cal B}(K_{e3})$ and ${\cal B}(K_{\mu3})$ by 25\% and 50\%, respectively, depending on the exact values assumed.

%%%%%%%%%%%%%%%%%%%%%

\subsection{Summary of physics sensitivity}

HIKE will make the first observation, with significance above $5\sigma$, and measurement, to a precision of about 15\%, of the ultra-rare golden-mode decays $K_L\to\pi^0 e^+e^-$ and $K_L\to\pi^0\mu^+\mu^-$. These results will lead to an independent determination of the CKM parameter ${\rm Im}~\lambda_t$ to 20\% precision.
 
Furthermore, HIKE Phase~2 will allow for a wide reach of precision measurements and searches for rare and forbidden $K_L$ decays. The broader physics programme is being developed, and includes (but is not limited to) a measurement of the $K_L\to\mu^+\mu^-$ rate to percent-level precision, searches for lepton flavour and lepton number violating decays (of which $K_L\to\mu^\pm e^\mp$ is the most prominent
channel), and first-row CKM unitarity tests addressing the origin of the Cabibbo-angle anomaly.

\section{Impact of the HIKE kaon physics programme}
\label{sec:k-impact}

Measurements of quantities well predicted by the SM, like  ${\cal B}(K\to\pi\nu\bar\nu)$, offer model-independent standard candles that can constrain any BSM scenario, present or future.
The status of BSM models in the future is hard to predict, but measurements made by the unrivaled HIKE experimental programme will be durable standards against which many of those models will be judged.

The HIKE project will bring the rare $K^+$ and $K_L$ decay programme to an unprecedented level of precision. 
HIKE will collect $2.0\times 10^{13}$ kaon decays in the decay volume per year ($8\times 10^{18}$~POT/year) in Phase~1 with a $K^+$ beam, and 
$3.8\times 10^{13}$ kaon decays in the decay volume per year ($1.2\times 10^{19}$~POT/year) in Phase~2 with a $K_L$ beam. 
Table~\ref{tab:hike-flavour} lists a selection of
the many unique measurements that HIKE can perform.  

In the following, we present examples of how HIKE can probe BSM physics, and 
how the physics potential of HIKE compares with other ongoing or planned experimental efforts. From the global fits performed using individual measurement projections as input, it is evident that HIKE's measurements of rare kaon decays are a very powerful tool to 
expose clear deviations from SM expectations or constrain BSM scenarios, greatly reducing the unexplored parameter space. We note that, beyond the measurements of the branching ratios used as inputs in these global fits, HIKE will also be able to establish the nature (scalar or vector) of BSM contributions to the $K^+\to\pi^+\nu\nu$ decay and perform an independent determination of the CKM parameter ${\rm Im}~\lambda_t$ to 20\% precision.

In each of the benchmark scenarios considered below, it is clear that HIKE will provide a significant step forward. HIKE sensitivity is better than or competitive with, and complementary to, projected bounds on these models from other experiments. 
In the case of a top-philic $Z^\prime$, HIKE sensitivity is better than that projected for ATLAS, and provides the strongest expected constraints at higher $Z^\prime$ masses.
For the scalar leptoquark model, HIKE is able to provide the best constraints across the entire parameter space considered, with only Belle~II (at full target luminosity) projected to give comparable, if weaker, constraints. For the vector leptoquark model, HIKE is able to cover almost all of the unexplored parameter space. 

Finally, both HIKE Phases~1 and 2 will have a considerable impact in elucidating the nature of the first-row CKM unitarity deficit.

%%%%%%%%%%%%%%

\begin{table}[htb]
{\small
\centering
\caption{Summary of HIKE sensitivity for flavour observables. The $K^+$ decay measurements will be made in Phase~1, and the $K_L$ decay measurements in Phase~2. The symbol $\cal B$ denotes the decay branching ratios.}
\vspace{-1mm}
\begin{tabular}{lll}
\hline
$K^+\to\pi^+\nu\bar\nu$ & $\sigma_{\cal B}/{\cal B}\sim5\%$ & BSM physics, LFUV \\
$K^+\to\pi^+\ell^+\ell^-$ & Sub-\% precision on form-factors & LFUV \\
$K^+\to\pi^-\ell^+\ell^+$, $K^+\to\pi\mu e$ & Sensitivity ${\cal O}(10^{-13})$ & LFV / LNV \\
Semileptonic $K^+$ decays & $\sigma_{\cal B}/{\cal B}\sim0.1\%$ & $V_{us}$, CKM unitarity \\
$R_K={\cal B}(K^+\to e^+\nu)/{\cal B}(K^+\to\mu^+\nu)$ & $\sigma(R_K)/R_K\sim {\cal O}(0.1\%)$ & LFUV \\
Ancillary $K^+$ decays & \% -- \textperthousand & Chiral parameters (LECs) \\
(e.g. $K^+\to\pi^+\gamma\gamma$, $K^+\to\pi^+\pi^0 e^+e^-$) \\
\hline
$K_L\to\pi^0\ell^+\ell^-$ & $\sigma_{\cal B}/{\cal B} < 20\%$ & ${\rm Im}\lambda_t$ to 20\% precision, \\
&& BSM physics, LFUV \\
$K_L\to\mu^+\mu^-$ & $\sigma_{\cal B}/{\cal B}\sim 1\%$ & Ancillary for $K\to\mu\mu$ physics \\
$K_L\to\pi^0(\pi^0)\mu^\pm e^\mp$ & Sensitivity ${\cal O}(10^{-12})$ & LFV \\
Semileptonic $K_L$ decays & $\sigma_{\cal B}/{\cal B}\sim 0.1\%$ & $V_{us}$, CKM unitarity \\
Ancillary $K_L$ decays & \% -- \textperthousand & Chiral parameters (LECs), \\
(e.g. $K_L\to\gamma\gamma$, $K_L\to\pi^0\gamma\gamma$) & & SM $K_L\to\mu\mu$, $K_L\to\pi^0\ell^+\ell^-$ rates\\
\hline
\end{tabular}
\label{tab:hike-flavour}
}
\end{table}

%%%%%%%%%%%%%%%

\subsection{Global sensitivity to new physics}
\label{sec:GlobalNPSensitivity}

The rare kaon decays investigated by HIKE offer the possibility to search for BSM physics with a global fit technique, for example, in the context of lepton flavour universality (LFU) tests.

In the SM, the three lepton flavours ($e$, $\mu$ and $\tau$) have exactly the same gauge interactions and are distinguished only through their couplings to the Higgs field and hence the charged lepton masses. BSM models, on the other hand, do not necessarily conform to the lepton flavour universality hypothesis and may thereby induce subtle differences between the different generations that cannot be attributed to the different masses. Among the most sensitive probes of these differences are rare kaon decays with electrons, muons or neutrinos in the final state.

Using a low-energy effective theory approach, the physics is described in terms of an effective Hamiltonian, which is an expansion series of fermion current operators.
The coefficients of this expansion are the Wilson coefficients $C_k$, where the index $k$ runs over the expansion terms.
The assumption is that BSM physics modifies the $C_k$ with respect to their SM values.
The Wilson coefficients are then fitted using experimental values of observables from rare kaon decays as input variables.

The effective Hamiltonian describing the FCNC decay $s\to d$ is given by~\cite{DAmbrosio:2022kvb} 
\begin{equation}\label{eq:Heff}
\mathcal{H}_{\rm eff}=-\frac{4G_F}{\sqrt{2}}V_{td}V_{ts}^*\frac{\alpha_e}{4\pi}\sum_k C_k^{\ell}O_k^{\ell}\,,
\end{equation}
where $G_F$ denotes the Fermi constant, $\alpha_e$ the fine-structure constant, and the Wilson coefficients $C_k^\ell$ multiply the effective operators $O_k^\ell$. 
For the study of BSM contributions $\delta C_k^\ell$, it is possible to reduce the set of operators by considering only scenarios where the neutral and charged leptons are related by SU(2)$_{\rm L}$ gauge symmetry, such that $\delta C_{L}^{\ell} \equiv \delta C_9^{\ell} = - \delta C_{10}^{\ell}$.

\begin{figure}[p]
\begin{center}
\includegraphics[width=0.495\textwidth]{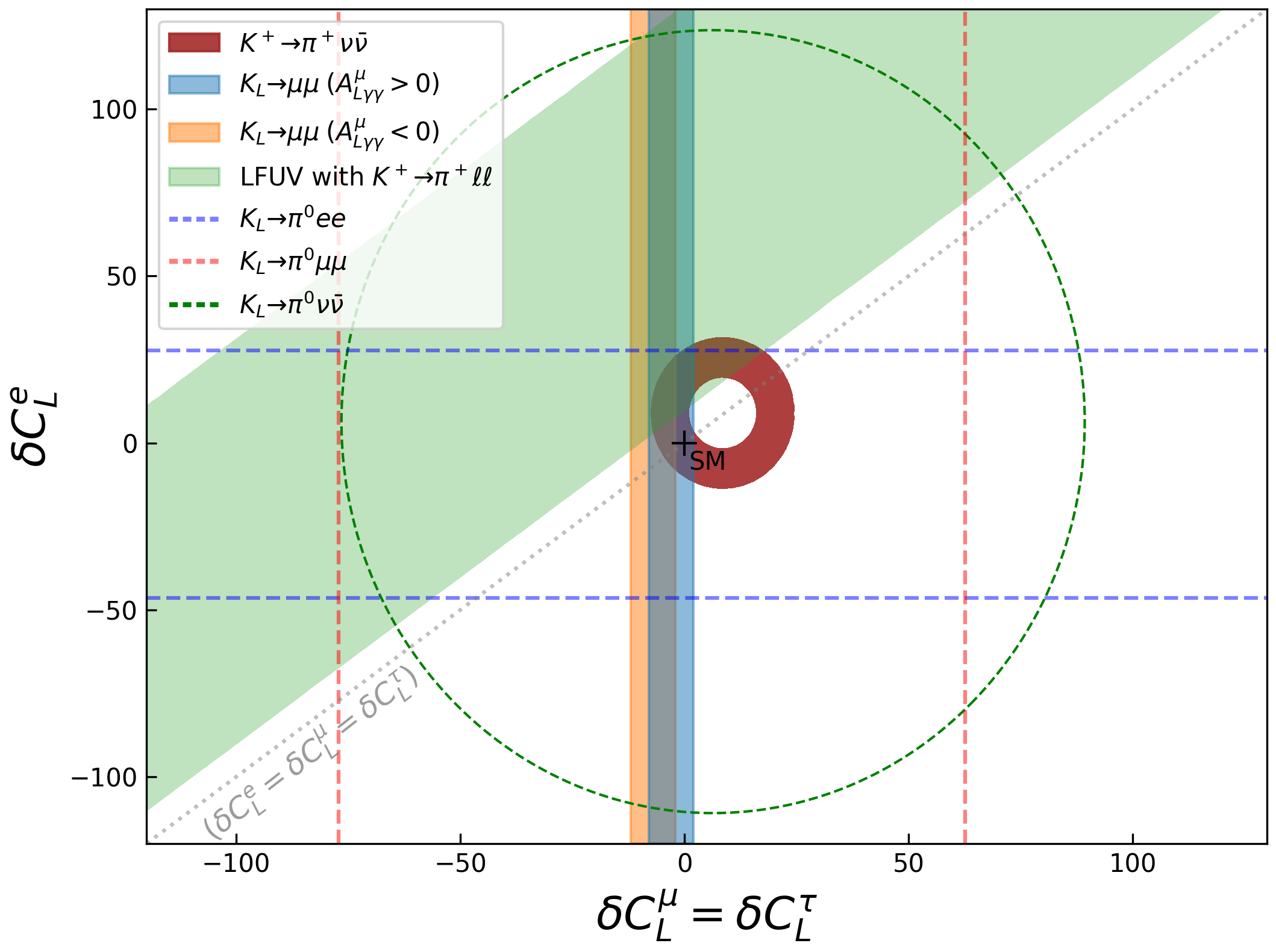}%
\includegraphics[width=0.495\textwidth]{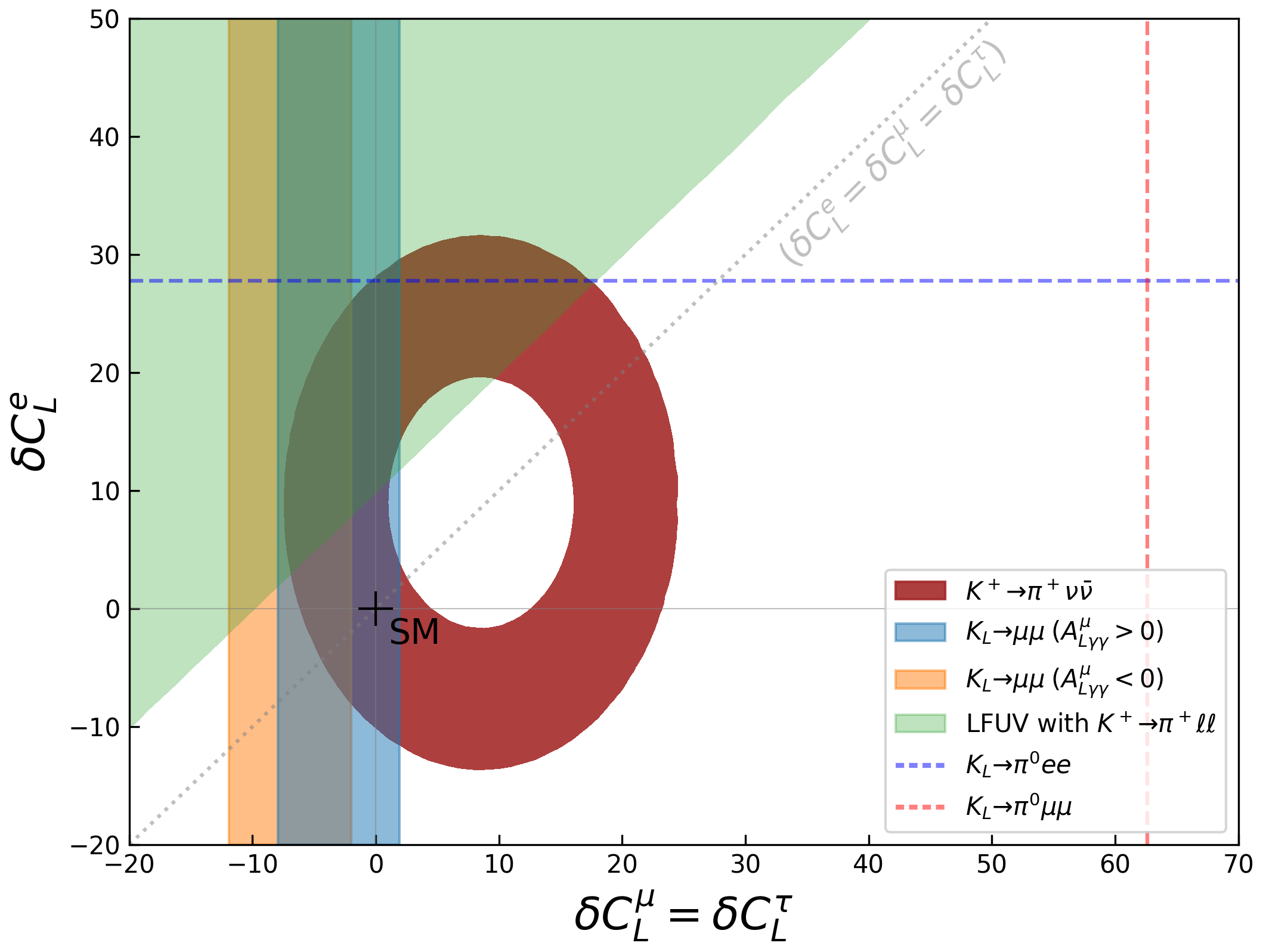}
\vspace{-3mm}
\caption{Bounds on LFU violating new physics contributions to Wilson coefficients from individual observables in the kaon sector. The right panel is the zoomed version of the left panel. See Fig.~7 in Ref.~\cite{DAmbrosio:2022kvb} for further information.
\label{fig:all_obs_individually}}
\end{center}
\end{figure}

For BSM scenarios with LFU violating effects, focusing
on the case where the NP effects for electrons are different from the those for muons and taus, the individual constraints on $\delta C_L^e$ and $\delta C_L^\mu = \delta C_L^\tau$ are shown in Fig.~\ref{fig:all_obs_individually} (taken from Ref.~\cite{DAmbrosio:2022kvb}, which also contains the list of constraints considered).
This shows that the main constraining observables are ${\cal B}(K^+\to\pi^+ \nu\bar\nu)$ and ${\cal B}(K_L\to\mu^+\mu^-)$, where for the latter, the unknown sign of the LD interference term plays an important role. 

%%%%%%%%%%%%%%%%%%%%%%%%%%%%%%
% A PAGE OF 3 FLOATING FIGURES

\begin{figure}[p]
\begin{center}
\vspace{-4mm}
\includegraphics[width=0.495\textwidth]{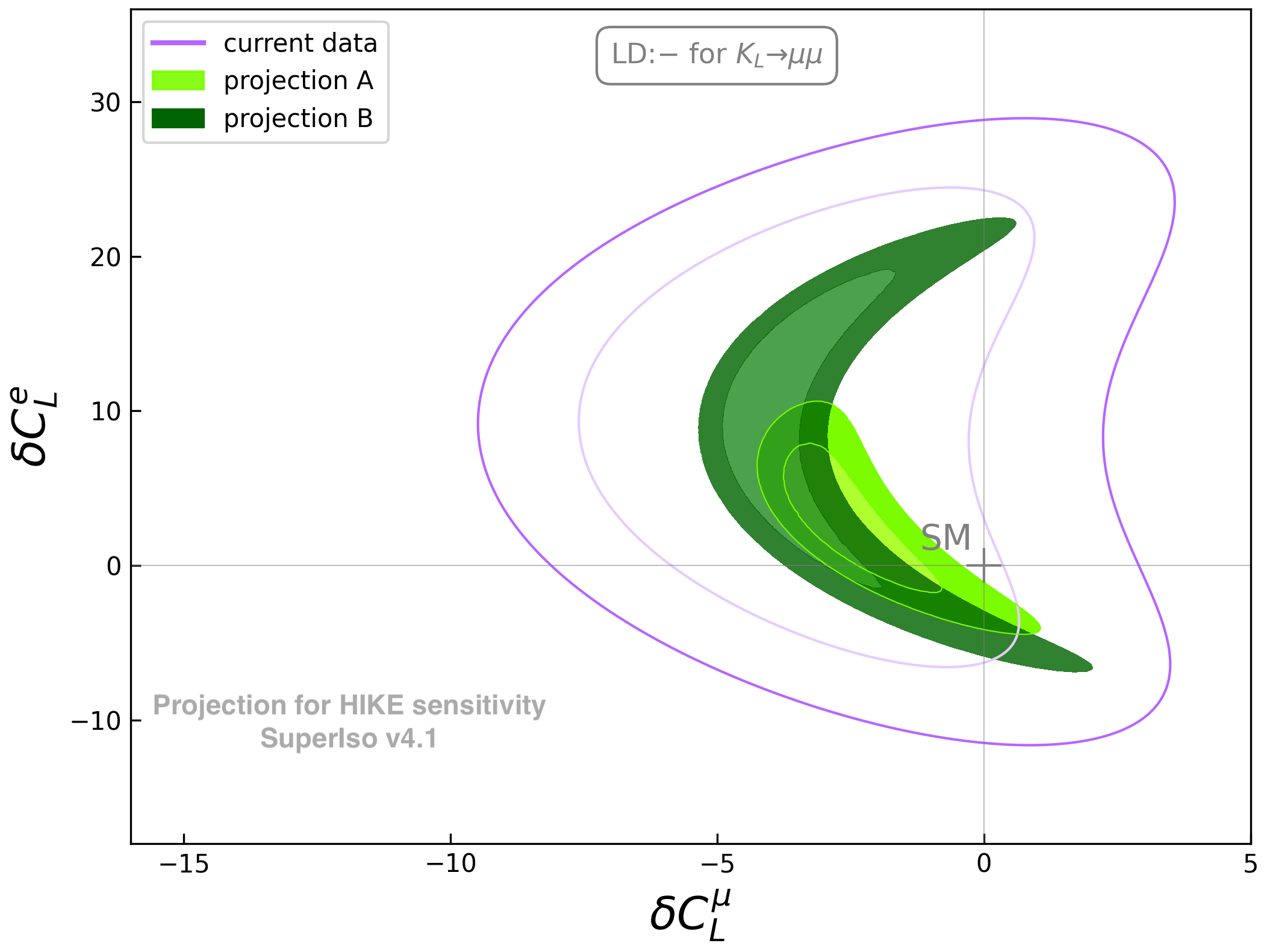}%
\includegraphics[width=0.495\textwidth]{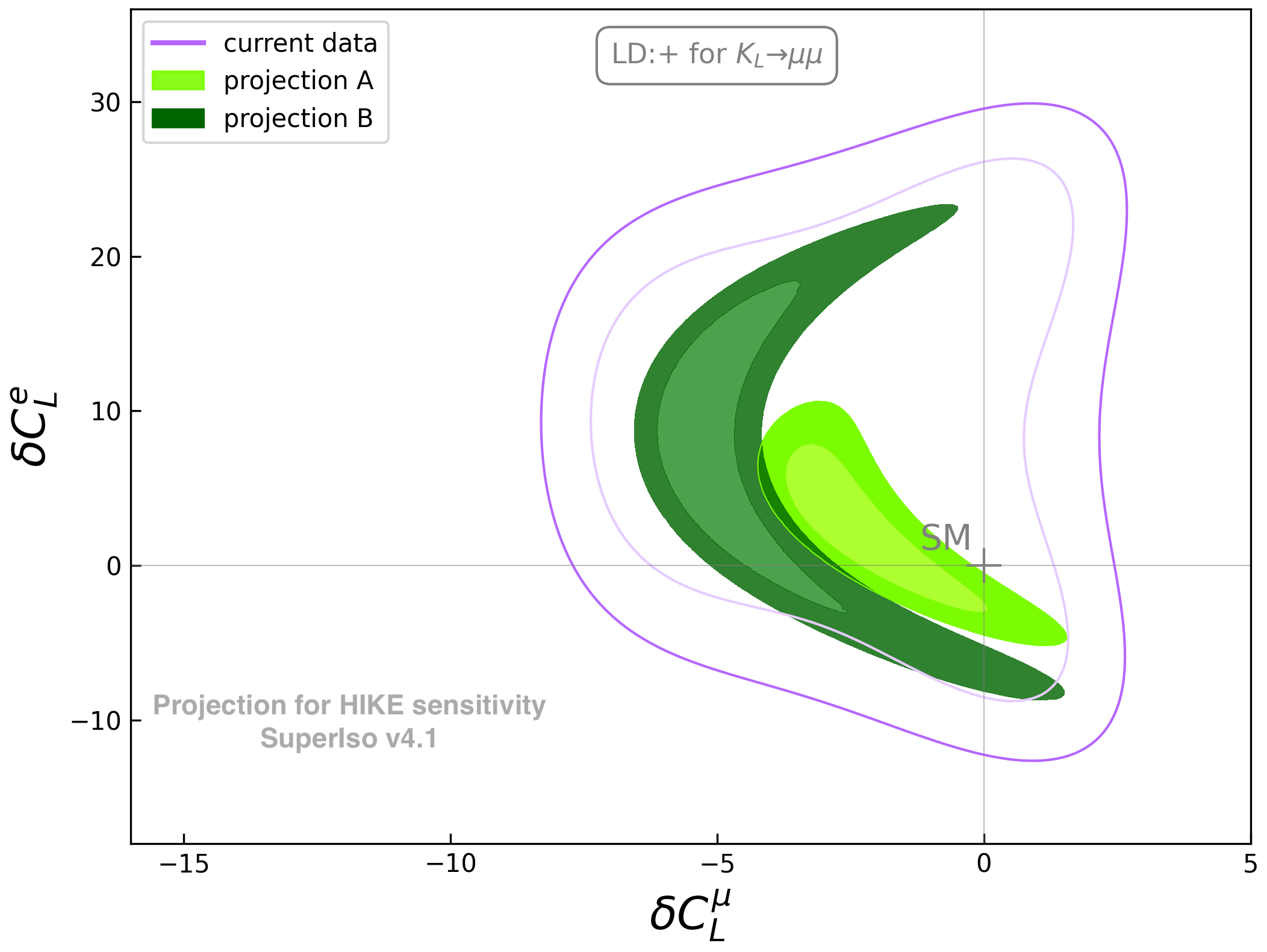}%
\vspace{-4mm}
\caption{Global fits in the $\{\delta C_L^{e}, \delta C_L^\mu(=\delta C_L^{\tau})\}$ plane with current data (purple contours) and the projected scenarios.
as described in the text (green regions). 
For further details and the list of inputs considered, see Ref.~\cite{DAmbrosio:2022kvb}. 
The curves reflect the status expected at the end of HIKE Phase~1.
\label{fig:fitcomparison}}
\end{center}
\end{figure}

\begin{figure}[p]
\begin{center}
\vspace{-7mm}
\includegraphics[width=0.495\textwidth]
{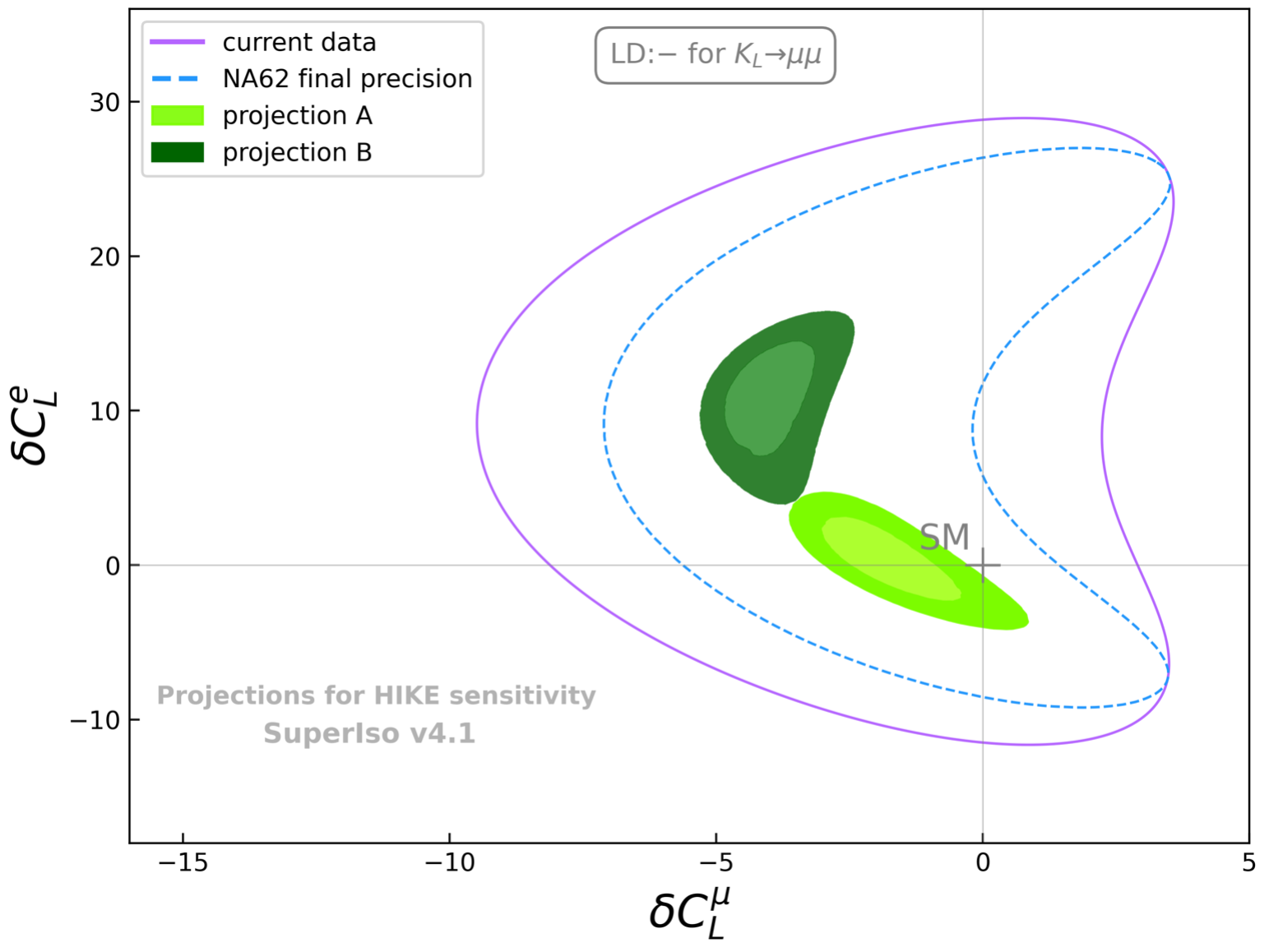}%
\includegraphics[width=0.495\textwidth]
{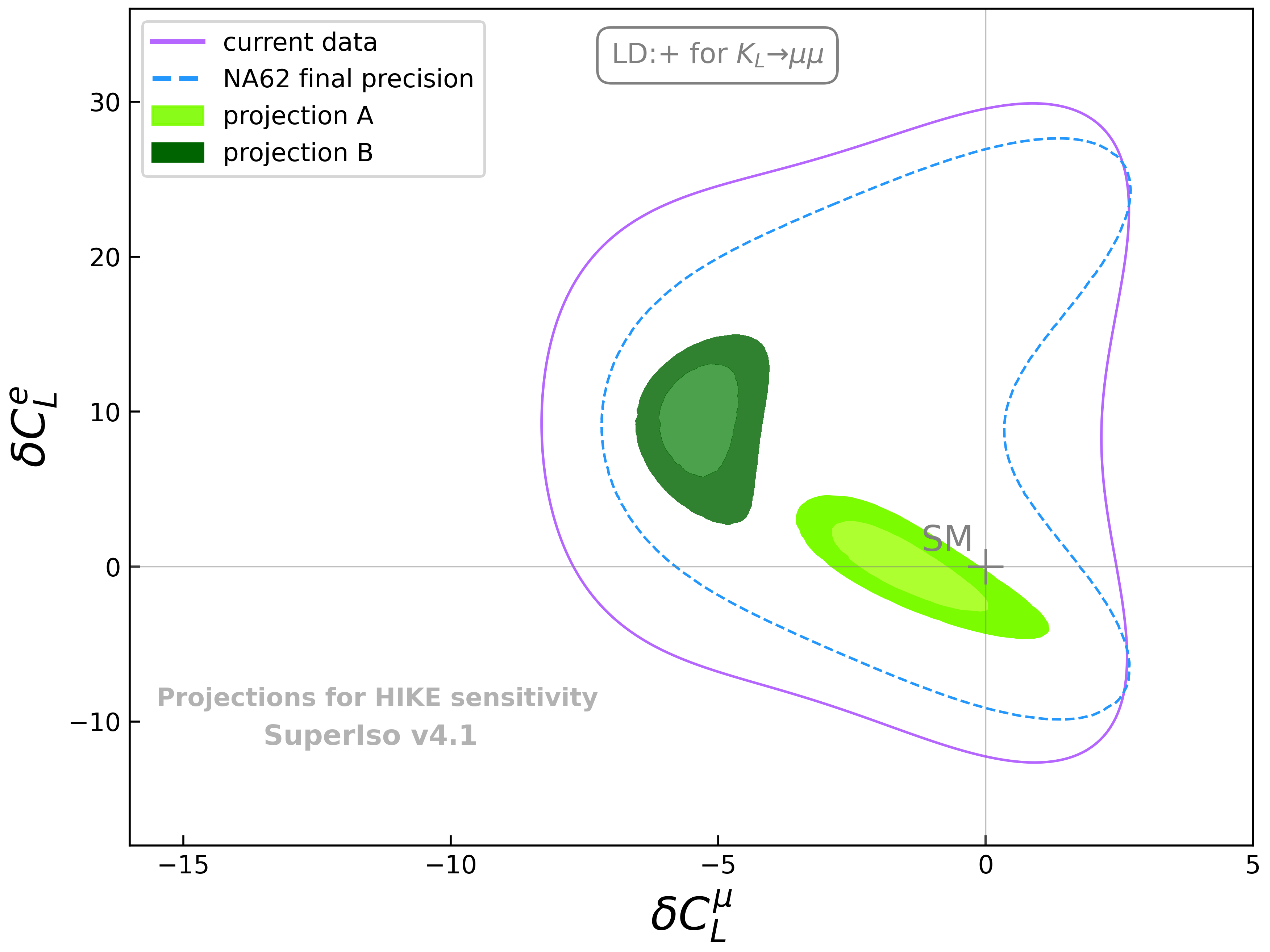}
\vspace{-4mm}
\caption{Global fits in the $\{ \delta C_L^{e}, \delta C_L^\mu(=\delta C_L^{\tau})\}$ plane with current data (purple contours) and the full projected scenarios (green regions) at the end of HIKE Phases~1 and~2. The list of inputs is summarised in Table~\ref{tab:GiancarloData}. The blue dotted curve represents the NA62 projection at the end of 2025. For further details of the theory approach, see Ref.~\cite{DAmbrosio:2022kvb}.
\label{fig:fitcomparison_new}}
\end{center}
\vspace{-7mm}
\end{figure}

% END OF PAGE OF FIGURES
%%%%%%%%%%%%%%%%%%%%%%%%

A combined fit of all the decay modes is then performed~\cite{DAmbrosio:2022kvb}.
Projections based on the fits require assumptions for both the possible future measured (central) values as well as the experimental precision. For the latter, the expected long-term experimental precisions are considered, while for the central values two scenarios are assumed:
\begin{itemize}
\item {\bf Projection A}: predicted central values for observables with only an upper bound available are taken to be the same as the SM prediction while for measured observables the current central values are used.
\item {\bf Projection B}: the central values for all of the observables are projected with the best-fit points obtained from the fits with the existing data.
\end{itemize}
The results of a combined fit of all the decay modes shown in Fig.~\ref{fig:all_obs_individually} in the $\{\delta C^e, \delta C^\mu(=\delta C^\tau) \}$ plane is displayed in Fig.~\ref{fig:fitcomparison}.
The projection inputs include a 20\% precision measurement of ${\cal B}(K_L\to\pi^0\nu\bar\nu)$ by KOTO-II, 1\% precision on ${\cal B}(K_L\to\mu^+\mu^-)$ and 100\% precision on ${\cal B}(K_L\to\pi^0\ell^+\ell^-)$~\cite{DAmbrosio:2022kvb}. No improvement in the theoretical precision is assumed for either projections. 
The light and dark solid purple contour outlines indicate the 68\% and 95\% confidence level (CL) regions, respectively, using current data, and the two panels correspond to the two possible signs of the LD contributions to $K_L\to\mu^+\mu^-$. 

The projected fits for the two scenarios are also displayed in Fig.~\ref{fig:fitcomparison}, where the 68\% and 95\% CL regions are shown with two shades of light green for projection A and the two shades of dark green for projection B. 
The two scenarios give different results: while projection A indicates overall consistency with the SM at the level of $3\sigma$ almost by construction, projection B clearly shows departures at $3\sigma$, especially for positive LD interference. Theory work is ongoing to determine the sign of the LD interference amplitude (Section~\ref{sec:k-decays}). Considering the specific projections taken as input, and the fact that here the $K^+$ future measurements have a dominant effect,
the curves reflect the status at the end of HIKE Phase~1. 

A fit for a modified projection made within the Physics Beyond Colliders initiative~\cite{Neshatpour:2022fak} following the strategy of Ref.~\cite{DAmbrosio:2022kvb} for the measurements to be performed in HIKE Phases~1 and~2 is shown in Fig.~\ref{fig:fitcomparison_new}. The inputs to the fit are summarised in Table~\ref{tab:GiancarloData}: notably, the precision on ${\cal B}(K_L\to\pi^0\ell^+\ell^-)$ is assumed to be 20\%.  The figure includes the projection for the NA62 final result on ${\cal B}(K^+ \to \pi^+ \nu\nu )$.
The significant impact of the much-improved precision of ${\cal B}(K_L\to\pi^0\ell^+\ell^-)$ is in line with the expectation of Ref.~\cite{DAmbrosio:2022kvb}. 
It is evident from Figs.~\ref{fig:fitcomparison} and~\ref{fig:fitcomparison_new} that the measurements foreseen at HIKE have the potential to show a clear deviation from the SM or to strongly constrain the parameter space available to BSM physics.

\begin{figure}[p]
\begin{center}
\vspace{-5mm}
\includegraphics[width=0.5\textwidth]
{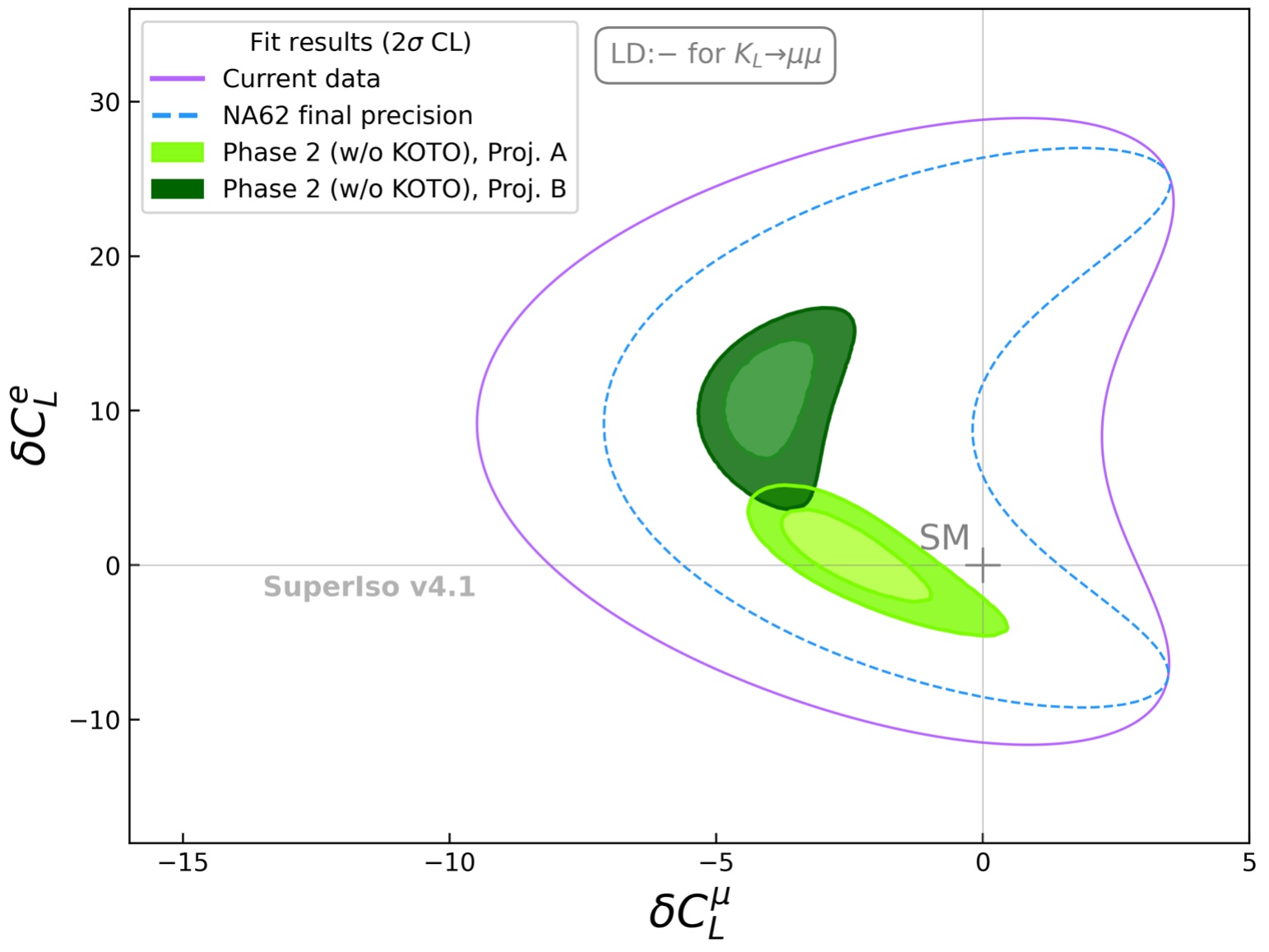}%
\includegraphics[width=0.5\textwidth]
{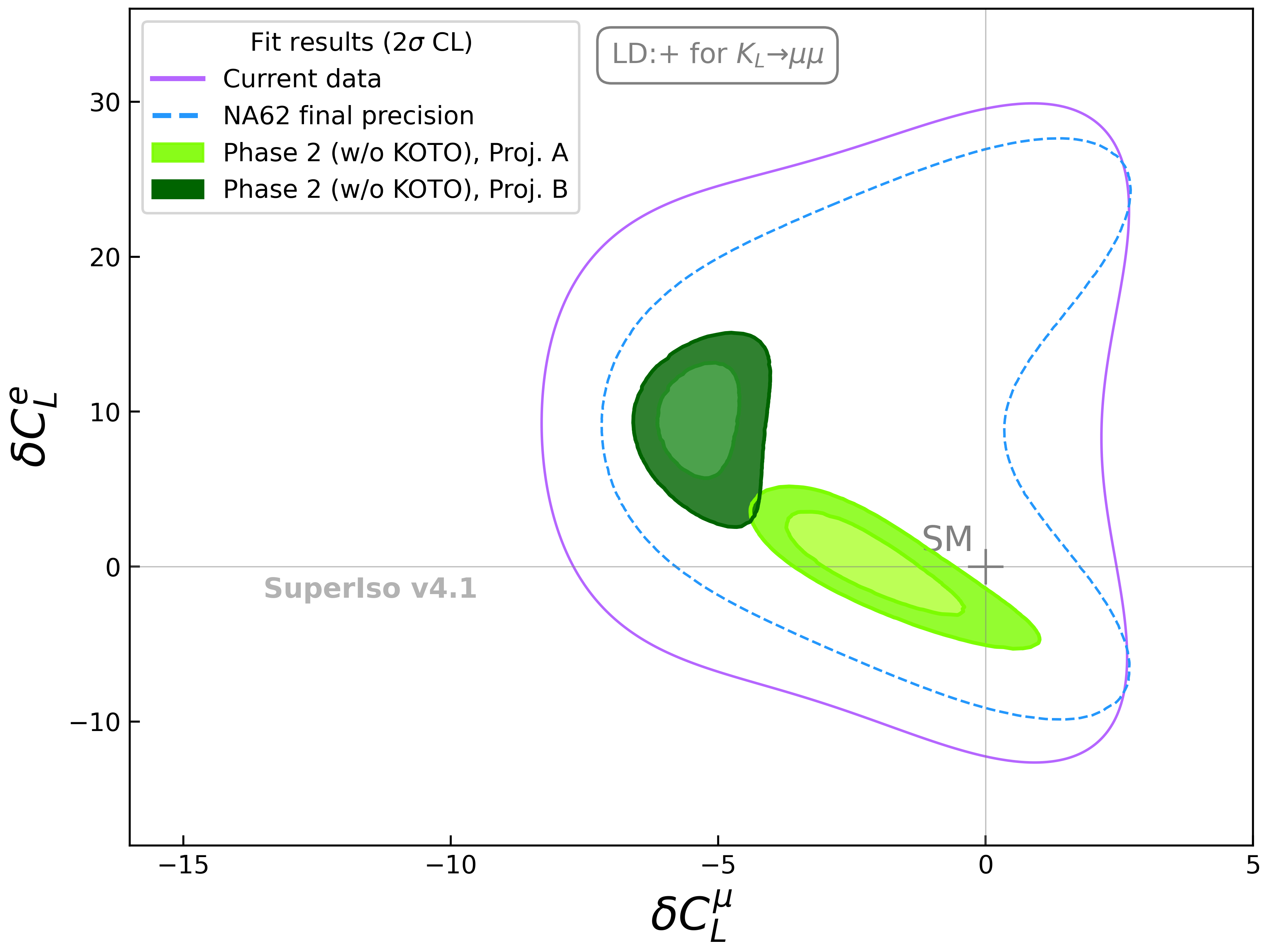}
\vspace{-7mm}
\caption{Global fits as in Fig.~\ref{fig:fitcomparison_new} but removing the measurement from KOTO-II.
\label{fig:fitcomparison_noKOTO}}
\end{center}
\end{figure}

\begin{figure}[p]
\begin{center}
\vspace{-5mm}
\includegraphics[width=0.5\textwidth]
{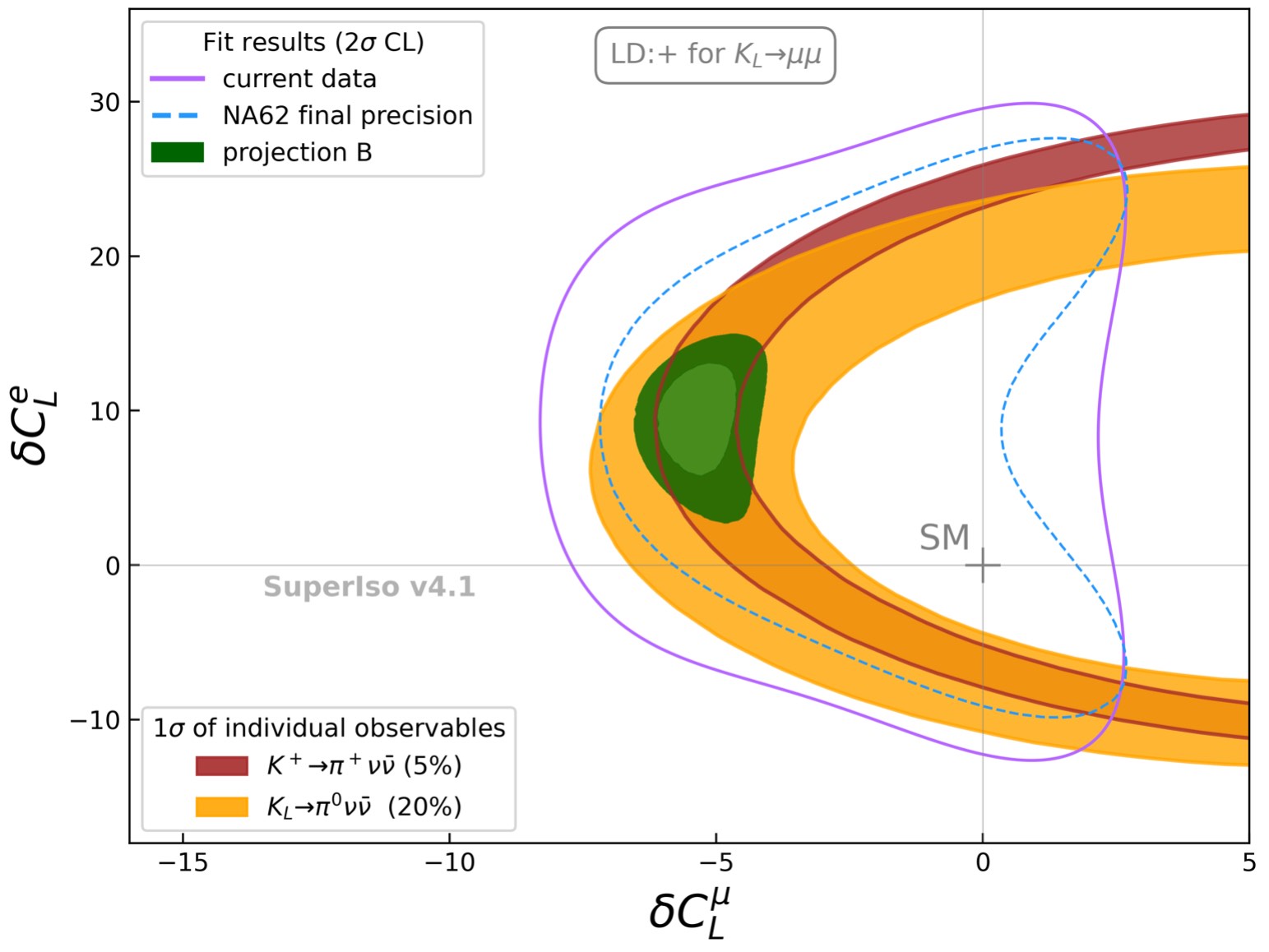}%
\includegraphics[width=0.5\textwidth]
{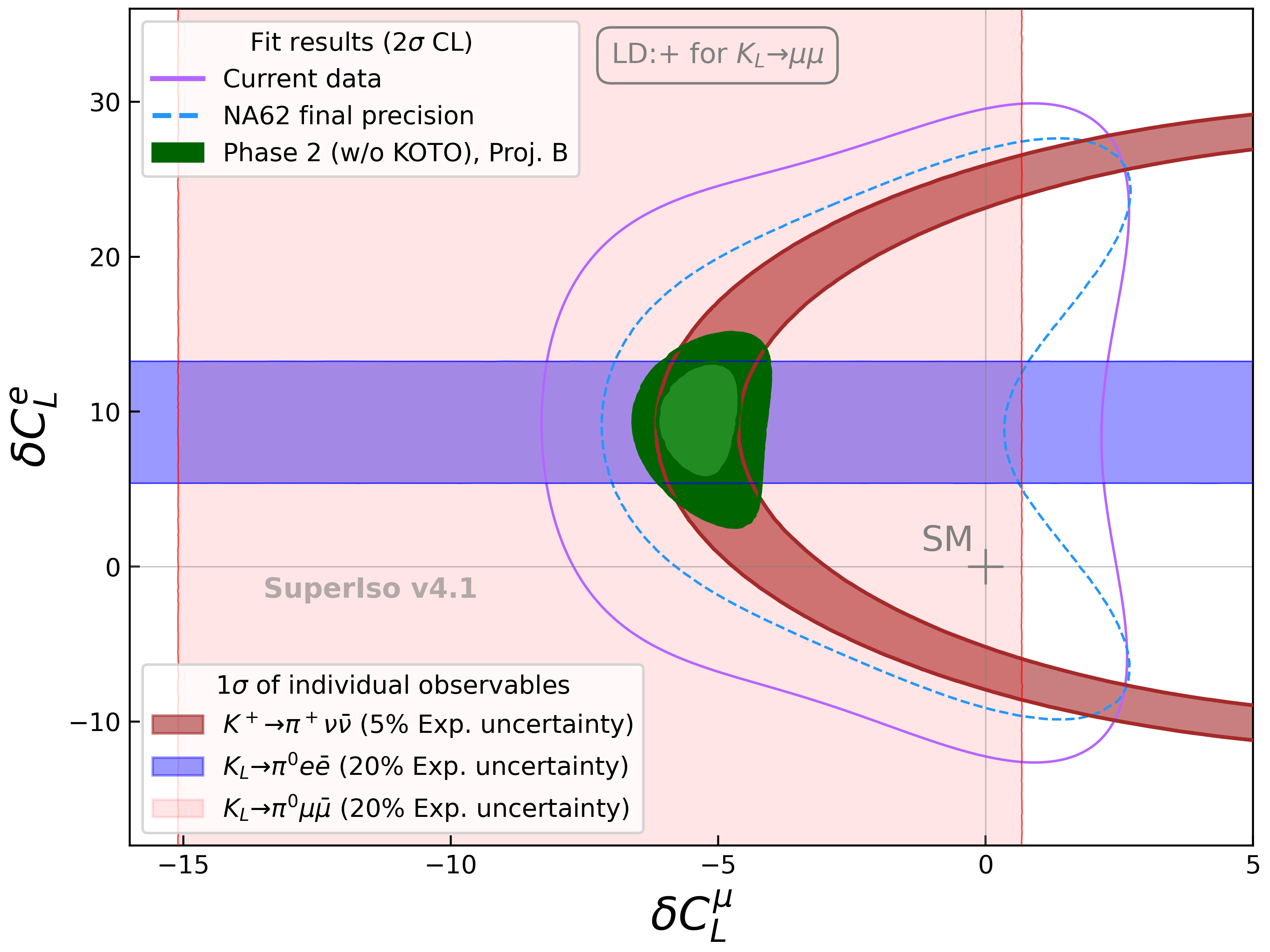}
\vspace{-7mm}
\caption{Global fits as in Fig.~\ref{fig:fitcomparison_new} (right), highlighting the contributions of single constraints in projection~B (with KOTO-II on left, without KOTO-II on the right).
\label{fig:fitcomparison_constributions}}
\end{center}
\end{figure}

Further global fits have been performed to highlight the effect of the potential absence of KOTO-II measurement~\cite{dambrosio2023}, as shown in Fig.~\ref{fig:fitcomparison_noKOTO}. Further studies exemplify the contribution of specific constraints in case of positive long-distance contribution to ${\cal B}(K_L \to\mu^+\mu^-)$ and projection~B~\cite{dambrosio2023}, as shown in Fig.~\ref{fig:fitcomparison_constributions}. It is clear that HIKE offers strong sensitivity to new physics, with the measurement of ${\cal B}(K_L\to\pi^0\ell^+\ell^-)$ especially in the electron mode giving excellent constraints even in absence of the KOTO-II measurement.

%%%%%%%%%%%%%%%%%%%%%%%%%%%%%%
% A PAGE OF 3 FLOATING FIGURES

\begin{table}[p]
\vspace{-5mm}
\caption{SM predictions, current experimental status and the expected sensitivities for key observables in the kaon sector. The ``($+$)'' and ``($-$)'' signs in the first column denote constructive and destructive interference of the amplitudes. These inputs are used to produce Fig.~\ref{fig:fitcomparison_new}.}
\vspace{-6mm}
\renewcommand{\arraystretch}{1.3}
\begin{center}
\setlength\extrarowheight{0.8pt}
\scalebox{0.83}{
\begin{tabular}{lllc}
\hline
Observable & SM prediction & Experimental status & Projections \\
\hline
%%%
${\cal B}(K^+\to \pi^+\nu\bar\nu)$ & $(7.86 \pm 0.61)\times 10^{-11}$  & $(10.6^{+4.0}_{-3.5} \pm 0.9 ) \times 10^{-11}$~\cite{NA62:2021zjw} & 5\%, HIKE Phase~1 \\
%%%
${\cal B}(K_L\to \pi^0\nu\bar\nu)$  & $(2.68 \pm 0.30) \times 10^{-11}$ & $ <300\times 10^{-11}$ @ 90\% CL~\cite{KOTO:2018dsc} & 20\%, KOTO-II \\
%%%
LFUV($a_+^{\mu\mu}-a_+^{ee}$)&\multicolumn{1}{c}{0}&$-0.031\pm 0.017$~\cite{DAmbrosio:2018ytt,NA62:2022qes}&$\pm0.007$ HIKE Phase~1\\
${\cal B}(K_L\to \mu\mu)$ ($+$) & $(6.82^{+0.77}_{-0.29})\times 10^{-9}$ & \multirow{2}{*}{$(6.84\pm0.11)\times 10^{-9}$~[PDG]} & \multirow{2}{*}{1\%, HIKE Phase~2}\\
%%%
${\cal B}(K_L\to\mu\mu)$ ($-$) &
$(8.04^{+1.47}_{-0.98})\times 10^{-9}$
\\
%%%
${\cal B}(K_S\to\mu\mu)$ & $(5.15\pm1.50)\times 10^{-12}$  & $ < 2.1\times 10^{-10}$ @90\% CL~\cite{LHCb:2020ycd} & {\small Upper bound at its current value}\\
%%%
${\cal B}(K_L\to \pi^0 ee)(+)$ & $(3.46^{+0.92}_{-0.80})\times 10^{-11}$ & \multirow{2}{*}{$ < 28\times 10^{-11}$ @90\% CL~\cite{KTeV:2003sls}} &  \multirow{2}{*}{20\%, HIKE Phase~2}\\
%%%
${\cal B}(K_L\to \pi^0 ee)(-)$ & $(1.55^{+0.60}_{-0.48})\times 10^{-11}$  &  & \\
%%%
${\cal B}(K_L\to \pi^0 \mu\mu)(+)$ & $(1.38^{+0.27}_{-0.25})\times 10^{-11}$ & \multirow{2}{*}{$ < 38\times 10^{-11}$ @90\% CL~\cite{KTEV:2000ngj}} &  \multirow{2}{*}{20\%, HIKE Phase~2} \\
%%%
${\cal B}(K_L\to \pi^0 \mu\mu)(-)$ & $(0.94^{+0.21}_{-0.20})\times 10^{-11}$ &  &  \\
\hline
\end{tabular}}
\label{tab:GiancarloData}
\end{center}
\end{table}

% END OF PAGE OF FIGURES
%%%%%%%%%%%%%%%%%%%%%%%%

%\subsection{Scalar leptoquark coupled to third generation}

\subsection{Specific BSM models}
\label{sec:specific-models}

Specific models are considered within the Physics Beyond Colliders initiative, to exemplify the impact of HIKE on BSM searches and the interplay with other experiments: 
\begin{enumerate}
\item a $Z^\prime$ model;
\item a scalar leptoquark model;
\item a vector leptoquark model.
\end{enumerate}
In all cases, the current NA62 exclusion limit is $0.42{\cal B}_\text{SM} < {\cal B}(K^+ \to\pi^+\nu\bar\nu) < 2.04{\cal B}_\text{SM}$~\cite{NA62:2021zjw}, while the NA62 and HIKE sensitivity projections assume that the SM value of ${\cal B}(K^+\to\pi^+\nu\bar\nu)$ is confirmed to 20\% and 5\% precision, respectively. 

%%%%%%%%%%%%%%%%%%%

\begin{figure}[t]
\begin{center}
\includegraphics[width=0.7\textwidth]{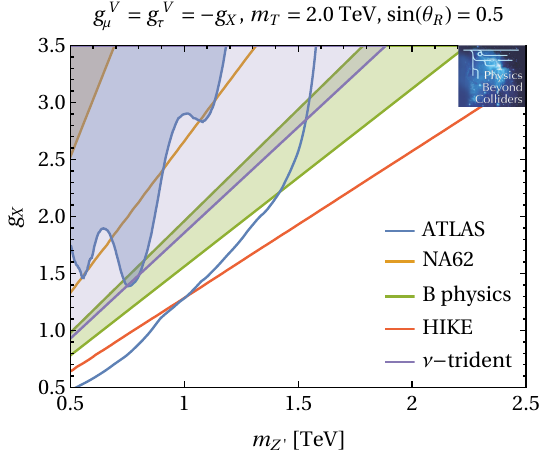}
\vspace{-2mm}
\caption{\small Constraints on a top-philic $Z'$ with mass $m_{Z'}$ and gauge coupling $g_X$~\cite{Kamenik:2017tnu,Fox:2018ldq}. We assume vector couplings to muons and tau leptons $g_\mu^V = g_\tau^V = - g_X$ and couplings to top quarks induced via mixing with a vector-like quark with mass $m_T = 2~\mathrm{TeV}$ and mixing angle $\sin \theta_R = 0.5$. The lepton couplings are chosen such that various anomalies in $b \to s$ transitions can be fitted (green shaded region), see Ref.~\cite{Alguero:2023jeh}. The purple shaded region is excluded by neutrino trident production~\cite{Altmannshofer:2014pba}. Blue shaded regions (blue lines) indicate the current exclusion with 139~fb$^{-1}$ (projection for 3~ab$^{-1}$) for ATLAS~\cite{ATLAS:2019erb}, while orange shaded regions (orange lines) indicate the current NA62 exclusion limit (projected sensitivity). The red line corresponds to the HIKE projected sensitivity.
\label{fig:Zprime}}
\end{center}
\end{figure}

%%%%%%%%%%%%%%%%%%%%

\subsubsection{$Z^\prime$ model}

The model is based on the top-philic $Z^\prime$~\cite{Kamenik:2017tnu,Fox:2018ldq}, but vector couplings to both muons and tau leptons are considered, giving rise to an interplay between the LHC which gives the dominant constraints for small $Z^\prime$ masses, and flavour physics which achieves sensitivity for large $Z^\prime$ masses.

Constraints on a top-philic $Z^\prime$ with mass $m_{Z^\prime}$ and gauge coupling $g_X$~\cite{Kamenik:2017tnu,Fox:2018ldq} are shown in Fig.~\ref{fig:Zprime}. Vector couplings to muons and tau leptons $g_\mu^V = g_\tau^V = - g_X$ are assumed, and couplings to top quarks induced via mixing with a vector-like quark with mass $m_T = 2~\mathrm{TeV}$ and mixing angle $\sin\theta_R = 0.5$. The lepton couplings are chosen such that various anomalies in $b \to s$ transitions can be fitted (green shaded region)~\cite{Alguero:2023jeh}. The purple shaded region is excluded by neutrino trident production~\cite{Altmannshofer:2014pba}. Blue shaded regions (blue lines) indicate the current exclusion with 139~fb$^{-1}$ (projection for 3~ab$^{-1}$) for ATLAS~\cite{ATLAS:2019erb}. Orange shaded regions and orange lines indicate the current NA62 exclusion limit and the projected limit by 2025, respectively. The red line corresponds to the HIKE projected sensitivity.

%%%%%%%%%%%%%%%%%%%%

\begin{figure}[p]
\centering
\vspace{-4mm}
\includegraphics[width=0.62\textwidth]{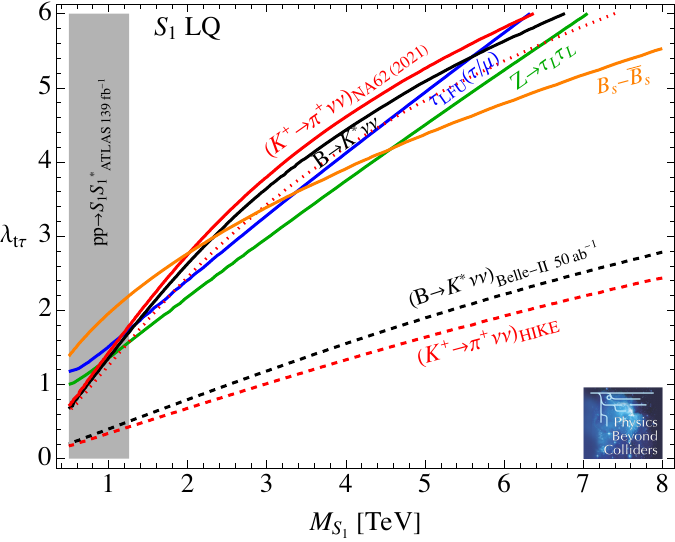}
\vspace{-3mm}
\caption{\small Constraints on the $\lambda_{t\tau}$ coupling of the $S_1$ leptoquark from flavour and electroweak observables, as function of the leptoquark mass $M_{S_1}$: the region above each line is excluded at 95\%CL by the corresponding observable. Shown here are constraints from NA62 (red), Belle~\cite{Belle:2017oht} (black), lepton-flavour universality in $\tau$ decays \cite{Pich:2013lsa} (blue), $Z$ boson couplings to tau leptons \cite{ALEPH:2005ab} (green), and from $B_s - \bar{B}_s$ mixing \cite{UTfit:2007eik} (orange) (other $\Delta F=2$ transitions provide similar but slightly weaker constraints). The shaded gray region is excluded by ATLAS from pair-production searches~\cite{ATLAS:2021jyv}. Also shown is the projected sensitivity for NA62 (dotted red) and for HIKE (dashed red), and for $B \to K^*\nu\bar\nu$ from Belle-II with 50~ab$^{-1}$ of luminosity~\cite{Belle-II:2018jsg} (dashed black). The constraints are derived using the complete one-loop matching of this leptoquark to the SMEFT derived in Ref.~\cite{Gherardi:2020det}, following the phenomenological analysis of Refs.~\cite{Gherardi:2020qhc,Marzocca:2021miv}.}
\label{fig:scalar_leptoquark}
\end{figure}

\begin{figure}[p]
\begin{center}
\vspace{-8mm}
\includegraphics[width=0.85\textwidth]{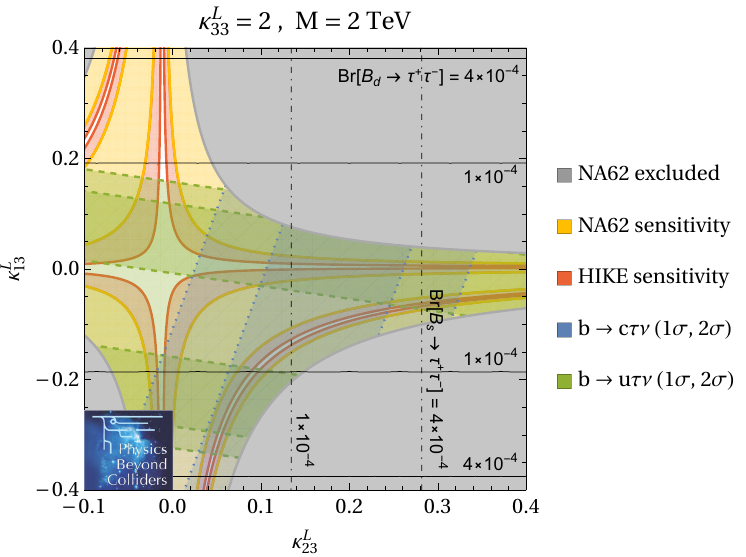}
\vspace{-3mm}
\caption{\small Constraints on a vector leptoquark $SU(2)$ singlet with mass $M = 2~\mathrm{TeV}$ and dominant coupling to the third generation ($\kappa_{33}^L$ = 2) as a function of the (real) flavour-changing couplings $\kappa_{13}^L$ and $\kappa_{23}^L$~\cite{Crivellin:2018yvo}. The grey shaded region is excluded by NA62, the blue (green) shaded regions are preferred by $b \to c(u) \tau \nu$ data~\cite{Charles:2004jd,HFLAV:2016hnz}. The NA62 (HIKE) sensitivity is indicated by  yellow (red) shading. Black lines represent the branching ratios for $B_s \to \tau^+ \tau^-$ and $B_d \to \tau^+ \tau^-$, which can be constrained by LHCb and Belle~II.
\label{fig:vector_leptoquark}}
\vspace{-15mm}
\end{center}
\end{figure}

%%%%%%%%%%%%%%%%%%%

\subsubsection{Scalar leptoquark model}

The model~\cite{Gherardi:2020det,Gherardi:2020qhc} introduces a scalar leptoquark $S_1 \sim ({\bf \bar{3}}, {\bf 1})_{3}$ coupled only to the third generation of quark and lepton SU(2)$_L$ doublets:
\begin{equation}
\mathcal{L} \supset \lambda_{t \tau} \bar{q}_3^c l_3 S_1 + \text{h.c.},
\end{equation}
where $q_3 = (t_L, \, V_{t d_j} d_L^j)$, $l_3 = (\nu_{\tau}, \, \tau_L)$. In this up-quark basis, the coupling to left-handed down quarks $d^i_L$ is proportional to the corresponding $V_{t d_i}$ CKM element. 

The present bounds on the $\lambda_{t\tau}$ coupling of the $S_1$ leptoquark from flavour and electroweak observables as a function of the leptoquark mass $M_{S_1}$ given by the model described above are shown in Fig.~\ref{fig:scalar_leptoquark}. The constraints are derived using the complete one-loop matching of this leptoquark to the SMEFT derived in Ref.~\cite{Gherardi:2020det}, following the phenomenological analysis of Refs.~\cite{Gherardi:2020qhc,Marzocca:2021miv}. Each observable shown in the plot excludes the region above the corresponding line at 95\%~CL. 
Constraints from $K^+\to\pi^+ \nu\bar\nu$~\cite{NA62:2021zjw} (red), $B\to K^*\nu\bar\nu$~\cite{Belle:2017oht} (black), lepton-flavour universality in $\tau$ decays~\cite{Pich:2013lsa} (blue), $Z$ boson couplings to tau leptons~\cite{ALEPH:2005ab} (green), and from $B_s - \bar{B}_s$ mixing~\cite{UTfit:2007eik} (orange) are shown (other $\Delta F=2$ transitions provide similar but slightly weaker constraints). The shaded gray region is excluded by ATLAS from pair-production searches~\cite{ATLAS:2021jyv}.
Also shown are the future expected constraints from $K^+ \to\pi^+\nu\bar\nu$ by NA62 by 2025 (dotted red) and HIKE (dashed red), and from $B\to K^*\nu\bar\nu$ by Belle~II with a 50~ab$^{-1}$ dataset~\cite{Belle-II:2018jsg} (dashed black).

%%%%%%%%%%%%%%%%%%%

\subsubsection{Vector leptoquark model}

The model~\cite{Crivellin:2018yvo} introduces a vector leptoquark SU(2) singlet with hypercharge -4/3 and dominant couplings to third-generation leptons:
\begin{displaymath}
\mathcal{L} \supset (\kappa_{fi}^L \overline{Q_f} \gamma_\mu L_i + \kappa_{fi}^R \overline{d_f} \gamma_\mu e_i) V^{\mu\dagger}_1 + \text{h.c.}
\end{displaymath}
Constraints on a vector leptoquark SU(2) singlet with mass $M = 2~\mathrm{TeV}$ and dominant coupling to the third generation ($\kappa_{33}^L$ = 2) as a function of the (real) flavour-changing couplings $\kappa_{13}^L$ and $\kappa_{23}^L$~\cite{Crivellin:2018yvo} are shown in Fig.~\ref{fig:vector_leptoquark}. The grey shaded region is excluded by NA62; the blue (green) shaded regions are preferred by $b \to c(u) \tau \nu$ data~\cite{Charles:2004jd,HFLAV:2016hnz}. The NA62 sensitivity by 2025 and the HIKE sensitivity are indicated by yellow and red shading, respectively. Black lines represent the branching ratios, enhanced in BSM, for $B_s \to \tau^+ \tau^-$ and $B_d \to \tau^+ \tau^-$, which can be constrained by LHCb and Belle~II.

%%%%%%%%%%%%%%%%%%

\subsection{First-row CKM unitarity}

\begin{figure}[ht]
\begin{center}
\includegraphics[width=\textwidth]{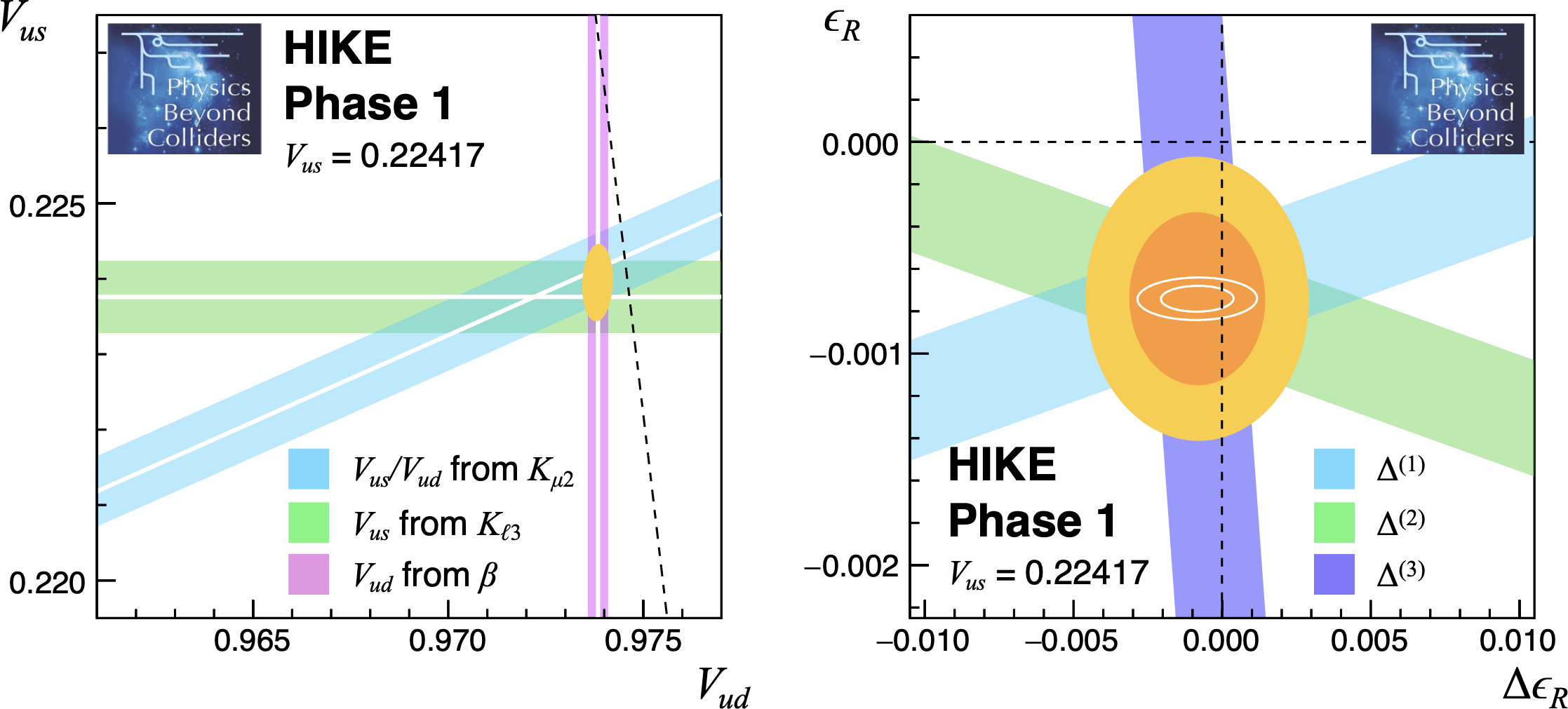}
\vspace{-7mm}
\caption{\small Status of first-row CKM unitarity in future scenario with measurements from HIKE Phase 1 confirming $V_{us} = 0.22417$. Left: measurements of $V_{us}$, $V_{us}/V_{ud}$, and $V_{ud}$ and relation to CKM unitarity. Right: constraints on right-handed currents from observed unitarity deficits.}
\label{fig:ckm_phase1}
\vspace{-7mm}
\end{center}
\end{figure}

As an example of the potential impact of the HIKE Phase 1 measurements on the understanding of first-row unitarity, 
Fig.~\ref{fig:ckm_phase1} shows the plots of Fig.~\ref{fig:ckm_2023}, updated assuming that HIKE Phase 1 obtains measurements of the five ratios as described in Section~\ref{sec:ckm_phase1} that consistently give a value for $V_{us}$ intermediate to those obtained currently from $K_{\ell3}$ and $K_{\mu2}$ decays.
The consistency of the values obtained for $V_{us}$ is seen in the intersection of the coloured bands. The unitarity deficit persists at approximately the current level, but the value of $\Delta^{(3)}$ is consistent with zero. Evidence for right-handed currents is decreased to the level of $2.3\sigma$.
The increase in sensitivity from the reduced uncertainties for the branching ratios can be appreciated in the smaller size of the white ellipses.  

\begin{figure}[ht]
\begin{center}
\includegraphics[width=\textwidth]{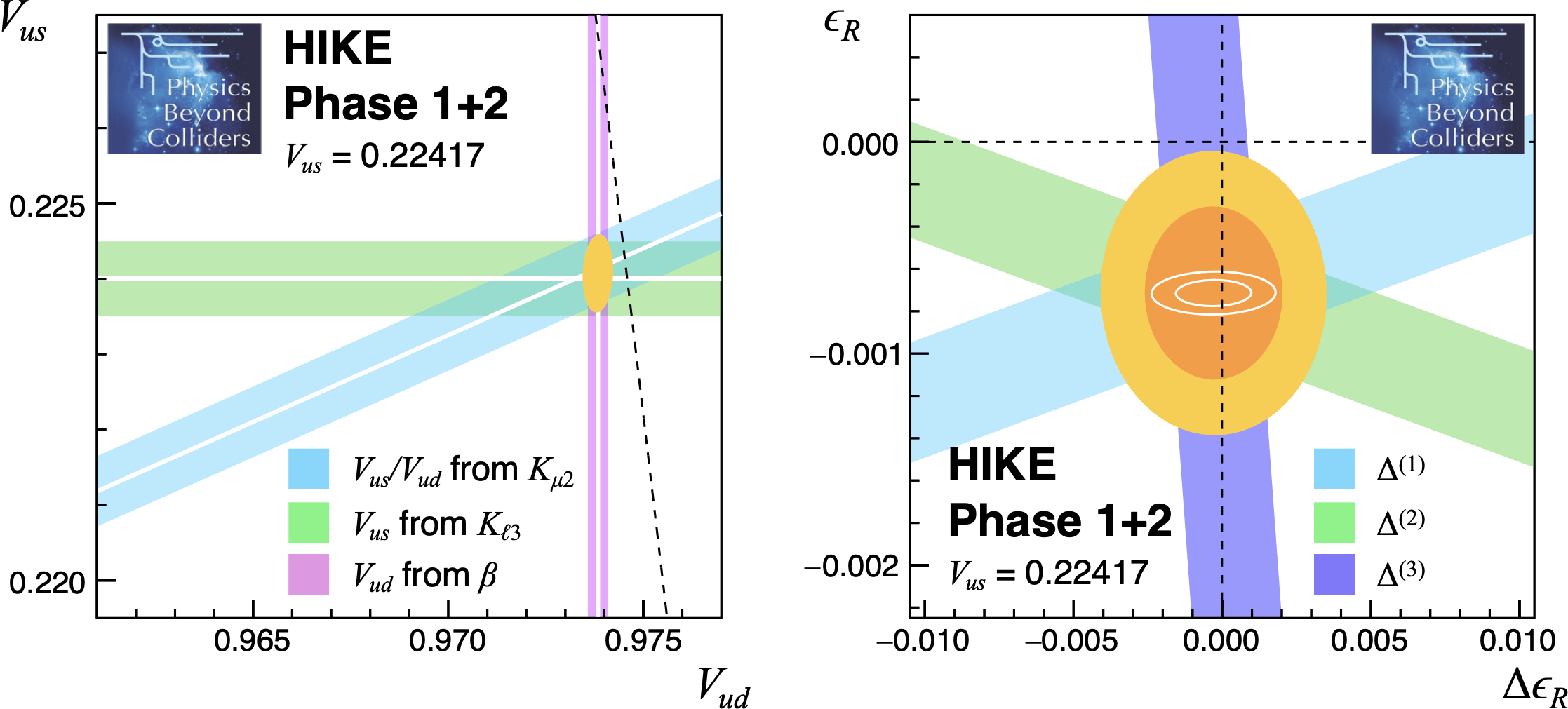}
\vspace{-7mm}
\caption{\small Status of first-row CKM unitarity in future scenario with measurements from HIKE Phases 1 and 2 confirming $V_{us} = 0.22417$.
Left: measurements of $V_{us}$, $V_{us}/V_{ud}$, and $V_{ud}$ and relation to CKM unitarity.
Right: constraints on right-handed currents from observed unitarity deficits.}
\label{fig:ckm_phase2}
\vspace{-10mm}
\end{center}
\end{figure}

The impact of adding the HIKE Phase~2 measurements (as detailed in Section~\ref{sec:ckm_phase2}) to the global fit in this scenario, together with the improvements from Phase~1, can be seen in Fig.~\ref{fig:ckm_phase2}.
Under the assumption that consistent results are obtained for $K_{\mu2}$ and $K_{\ell3}$, the values obtained for $V_{us}$ are perfectly consistent, indicating that if the unitarity deficit is attributed to right-handed currents, they must be SU(3) flavor universal. 
The level of exclusion of the point $\epsilon_R = \Delta\epsilon_R = 0$ is greatly decreased: the current $3.1\sigma$ evidence for right-handed currents is reduced to a mere $2.2\sigma$ curiosity.
In this scenario, while the kaon measurements are consistent, the unitarity deficit remains, and the precision obtained in the kaon sector 
strongly motivates further progress on the determination of $V_{ud}$, especially in the theoretical calculation of the radiative corrections.

%%%%%%%%%%%

\section{HIKE feebly-interacting particle programme}
\label{sec:fips}

\subsection{Benchmark scenarios accessible at HIKE}
\label{sec:FIPS:benchmarks}

Search for feebly-interacting particles (FIPs), i.e. manifestations of weakly-interacting NP potentially at low energy scales, represents an alternative paradigm to the search for NP with high-mass scales.
A number of SM extensions aimed at explaining the abundance of dark matter in the universe predict ``hidden sectors'' with mediator fields or matter fields in the MeV--GeV mass range. The realisations of such scenarios are usually classified in terms of the Lagrangian operators (``portals'') connecting SM particles to the new mediators. The related coupling constants are small, thus evading the existing exclusion bounds and justifying the term ``feebly interacting'' given to these new-physics models~\cite{Beacham:2019nyx,Agrawal:2021dbo,Antel:2023hkf}. The masses of hidden-sector field mediators and the couplings are free parameters of the models.
The simplest portals are summarised in Table~\ref{tab:PortalsTable}.

\begin{table}[h]
\centering
\vspace{-3mm}
\caption{Summary of generic portal models with FIPs~\cite{Beacham:2019nyx,Agrawal:2021dbo,Antel:2023hkf}.}
\vspace{-2mm}
\begin{tabular}{l|l}
\hline
Portal & FIPs \\ 
\hline
Vector & Dark photon, $A^{\prime}$ \\ 
Scalar & Dark Higgs/scalar, $S$ \\ 
Fermion & Heavy neutral lepton (HNL), $N$ \\ 
Pseudoscalar & Axion/axion-like particle (ALP), $a$ \\ 
\hline
\end{tabular}
\vspace{-2mm}
\label{tab:PortalsTable}
\end{table}

HIKE offers world-leading sensitivity to following the PBC benchmark scenarios, following the classification proposed in Ref.~\cite{Beacham:2019nyx}. Datasets to be collected in the kaon and beam-dump modes will provide sensitivities in complementary regions of FIP phase space.
\begin{itemize}
\item[BC1] Dark photon: a vector state $A^\prime$ interacting with electromagnetic current. The free parameters of the model are the mass $m_{A^\prime}$ and the coupling constant $\varepsilon$. In the kaon mode, HIKE will improve on the upper limits of $\varepsilon$ established by the NA48/2 experiment considering the prompt $K^+\to\pi^+\pi^0$, $\pi^0\to\gamma A^\prime$, $A^\prime\to e^+e^-$ decay chain~\cite{NA482:2015wmo}, reaching the values $\varepsilon\sim10^{-4}$. HIKE sensitivity for even lower $\varepsilon$ values using a displaced $A^\prime\to e^+e^-$ vertex analysis is yet to be evaluated. Another production channel available to HIKE is $K^+\to\mu^+\nu A'$, followed by either prompt or displaced $A'\to e^+e^-$ decays, allowing for improvement of the existing limits above the $\pi^0$ mass. In the beam-dump mode, the dark photon can be produced by proton strahlung and through secondary meson decays ($\pi^0$, $\eta^{(\prime)}$, etc.). For $m_{A'}<700$~MeV/$c^2$, the decay width is dominated by di-lepton final states. The HIKE projection is obtained using simulations developed for the analysis of the NA62 dataset~\cite{NA62:2023qyn}.
\item[BC2] Dark photon decaying to invisible final states. This scenario will be probed in the kaon mode by searching for the decay $K^+\to\pi^+\pi^0$ followed by $\pi^0\to\gamma A^\prime$. The technique for the search has been established by the NA62 experiment~\cite{NA62:2019meo}.
The HIKE projection is obtained by extrapolation of the established NA62 analysis.
\item[BC4] Dark scalar: a scalar state $S$ that interacts with the SM Higgs. The free parameters of the model are the mass $m_S$ and the mixing parameter $\theta$. In the kaon mode, the long-lived dark scalar is produced in the $K^+\to\pi^+S$ decay. The search in the mass range $m_S<2m_\pi$ proceeds by direct extension of the $K^+\to\pi^+\nu\bar\nu$ measurement for $m_S$ values not in the vicinity of $m_{\pi^0}$ (with a very low background, mainly from the $K^+\to\pi^+\nu\bar\nu$ decay itself), and by a dedicated analysis for $m_S\approx m_{\pi^0}$, using techniques established by NA62~\cite{NA62:2021zjw,NA62:2020xlg,NA62:2020pwi}.
The gap at $m_S \approx m_{\pi^0}$ is due to the $K^+\to \pi^+\pi^0$ background with both photons from the $\pi^0\to\gamma\gamma$ decay undetected. 
HIKE Phase~2 will provide sensitivity in the gap via the $K_L\to\pi^0 S$ decay, which is however limited by the $K_L \to\pi^0\pi^0$ background; the ultimate reach is provided by the possible HIKE Phase~3~\cite{HIKE:2022qra}. In the beam-dump mode, the dark scalar is produced mainly through flavour-changing neutral-current $B\to K^{(\ast)}S$ decays. The effective coupling constant is proportional to the fermion mass. Hence, when scanning the $S$ mass from MeV to GeV and above, each new decay mode kinematically accessible ($e^+e^-$, $\mu^+\mu^-$, $\pi^+\pi^-$, etc.) tends to either saturate or significantly increase the $S$ decay width. HIKE sensitivity evaluation currently stops at $m_S\approx2$~GeV due to the lack of a clean hadronic decay width calculation for higher masses~\cite{Winkler:2018qyg}, though there is no similar limitation for the leptonic decay width. HIKE projections rely on the simulation technique presented in Ref.~\cite{Jerhot:2022chi} and the associated public code repository~\cite{jan_jerhot_2023_7963458}.
\item[BC5] Dark scalar $S$ with enhanced pair-production. HIKE sensitivity in the beam-dump mode is similar to that of the scenario BC4; the reach to low coupling values in the mass region around 1~GeV however visibly exceeds that for BC4. Two reference values of the $\lambda_S$ parameter which quantifies the coupling of the Higgs boson to a pair of dark scalars are considered, corresponding to the current bound of the Higgs to invisible branching ratio~\cite{ATLAS:2023tkt} and the projected branching ratio considered in Ref.~\cite{Antel:2023hkf}.
\item[BC6--8] Heavy neutral leptons (HNLs): fermions $N$ that interact with the SM lepton doublets, and mix with the SM neutrinos through a matrix denoted $U$. The models BC6, BC7 and BC8 assume the presence of a single HNL of Majorana type with dominant $e$, $\mu$, or $\tau$ neutrino couplings, respectively. The free parameters are the HNL mass, $m_N$, and coupling parameter, $|U_{\ell 4}|^2$. In the kaon mode, the scenarios BC6 and BC7 are probed mainly via the $K^+\to\ell^+N$ decays ($\ell=e,\mu$). The $\pi^+\to e^+N$ decay of the beam pions provides sensitivity to BC6 in a complimentary $m_N$ range. Unlike beam-dump data, $K^+$ and $\pi^+$ decays provide direct sensitivity to the $|U_{e4}|^2$ and $|U_{\mu4}|^2$ couplings without any assumptions about the other coupling parameters. HIKE projections are obtained on the basis of the NA62 analysis~\cite{NA62:2020mcv,NA62:2021bji} assuming similar background conditions and a non-downscaled software trigger. In the beam-dump mode, HNL production is dominated by leptonic and semileptonic decays of charmed mesons and $\tau$ leptons (each model favouring the corresponding lepton flavour),
while the HNL decay width is dominated by two-body (meson-lepton and meson-neutrino) final states. HIKE-dump projections are obtained using the assumptions of Ref.~\cite{Drewes:2018gkc}.
\item[BC9] A pseudoscalar state (axion-like particle, ALP) $a$ with photon coupling. The free parameters of the model are $m_{a}$ and the coupling $C_{\gamma\gamma}$. 
The kaon mode is not sensitive to this scenario as the $K^+\to\pi^+a$ decay occurs at one-loop level in QED. In the beam-dump mode, ALP production is dominated by the interaction of photons from decays of secondary mesons ($\pi^0$, $\eta^{(\prime)}$) with the EM fields of the nuclei. The decay width is dominated by  $a\to\gamma\gamma$ decays in the accessible $m_a$ range. 
HIKE projections rely on the simulation technique discussed in Ref.~\cite{Jerhot:2022chi} and the associated public code repository~\cite{jan_jerhot_2023_7963458}.
\item[BC10] A pseudoscalar state $a$  with fermion coupling. The free parameters of the model are $m_{a}$ and the Yukawa coupling $g_Y$ to the SM fermionic fields. In the kaon mode, improvements will be obtained directly from interpretation of the $K^+\to\pi^+\nu\bar\nu$ measurement using techniques established by NA62~\cite{NA62:2021zjw,NA62:2020xlg,NA62:2020pwi}. The sensitivity provided by $K_L\to\pi^0\ell^+\ell^-$ measurements~\cite{Beacham:2019nyx} at HIKE Phase~2 is yet to be evaluated. In the beam-dump mode, the production is dominated by decays of secondary $B$ mesons ($B\to K^{(\ast)} a$). For $m_a<700$~MeV/$c^2$, the decay width is dominated by di-lepton final states; HIKE projections rely on the simulation technique discussed in Ref.~\cite{Jerhot:2022chi} and the associated public code repository~\cite{jan_jerhot_2023_7963458}.
\newpage
\item[BC11] A pseudoscalar state $a$ with gluon coupling. In the kaon mode, the search proceeds by interpretation of the $K^+\to\pi^+\nu\bar\nu$ and $K^+\to\pi^+\gamma\gamma$ measurements (considering also displaced $a\to\gamma\gamma$ vertices in the latter case) following the theoretical description of Refs.~\cite{Aloni:2018vki,Bauer:2021wjo}. The technique has been established by the NA62 experiment, and the NA62 Run~1 results have been published~\cite{NA62:2023olg}. In the beam-dump mode, the broad phenomenology can reproduce either of the models BC9 or BC10, depending on the ALP mass: one-loop corrections lead to an effective coupling to photons, and an effective coupling to quarks is generated. The former case leads to the ALP di-photon decay; the latter to hadronic decays, e.g. $a\to\pi^+\pi^-\gamma$, but not leptonic decays. HIKE projections in the dump mode rely on the simulation technique discussed in Ref.~\cite{Jerhot:2022chi} and the associated public code repository~\cite{jan_jerhot_2023_7963458}.
\end{itemize}

%%%%%%%%%%%%%%%%%%%%%%%%%

\subsection{FIP production in kaon decays}
\label{sec:fips:kaons}

Due to the availability of large datasets and the suppression of the total kaon decay width, kaon decays represent uniquely sensitive probes of light hidden sectors via searches for FIP production. A discussion of the search strategies and sensitivities in kaon decays can be found in a recent review~\cite{Goudzovski:2022vbt}. Experimentally, $K^+$ decays typically provide better sensitivity than $K_L$ decays due to the availability of the $K^+$ momentum and time measurements. Therefore the production processes and projections reported below are limited to those from $K^+$ decays. The reach of the $K_L$ phase will be examined in detail for the Technical Design Report.

In Sections~\ref{sec:KpiXSensitivity}, \ref{sec:K-HNL-Sensitivity} and \ref{sec:BC2sensitivity}, we introduce the principal $K^+$ decay channels to be exploited at HIKE in the kaon mode to address the PBC benchmarks. This list of channels is not exhaustive, and is limited to those with experimental strategies established by the NA62 experiment.

It should be stressed that kaon decays additionally provide sensitivity to a large number of non-minimal scenarios which evade detection in beam-dump experiments~\cite{Harris:2022vnx,Gori:2022vri,Ballett:2019pyw}. Examples of such scenarios accessible at HIKE are: short-lived Majorana HNLs decaying via a displaced-vertex topology $K^+\to\ell_1^+N$, $N\to\pi^-\ell_2^+$~\cite{Atre:2009rg}; dark neutrino produced and decaying via the $K^+\to\ell^+N$, $N\to\nu Z'$, $Z'\to e^+e^-$ chain~\cite{Ballett:2019pyw}; a muonphilic force scenario leading to $K^+\to\mu^+\nu X$ decays~\cite{Krnjaic:2019rsv}. Further signatures accessible in kaon decays include pair-production of FIPs~\cite{Hostert:2020xku}; the search strategy in this case is established by the NA62 experiment~\cite{NA62:2023rvm}.

%%%%%%%%%%%%%%%%%%%%%

\subsubsection{$K^+\to\pi^+X_{\rm inv}$ decays}
\label{sec:KpiXSensitivity}

The search for the $K^+\to\pi^+X_{\rm inv}$ decay, where $X_{\rm inv}$ is an invisible particle, provides sensitivity to the benchmark scenarios BC4 (dark scalar), BC10 (ALP with fermion coupling) and BC11 (ALP with gluon coupling). The search at HIKE Phase~1 will be performed by direct extension of the $K^+\to\pi^+\nu\bar\nu$ measurement, by peak search in the spectrum of the reconstructed missing mass $m_{\rm miss}^2=(P_{K^{+}}-P_{\pi^{+}})^{2}$.
The accessible $m_X$ ranges are approximately 0--110~MeV/$c^2$ and 150--260~MeV/$c^2$, corresponding to the $K^+\to\pi^+\nu\bar\nu$ signal regions. The principal background comes from the $K^+\to\pi^+\nu\bar\nu$ decay itself. The search strategy has been established by the NA62 experiment~\cite{NA62:2020xlg}, and the full NA62 Run~1 (2016--2018) dataset has been analysed~\cite{NA62:2021zjw}.
The HIKE Phase~1 sensitivity projection has been performed by extension of the NA62 analysis, assuming a 40-fold increase in the size of the data sample with respect to NA62 Run~1.

The above scenarios will be also addressed by a dedicated search for the $\pi^0\to X_{\rm inv}$ decay using a technique established by NA62~\cite{NA62:2020pwi}. This search covers the $m_X$ region in the vicinity of the $\pi^0$ mass. The region $m_X>260~{\rm MeV}/c$ for the scenarios BC4 and BC11 will be covered by searches for $K^+\to\pi^+X$ decays followed by displaced $X\to\mu^+\mu^-$ or $X\to\gamma\gamma$ decays, respectively.

%%%%%%%%%%%%%%%%%%%%%%%%%

\subsubsection{$K^+\to\ell^+N$ decays}
\label{sec:K-HNL-Sensitivity}

Searches for the $K^+\to\ell^+N$ decays ($\ell=e,\mu$), where $N$ is an invisible particle, provide sensitivity to the benchmark scenarios BC6 (HNL with electron coupling) and BC7 (HNL with muon coupling). The technique has been established by the NA62 experiment, which has obtained world-leading exclusion on the HNL mixing parameters $|U_{\ell 4}|^2$ over much of the accessible mass range of 144--462~MeV/$c^2$ with the Run~1 dataset~\cite{NA62:2020mcv,NA62:2021bji}. Both searches are limited by background. In particular, the $K^+\to\mu^+\nu$ decay followed by $\mu^+\to e^+\nu\bar\nu$ decay in flight, and the $\pi^+\to e^+\nu$ decay of the pions in the unseparated beam, represent irreducible backgrounds to the $K^+\to e^+N$ process.
The peaking nature of the $K^+\to\ell^+N$ signal in terms of the reconstructed missing mass allows for data-driven background evaluation, reducing the systematic uncertainties in the background estimates.

The HIKE sensitivity projection is obtained by extension of the NA62 analysis assuming the similar resolution and background. In the $K^+\to\mu^+N$ case, it is assumed additionally that, unlike NA62, the trigger line is not downscaled, which is possible for a fully software trigger. HIKE sensitivity to $|U_{e4}|^2$ in the mass range 144--462~MeV/$c^2$ approaches the seesaw neutrino mass models~\cite{Abdullahi:2022jlv}. For $m_N<140~{\rm MeV}/c^2$, HIKE will improve the PIENU limits~\cite{PIENU:2017wbj} on $|U_{e4}|^2$ via the $\pi^+\to e^+N$ decays of pions in the unseparated beam, and has a further potential via the $K^+\to\pi^0e^+N$ decay~\cite{Tastet:2020tzh}. HIKE will also approach the seesaw neutrino mass models for $|U_{\mu4}|^2$.

%%%%%%%%%%%%%%%%%%

\subsubsection{$\pi^0\to\gamma A'$ decay}
\label{sec:BC2sensitivity}

The search for the $K^+\to\pi^+\pi^0$, $\pi^0\to\gamma A^\prime$ decay chain with a technique established by the NA62 experiment~\cite{NA62:2019meo} addresses the benchmark scenario BC2. The HIKE sensitivity, estimated in a dedicated study, assumes improvements at the trigger level, with Cherenkov rings reconstructed by the RICH providing fast measurement of $\pi^+$ momentum and direction downstream of the spectrometer magnet. Including the EM calorimeter information, the missing energy $E_\mathrm{miss}$ is measured at the trigger level and the condition $E_\mathrm{miss}>20$~GeV is required.

At NA62, the background is dominated by $\pi^0\to\gamma\gamma$ decays with one photon lost, and the other detected photon converting into a $e^+e^-$ pair before reaching the LKr calorimeter, thus leading to a systematically lower reconstructed energy; the background is estimated using a control data sample~\cite{NA62:2019meo}. The HIKE projection is evaluated under the assumption that the NA62 background scales linearly with the accumulated statistics. However an intrinsically lower background is expected due to the reduced material budget: 40\% less material than in NA62 for the straw-tube tracker and absence of passive material just in front of the ECAL (as opposed to the cryostat in front of the LKr calorimeter). Beyond the increased sensitivity guaranteed by two order of magnitudes more statistics acquired thanks to the trigger improvements, the background reduction allows a further sensitivity improvement by a factor of 2.4 (in terms of couplings) in the region of dark photon masses above 50~MeV. Further improvements in sensitivity can be obtained by optimising the $\pi^+$ momentum range,
the veto conditions in the EM calorimeter, the minimum photon energy requirement, and the quality conditions for the reconstruction of the $K^+$ and $\pi^+$ momenta.

%%%%%%%%%%%%%%%%

\newpage

\subsection{FIP production and decays in dump mode}

\subsubsection{Experimental layout}

Thanks to the high-intensity 400~GeV/$c$ proton beam extracted from the CERN SPS and high-performance detectors,
HIKE will achieve the sensitivities required for competitive searches for production and decay of long-lived light mediators in a variety of new-physics scenarios.
FIPs can be produced in a plethora of open production channels when the proton beam is dumped onto the upstream HIKE collimators with the kaon production target removed. Long-lived FIPs can traverse the beam dump without being stopped, and, with typical decay lengths of several meters to kilometers, reach the decay volume. 

Due to the forward geometry, HIKE beam-dump data is particularly sensitive to specific cases in which a hidden-sector mediator is produced by Primakoff scattering or light-meson mediated decays. Prominent examples are the vector and axion-like particle scenarios. For all benchmark cases in which exotics are generated from meson decays or meson mixing, meson production is simulated using PYTHIA~8.2. For some of the light mesons, the kinematics, cross-sections and yield have been validated against available data where possible~\cite{Dobrich:2019dxc}; for heavier mesons, where the experimental picture is less clear, we have partially validated the meson spectra against simulations used in other groups~\cite{Dobrich:2018jyi}.
Note that no ``cascades''~\cite{CERN-SHiP-NOTE-2015-009} are implemented in our simulation. We therefore expect our signal projections in which forward exotics production plays an important role to be on the conservative side. The most relevant features of the hidden-sector phenomenology for searches in beam-dump mode are:
\begin{itemize}
\item The mediators can be produced in proton-nucleus interactions through a number of mechanisms, the most important being direct Primakoff production and meson-mediated tertiary production. The specific production mechanism for each mediator type differs in terms of cross section and momentum-angle spectra of the mediator emitted.
\item At the SPS proton energy of 400~GeV, the decay lengths of mediators with momenta above 10~GeV/$c$ range from tens of meters to tens of kilometres in the feeble coupling interval of interest.
\item Due to the feeble interaction with the SM particles, the emitted mediators can reach the decay volume after punching through tens of meters of traversed material before decaying.
\end{itemize}

HIKE operation in the beam-dump mode will build upon the experience accumulated at NA62. By the end of 2023, a sample of $3.9\times 10^{17}$~POT has been collected in the beam-dump mode at NA62 in 20~days of operation. The T10 target used to generate the NA62 secondary hadron beam was removed from the beamline, and the proton beam was made to interact in the NA62 movable collimators, called TAX, located 23~m downstream of T10 within two pair of dipoles, and 80~m upstream of the decay volume. An ad-hoc setting of the dipoles allows a substantial reduction of the rate of muons emitted by pion decays in the proton-induced hadronic showers in the TAX (the so called halo muons). This above optimisation reduces the background from secondary interactions of halo muons by at least one order of magnitude~\cite{Gatignon:2650989}. The first NA62 physics results based on $1.4\times 10^{17}$~POT demonstrate a low background level~\cite{NA62:2023qyn}. 

Operation with an intensity of $2\times10^{13}$ protons per 4.8~s spill, and a total sample of $5\times10^{19}$ POT are foreseen in the beam-dump mode. When compared to NA62, these figures correspond to an intensity and statistics increase by factors of 4 and 50, respectively.
However, higher beam intensities up to twice as that can be tolerated in beam-dump mode since 
the detector rate foreseen during beam-dump operation at the above intensity amounts to less than 200~kHz.
Considering limitations arising from the current design of the future target and TAX and radio-protection consideration, operation at $2.4\times10^{13}$~POT for 4.8~s spills is within the TAX design margins (Section~\ref{sec:beam}).

Using a movable dump allows for a quick switch between a kaon-beam and a beam-dump operation, making it possible to perform specific online calibration procedures in kaon-beam mode during beam-dump operation periods. The importance of this aspect was demonstrated in NA62 in 2021, significantly improving the overall data quality. 
As demonstrated by the NA62 beam-dump data, regular accurate calibration of the beam secondary-emission intensity monitors (BSI) is necessary, to ensure an overall uncertainty in the integrated proton flux at the 5\% level.  
Monitors based on the current BSI technology are 
adequate to withstand the projected beam intensity. Exploratory experimental studies by the CERN SY-BI and BE-EA experts show that the goal of achieving a 5\% uncertainty in the proton flux measurement is within reach~\cite{VanDijk:2023}. 

%%%%%%%%%%%%%%%%

\subsubsection{Background evaluation}

The results from the analyses already completed with NA62 data sets allow a quantitative extrapolation of the background level for a statistics corresponding to $5\times10^{19}$~POT.

\begin{figure}[tb]
\centering
\includegraphics[width=.66\linewidth]{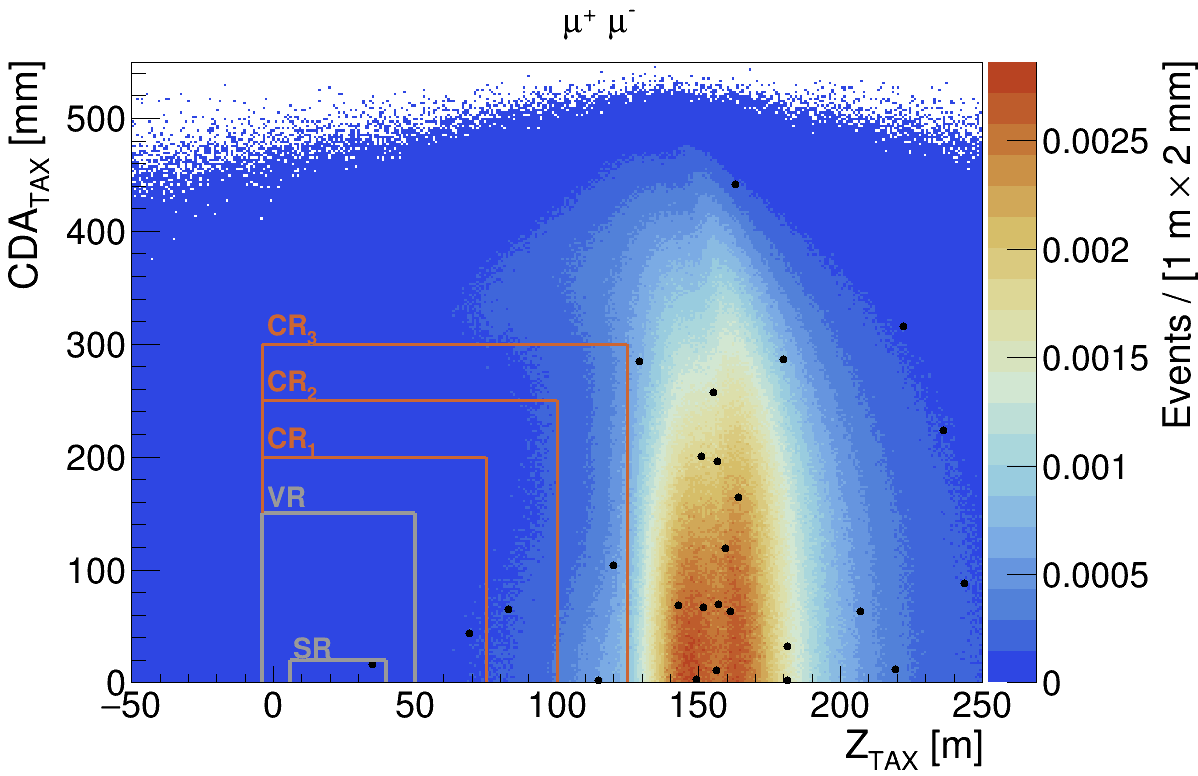}
\vspace{-3mm}
\caption{\label{fig:SREvent} Distributions of $\mathrm{CDA}_\mathrm{TAX}$ vs $\mathrm{Z}_\mathrm{TAX}$: data (markers); expected background (colour-scale).}
\end{figure}

A search for dark photons decaying in flight to $\mu^+\mu^-$ pairs based on a sample of $1.4\times 10^{17}$ protons on dump collected in 2021 is reported in Ref.~\cite{NA62:2023qyn}.
Fig.~\ref{fig:SREvent} reproduced from this publication shows the remaining data events that pass selection criteria for decays from FIPs originating at the TAX and decaying into a $\mu^+\mu^-$ pair in the NA62 fiducial volume.
These pairs, in order to constitute a possible signal event, must come from a small region, labelled ``SR'', in which the reconstructed FIP trajectory has its closest distance of approach to the TAX in a position along the beam axis around 23~m. The data is overlayed with a colour-scale that reflects the combinatorial background expectation obtained from an independent single-track sample which was statistically combined~\cite{NA62:2023qyn}. Expectation and observations are in good agreement.

For searches including other final states with two charged particles (electrons, pions, muons), the background is dominated by interactions of a single beam-halo muon in material upstream of the decay volume, such as photo-production of pions or electromagnetic showers. The background is constituted by pairs of in-time tracks. A data-driven simulation has been developed to evaluate the expected background: single muon tracks reconstructed in data are extrapolated backwards to a transversal plane upstream of the beam-achromat system hosting the GTK stations, correcting for the average energy loss in the material traversed, and then used as input to the standard NA62 Monte Carlo simulation.  
A control sample used to cross-check the reliability of such simulation is constituted by final states with electron-muon, positron-muon pairs identified within the 2021 beam-dump dataset. Data agree with the expectation from the simulation, not only after the full event selection is applied, but also at the various analysis steps~\cite{StefanMoriond}, as shown in Table~\ref{tab:emu_CR_SR_rejections}. Here ``PID'' refers to the particle-identification conditions, whereas ``ANTI-0'' and ``LAV'' refer to the veto conditions applied using the information from the related detectors; finally, ``CR'' and ``SR'' refer to the control and signal regions in the plane shown in Fig.~\ref{fig:SREvent}.

%%%%%%%%%%%%%%%%%%%%

\begin{table}[tb]
\caption{Summary of the expected number of $e\mu$ vertices from the simulation ($N_{\text{exp}}$) before and after requiring the ANTI-0, LAV, and LAV+ANTI-0 veto conditions, the related total uncertainty ($\delta N_{\text{exp}}$), the observed events in data ($N_{\text{obs}}$) and the $p$-values representing the probability to obtain a likelihood $L$ for data-MC compatibility equal or worse than that corresponding to $N_{\text{obs}}$. The entry with an asterisk is obtained from a Fisher test on proportions assuming integer counts.}
\vspace{-2mm}
\centering
\begin{tabular}{l|c|c|c}
\hline
Condition & $N_{\text{exp}}\pm\delta N_{\text{exp}}$ & $N_{\text{obs}}$ & $p(L< L_{\text{obs}})$\\
\hline
$e\mu$ PID & $2905 \pm 1455$ & 2896 & 0.97\\
$e\mu$ PID, ANTI-0 & $8.6 \pm 6.1$ &  12 & 0.61\\
$e\mu$ PID, LAV & $728 \pm 365$ &  645 & 0.94\\
$e\mu$ PID, LAV+ANTI-0 & 0  &  2 & 0.25$^\ast$  \\
$e\mu$ PID, CR & $50\pm26$    &  49 & 0.98 \\
$e\mu$ PID, SR & $2.5\pm1.8$  &  3  & 0.83 \\
$e\mu$ PID, LAV+ANTI-0, CR & 0  &  0 & -- \\
$e\mu$ PID, LAV+ANTI-0, SR & 0  &  0 & -- \\
\hline
\end{tabular}
\label{tab:emu_CR_SR_rejections}
\end{table}

%%%%%%%%%%%%%%%%%

\begin{table}[t]
\caption{\label{tab:bkg}
Background estimates for an expected HIKE beam-dump dataset of $5\times10^{19}$~POT, obtained as projections of the analyses of $1.4\times10^{17}$~POT collected in beam-dump mode by NA62 in 2021. Upper limits are quoted at 90\% CL.}
\vspace{-2mm}
\centering
\begin{tabular}{c|c}
\hline
Final state & Expected background \\
\hline
$\mu^+\mu^-$ & $0.0024$ \\
$e^+e^-$ & $<0.9$ \\
$\pi^+\pi^-(\gamma)$ & $<0.09$ \\  
$\mu^\pm \pi^\mp$, $e^\pm\pi^\mp$ & $<0.1$  \\  
$\gamma\gamma$ & work in progress \\
\hline
\end{tabular}
\end{table}

%%%%%%%%%%%%%%%%%

From these analyses, a prospective level of expected background can be derived for HIKE. Expected backgrounds extrapolated from the completed and ongoing analyses are summarised in Table~\ref{tab:bkg}. The bullets below provide details for each decay mode.
\begin{itemize}
\item $\mu^+\mu^-$ final state: random pairing of two unrelated beam-halo muons in a $\pm5$~ns time window produces a background below 0.02~events in a sample of $1.4\times10^{17}$~POT,  with no ANTI-0 veto condition applied. In HIKE baseline conditions, i.e.\ with $2\times10^{13}$ ppp with a 4.8 s flat top, a factor of four higher intensity is foreseen, and is compensated by the improved detector time resolution of the forward hodoscope, so the combinatorial background probability is substantially unchanged. For a total HIKE dump-mode data set of $5\times10^{19}$ POT, 6 combinatorial background events are expected without applying the ANTI-0 veto. The ANTI-0 veto condition is capable of rejecting any charged particle traversing it, with at most a 2\% inefficiency per particle. Applying the ANTI-0 veto, 0.0024 combinatorial background events are expected in the HIKE baseline conditions.\footnote{In principle, HIKE would be able to take also the full SPS beam (corresponding to $4\times10^{13}$ ppp) from the detector and analysis point of view. With a 4.8 s flat top, it would lead to a $0.0024 \times 2^2 \approx 0.01$ combinatorial background events expected, while with a 1.2 s flat top it would lead to $0.01 \times 4^2 = 0.16$ combinatorial background events expected.}
\item $e^+e^-$ final state: in-time background due to interactions of beam-halo muons produces a background below 0.04~events at 90\% confidence level in a sample of $1.4\times10^{17}$~POT. Statistical scaling yields an upper limit below 14 background events. This figure is reduced to below 0.9~events when invariant masses above 40~MeV/$c^2$ are considered.
\item $\pi^+\pi^-(\gamma)$ final state: analysis of $1.4\times10^{17}$~POT is in progress. Control data samples indicate that the background (if any) is dominated by interactions of beam-halo muons in the material traversed. Moreover, the identification of pion tracks from a preliminary analysis has shown that pion production is a factor of 10 less abundant than electron/positron production. The overall background level is expected to be below 0.09~events at $5\times10^{19}$~POT.
\item $\pi^\pm e^\mp$ final states: analysis of $1.4\times10^{17}$~POT is in progress. Control data samples indicate that the background is dominated by interactions of beam-halo muons in the material traversed. Considerations similar to those done for the pion-pion final states apply in this case. The overall background level is expected to be below 0.1~events at $5\times10^{19}$~POT.
\item $\gamma\gamma$ final state: analysis of $1.4\times10^{17}$~POT is in progress. The loss of rejection power due to the lack of extrapolation to the production point (a factor of 100) can be recovered by raising the minimum energy threshold for photons in the final state: as seen for the $e^+e^-$ final state, the spectrum of secondaries produced by interactions of halo muons is dominantly soft. It has been shown that a threshold of 30~GeV on the total energy reduces the background component by a factor of 100. Background due to production and decay of $K_S$ and $\Lambda$ has been studied (as reported to the SPSC in 2022) and reduced by a factor of~200 by increasing the sweeping power in the beamline. A reliable scaling to $5\times10^{19}$~POT requires the completion of the $\gamma\gamma$ search with the 2021 dataset.
\end{itemize}

For neutrino-induced backgrounds, a dedicated discussion is worth. Simulations predict the overall number of neutrino interactions in the various detector materials. The background expected in HIKE from neutrino interactions turns out to be negligible when searching for di-lepton final states. The effect of neutrino interactions has been cross-checked for $\pi^\pm\mu^\mp$ final states. Estimates based on a sample of MC events folded with the composition and material budget of the main components that could contribute to induce a neutrino background have been performed. In particular, we have considered background sources from neutrino interactions in the final collimator just upstream of the decay volume (thickness of 1~m, steel), in the upstream end cap of the vacuum vessel (9~mm, iron), in the residual air in the fiducial volume (50~m, air at $10^{-6}$~mbar) and in the first straw tracker station (2.2~mm, equivalent material with a density of 1~g$/$cm$^3$). Considering neutrino momenta above 20~GeV, the expected numbers of events after $1.5\times10^{17}$ dumped protons with a neutrino interaction producing a muon in the final state are 0.008, 1.1, $10^{-6}$, and 0.04 for the final collimator, vessel end cap, residual air and the first straw chamber, respectively. These can be considered upper limits for the number of background events, since we considered any interaction producing a muon in the final state, while the background will be due to the simultaneous production of a charged pion and a muon. 

From these results, it seems likely that the neutrino-induced background from the upstream end cap of the vacuum vessel would become visible at HIKE-dump fluxes. To a lesser extent, this holds for the first straw chamber, as well. It must be noted that with a suitable restriction of the fiducial volume, the background from neutrino interactions from the final collimator, the vessel end cap, and the first straw chamber can be significantly reduced with modest losses in terms of signal acceptance. This statement holds for any final state with charged particles, while it remains to be proved for the $\gamma\gamma$ search.
%%%%%%%%%%%%%%%%%

The future HIKE setup is foreseen to coexist with the SHADOWS detector. A dedicated simulation of the muon halo in presence of the SHADOWS setup has been performed, based on the BDSIM package. The equivalent statistics of such simulation is $1.1\times 10^9$ protons on dump. A comparison of the muon halo flux at the first straw station with the setup in absence of SHADOWS has been derived. 
From this study, we have no indication of any evident degradation of the expected sensitivity for HIKE operation in dump mode in presence of SHADOWS.

\subsection{Summary of physics sensitivity}

With a combination of data taken in kaon and beam-dump modes, HIKE will be able to reach unprecedented sensitivities for all PBC benchmarks~\cite{Beacham:2019nyx}, with the exception of BC3 for which the sensitivity is yet to be evaluated. The analysis methods and background estimates for HIKE rely solidly on the data collected and analysed by NA62; simulation techniques and FIP production models are well established. The sensitivity curves presented in Section~\ref{sec:fips_sensitivity} show marked improvements with respect to present experimental status for FIP masses below 2~GeV/$c^2$, reaching ${\cal O}(10^{-5})$ to ${\cal O}(10^{-10})$ in the FIP coupling depending on the scenario.

\section{Impact of the HIKE feebly-interacting particle programme}
\label{sec:fips_sensitivity}

Operation in both kaon and beam-dump modes will allow HIKE to address a uniquely broad range of hidden-sector scenarios. For many scenarios, kaon and beam-dump datasets are sensitive to complementary FIP mass ranges. Therefore HIKE will provide coverage in a broad FIP mass region spanning from about 10~MeV to a few GeV. Furthermore, operation in kaon mode provides excellent sensitivity to non-minimal dark sector scenarios involving short-lived FIPs which evade detection in beam-dump experiments~\cite{Harris:2022vnx,Gori:2022vri,Ballett:2019pyw}. 

\begin{figure}[p]
\centering
\vspace{-8mm}
\includegraphics[width=0.92\textwidth]{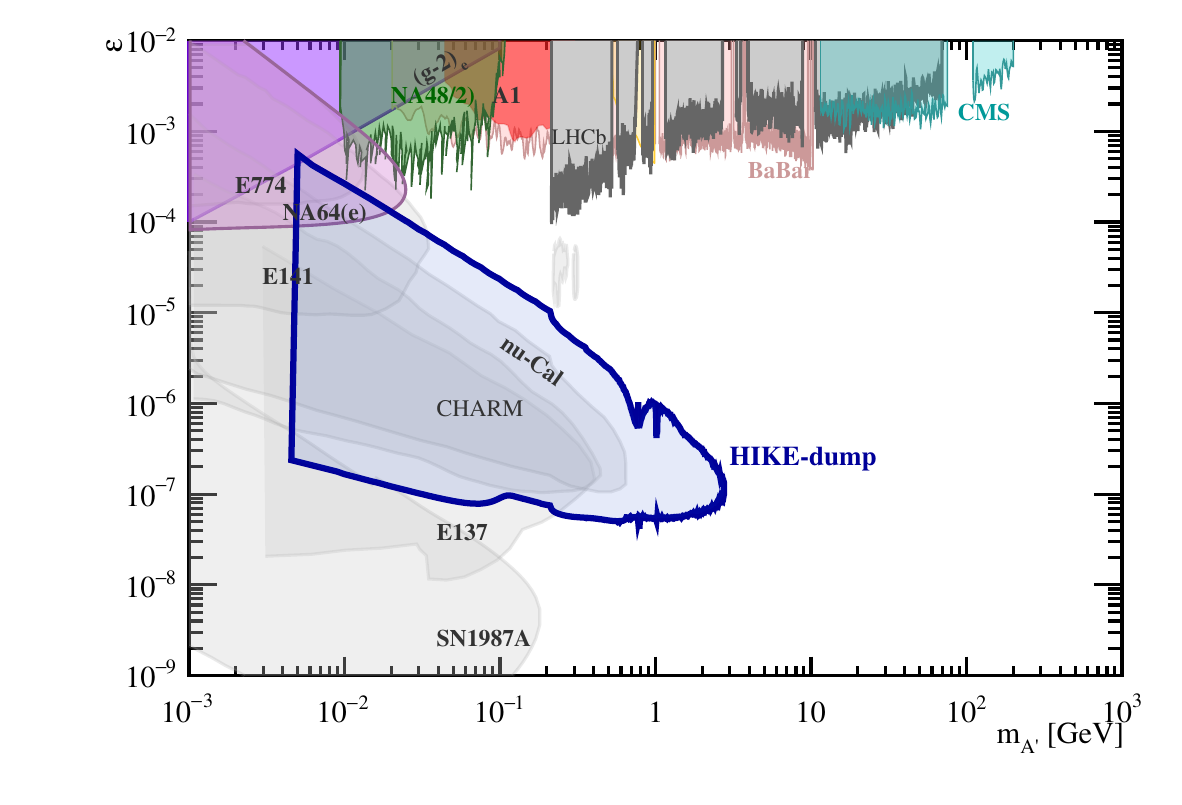}%
\put(-408,135){\LARGE BC1}
\vspace{-7mm}
\caption{HIKE exclusion region in the BC1 scenario (dark photon) in the dump mode. In the kaon mode, HIKE will improve the NA48/2 upper limits of $\varepsilon$ to the $10^{-4}$ level in the region $m_{A'}<m_{\pi0}$ (which is not however indicated in the plot). Existing experimental limits are also shown.}
\label{fig:bc1}
\end{figure}

\begin{figure}[p]
\centering
\vspace{-17mm}
~~~~~~~~~~~~~~~~~~\includegraphics[width=0.91\textwidth]{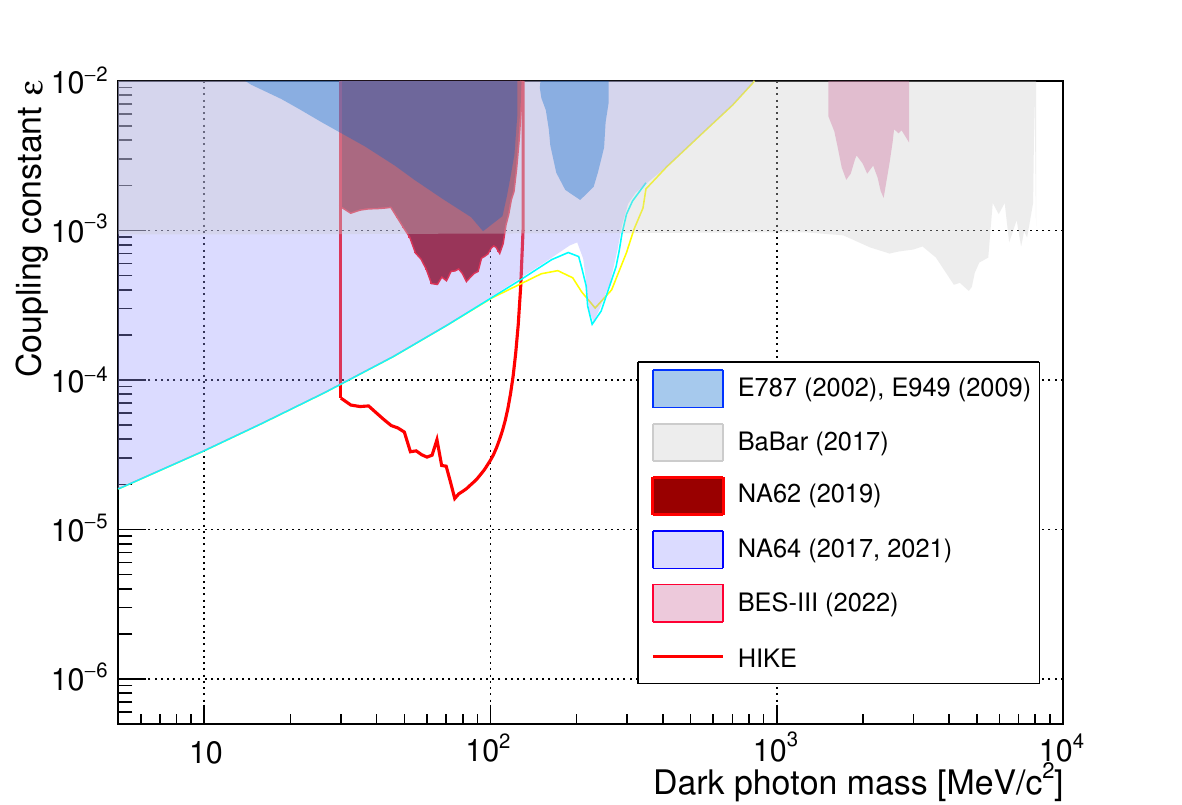}%
\put(-425,135){\LARGE BC2}
\vspace{-2mm}
\caption{HIKE exclusion region in the BC2 scenario (invisible dark photon) in the kaon mode. Existing experimental limits are also shown.}
\label{fig:bc2}
\end{figure}

\begin{figure}[p]
\centering
\vspace{-8mm}
\includegraphics[width=0.935\textwidth]{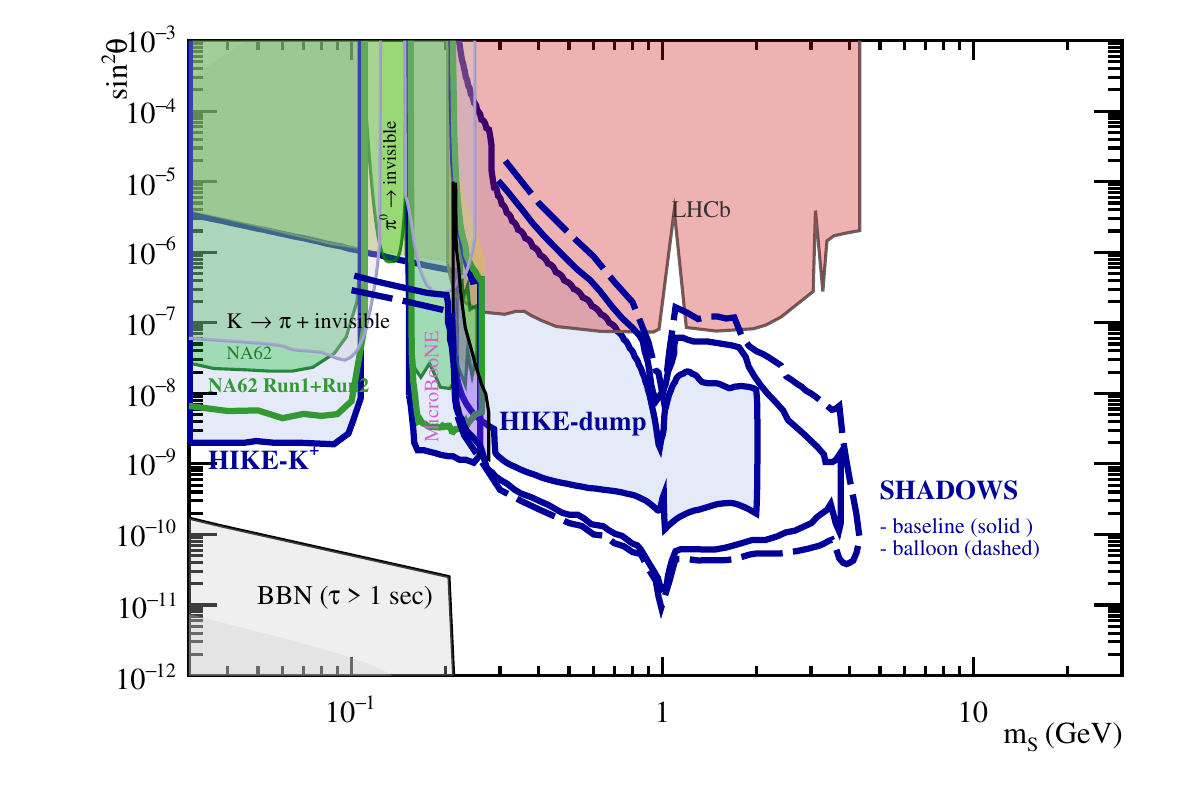}%
\put(-406,140){\LARGE BC4}
\vspace{-8mm}
\caption{HIKE (filled blue contours) and SHADOWS (empty contours) exclusion regions in the BC4 scenario: dark scalar. Existing experimental limits are also shown.}
\label{fig:bc4}
\end{figure}

\begin{figure}[p]
\centering
\vspace{-14mm}
~~~~\includegraphics[width=0.85\textwidth]{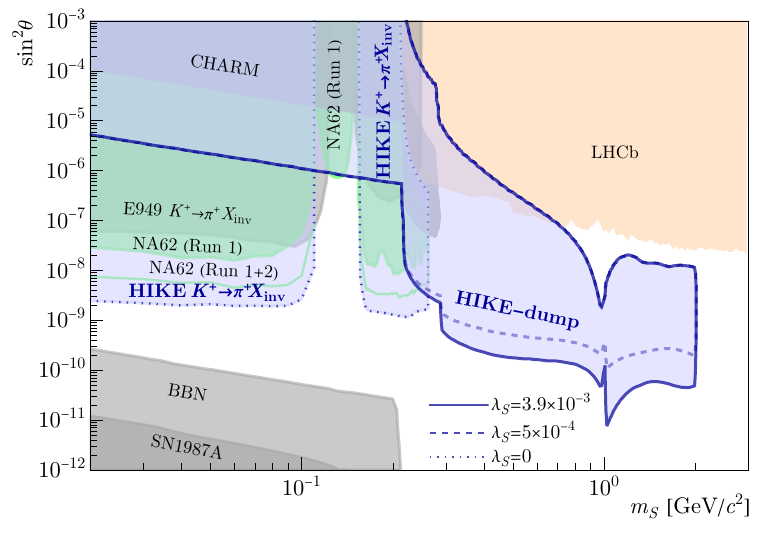}%
\put(-390,140){\LARGE BC5}
\vspace{-6mm}
\caption{HIKE exclusion regions in the BC5 scenario: dark scalar with enhanced pair-production  (filled blue contours). HIKE sensitivity in the dump mode is evaluated for two reference values of the $\lambda_S$ parameter, $\lambda_S=3.9\times 10^{-3}$ and $\lambda_S=5\times 10^{-4}$~\cite{ATLAS:2023tkt,Antel:2023hkf}. HIKE sensitivity in the kaon mode ($K^+\to\pi^+X_{\rm inv}$) i shown for $\lambda_S=0$, which corresponds to the scenario BC4. Existing experimental limits are also shown.}
\label{fig:bc5}
\end{figure}

\begin{figure}[p]
\centering
\vspace{-12mm}
\includegraphics[width=0.75\textwidth]{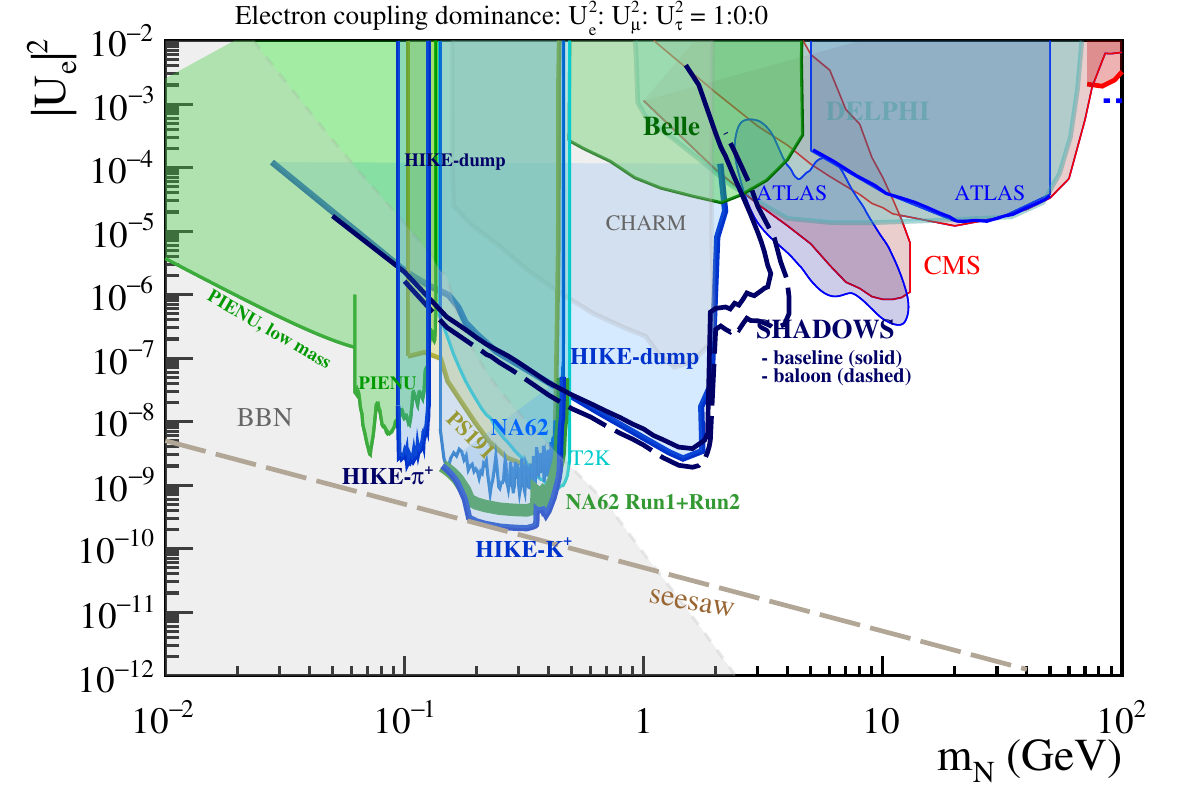}%
\put(-350,115){\LARGE BC6}\\
\vspace{-2mm}\includegraphics[width=0.75\textwidth]{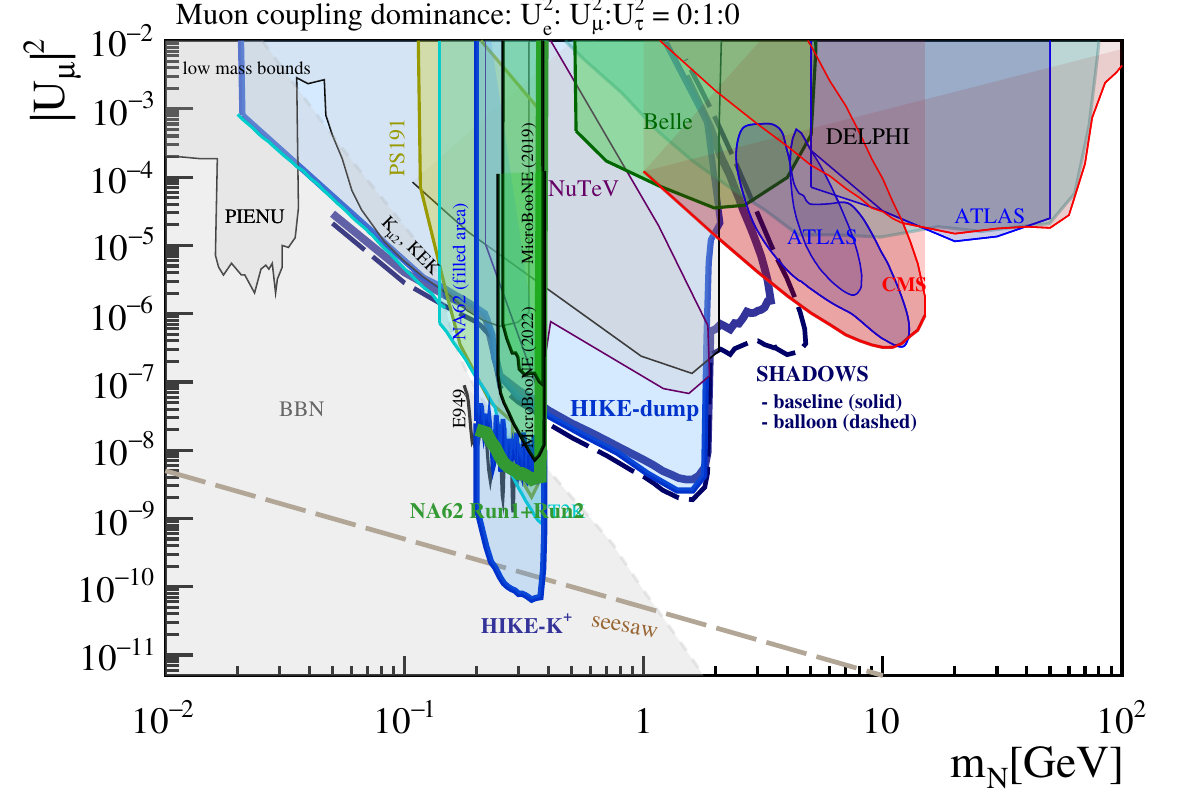}%
\put(-350,115){\LARGE BC7}\\
\vspace{-2mm}
\includegraphics[width=0.75\textwidth]{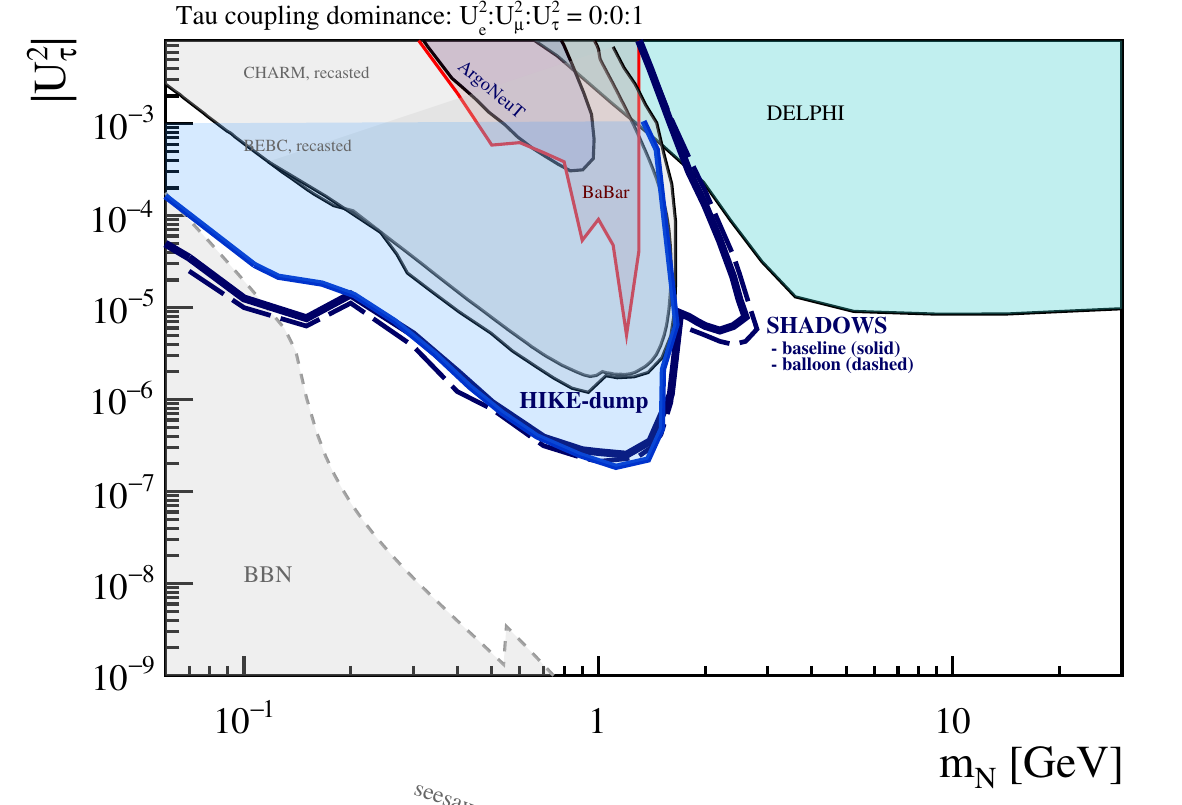}%
\put(-350,115){\LARGE BC8}
\vspace{-4mm}
\caption{HIKE (filled blue contours) and SHADOWS (empty contours) exclusion regions in the scenarios BC6,7,8: HNL with $e$, $\mu$ and $\tau$ coupling. Existing experimental limits are also shown.}
\label{fig:bc6to8}
\end{figure}

\begin{figure}[p]
\centering
\vspace{-10mm}
\includegraphics[width=0.8\textwidth]{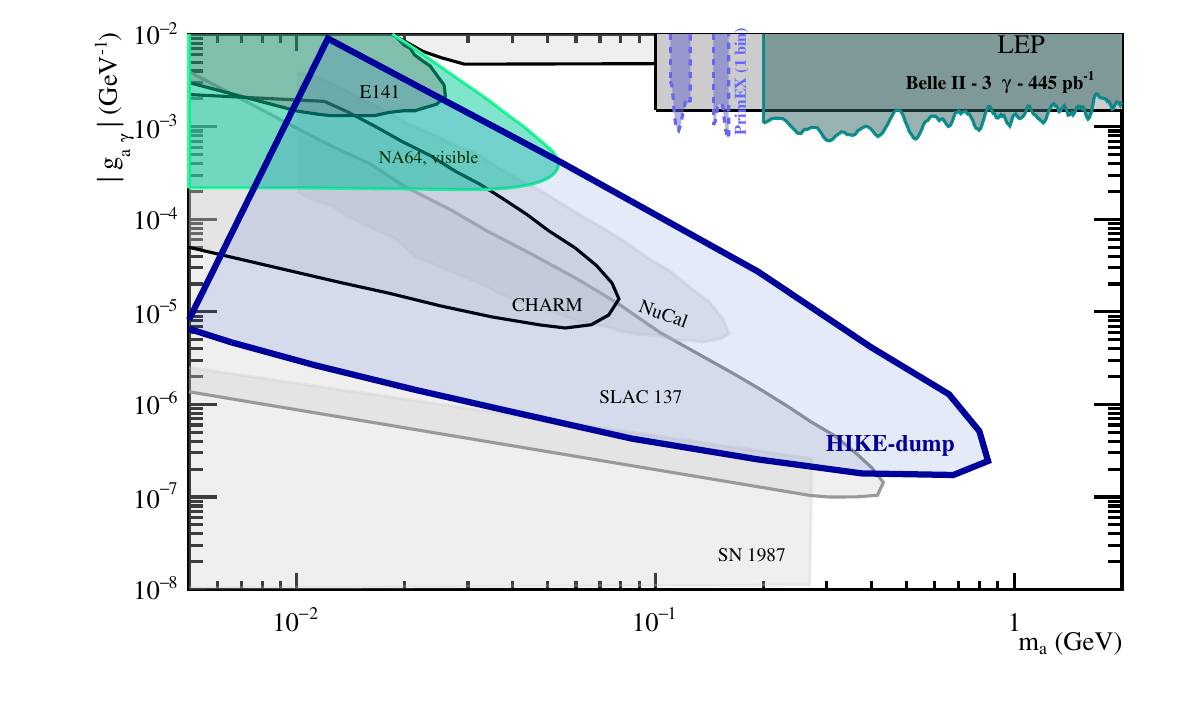}%
\put(-365,115){\LARGE BC9}\\
\vspace{-5mm}
\includegraphics[width=0.8\textwidth]{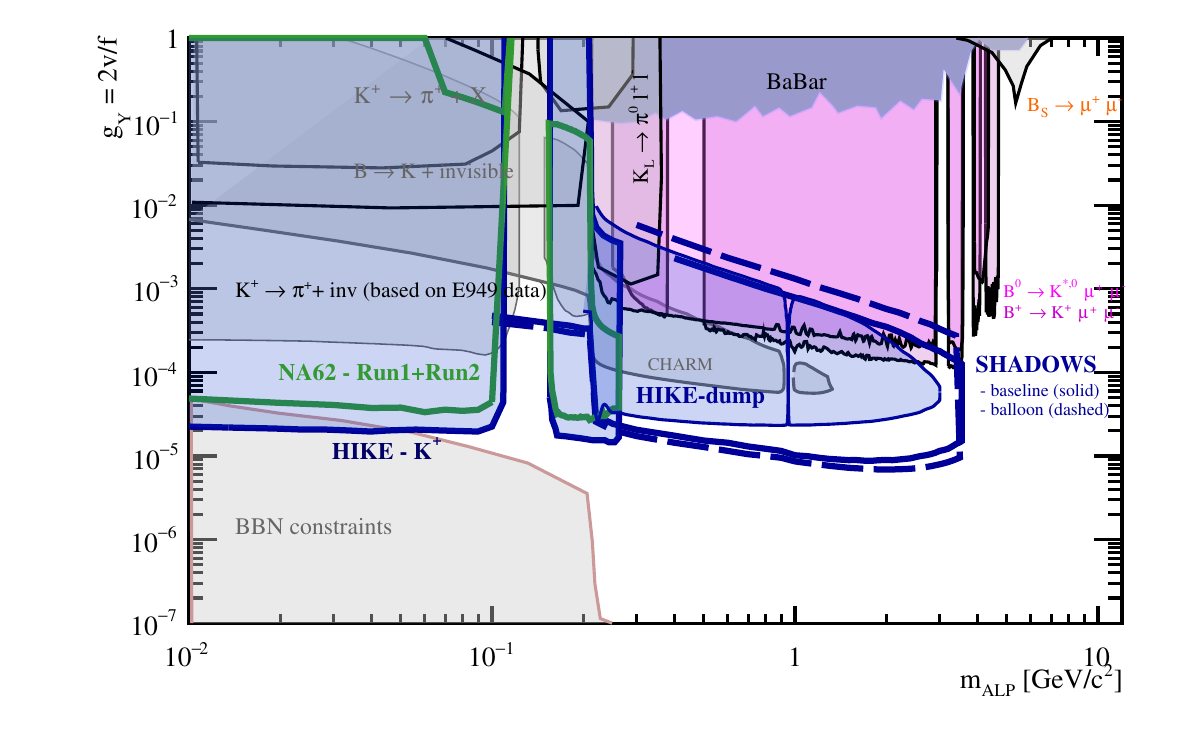}%
\put(-365,115){\LARGE BC10}\\
\vspace{-5mm}
~~~\includegraphics[width=0.72\textwidth]{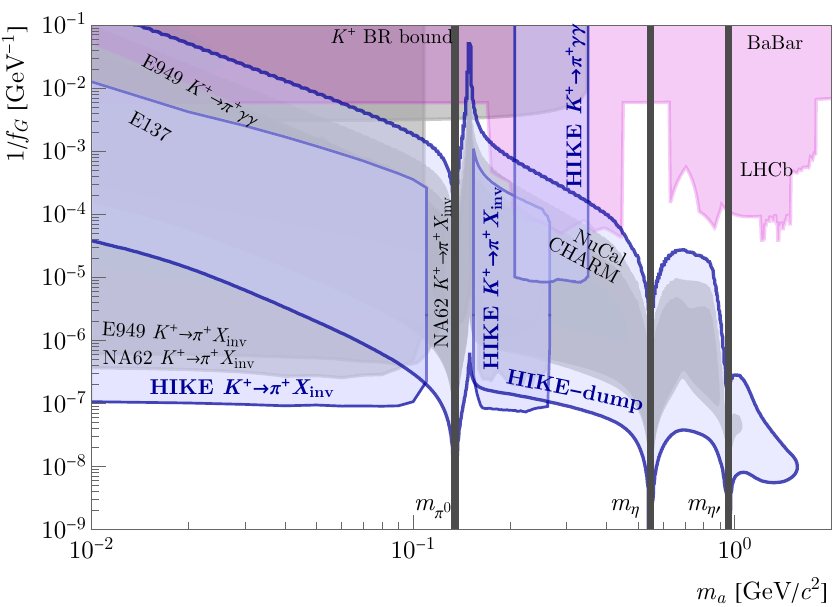}%
\put(-350,115){\LARGE BC11}\\
\vspace{-2mm}
\caption{HIKE (filled blue contours) and SHADOWS (empty  contour, scenario BC10 only) exclusion regions in the scenarios BC9,10,11: ALP with photon, fermion and gluon coupling. Existing experimental limits are also shown.}
\label{fig:bc9to11}
\end{figure}

The HIKE sensitivity for the PBC benchmark scenarios BC1--11 described briefly in Section~\ref{sec:FIPS:benchmarks} (with the exception of BC3 for which the sensitivity is yet to be evaluated) is summarised in Figs.~\ref{fig:bc1}--\ref{fig:bc9to11}. The PBC conventions for the coupling constants~\cite{Beacham:2019nyx} are used in all plots. Note that the PBC convention for the ALP couplings is different from the widely used convention of Refs.~\cite{Aloni:2018vki,Afik:2023mhj}.
For the scenarios accessible in both kaon and beam-dump modes of operation, HIKE sensitivity in both modes is shown. Also shown are the reach of the proposed SHADOWS experiment which will be collecting data in the beam-dump mode simultaneously with HIKE, and the existing limits from past experiments. Note that the Big Bang Nucleosynthesis (BBN) and supernova (SN1987A) exclusion regions, shown as grey areas in the plots, are strongly model-dependent in contrast to the HIKE and SHADOWS projections.

The combined sensitivity of HIKE and SHADOWS is evaluated assuming `Scenario~C' for the sharing of beam time between the kaon and beam-dump modes (see Section~\ref{sec:schedule} for details). This scenario is common between the two experiments, and assumes collection of $5\times 10^{19}$~POT in the dump-mode in 8~years from the arrival of the high-intensity beam. The details of SHADOWS sensitivity and the assumptions behind it are discussed in the SHADOWS proposal.

%%%%%%%%%%%%%%%%

%%%%%%%%%%%%%%%%%%%%%%%%%

\clearpage
 
%%%%%%%%%%%%%%%%%%%%%%%%%

\clearpage

%%%%%%%%%%%%%%

\section{High-intensity beams}
\label{sec:beam}

As the acronym HIKE suggests, the proton beamline must provide significantly higher beam intensities than those delivered currently to the NA62 experiment, on the order of 4 to 6 times of that currently provided, which would correspond to between 1.3 and $2 \times 10^{13}$ protons on the T10 target per SPS extraction. Here, we assume that a typical running year has 200~days with 3000 spills each, including an SPS uptime of about 80\%, and a flat top length of 4.8~seconds.

The proposed HIKE programme includes both a charged beam ($K^{+}$) phase and a neutral beam ($K_{L}$) phase. For both parts of the programme, protons at much higher intensity have to be transported via the P42 beam line from the T4 target to the kaon production target T10. From the beam transport and optics point of view, the $K^+$ beam is proposed to be the same as the present beam line given the optimisations and improvements done thanks to the experience of running NA62, but consolidation and upgrades are needed to handle the higher beam intensities, in particular for the technical infrastructure, including a service strategy to guarantee reliable beam operation during the proposed running time. Improved beam instrumentation will provide better diagnostics, help to monitor the beam online, and reduce beam losses. A more precise calibration of the primary beam intensity measurements by the secondary emission monitors at T4 and T10 is important for beam optimisation and for normalisation purposes in dump mode. Efforts in this respect have already been started by the SY-BI, SY-ABT, and BE-EA groups at CERN with the help of the ECN3 beam delivery task force established in the framework of Physics Beyond Colliders. The neutral beam line is a new design that would replace the current K12 beam line, profiting from the experience with the NA48 experiment, and requires relatively minor local modifications to the last section of the P42 line.

%%%%%%%%%%%%%%

\subsection{Beam delivery to the kaon production target}
\label{sec:proton_beam}

The P42 beam line started operation in 1980, and has hardly been modified since then~\cite{Banerjee:2774716}. Most of the equipment dates from then and is showing strong effects of ageing. Many issues and suggestions for mitigation have been described in the reports from the PBC Conventional Beams working group~\cite{Gatignon:2650989}. The first phase of a vast consolidation programme for the CERN North Area (NA-CONS) has been approved and funded, and the second phase has been prepared~\cite{Kadi:2018,Kadi:2019,Kadi:2021}. This will allow restoration of the equipment and beam infrastructure to a reliable state. Some upgrades are required to cope with the up to six times higher beam intensity, including operational requirements from the rest of the North Area complex and with stricter safety and radiation protection requirements compared to the original installation. Many studies have been performed by the SPS Losses and Activation working group (SLAWG)~\cite{Balhan:2668989, Fraser:ipac17-mopik045, Fraser:ipac19-wepmp031}, the Physics Beyond Colliders (PBC) Proton Delivery working group~\cite{KoukoviniPlatia:2675225}, the PBC Conventional Beams working group~\cite{Gatignon:2650989}, and more recently, by the PBC ECN3 beam delivery task force~\cite{Brugger:2022}, focusing on the possibility to extract and deliver such high intensities towards TCC8 and ECN3. A comprehensive summary can be found in the PBC Report ``Post-LS3 Experimental Options in ECN3''~\cite{Arduini:2023}.

\begin{figure}[tb]
\centering
\includegraphics[width=\textwidth,height=8.5cm]{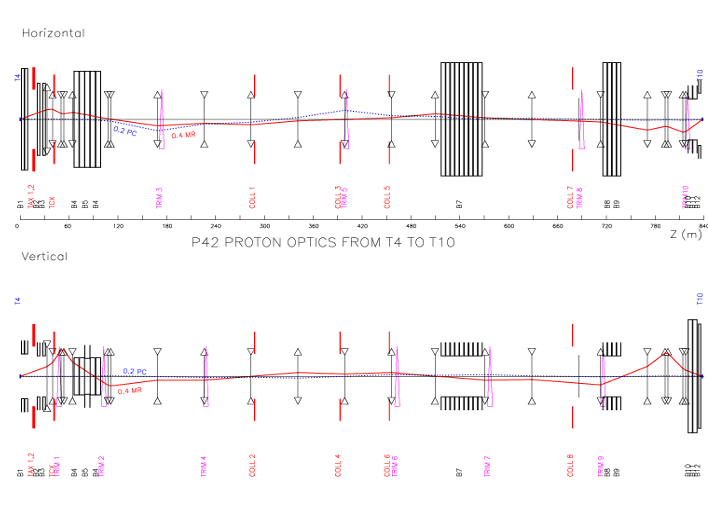}
\vspace{-11mm}
\caption{Beam optics for the P42 primary beam line, delivering protons from the T4 production target to the T10 kaon production target. The red line depicts the local values of the R12 (horizontal) and R34 (vertical) beam transport matrix terms. The dotted blue line shows the local dispersion terms R16 (horizontal) and R36 (vertical).}
\label{fig:P42optics}
\end{figure}

The present optics of the P42 beam is shown in Fig.~\ref{fig:P42optics}. In order to reduce the attenuation of the primary beam in the T4 target, several options have been considered. According to the aforementioned studies, the preferred option is to run dedicated cycles for ECN3 in which the beam is bumped to pass above or below the target plate. In principle, this could allow even higher intensities than $2 \times 10^{13}$ protons-per-pulse (ppp) on the T10 target, and intensities for beam-dump running up to $2.4 \times 10^{13}$ could be envisaged with the current design of the TAX (described below).

\begin{figure}[tb]
\centering
\includegraphics[width=0.62\textwidth]{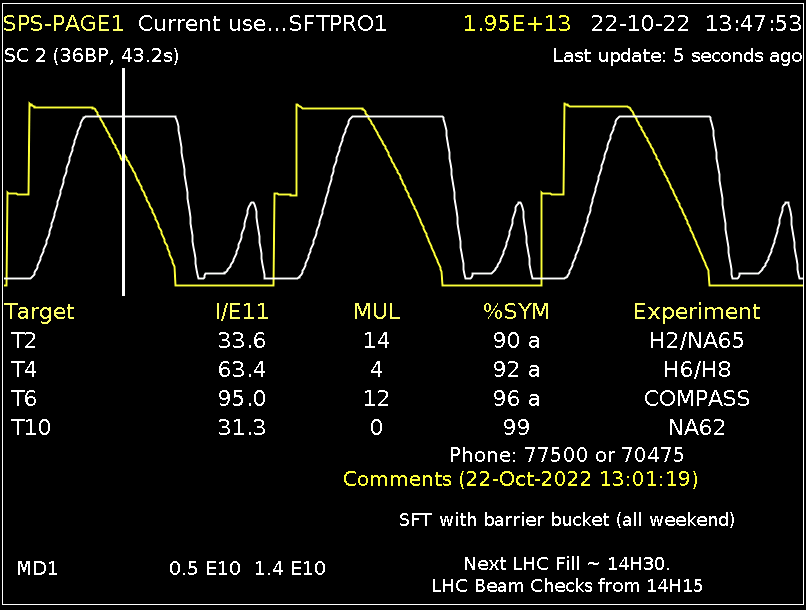}
\caption{The spill configuration used on 22 October 2022, as seen from SPS Page~1.}
\label{fig:screenshot}
\end{figure}

At present, the beam is provided in spills of 4.8~seconds, typically twice per super-cycle of about 40 to 45~seconds during day time, and two equally-spaced spills during 33.6~seconds during nights. One spill every 14.4~seconds is also technically possible, and is the standard cycle adopted during the night in 2023 (Fig.~\ref{fig:screenshot}). 

The PBC ECN3 Beam Delivery Task Force has concluded that, in a dedicated beam delivery scenario without beam splitting in TDC2 and without impacting the targets in TCC2 (including T4 serving H6 and H8, the requirements of all PBC LoIs (HIKE, SHADOWS, and SHiP) can be met while continuing to service the other NA experiments and users in EHN1 and EHN2~\cite{Brugger:2022}. 
More specifically, the preferred beam delivery scenario identified by the task force is a mixed super-cycle with the present-day SFTPRO cycles for EHN1 and EHN2 along with dedicated high-intensity cycles for ECN3. HIKE running with $2\times10^{13}$~ppp and a 4.8~s flat top corresponds then to about $1.2\times10^{19}$~POT/year in ECN3 (POT = protons on target).

In recent years, on average 3000~spills per day were delivered effectively, including the typical availability of SPS beams of about 80\%. The figure of merit is the time on flat top per year, which is proportional to the duty cycle and the running time. It is also important that the beam intensity is delivered as uniformly as possible. At the time of the NA62 proposal, one assumed an effective spill of 3.3~seconds, but thanks to recent improvements, we anticipate that it can be significantly better, ideally better than 4~seconds for a 4.8~second spill. The spill length can be modified to some extent, but it is important to maintain the duty cycle as well as the instantaneous intensity, i.e. extracted protons per second during the spill.

\begin{figure}[tb]
\centering
\includegraphics[width=0.9\textwidth]{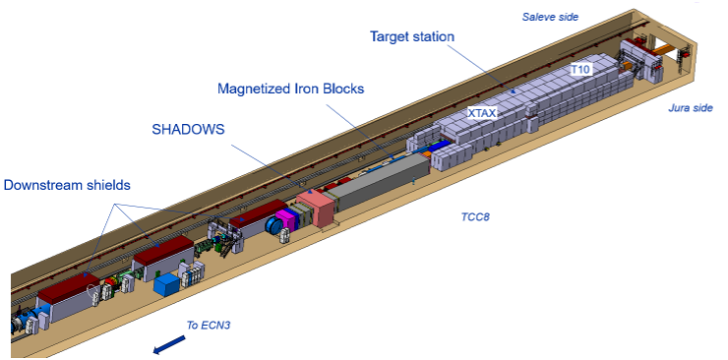}
\includegraphics[width=0.9\textwidth]{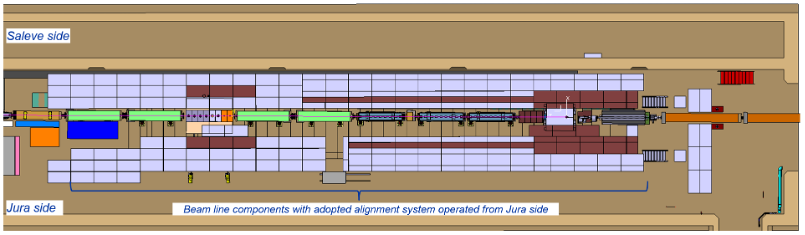}
\caption{Integration model of the TCC8 target-TAX region. On the top, the main features and locations of target and TAX are visible, surrounded by the massive RP shielding. Infrastructure and potential location of the SHADOWS experiment can be seen, as well. On the bottom, details within the area shielding are depicted, with the P42 proton beam entering from the right side.}
\label{fig:integration_target_TAX}
\end{figure}

Studies by the SY-STI group in the PBC context suggest that the TAXs in P42 and K12 are already running close to the maximum intensity allowed~\cite{Gatignon:2650989}. An upgrade is being studied at the technical level in the framework of the ECN3 task force. This implies a substantial upgrade of the T10 target-TAX complex to withstand the higher beam power and to be compatible with RP and maintenance requirements. A first integration for this new complex has been successfully concluded, also addressing the requirements and constraints of potentially adding the SHADOWS experiment in TCC8. An overview of the integration model is depicted in Fig.~\ref{fig:integration_target_TAX}, featuring a massive new shielding configuration surrounding the new complex and indicating the potential position of SHADOWS.  

A new target design was inspired by the concepts of the CNGS target and is currently being studied.
The target material is proposed to be changed from beryllium to carbon for better handling and cooling. However, more dense target materials can also be considered as an option outside of the present baseline, as they could be advantageous in particular for the neutral beamline, in preparation of a later phase of the experiment. TAX holes (apertures) will be defined to ensure compatibility with the charged beam as well as the neutral beam.  A different choice of material in the TAX blocks (Cu+Fe) and target collimator will be required, as well as cooling closer to the beam impact point for the TAX.

The present interlocks protect both the beam line equipment and the NA62 experiment against incorrect magnet currents, T10 target and K12 TAX cooling problems, and closed vacuum valves in the K12 beam line. For the higher intensities in the future, this system must be sped up. This is planned within the NA-CONS programme baseline, independently of the proposed high-intensity upgrade for ECN3, profiting from a new generation of power converters. On the shorter time scale, some improvements are already under consideration, for instance using the software interlock system to dump the beam directly in the SPS, compatible with the presented HIKE timeline.  
Many studies are on-going concerning radiation dose levels to equipment and on the surface, which must remain under control also at the higher intensities. All these issues are addressed in the report from the ECN3 beam delivery task force~\cite{Brugger:2022}.

For HIKE Phase 2, the kaon beam production angle has to be increased to 2.4~mrad from the zero angle used for the charged mode. This can be achieved without modifications of the P42 beam line. In case a higher production angle of up to 8~mrad is needed, the three last dipoles will need to be realigned, which is a minor modification as documented in the Conventional Beams Working Group Report~\cite{Gatignon:2650989}. The high-intensity proton beam will ultimately impinge on the kaon production target, T10, which should be equipped with another target head with larger diameter of 4~mm in order to accommodate such a possibility in addition to the one for Phase 1 and production angles smaller than 2.4~mrad, which has a diameter of 2~mm.
The target is then followed by the TAX beam dumps. Only for Phase 2, an additional proton beam dump closer to the T10 target is envisaged as it would provide better mitigation of muonic backgrounds. The detailed design of this object would be subject to the upcoming TDR design and preparation phase.

%%%%%%%%%%%%%%

\subsection{Charged kaon beamline}

\begin{figure}[tb]
\centering
\vspace{-6mm}
\includegraphics[width=0.8\textwidth]{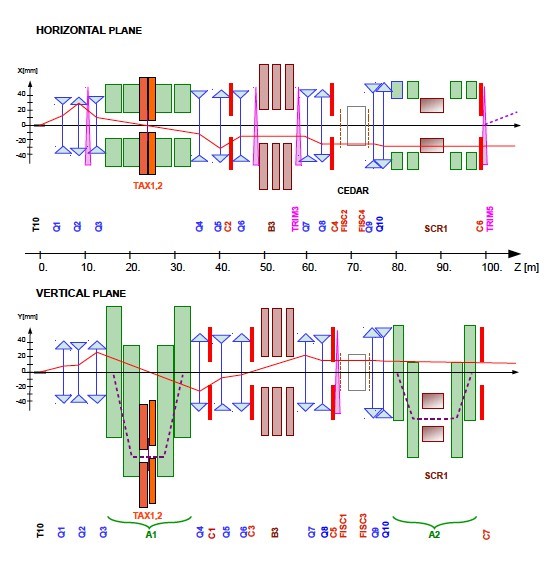}
\vspace{-6mm}
\caption{Beam optics for the K12 beamline. The red line depicts the local values of the R12 (horizontal) and R34 (vertical) beam transport matrix terms. The dotted black line shows the local dispersion terms R16 (horizontal) and R36 (vertical).}
\label{fig:K12optics}
\end{figure}

The HIKE charged kaon beamline will remain conceptually as it is now, with essentially the same beam properties apart from the intensity (Fig.~\ref{fig:K12optics}). Consolidation of some equipment is required in the framework of NA-CONS, such as the K12 machine protection system in synergy with the P42 interlock upgrade. The main upgrades concern the T10 production target, the K12 TAX, including the needed shielding as well as the beamline elements between target and TAX, which will have to be rebuilt with new, radiation-hard magnets with provision of full remote handling and busbar connections. A pre-study presented in the Conventional Beams Working Group report~\cite{Gatignon:2650989} has been followed up further. The beamline will allow for a beam-dump mode similar to what is available now, i.e. moving out the T10 target head and moving the TAX into the dump position. This mode change is done fully remotely and takes 15--20 minutes.

The K12 beam is a mixed, positively charged beam containing 6\% of kaons at 75~GeV/$c$. A study has been performed to evaluate if the beam can be enriched by RF separation. This requires a much longer beamline, implying that more kaons decay before the fiducial volume, which decreases both the kaon flux and the initial kaon fraction in the beam before enrichment. An RF-separated beam cannot fulfil the HIKE requirements, even with state-of-the-art RF systems~\cite{Appleby:2765940}.

%%%%%%%%%%%%%%%%%

%%%%%%%%%%%%%%

\subsection{Neutral kaon beamline}
\label{sec:neutral_beam}

\begin{figure}[p]
\centering
\includegraphics[width=0.95\textwidth]{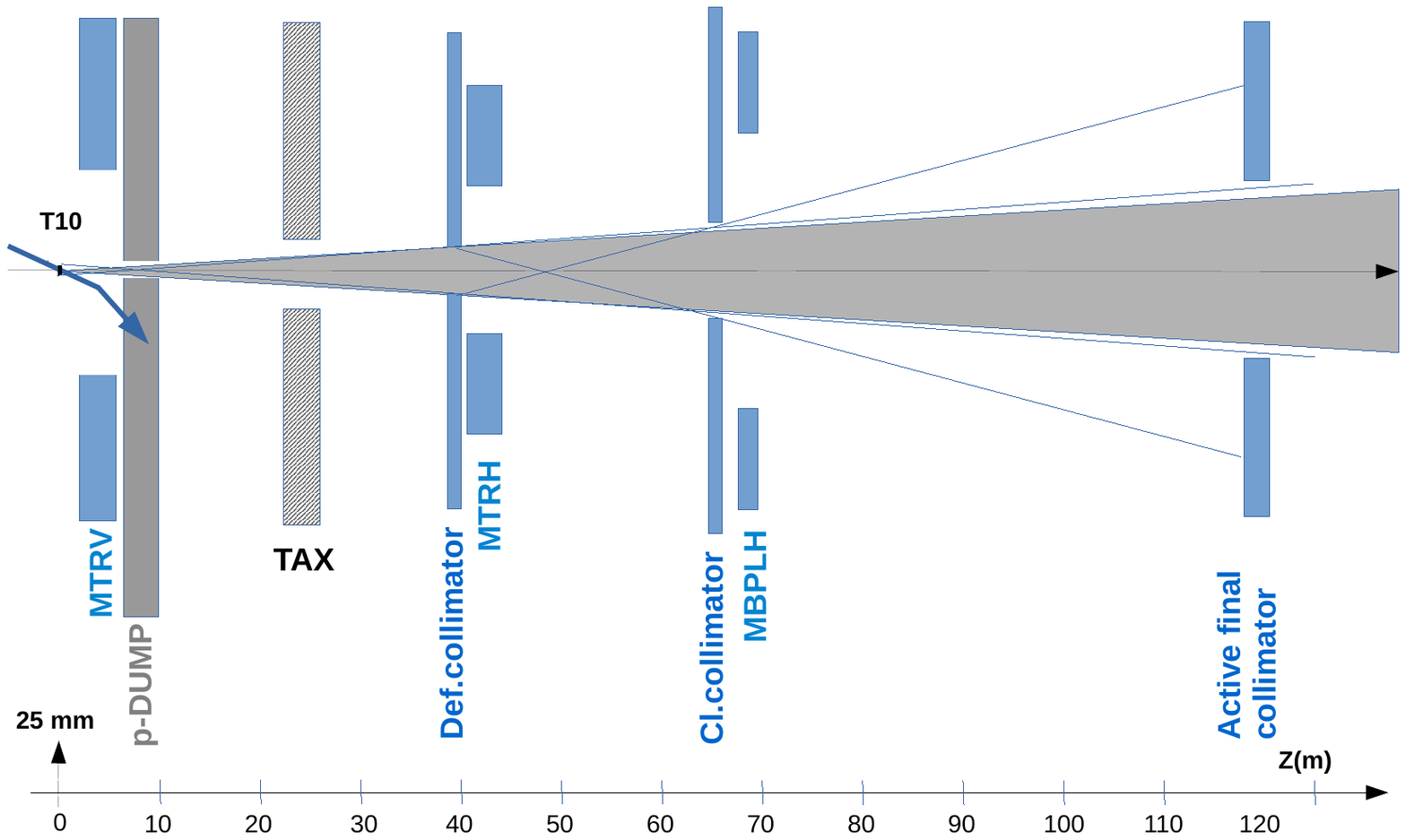}
\vspace{-14mm}
\caption{Baseline concept of the HIKE Phase~2 neutral beam. Several collimation steps define the beam and then clean it from debris stemming from scattering off the collimator edges.}
\label{fig:phase2_beam}
\end{figure}

\begin{figure}[p]
\centering
\includegraphics[width=0.7\textwidth]{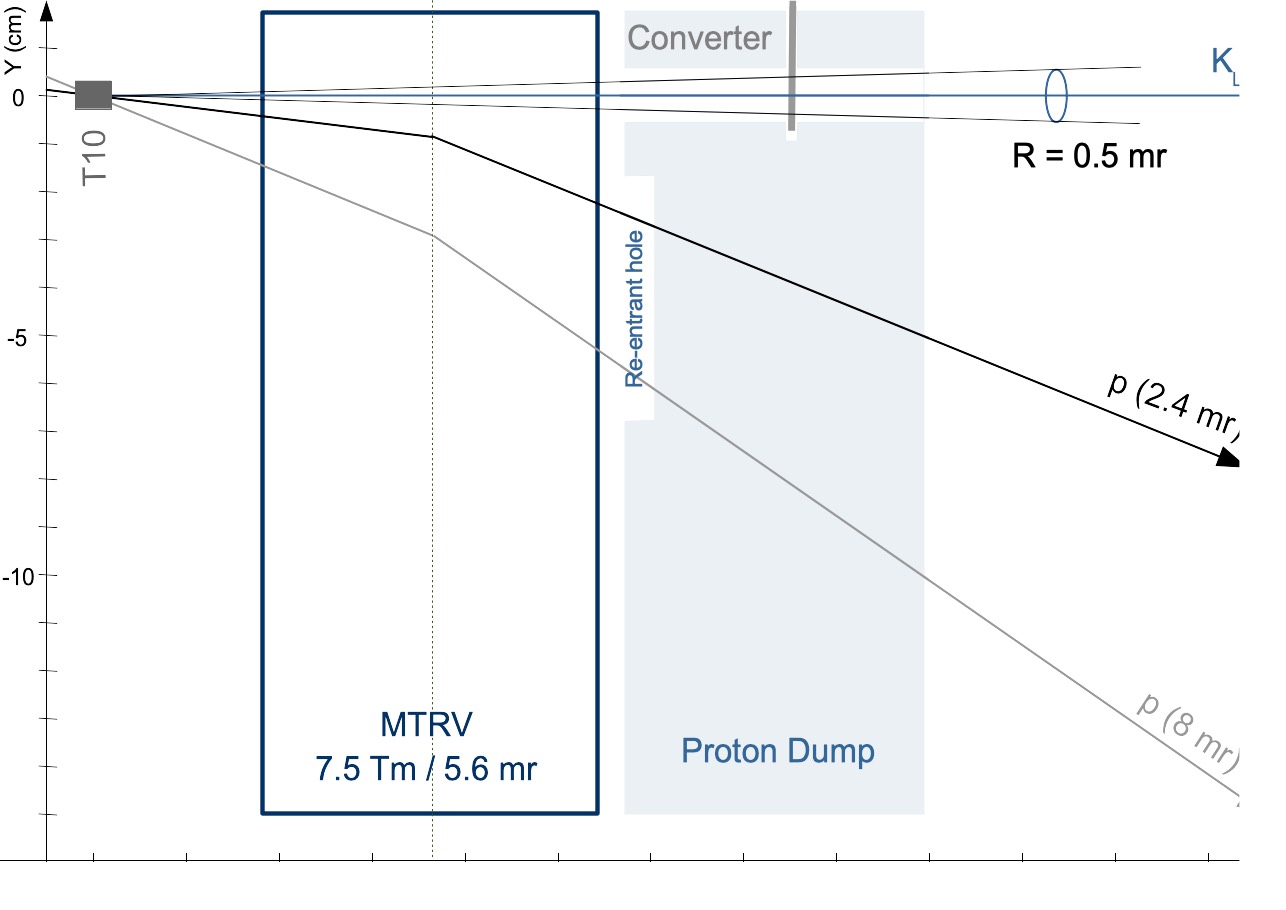}
\vspace{-2mm}
\caption{Conceptual view of the beam front-end for HIKE Phase~2.}
\label{fig:phase2_target_region}
\end{figure}

The design and simulations for the neutral beamline were performed by the Conventional Beams Working Group~\cite{Gatignon:2650989}. The design (Figs.~\ref{fig:phase2_beam}, \ref{fig:phase2_target_region}), based on the experience with the $K_L$ beamlines for NA31 and NA48 following the guidelines laid out in Ref.~\cite{Gatignon:2730780},
forms the basis for the HIKE Phase~2 beamline, and can be extended in length for a measurement of the $K_L\to\pi^0\nu\bar\nu$ decay (HIKE Phase~3) in the future. Further details can be found in the HIKE Letter of Intent~\cite{HIKE:2022qra}.
After removal of the charged beamline elements, the neutral beamline would start from the present T10 location. Its length is about 120~m, which is 18~m longer than the charged beamline. The primary proton beam impinges on the T10 target with a downward vertical angle of 2.4~mrad, which can be increased to 8~mrad as and if required for tests towards a possible future HIKE Phase~3. This choice balances several factors for the $K_L\to\pi^0\ell^+\ell^-$ measurements (Section~\ref{sec:phase2}).

The target is immediately followed by a first collimator that stops hadrons outside the beam acceptance before they decay into muons. The non-interacting protons are swept further downward by a strong dipole magnet and dumped in a dedicated proton dump. This dump would be an addition to the TAX of Phase~1. Alternatively, the proton beam could also be transported to the Phase~1 TAX in order to dump the beam protons, for which studies are currently being conducted. For background estimates, we assume having a proton dump as the baseline.
In both cases, a photon converter made of 9$X_0$ of high-$Z$ material, positioned either at the centre of the proton dump in a dedicated compartment or in the centre of the TAX, reduces the flux of high-energy photons ($E_\gamma > 5$~GeV) in the neutral beam by two orders of magnitude. A thinner, oriented crystal converter is an optional possibility under study. In a next step, three collimators define the neutral beam acceptance in a clean way. The defining collimator is located at $1/3$ of the distance to the final collimator and defines the beam angular acceptance of $\pm$0.4~mrad, matching the size of the central bore in the proposed HIKE calorimeter (Section~\ref{sec:calorimeter}).
A cleaning collimator stops debris from scatterings in the jaws of the defining collimator, and a final collimator stops scattering products from the cleaning collimator. All collimators comprise surfaces that do not face both target and detectors in order to reduce background from scattering off these surfaces as much as possible. Charged background from inelastic interactions at the collimators is reduced further by introducing strong sweeping magnets with apertures larger than the beam acceptance. The active final collimator defines the start of the fiducial volume. Sufficient space is available to add more collimation stages, depending on future design iterations within the Conventional Beams Working Group. In addition, the current design allows to install a dedicated veto detector either in front of or surrounding the active final collimator if deemed necessary (Section~\ref{sec:neutral_detectors}).
Integration studies for the neutral beam line are in progress, with a detailed document in preparation, including possible 
modifications of the Phase-2 beam layout and of the cost estimate.

%%%%%%%%%%%%%%%%%%

\section{HIKE detectors}
\label{sec:detectors}

The general aim of the HIKE detectors is to achieve the NA62 detector performance at a factor 4--6 higher beam intensity, with suitable time resolution and rate capability. 

The setup and detectors in Phase~1 (the $K^+$ phase) will be optimised for the $K^+\to\pi^+\nu\bar\nu$ branching ratio measurement to $\mathcal{O}$(5\%) precision.  While the conceptual layout is based on the successful one of NA62, new detectors will replace those of NA62 with the goal of improving the performance and sustaining higher rates. The prime examples are the beam tracker and the straw spectrometer. Thanks to the relatively compact detector, Phase~2 with the neutral beam allows for a 90 m long decay volume to be accommodated in the present ECN3 experimental hall, with no major civil engineering work. Phase~2 will use an experimental setup with minimal modifications with respect to the $K^+$ phase. The beam tracker, kaon-identification, pion-identification will be removed;
the small-angle-calorimeter will be moved; 
the main tracking spectrometer will be shortened, and central holes of the chambers will be realigned on the neutral beam axis; the large-angle veto detectors will be repositioned and possibly reduced in number. 
The EM calorimeter for the neutral phase must be able to operate in an environment with intense soft-photon and neutron fluxes. Other detectors will remain unchanged.
A summary of detector configurations is presented in Table~\ref{tab:hike-detectors}.
Rates in the main detectors due to the principal kaon decays in HIKE Phase 2 estimated using detailed beamline simulations  (Section~\ref{sec:beam}) are summarised in Table~\ref{tab:ratesKL}.

\begin{table}[b]
\caption{Average and maximum particle rates for specified checkpoints at the detector entry, along the length of the setup in HIKE Phase~2. The particle rates are collected from the cross-section of the outer volume for the subdetector. The last column represents the rate from electrons, muons and pions together.
A geometrical cut is applied to remove particles that pass though the checkpoints within the beam envelope.}
\vspace{-7mm}
\begin{center}
\begin{tabular}{|l|clllllllll|}
\hline
\multicolumn{1}{|l|}{\multirow{3}{*}{Checkpoint}} & \multicolumn{10}{c|}{Particle rates, kHz/cm$^{2}$} \\ \cline{2-11} 
\multicolumn{1}{|c|}{} & \multicolumn{2}{c|}{$e^{+/-}$} & \multicolumn{2}{c|}{$\mu^{+/-}$} & \multicolumn{2}{c|}{$\pi^{+/-}$} & \multicolumn{2}{c|}{$\gamma$} & \multicolumn{2}{c|}{Charged particles} \\ \cline{2-11} 
\multicolumn{1}{|l|}{} & \multicolumn{1}{c|}{Max} & \multicolumn{1}{c|}{Avg} & \multicolumn{1}{c|}{Max} & \multicolumn{1}{c|}{Avg} & \multicolumn{1}{c|}{Max} & \multicolumn{1}{c|}{Avg} &  \multicolumn{1}{c|}{Max} & \multicolumn{1}{c|}{Avg} & \multicolumn{1}{c|}{Max} &
\multicolumn{1}{c|}{Avg} \\
\hline
STRAW1 entry$\!$ & \multicolumn{1}{c|}{264} & \multicolumn{1}{c|}{0.26} & \multicolumn{1}{c|}{224} & \multicolumn{1}{c|}{1.58} & \multicolumn{1}{c|}{117} & \multicolumn{1}{c|}{0.41} & \multicolumn{1}{c|}{13426} & \multicolumn{1}{c|}{13.87} & \multicolumn{1}{c|}{360} & \multicolumn{1}{c|}{2.56} \\ \hline
MEC entry & \multicolumn{1}{c|}{85} & \multicolumn{1}{c|}{1.76} & \multicolumn{1}{c|}{174} &  \multicolumn{1}{c|}{1.12} & \multicolumn{1}{c|}{109} & \multicolumn{1}{c|}{0.24} & \multicolumn{1}{c|}{166} & \multicolumn{1}{c|}{0.30} & \multicolumn{1}{c|}{238} & \multicolumn{1}{c|}{3.25}\\ \hline
HCAL entry  & \multicolumn{1}{c|}{5} & \multicolumn{1}{c|}{0.01} & \multicolumn{1}{c|}{190} & \multicolumn{1}{c|}{1.04} & \multicolumn{1}{c|}{32} & \multicolumn{1}{c|}{0.03} & \multicolumn{1}{c|}{29} & \multicolumn{1}{c|}{0.01} & \multicolumn{1}{c|}{190} & \multicolumn{1}{c|}{1.07}\\ \hline
MUV entry  & \multicolumn{1}{c|}{5} & \multicolumn{1}{c|}{$10^{-4}$} & \multicolumn{1}{c|}{115} & \multicolumn{1}{c|}{0.89} & \multicolumn{1}{c|}{3} & \multicolumn{1}{c|}{$10^{-4}$} & \multicolumn{1}{c|}{3} & \multicolumn{1}{c|}{$10^{-3}$} & \multicolumn{1}{c|}{115} & \multicolumn{1}{c|}{0.88} \\
\hline
\end{tabular}
\end{center}
\vspace{-3mm}
\label{tab:ratesKL}
\end{table}

HIKE, with new or upgraded detectors and readouts to profit from the increased beam intensity, will improve the acceptance of kaon decays and keep the random veto under control at much higher intensity. 
The basic technologies for HIKE detectors all exist already, or a prototype is foreseen in the very near future. The detectors are challenging but smaller in size or quantity compared to many other experiments. Several detectors are synergetic to on-going developments for High-Luminosity LHC~(HL-LHC) experiments.
A beam intensity uniformly distributed over the 4.8~s
spill is essential in all phases, to optimally collect high statistics while effectively managing detector rates and spurious intensity effects.

The HIKE beam dump operation will build upon the NA62 experience. 
Proton beam will interact in the T10 TAX; switching between kaon and dump mode will be possible during a 8-hour SPS Machine Development~(MD) time slot. The detectors used are the same as the kaon mode (with a much lower occupancy), plus the ANTI-0 to veto beam-halo muons.

Details of specific technologies envisaged for detectors and readout systems are provided in this section.
While the current proposal refers to HIKE Phases~1 and~2 only, the detector design should remain compatible with a possible future Phase~3 (please refer to the HIKE Letter of Intent~\cite{HIKE:2022qra}).

\subsection{Detectors upstream of the decay volume}

%%%%%%%%%%%%%%%%%%%%%%
\subsubsection{Cherenkov kaon tagger (KTAG)}
\label{sec:ktag}

A differential ring-focusing Cherenkov detector used for $K^+$ tagging in the NA62 experiment at a kaon rate of 40~MHz provides tagging efficiency above 95\%~\cite{NA62:2017rwk,Goudzovski:2015xaa}. The detector uses hydrogen at a pressure of 3.8~bar as the radiator gas, and has a total thickness of 0.73$X_0$ in the beam. A diaphragm placed in the focal plane transmits only the Cherenkov photons emitted by kaons. Ring-imaging optics focuses these photons onto eight spherical mirrors reflecting them onto eight photodetector arrays. Each array is equipped with a matrix of 48 Hamamatsu R7400-U03 or R9880-210 single-anode phototubes (PMTs) with peak quantum efficiency (QE) of 20--40$\%$. The average rate of detected photons per PMT depends on the QE and PMT position and varies between 3--5~MHz, for an effective area of 2.5~cm$^2$. The single-photon time resolution is 300~ps, and the 22 Cherenkov photons are detected per kaon on average, leading to a 65~ps kaon time resolution.

For HIKE Phase~1, the KTAG must to provide a time resolution of 15--20~ps with a tagging efficiency above 95\%. The vessel containing the radiator gas and ring-imaging optics, along with the gas distribution system (both designed and constructed specifically for NA62) are fully suitable for HIKE operation. However the expected $K^+$ rate in the HIKE beam is about 200~MHz, corresponding to a 10~MHz/cm$^2$ maximum rate of detected photons, including a safety margin.
New photodetection, front-end and readout systems are required to satisfy the rate requirements.

Photodetection devices under consideration for the HIKE KTAG detector are micro-channel plate photomultipliers (MCP-PMTs) which are compact devices providing a single-photon time resolution of 50--70~ps with a low dark count rate (below 1~kHz/cm$^2$), able to operate at the expected maximum photon rate of 10~MHz/cm$^2$. 
Assuming a gain of $10^6$, 200~days of runtime and 3000~spills per day, the above photon rate leads to an Integrated Anode Charge (IAC) of 5~C/cm$^2$ per year of data taking.

%%%%%%%%%%%%%%%%%

\begin{figure}[p]
\centering
\includegraphics[width=0.5\linewidth]{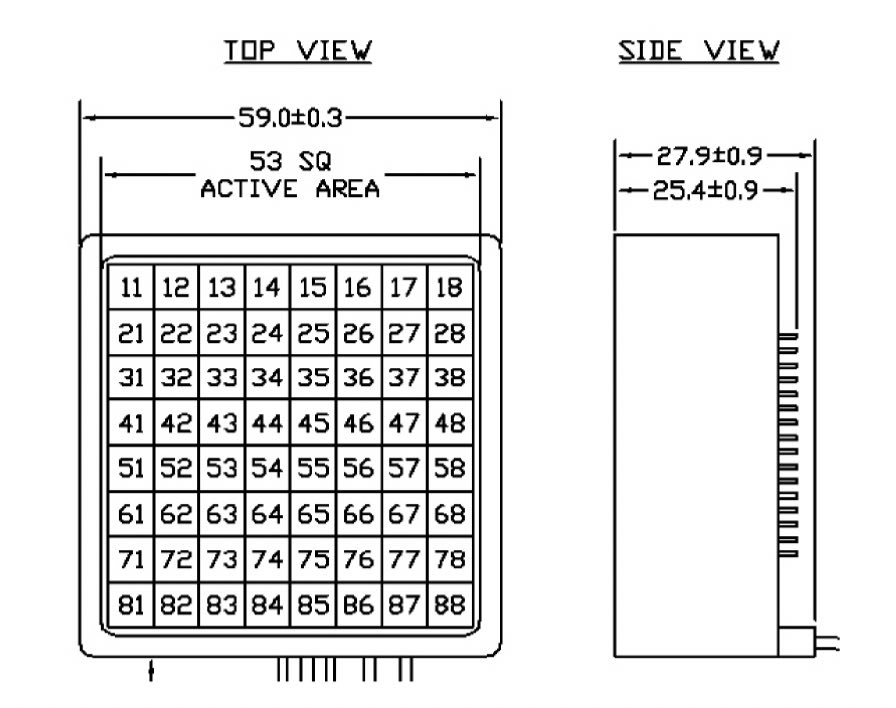}
\caption{Schematic drawing of a Photonis Planacon square MCP-PMT array.}
\label{fig:mcp-pmt}
\end{figure}

\begin{figure}[p]
\centering
\includegraphics[width=0.4\textwidth]{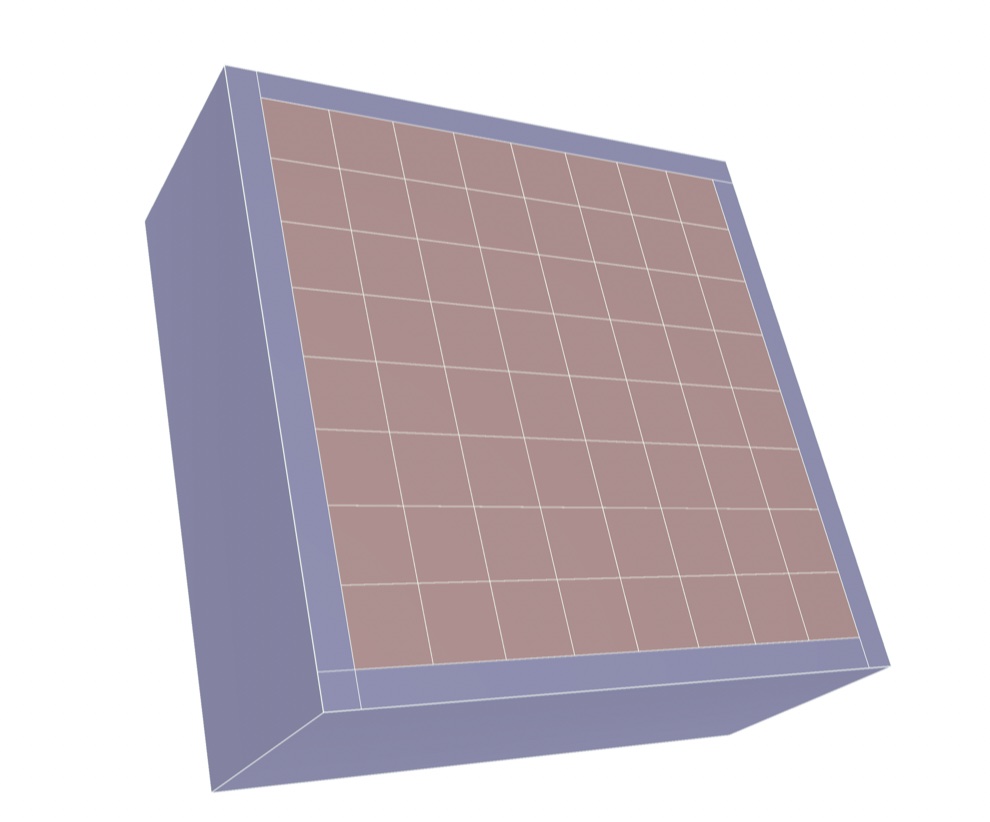}%
\includegraphics[width=0.4\textwidth]{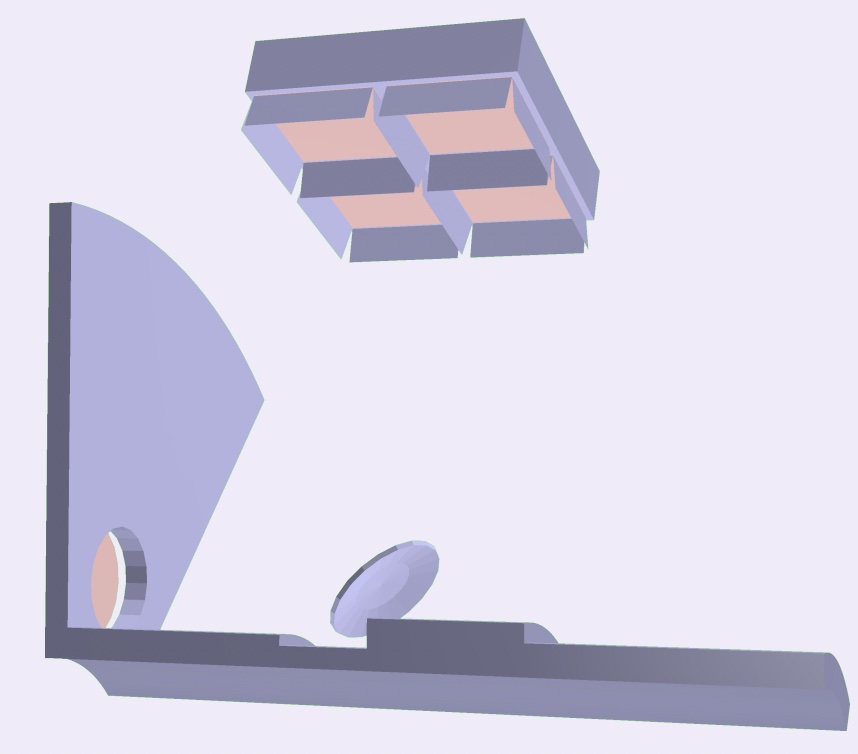}
\caption{Sketches of a single MCP-PMT array (left) and a matrix of four MCP-PMTs (right) to be used to detect photons in each of the eight octants of the HIKE KTAG detector.}
\label{fig:octant}
\end{figure}

\begin{figure}[p]
\centering
\includegraphics[width=0.5\linewidth]{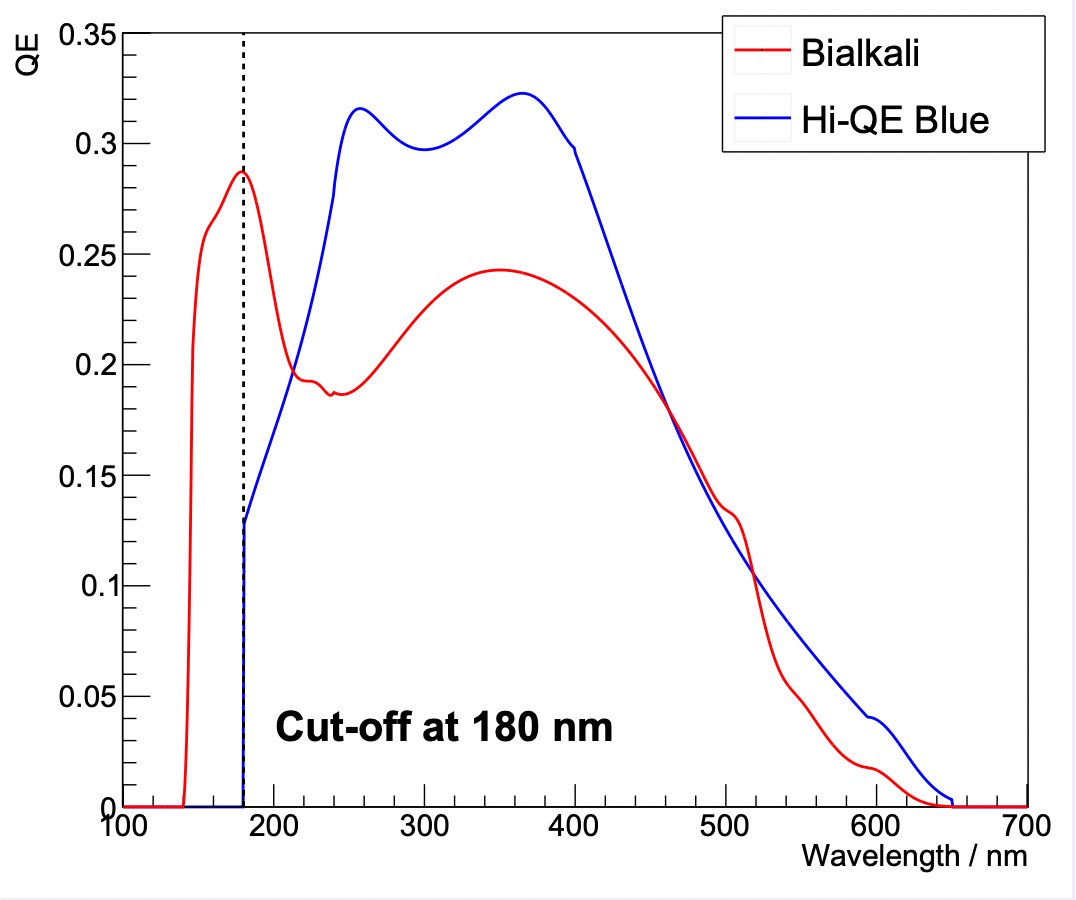}
\caption{QE of the Photonis Planacon MCP-PMTs as functions of wavelength for the two photocathode types considered (standard bialkali and Hi-QE Blue). A cutoff at 180~nm corresponds to the minimum wavelength of the Cherenkov photons reaching the photodetection system.}
\label{fig:QE}
\end{figure}

%%%%%%%%%%%%%%%

A limitation of standard MCP-PMTs is their limited lifetime, caused by a rapid decrease in the QE of the photocathode (PC) with increasing IAC. This ageing of the PC is caused by heavy feedback ions from the residual gas. For standard MCP-PMTs the QE was observed to drop by 50\% for IAC of 0.2~C/cm$^2$~\cite{Britting:2011zz}. 
MCP-PMT treatment with atomic layer deposition (ALD) coating increases the lifetime dramatically~\cite{Pfaffinger:2018huh}. The Planacon XP85112 from Photonis, a square-shaped MCP-PMT treated with an improved ALD technique involving two layers, has been demonstrated to reach a lifetime of 33~C/cm$^2$ without any sign of ageing~\cite{LehmanTBC}. This represents an improvement in lifetime by a factor of more than~150 with respect to standard MCP-PMTs, and a factor of at least~10 compared to earlier lifetime optimisation techniques.

A Photonis Planacon MCP-PMT of 2''$\times$2'' size consisting of $8\times 8$ pixel sensors (Fig.~\ref{fig:mcp-pmt}), treated with two-layer ALD coating, is a viable solution for the photodetection system of the KTAG in high intensity applications. 
The Cherenkov photons at each photodetector plane can be detected with a matrix of four MCP-PMTs (Fig.~\ref{fig:octant}). The MCP-PMT is available with standard bialkali and Hi-QE Blue photocathode types; the two QE parameterisations provided by the manufacturer are shown in Fig.~\ref{fig:QE}. Simulations considering a geometrical filling factor of 75\%
show that the expected number of Cherenkov photons detected per kaon in the HIKE conditions is similar to that of NA62, and a 15--20~ps kaon time resolution is achievable. The linearity of MCP-PMTs with rate at high gain is under investigation. However the rate stability can be adjusted by lowering the resistance and/or capacitance of the MCP, or by operating at lower gain ($5 \times 10^5$ instead of $10^6$).

Performance studies and characterisation of MCP-PMTs intended for use at HIKE
are starting at the University of Birmingham in September 2023. The complete setup for the measurements and a Photonis XP85112 Planacon MCP-PMT with double ALD coating layers (R2D2) and low-resistance MCP have been procured. Measurements of the analog signal shape, charge spectrum distribution, single-electron response (SER), single-photon time resolution (SPTR), QE, rate capability and lifetime for an integrated anode charge up to 50~C/cm$^2$ are foreseen. The MCP-PMT pulse shape and the light intensity will be monitored against changes in gain and light source temperature. Gain and dark count rates will be measured at regular intervals. QE measurements relative to the nominal value will be performed to assess stability vs integrated anode current. Gain measurements relative to the nominal value will be used to study linearity vs hit rates. 

The development of long lifetime and high-rate capability vacuum photodetector is supported by the detector R$\&$D collaboration at CERN (DRD4) and has technology synergies with the requirements of next generation experiments foreseen at HL-LHC. The R$\&$D activities on MCP-PMTs planned at the University of Birmingham, and finalised to ultra-fast timing single-photon detection capability with extended lifetime, are included into the DRD4 work plan.

The above HIKE kaon time resolution figures do not include contributions from front-end and readout systems. Development of read-out electronics capable to reach a few ps timing resolution is also supported within the DRD4 collaboration at CERN. The fastIC~\cite{Gomez_2022} and picoTDC~\cite{picoTDC2015Talk} electronics (Section~\ref{sec:ro-boards}) being currently developed for other applications are viable options. The FastIC technology provides a 20~ps time resolution and better linearity for signals from SiPM and MCP-PMTs. PicoTDC chips with 64~channels and 12~ps binning are able to provide a 4~ps time resolution. Both time contributions will only have a marginal effect on the KTAG timing capability.

\subsubsection{Beam tracker} 
The GigaTracker~\cite{Rinella_2019} is the beam tracker used by the NA62 experiment to measure the momentum and time of the beam particles with high momentum, angular and time resolutions.
The detector is made by planes of hybrid pixel silicon detectors, with a dipole magnets between the planes to allow the momentum measurement. For NA62 Run~1 (2016--2018) three silicon planes were used, then for NA62 Run~2 (2021--LS3) a fourth plane was added to further improve the tracking efficiency.
Each plane is made of a n-in-p planar sensor, with a thickness of \SI{200}{\micro \meter} and a sensitive area of $60 \times 27 ~\rm{mm^2}$, bump-bonded to 10 readout ASICs (TDCPix). The ASICs need to operate with a cooling system: the NA62 GigaTracker used a silicon microchannel cooling plate that represented the first use of this technology in high-energy experiments.
In order to minimise the probability of interactions between the beam particles and the detector, the material budget must be kept as low as possible: the total thickness of a single plane is around \SI{500}{\micro\meter}, namely $0.5\%X_0$.
The NA62 GigaTracker has achieved very good performances, namely a track time resolution of
${\cal O}(100~{\rm ps})$, an angular resolution of 16~$\mu$rad and a momentum resolution of 0.2\%.

Among all the above characteristics of the detector, the time resolution represents a limiting parameter for its operations in the context of HIKE. With more than four-fold instantaneous intensity increase, the time resolution should scale accordingly, namely from less than 200~ps to below 50~ps for a single hit. The radiation hardness of the new detector needs to be improved by at least a factor four if a replacement approach is to be preserved, or more in case of a single installation for the entire HIKE data taking campaign.
Table~\ref{tab:gtk_comparison} summarises the comparison between the current NA62 GigaTracker and the new detector needed for HIKE.

\begin{table}[tb]
\centering
\caption{Comparison between the NA62 Gigatracker main characteristics and the requirements for the HIKE beam tracker. One year corresponds to 200 days of beam.}
\vspace{-2mm}
\begin{tabular}{l|c|c}
\hline
&NA62 GigaTracker& New beam tracker \\
\hline
Single hit time resolution & < 200 ps & < 50 ps \\
\hline
Track time resolution & < 100 ps & < 25 ps \\
\hline
Peak hit rate & $2 ~\rm{MHz/mm^2}$ & $8 ~\rm{MHz/mm^2}$\\
\hline
Pixel efficiency & > 99\% & > 99\% \\
\hline
Peak fluence / 1 year [$10^{14} ~ 1~\rm{MeV~n_{eq}/cm^2}$] & 4 & 16 \\
\hline

\end{tabular}
\label{tab:gtk_comparison}
\end{table}

The interest for silicon detectors with fast timing information capable to operate in a high-radiation environment is shared among different particle-physics collaborations, including the LHC experiments
for the high luminosity phase of the collider. 
Following the 2020 Update to the European Strategy for Particle Physics, 4D tracking at high fluences was identified as one of the most urgently needed technologies for future detectors (2021 European Collaboration for Future Accelerators; Detector Research and Development Roadmap DRDT 3.2 and DRDT 3.3). 
Therefore several promising R\&D projects are ongoing and their outcome could be adopted for HIKE.

\begin{figure}[tb]
\centering
\includegraphics[width=0.75\textwidth]{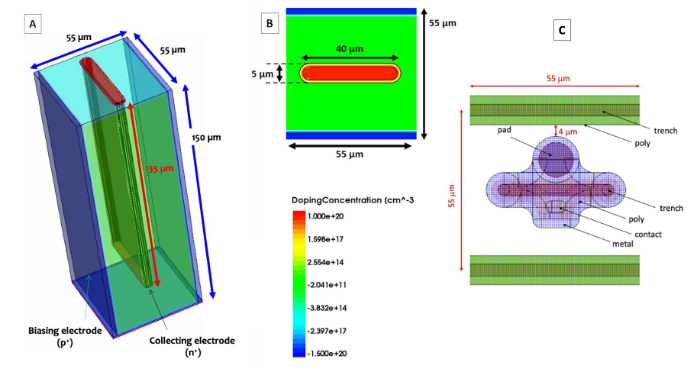}
\caption{Elementary cell of a 3D-trench silicon detector, as designed by the TIMESPOT project~\cite{Anderlini_2020}. Doping profiles are shown (n++ in red, p-- in green, p++ in blue). A) 3D rendering; B) pixel section; C) pixel layout.}
\label{fig:3dtrench}
\end{figure}
\begin{figure}[h]
\begin{center}
\includegraphics[width=0.75\textwidth]{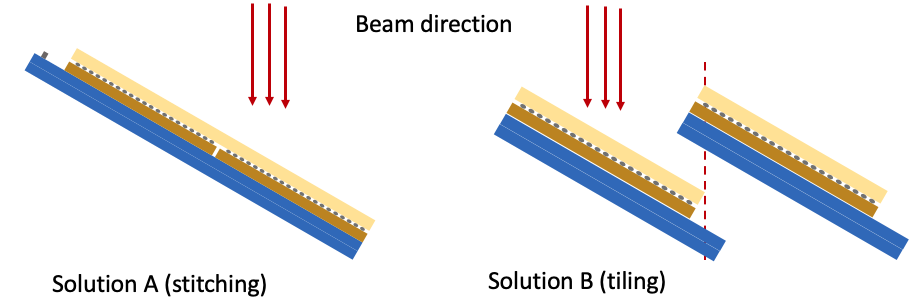}
\caption{Die stitching and module tiling solutions for HIKE beam tracker.}
\label{fig:timespot}
\end{center}
\vspace{-4mm}
\end{figure}

A strong option is the TimeSPOT project~\cite{hep-ph_Lai_2018,hep-ph_ignite_ALai_2022} 
which develops a technology called hybrid 3D-trenched pixels in which the pixel electrode geometry is optimised for timing performance.  A representation of one cell of a 3D-trench sensor is reported in Fig.~\ref{fig:3dtrench}. As for standard 3D-pixels, the sensors are able to withstand very large irradiation. The project has experimentally demonstrated that sensors with a pitch of \SI{55}{\micro\meter} provide a time resolution of \SI{10}{ps} up to fluence of \SI{2.5e16}{MeV~n_{eq}/cm^2}~\cite{hep-ph_Lampis_2022}. The TimeSPOT collaboration is planning to extend these tests up to fluences of \SI{1e17}{MeV~n_{eq}/cm^2}, that has never been achieved so far. Excellent detection efficiencies were also measured~\cite{hep-ph_Lampis_2022} by operating the sensor inclined by an angle of \SI{20}{\degree} with respect to the beam incidence. As the electrode itself is not sensitive, the inclination ensures that particle always cross some active pixel region. Sensor with a size of $2\times2$~cm$^2$ can be produced and technical solutions such as die stitching or module tiling can be used to obtain a detector plane with the dimensions required.
These two solutions are sketched in Figure~\ref{fig:timespot}. Die stitching has been already implemented at FBK on 3D devices; although dedicated  development for HIKE will be necessary, the feasibility of this technique has been demonstrated~\cite{Boscardin:2020two}. Module tiling requires precise mechanics for the alignment of sub-modules (as shown in the figure), and exploits the need of module tilting for 3D, in order to gain geometrical efficiency. Again, while the technique has to be perfected for HIKE, no significant technical difficulties are expected.

A first prototype of the ASIC, Timespot1, has been realised~\cite{hep-ph_CadedduEtAl_2022}. The chip contains a pixel matrix of $32\times 32$ elements, with a pixel pitch of $55 ~\rm{\mu m}$. Each pixel in the matrix contains an analogue and a digital circuit, with its TDC to measure the time of the arrival of the signals from the sensor. Digitised signals are then sent to the periphery of the matrix where 8~multiplexed output links send the data out at 1.28~Gb/s speed. Efforts are on-going to scale the size of the ASIC, with the aim of reaching an integrated ASIC on a $2\time2 ~\rm{cm}^2$ by 2026.
The ASIC specifications in terms of a power consumption is \SI{<1.5}{W/cm^{2}} which is comparable to the one of the existing GTK. Hence, the solution developed for the GTK cooling could be reemployed.
The chip and sensor production has been made by FBK, on an original design by the TimeSPOT project (there are no issues related to patenting).
All the above considerations make TimeSPOT with its ASIC Timespot1 a viable option for the HIKE beam tracker.

Other projects involving different technologies, such as monolithic pixel sensors~\cite{hep-ph_IacobucciEtAl_2022} or LGADs~\cite{hep-ph_PellegriniEtAl_2015,hep-ph_PaternosterEtAl_2020}, are also taken in consideration. Monolithic detectors can be made with very small thickness and excellent time resolutions, however they are expected to have a lower radiation tolerance compared to the 3D-trench sensors. The same concerns apply to the LGAD technology.

\subsubsection{Veto counter (VC)} 
\label{sssec:vetocounter}

One of the main sources of background to the $K^+\to\pi^+\nu\bar\nu$ measurement is induced by $K^+\to\pi^+\pi^0$ and $K^+\to\pi^+\pi^+\pi^-$ decays upstream of the fiducial volume. The most dangerous of these decays occur between the last two GTK stations. The veto counter was installed in 2021, surrounding the beam pipe, with the aim of detecting the particles ($\pi^0\to \gamma\gamma$, $\pi^\pm$) produced in addition to the $\pi^+$ in the upstream region and has helped reduce the background. However, a significant improvement can be achieved by enhancing the capabilities of the VC to detect these early $K^+$ decay products passing at very small angles inside the holes accommodating the beam pipe. In addition, the random veto due to the usage of the VC is less than 2\% at NA62 nominal intensity but is expected to increase with the beam intensity, therefore an improved time resolution to less than 200~ps is needed. In events with more than one particle present, the granularity of the current detector does not allow to reliably separate them and distinguish between accidental activity due to halo muons and genuine background candidates: this facility would be welcome in HIKE.

\begin{figure}[tb]
\centering
\includegraphics[width=0.75\linewidth]{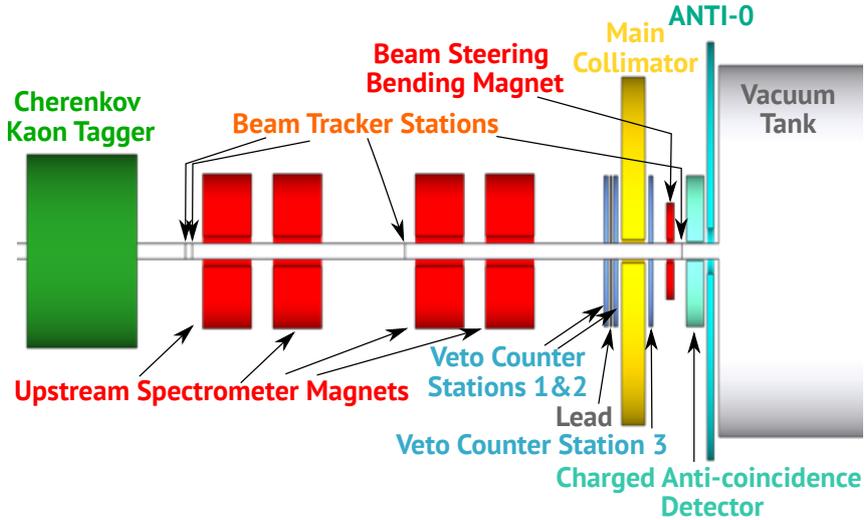}
\vspace{-1mm}
\caption{Zoomed view of the upstream region of the HIKE Phase~1 setup (see Fig.~\ref{fig:phase1-layout}), highlighting the location of the veto counter stations.}
\label{fig:UpstreamRegionZoom}
\end{figure}

An improved design based on scintillating fibre (SciFi) technology, developed for the SciFi tracking detector at LHCb~\cite{Joram2015LHCbSF}, is proposed to tackle these limitations. The detector will consist of three stations: two located in front of the main collimator, separated by a layer of lead acting as a photon converter, and a third one located immediately behind the collimator (Fig.~\ref{fig:UpstreamRegionZoom}). Each station must extend at least up to 10~cm above and 30~cm below the beamline to cover the full range where upstream decays are expected to cross the detection planes. The horizontal coverage needs to extend up to 6~cm on either side of the beam. A central hole will accommodate the reduced-size beam pipe, allowing it to reach only 2~cm from the beam centre. The thin SciFi modules would then be attached to the beam pipe allowing for an increased detection efficiency for interactions of photons and charged particles with the beam pipe. Following these specifications, the active area of the new veto counter will cover a rectangular surface area of $26\times $\SI{78}{\cm^2} with a hole at the centre to accommodate the passing beam. With the present state of the technology as used in LHCb, each station is made of scintillating fibres of diameter \SI{250}{\um}, read out by silicon photo-multipliers (SiPMs). The fibres are arranged in mats of multiple layers forming a detection plane. Two orthogonal SciFi planes form a single veto counter station. The described scheme will provide X--Y reading of charged particles crossing a detection station with a spatial resolution $\sigma_{x,y}\sim\SI{200}{\um}$, a time resolution below \SI{200}{\ps}, and detection efficiency above 99\%. A solution with the complete SciFi stations placed in vacuum can enhance the veto capabilities further, and will be investigated for the HIKE TDR. The number of readout channels for the above design will be 4096 channels per station. If needed, several SiPM channels could be coupled together into a single readout channel to reduce the number of channels at the expense of some spatial resolution. The thickness of the detection planes can also be modified to increase the light yield resulting in better detection efficiency and time resolution. The small granularity and high spatial resolution of each detection plane will provide tracking capabilities by combining information from consecutive stations, further improving the time resolution and reducing random veto. 

The estimated dose of non-ionising radiation, by extrapolation from NA62 at four times the intensity, is about 1--$2 \times 10^{11}$ 1 MeV $n_{\rm eq}$ per year over the entire surface of the SciFi VetoCounter. Conservatively estimating that all of this rate is concentrated close to the beam pipe, the maximum flux is $4 \times 10^9$ 1 MeV $n_{\rm eq}$/cm$^2$/year. This is about equivalent to the flux seen by the LHCb SciFi tracker about 1 m away from the beam, and is about 3-4 orders of magnitude less than that close to the beam pipe. Therefore, the use of a SciFi veto counter in the HIKE environment should definitely be safe for both the fibres and SiPMs without significant performance degradation.

The expected hit rate in the detector is extrapolated from the rates at the nominal NA62 beam intensity, assuming it grows linearly with the beam intensity.
The hit rate at four times the NA62 beam intensity is expected to be 30~MHz per detection plane, which is too high for the existing readout electronics of the SciFi detector but can be addressed with the TDC-Felix electronics commissioned in NA62 and foreseen in HIKE (Section~\ref{sec:readout}). The front-end electronics must also be adapted for HIKE. The present front-end ASIC used for the LHCb SciFi receives signals from the SiPMs, which are then shaped and digitised by a 2-bit ADC with a sampling rate of 40~MHz. Due to the time resolution requirements for HIKE, new front-end electronics will be used, based on ToT or FADC, similar to the solutions for other detector systems.

A different technology option is an update on the design currently used in NA62 using tiles of plastic scintillator. In this case, the use of tiles with smaller heights (4~cm in NA62) and the usage of a double plane for each station (a plane with tiles placed horizontally and a plane with tiles placed vertically) will allow to achieve a smaller granularity and separate signals left by multiple particles. An increased acceptance coverage will be achieved by adding tiles also on the left and right of the beam. A rectangular hole of $0.42\times 0.78$~cm$^2$ size will allow the beam pipe to pass through the detector. In addition, some tiles could be placed in the perpendicular direction along the beam pipe to increase the detection of particles escaping the beam pipe between stations.

Further improvements in upstream background rejection power could be achieved by rearranging the beamline elements in the VC region. Combined with the proposed hardware improvements, the new veto counter detector will reduce the upstream background by at least a factor of three with respect to the present system.

\subsubsection{ANTI-0}
\label{sec:anti0}

The baseline option for the charged particle veto hodoscope located at the entrance of the decay volume (ANTI-0) is a cell-structure hodoscope covering an area of 2.2~m diameter around the beampipe. The NA62 ANTI-0 hodoscope design is based on rectangular scintillating tiles with transverse size of $120\times120$~mm$^2$ read by SiPMs through short lightguides, arranged in a chessboard style at both sides of the central foil~\cite{Danielsson:2020opy}. The general design of the HIKE ANTI-0 detector follows that of the NA62 ANTI-0 hodoscope with the two major changes: finer granularity to sustain up to six times higher intensity, and two active layers to ensure higher efficiency. 
Since the rate should be kept below 1~MHz per tile, it seems reasonable to use tiles of 55 $\times$ 55 $\times$ 10 mm$^3$ dimensions. The final choice of tile size will be informed by further simulation studies.
The possibility of adding active tiles in dump mode in the area currently passive for beam passage is under investigation to increase the efficiency.

The ANTI-0 timing requirements are not very stringent in dump mode given the low particle rate, making the time resolution overall not particularly challenging. The time resolution of the ANTI-0 in NA62 is at the level of 1~ns, corresponding to a $5\sigma$ coincidence window of 5~ns. At the full HIKE beam-dump intensity, the use of a 5~ns coincidence window would contribute at most 2.5\% to the random veto efficiency.

A possible option for the tiles is the plastic scintillator UPS-923A, used for the CDF detector at the Fermilab Tevatron and for the ATLAS tile calorimeter~\cite{Artikov:2005mg}.
Specifications for UPS-923A scintillator are listed in Table~\ref{tab:ups923a}.

Another possibility, currently under investigation, is reabsorption-free fast nanocrystal-based plastic scintillators~\cite{Gandini:2020aaa}. The new scintillator already now can sustain a higher radiation dose, having an emission peak shifted to $\lambda>550$~nm, tolerating transparency losses in UV and blue regions. Some perovskite-based scintillators show very fast decay time, with the first time component of the order of 0.3~ns~\cite{JANA2022110}, which could be crucial for building fast future detectors. More details can be found in the description of the main calorimeter  (\Sec{sec:klever_mec}). These technological possibilities will be investigated further for the HIKE Technical Design Report.

\begin{table}[h]
\centering
\caption{Specifications for UPS-923A plastic scintillator from the Institute for Scintillation Materials (ISMA), Kharkiv, Ukraine.}
\vspace{-2mm}
\label{tab:ups923a}
\begin{tabular}{l|c} 
\hline
Scintillator & UPS-923A \\
\hline
Base & Polystyrene \\
Rise time (ns) &  0.9 \\
Decay time (ns) & 2--3 \\
Wavelength, max emission (nm) & 418 \\
Light output, \% of anthracene & 56 \\
\hline
\end{tabular}
\end{table}

\subsubsection{Detectors for the neutral beamline}
\label{sec:neutral_detectors}

As discussed in \Sec{sec:neutral_beam}, the last stage of collimation consists of an active final collimator (AFC).
The AFC is a collar counter made of shaped crystals with angled inner surfaces to
provide the last stage of beam collimation; signals from the AFC will be used to veto inelastic interactions from scattered beam particles, as well as photons and charged particles from beam particles decaying in transit through the collimator or immediately upstream of it.
The ideal material is a dense, high-$Z$, radiation-hard crystal with excellent light output and very fast time response.
Engineered LYSO\footnote{Luxium, formerly Saint-Gobain Crystals, \url{luxiumsolutions.com}} has a light output of 33000 photons/MeV and an emission decay time of 36~ns. LYSO is known to be extremely radiation resistant, especially to neutrons~\cite{Hu:2020aem}.
The AFC consists of 24 crystals of trapezoidal cross section, forming a detector with an inner radius of 60 mm and an outer radius of 100 mm.
The AFC is 800 mm in depth. The maximum crystal length for
a practical AFC design is about 250 mm, so the detector consists of 3 or 4 longitudinal segments.
Each crystal is read out on the downstream side with two avalanche photodiodes (APDs).
These devices couple well with LYSO and offer high quantum efficiency and
simple signal and HV management. Studies indicate that a light yield in excess of 4000~p.e./MeV should be easy to achieve.

As described in the HIKE Letter of Intent, in a future Phase 3 experiment, the AFC would be inserted into a hole in centre of an upstream-veto calorimeter (UV), a shashlyk calorimeter with the same basic structure as the MEC, but with a coarser readout granularity. With a sensitive area extending out to a radius of 100~cm and the ability to detect both charged particle and photons, the UV is effective at vetoing decays of neutral beam particles up to 40~m upstream of the final collimator, provided that it has an unobstructed view. The UV and upstream extension of the vacuum tank are not included in the baseline Phase~2 layout. However, they are compatible with the beamline layout and could be installed at a later time if sensitivity studies indicate a need for them. This could be the case, for example, if the HIKE Phase-2 programme is extended to the study of $K_L$ decays with all-neutral final states.

\subsection{Decay volume and its detectors}

\subsubsection{Charged particle anti-coincidence detector (CHANTI)}

The charged particle anti-coincidence detector provides veto for events with particles scattered inelastically off the last station of the beam tracker and halo particles originating from upstream decays, entering the decay region close to the beam. 

The NA62 CHANTI comprises six stations, each station covering an area of $30\times30$~cm$^{2}$ with a $9\times\SI{5}{\cm^2}$ inner opening for the beam. The CHANTI covers hermetically the angular region between \SI{34}{\milli rad} and \SI{1.38}{rad} with respect to the beam axis, and has an efficiency of about 90--95\%. The CHANTI, with its time resolution of about \SI{800}{\pico s}, induces a signal loss due to accidental activity of about 5$\%$ at the nominal intensity but is expected to become more significant with increasing intensity.
At the HIKE intensity, in order to keep the random veto and the efficiency at the same level, the time resolution has to be about four times better than the present one, i.e. around 200~ps. 

\begin{figure}[tb]
\centering
\includegraphics[width=0.42\textwidth]{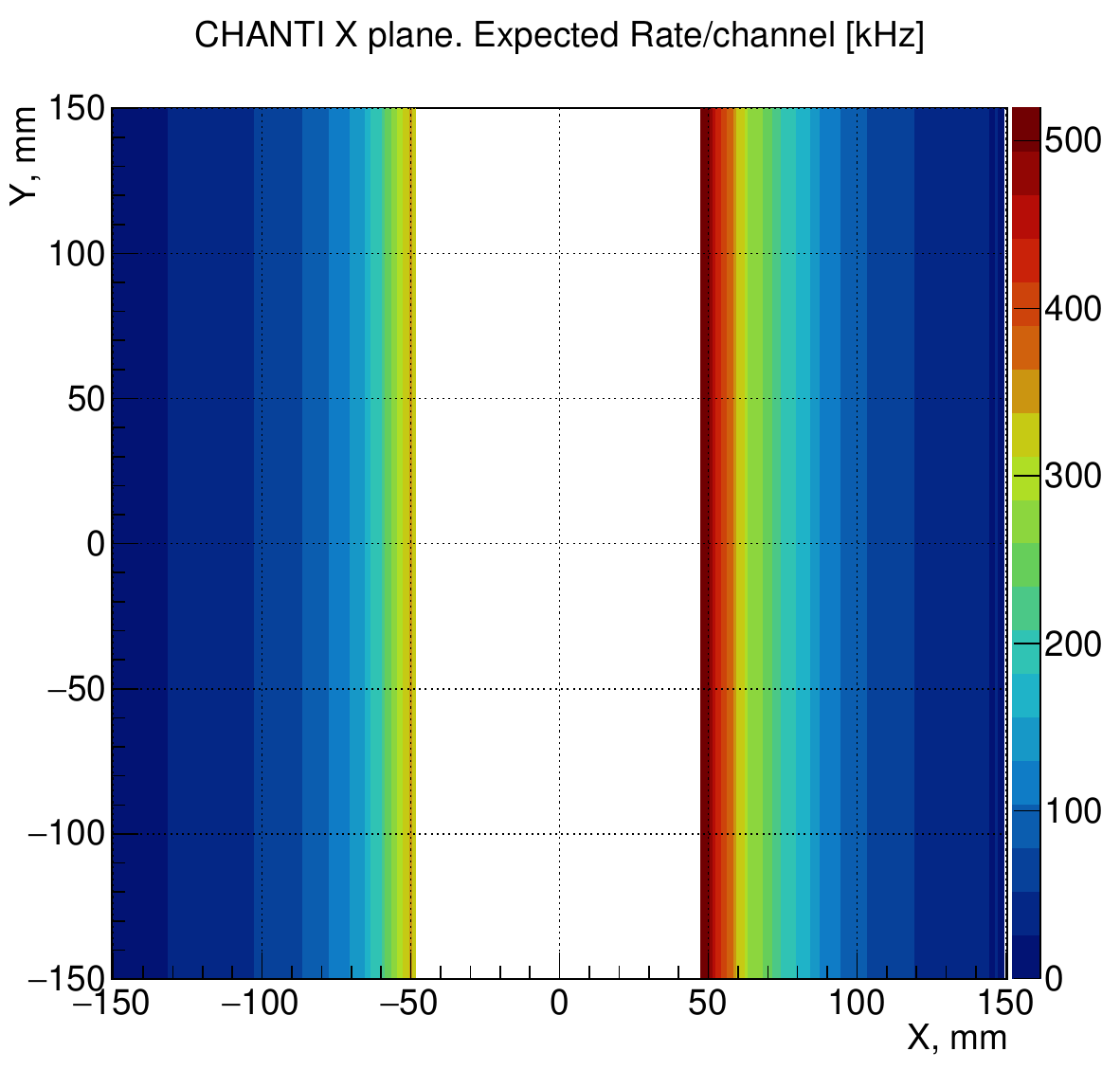}%
\includegraphics[width=0.42\textwidth]{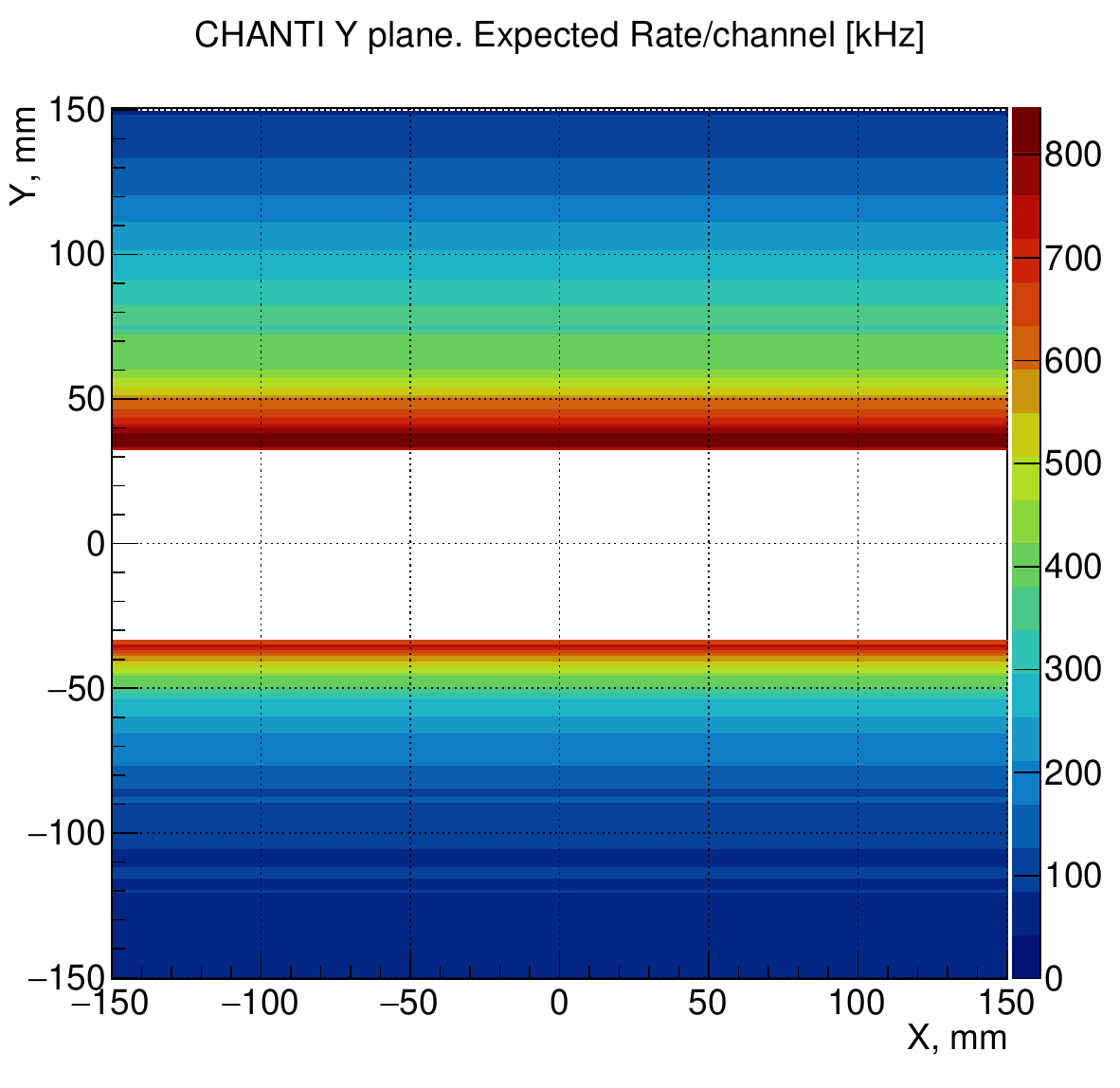}
\vspace{-4mm}
\caption{CHANTI expected rate per fibre for the vertical and horizontal planes.}
\label{fig:CHANTI_Exp_Rate}
\end{figure}

The design of the HIKE CHANTI stations is based on the usage of thick plastic scintillating fibres (from $\oslash 2$ up to $\oslash 3$~mm) read by SiPMs from both ends.
The expected particle rate has been estimated using the NA62 beamline simulation, assuming a hadron beam intensity six times higher than the nominal 750~MHz of NA62. The expected rate per channel in kHz is shown in Fig.~\ref{fig:CHANTI_Exp_Rate} for the option with $\oslash 2$~mm thick fibres overlapping by 0.5~mm. The option with $\oslash 2$~mm fibres, that is the maximum diameter currently widely available on the market, assumes 464 fibres per station (208 for vertically-oriented X station, and 256 for horizontally-oriented Y station) and about 1000 SiPMs/station. The required number of electronic channels could be reduced by connecting signals from the peripheral fibres into an analogue OR before digitization.

\paragraph{Alternative detector design using SciFi technology}

A more ambitious option would be to pursue the SciFi technology for the new CHANTI detector.
The new detector would consist of six stations with $39\times $\SI{39}{\cm^2} transverse dimensions, positioned to hermetically cover the required angular region with respect to the beam axis. Each station is made of scintillating fibres ($\oslash \SI{250}{\um}$) arranged in two planes that provide X-Y information about the incoming charged particles. Each plane is made of mats of several fibre layers, which are read out by SiPMs. The layout allows a time resolution below \SI{200}{\pico s} per station and a spatial resolution $\sigma_{x,y}\sim\SI{200}{\um}$. The time resolution would be at least a four-fold improvement, and the exceptional spatial resolution would provide tracking capabilities. The stations would be thicker than the ones currently used at LHCb~\cite{Joram2015LHCbSF}, bringing the detection efficiency above 99$\%$. 

The new CHANTI detector could also be used in conjunction with the new VetoCounter
(Section~\ref{sssec:vetocounter}), both made of SciFi technology, to separate halo muons from charged particles produced in interactions in the beam-tracker. Owing to their identical spatial and time resolutions and largely overlapping sensitive regions, a combination of the information between the two detectors can reduce significantly the accidental veto due to halo-muon activity.

Using a SciFi detector for the new CHANTI presents a technological challenge. At present detectors using SciFi technology are operated in air and a dedicated R$\&$D programme is required to enable their operation in vacuum. A solution to this challenge will allow the cooling of the SiPM to liquid nitrogen temperatures, eliminating the noise completely and improving radiation tolerance. This is an important improvement over the present CHANTI, which uses SiPMs without cooling. The SiPMs currently in operation at NA62 are heavily impacted by the effect of accumulated radiation dose; such an issue would not be present for a SciFi detector with cooled SiPMs in vacuum. 

The number of readout channels for the above design would be about 6000 channels per station with the present design of a single SciFi station. However, the spatial resolution of $\sigma_{X,Y}=200~\mu$m can be reduced by a factor of two without any impact on the veto performance. Therefore, pairs of neighbouring read-out channels can be merged, halving the number of read-out channels and reducing the data throughput. Similarly to the solution for the HIKE Veto counter, the readout is planned to be based on the TDC-Felix electronics (Section~\ref{sec:readout}), which is able to sustain the hit rate presented in Fig.~\ref{fig:CHANTI_Exp_Rate} and provide the necessary time resolution.

\subsubsection{Large-angle vetos (LAV)} 
\label{sec:lav}

The present NA62 photon veto system largely relies on the large-angle veto system based on the OPAL lead glass~\cite{OPAL:1990yff,NA62:2017rwk} consisting of 12 stations, of which 11 are incorporated into the vacuum tank, and one is located in air just upstream of the EM calorimeter. Some parameters of the NA62 LAV stations are listed in Table~\ref{tab:na62_lav}. The experience accumulated in NA62 in terms of LAV performance and operation is essential for designing the equivalent system for HIKE.

\begin{table}[tb]
\centering
\caption{Parameters of the present NA62 large angle veto stations.}
\vspace{-2mm}
\begin{tabular}{lccccc}
\hline
Stations & Diameter [mm] & \multicolumn{2}{c}{Block radius [mm]} & Layers & Blocks\\
& Outer wall & Inner & Outer & & \\
\hline
LAV1--LAV5  & 2168  &   537 &  907 & 5 & 160\\
LAV6--LAV8  & 2662  &   767 & 1137 & 5 & 240\\
LAV9--LAV11 & 3060  &   980 & 1350 & 4 & 240\\
LAV12       & 3320  &  1070 & 1440 & 4 & 256\\
\hline
\end{tabular}
\label{tab:na62_lav}
\end{table}

In the analysis of NA62 Run~1 data, the overall $\pi^0\to\gamma\gamma$ decay detection inefficiency was found to be close to $10^{-8}$~\cite{NA62:2020fhy}.
The inefficiency estimate is validated by studies of the single-photon detection efficiencies for each veto subsystem~\cite{NA62:2020pwi}.
The inefficiency of the NA62 LAVs for single photons was found to rapidly decrease with photon energy to a value of about $3\times10^{-3}$ at 300~MeV, and thereafter to slowly decrease with energy to values of a few $10^{-4}$ for photons of 4--6~GeV. This refers to the LAV system in operation, including the effects of the non-hermeticity of the retrofitted arrangement of lead-glass blocks. The actual detection inefficiency for the lead-glass detectors themselves was found with electrons to be an order of magnitude lower~\cite{Ambrosino:2007ss}.

In the decay-in-flight technique for the $K^+\to\pi^+\nu\bar\nu
$ measurement, the $K^+$ decay vertex is accurately reconstructed, and the momentum requirements on the secondary $\pi^+$ guarantee that the $\pi^0$ in $K^+\to\pi^+\pi^0$ decays has at least 40~GeV of energy. As a result, the LAV system only has to cover out to 50~mrad in the polar angle as seen from the decay volume. Moreover, the two photons from $K^+\to\pi^+\pi^0$ decay are correlated such that for events with one low-energy photon heading into the LAVs, the second photon has high energy and is detected in the EM calorimeter. 
These conditions relax the LAV single-photon inefficiency requirements for the $K^+$ programme.

On the other hand, the time resolution obtained with the existing LAVs is insufficient for HIKE.
The Cherenkov light produced in the large lead-glass blocks is highly directional, and the light propagates to the PMT photocathodes via complicated paths with multiple reflections. There is considerable time spread arising from the angle of particle incidence, and perhaps for different particle species, the Cherenkov characteristics of mips and electromagnetic showers being quite different. Above all, the light yield of the lead-glass blocks for electromagnetic showers is only 0.2~p.e./MeV. As a result, the NA62 LAVs have a time resolution of about 1~ns with significant extra-gaussian tails~\cite{NA62:2017rwk}. 

The HIKE time-resolution requirements are driven by the beam intensity. To mitigate the random veto effects at HIKE, the LAV time resolution needs to be improved by at least a factor of four, to about 250~ps, and the tails of the time distribution must be substantially reduced or eliminated. A preliminary estimate of the maximum total hit rate on the HIKE LAVs is 14~MHz, implying a 3.5\% random veto probability for a $\pm5\sigma$ coincidence window with a full width of 2.5~ns. The LAVs for HIKE should provide adequate photon detection efficiency for all phases of the programme, including a possible Phase~3 extension for the $K_L\to\pi^0\nu\bar\nu$ measurement. Indicatively, the photon detection inefficiency must be less than 5\% at 10~MeV, less than $2.5\times10^{-4}$ at 100~MeV, and less than $2.5\times10^{-6}$ for energies above 2.5~GeV. The positions and radii of the LAV stations for HIKE Phases~1 and 2 are expected to be similar to those for NA62.

%%%%%%%%%%%%%

\paragraph{LAV design and expected performance}

\begin{figure}[tb]
\centering
\includegraphics[width=0.5\textwidth]{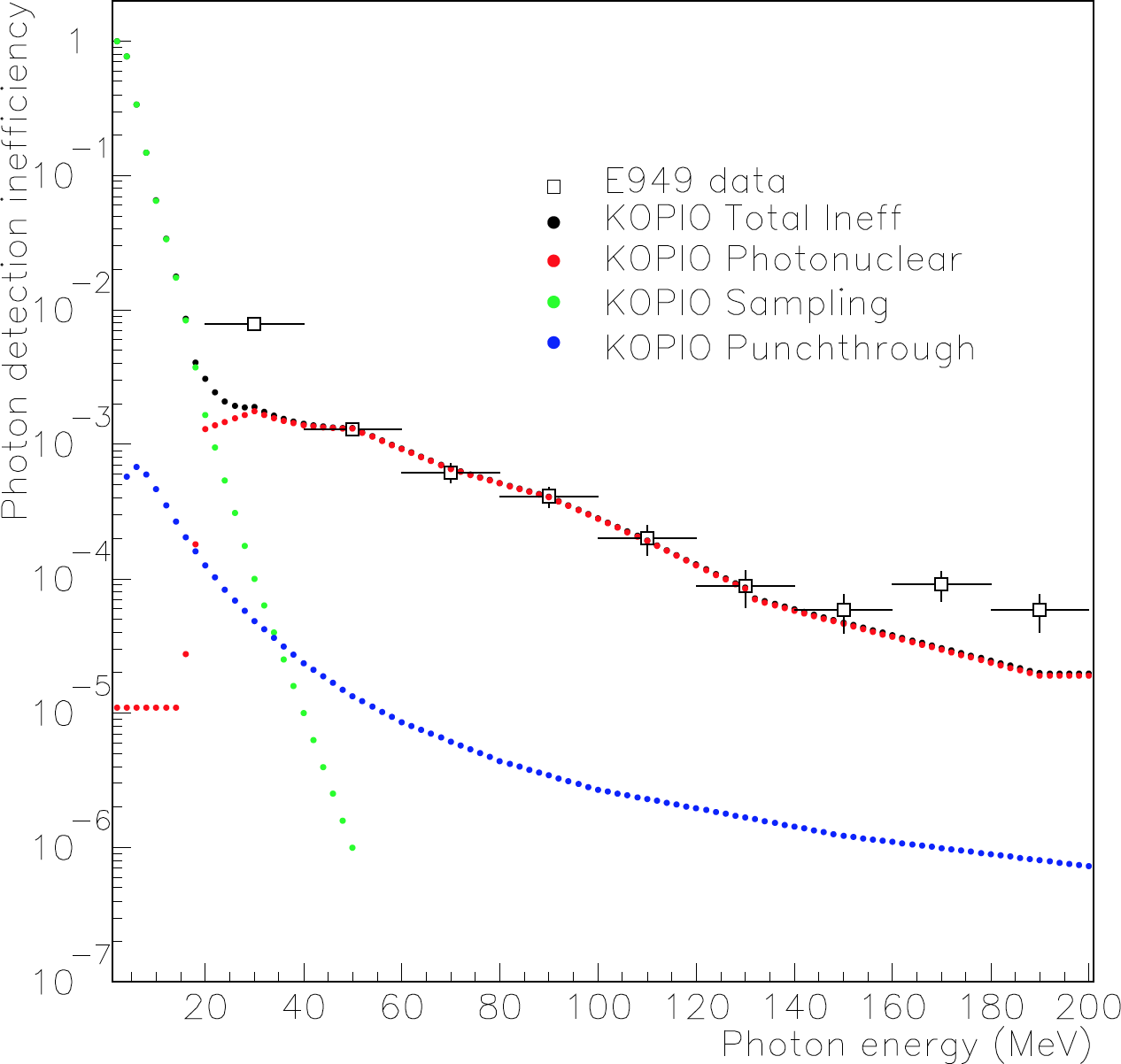}
\caption{Photon detection inefficiency parameterization from KOPIO~\cite{KOPIO+05:CDR}, broken down by source. Measured
inefficiencies for the E949 barrel veto~\cite{Atiya:1992vh} are also plotted.}
\label{fig:kopio_eff}
\end{figure}

As a reference for low-energy photon detection efficiency
achievable for the HIKE LAV, Fig.~\ref{fig:kopio_eff} shows the inefficiency
parameterization used for the KOPIO proposal~\cite{KOPIO+05:CDR}.
For the energy range
$50~{\rm MeV} < E < 170~{\rm MeV}$,
the overall parameterization is based on detection
inefficiencies measured for the
E787/949 barrel photon veto~\cite{Atiya:1992vh} using $K^+\to\pi^+\pi^0$
events; these data are also shown in the figure.
Outside of this range, the parameterization
is guided by FLUKA simulations with different detector designs, and
the overall result is (slightly) adjusted to reflect the segmentation
of the KOPIO shashlyk calorimeter, but for most of the interval
$E < 200$~MeV, the results do not differ much from the E949 measurements.
The contributions to the inefficiency from photonuclear interactions, sampling fluctuations and punch-through were estimated from known cross sections, statistical considerations and mass attenuation coefficients.

\begin{figure}
\centering
\includegraphics[width=0.5\textwidth]{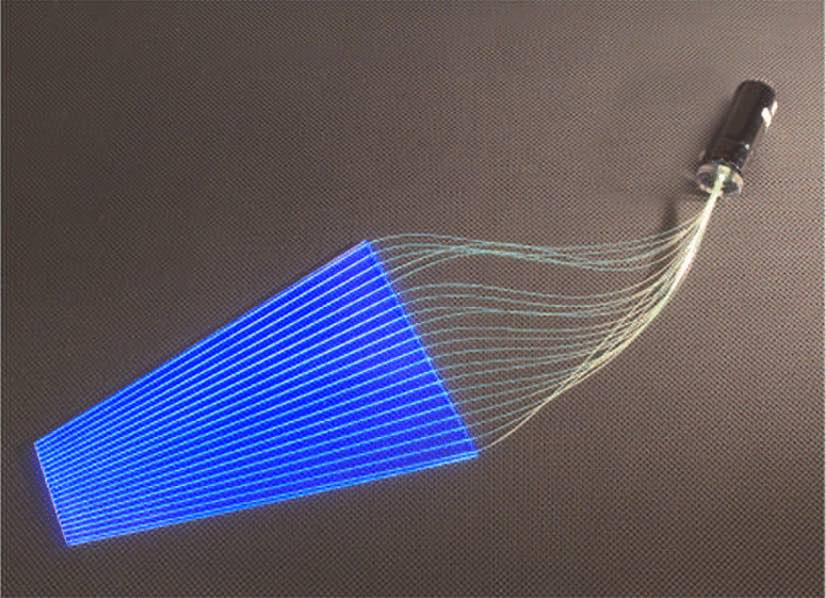}
\caption{Prototype tile with WLS fiber readout for the CKM VVS detector~\cite{Frank:2001aa}.}
\label{fig:ckm_tile}
\end{figure}

One possible design for the HIKE LAVs would be similar to the Vacuum Veto System (VVS) detectors planned for the CKM experiment at Fermilab~\cite{Frank:2001aa}. The CKM VVS is a lead/scintillator-tile detector with a segmentation of 1~mm Pb and 5~mm scintillator, for an electromagnetic sampling fraction of 36\%. This segmentation is the same as for the E787/949 barrel photon veto, so the same low-energy efficiencies might be expected. The wedge-shaped tiles are stacked into modules and
arranged to form a ring-shaped detector. The scintillation light is collected and transported by 1-mm-diameter WLS fibers in radial grooves, as seen in Fig.~\ref{fig:ckm_tile}.
In the approximate geometry for HIKE, the LAV modules would consist of 96 layers, for a total thickness of $\sim$60~cm, corresponding to $\sim$18 $X_0$. There are
50--65 modules per detector, each with 20 fibers per tile. 
In the original VVS design, the fibers brought the light to optical windows for readout by PMTs outside of the vacuum. Readout by SiPMs inside the vacuum would make for shorter fibers and would facilitate the mechanical design---the availability of economical SiPM arrays with large effective area makes this an attractive option. In the HIKE geometry, the fibers in a module would be bundled for readout by eight SiPM arrays. As in the VVS design, alternating fibers from each tile would be read by different SiPMs to provide redundancy, and there would be four readout layers in depth.

\begin{figure}[tb]
\centering
\includegraphics[width=0.61\textwidth]{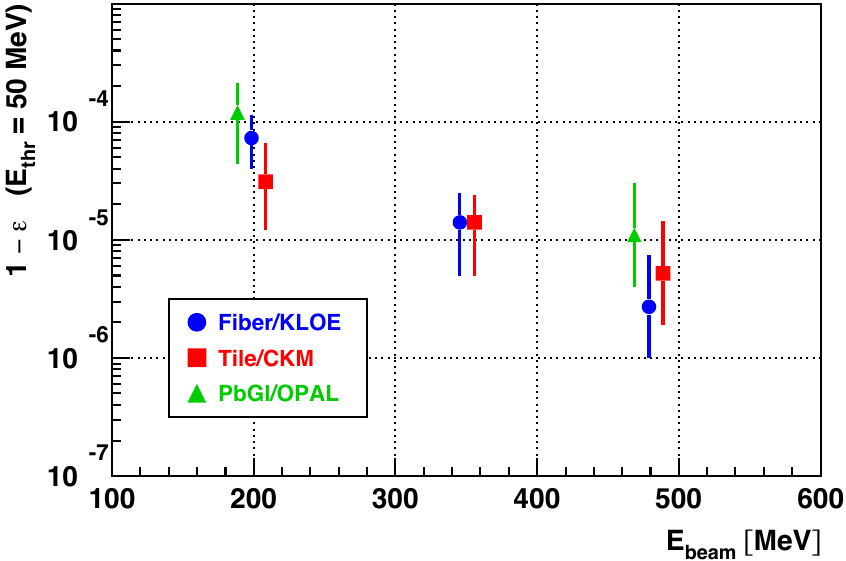}
\vspace{-2mm}
\caption{Measurements of detection inefficiency for tagged electrons with 203, 350, 483~MeV energies for three veto prototypes, made at the Frascati BTF in 2007. The red squares are for the CKM VVS prototype.}
\label{fig:hawaii}
\end{figure}

In the original NA62 proposal~\cite{Anelli:2005xxx}, before the OPAL lead glass became available,
very similar detectors were the baseline solution for the existing LAVs,
and in 2007, the efficiency of the CKM VVS prototype in the energy range 200--500~MeV was measured using a tagged electron beam at the Frascati
Beam-Test Facility (BTF) \cite{Ambrosino:2007ss}.
The results are shown in Fig.~\ref{fig:hawaii}. Earlier, the efficiency of the same prototype was measured at the
Jefferson National Laboratory with tagged electrons in the interval 500--1200~MeV. An inefficiency of $3\times10^{-6}$ was found at 1200~MeV,
even with a high threshold (80~MeV, or 1 mip)~\cite{Ramberg:2004en}.
The efficiency parameterization for the LAVs used in our simulations is based on the
KOPIO efficiencies up to the point at 129~MeV, and then extrapolated
through the three points measured at the BTF, to $2.5\times10^{-6}$ for photons
with $E > 2.5$~GeV. This is not unreasonable,
considering that the Jefferson Lab measurement shows that nearly this
inefficiency is already obtained at 1.2~GeV.
The tests of this prototype at the BTF were not optimized for the measurement of the time resolution, but indicated a time resolution of better than 250~ps for 500 MeV electrons, which is sufficient for HIKE.

%%%%%%%%%%%%%

\paragraph{LAV readout}

A leading contribution to the random veto inefficiency for the LAVs in NA62 is from halo muons. If the interactions of halo muons could be reliably distinguished from photon showers, the random veto inefficiency for the LAVs would be decreased, potentially relaxing the requirements on the LAV time resolution. In NA62, the LAV signals are discriminated against two thresholds, and signal amplitudes are obtained using the time over threshold technique. 
While some use of this information has been made in NA62 to partially recover the LAV random veto inefficiency and efforts are continuing in this direction, in practice, the separation between mips and low-energy photon showers has not so far proved to be reliable. To improve upon this situation, we are investigating the gains to be had with a fully digitizing FADC readout for the LAVs. In all, the 12 LAV modules will have about 5000 readout channels.

\subsubsection{Spectrometer}
\label{sec:straw}

\begin{figure}[t!]
\centering
\includegraphics[width=0.75\textwidth]{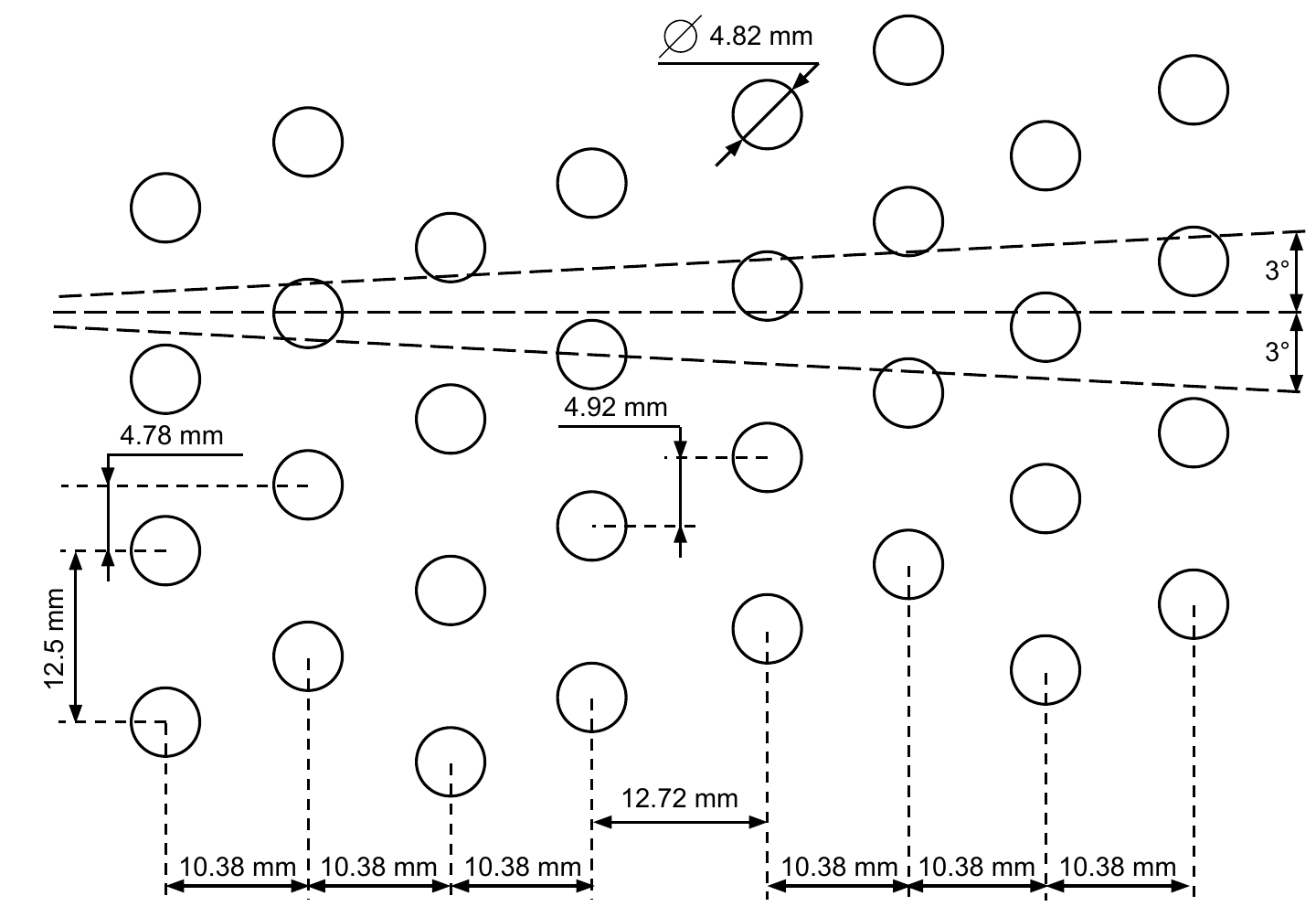}
\vspace{-2mm}
\caption{Optimised layout of straw tubes with a diameter of 4.82~mm in a view.}
\label{fig:straw_new_layout}
\end{figure}

A spectrometer similar to the NA62 straw tracker~\cite{NA62:2017rwk}, comprising four straw chambers and a dipole magnet, is planned to reconstruct the momentum and direction of charged particles in the final state. The straw tubes are expected to remain in the vacuum tank containing the decay region, profiting from the successful technology developed for NA62. Straws with diameter less than 5~mm are necessary to handle the expected particle rates at the higher beam intensity. Straw diameter reduction by a factor $\sim 2$ with respect to the NA62 straws will lead to shorter drift times and an improvement in the resolution of the trailing edge time from the current 30~ns to 6~ns. A smaller diameter of the straw also requires a change in the geometric placement of the straws in a single view.
Design work based on Monte Carlo simulations was performed, and the straw layout was optimised taking into account realistic spacing and dimension requirements, resulting in a choice of eight straw layers per view, shown in Fig.~\ref{fig:straw_new_layout}.

%%%%%%%%%%%%%%%%%%%%%

\begin{table}[t!]
\centering
\caption{Comparison of the NA62 and HIKE straw spectrometers.}
\vspace{-2mm}
\begin{tabular}{l|r|r}
\hline
&Current NA62 spectrometer& HIKE spectrometer \\
\hline
Straw diameter & 9.82~mm & 4.82~mm \\
Straw length & 2100~mm  & 2100~mm \\
Planes per view & 4 & 8 \\
Straws per plane & 112 & ${\sim}160$ \\
Straws per chamber & 1792 & ${\sim}5200$ \\
\hline
Mylar thickness & 36~$\mu$m  &  (12 or 19)~$\mu$m \\ 
Anode wire diameter & 30~$\mu$m & (20 or 30)~$\mu$m \\
Total material budget & 1.7\%\,$X_0$ & (1.0 -- 1.5)\%\,$X_0$ \\
\hline
Maximum drift time &  ${\sim}150~$ns &  ${\sim}80~$ns \\
Hit leading time resolution & (3 -- 4)~ns & (1 -- 4)~ns\\
Hit trailing time resolution & ${\sim}30~$ns & ${\sim}6~$ns \\
Average number of hits hits per view & 2.2 & 3.1 \\
\hline
\end{tabular}
\vspace{-2mm}
\label{tab:straw_comparison}
\end{table}

%%%%%%%%%%%%%%%%%%%%%%

The material used to make new straws using the ultrasonic welding technique will be the same as in the current spectrometer, namely Mylar coated with 50~nm of copper and 20~nm of gold on the inside.
To reduce the detector material budget, the thickness of the mylar will be reduced from $36~\mu$m to either $12~\mu$m or $19~\mu$m.
The diameter of the gold-plated tungsten anode wires might be reduced from $30~\mu$m to $20~\mu$m.
The final decision on the mylar thickness and the wire diameter will be made based on mechanical stability tests.

\newpage

Straws with reduced mylar thickness (either $19~\mu$m or $12~\mu$m) and the same metallisation thickness as in NA62 will be compatible with use in vacuum. The gas permeation is mainly determined by the thickness of the metal layer and the width of the welded seam in the straw, which will not change with respect to NA62. The gas tightness of the welded seam depends on the stress concentration, which scales linearly with the wall thickness and straw diameter, and which for the 19~$\mu$m option is the same for NA62, while for the 
12~$\mu$m option will increase by 50\%, remaining well below the theoretical limit. The total straw surface seen by the vacuum system (and hence the gas load), will increase by 50\% due to the increased number of straws, but there is already spare capacity in the NA62 vacuum system to keep the vacuum for a HIKE-type straw detector at the same level as in NA62 ($\sim 10^{-6}$~mbar).
More R\&D is needed in order to guarantee straws with sufficient quality for a large-scale detector with $\sim$20k straws.
The development of small-diameter thin-walled straws has synergies with R\&D work for COMET phase II at J-PARC~\cite{Nishiguchi:2017gei}, and is included in the ECFA detector R\&D roadmap~\cite{ecfareport:2021}.

\begin{figure}[p]
\centering
\includegraphics[width=0.83\textwidth]{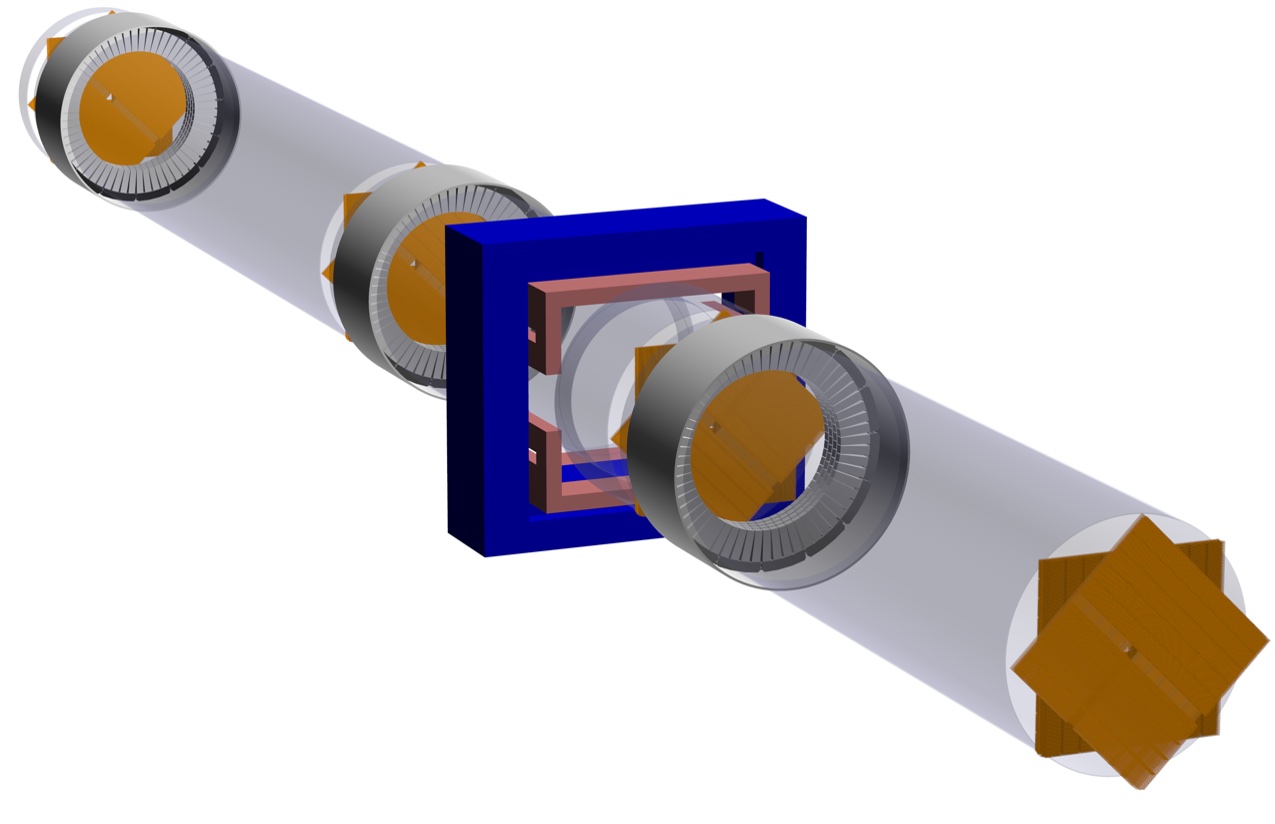}
\vspace{-6mm}
\caption{
\label{fig:straw_visualization}
Geant4 visualisation of the new straw spectrometer.}
\end{figure}

Based on the results of the design study, a Geant4-based simulation of the new spectrometer was developed using the same dimensions and positions of the straw chambers, the number and orientation of views in the chamber, the gas composition (Ar+CO$_2$ with 70:30 ratio) and the properties of the dipole magnets as in the current NA62 layout. 
A comparison between the two straw detectors is given in Table~\ref{tab:straw_comparison}, and a Geant4 visualisation of the new spectrometer is shown in Fig.~\ref{fig:straw_visualization}.
The new spectrometer is planned to have the capability of aligning the central holes of the straw chambers on the beam axis for both $K^+$ (Section~\ref{sec:phase1}) and $K_L$ (Section~\ref{sec:phase2}) modes of operation.

The NA62 track reconstruction algorithm was adapted for the new detector, and a preliminary resolution comparison indicates that the new spectrometer could improve the resolution for the reconstructed track angles and momenta by 10--20\% with respect to the existing NA62 spectrometer while maintaining the high track reconstruction efficiency.

\begin{figure}[p]
\centering
\includegraphics[width=0.45\textwidth]{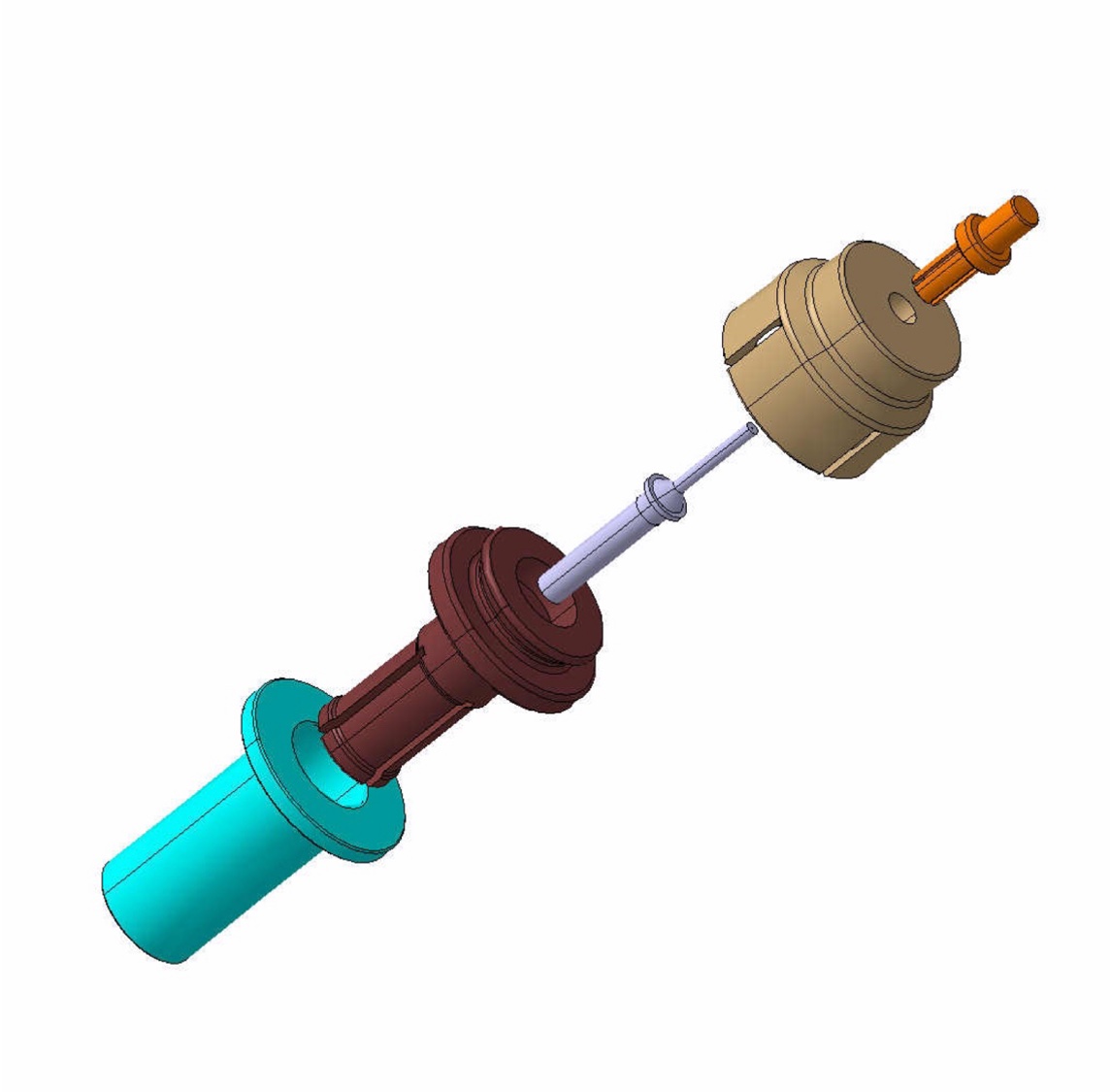}
\includegraphics[width=0.45\textwidth]{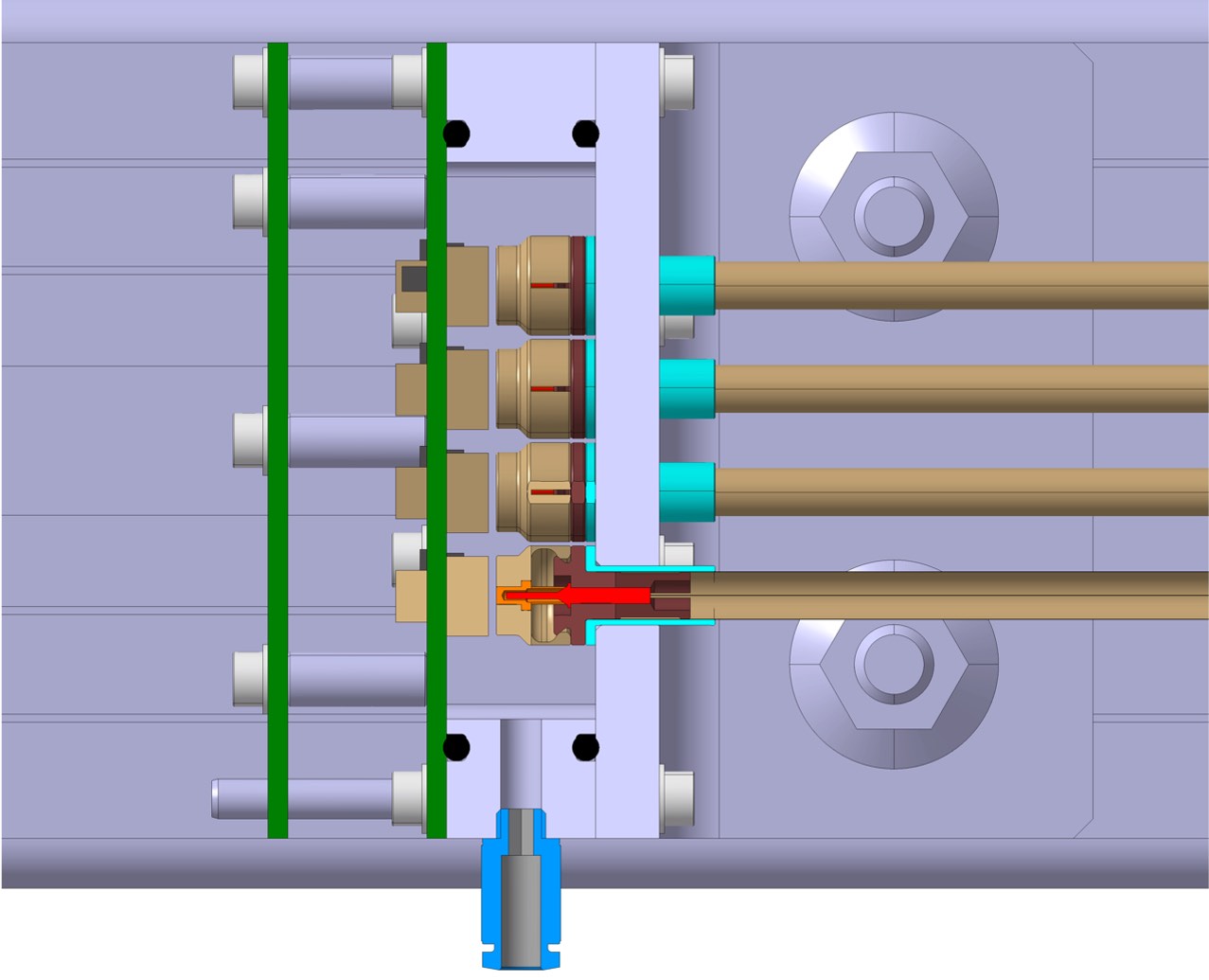}
\vspace{-2mm}
\caption{Left: detail of the connectivity elements (coaxial) in order to minimise the signal path to the front-end electronics.
Right: cross-section of the prototype to validate connectivity and basic performance of a 4.82~mm diameter straw using new front-end electronics with the capability to measure the trailing edge.}
\label{fig:spectrometer:connectivity}
\end{figure}

\begin{figure}[p]
\centering
\includegraphics[width=0.46\textwidth]{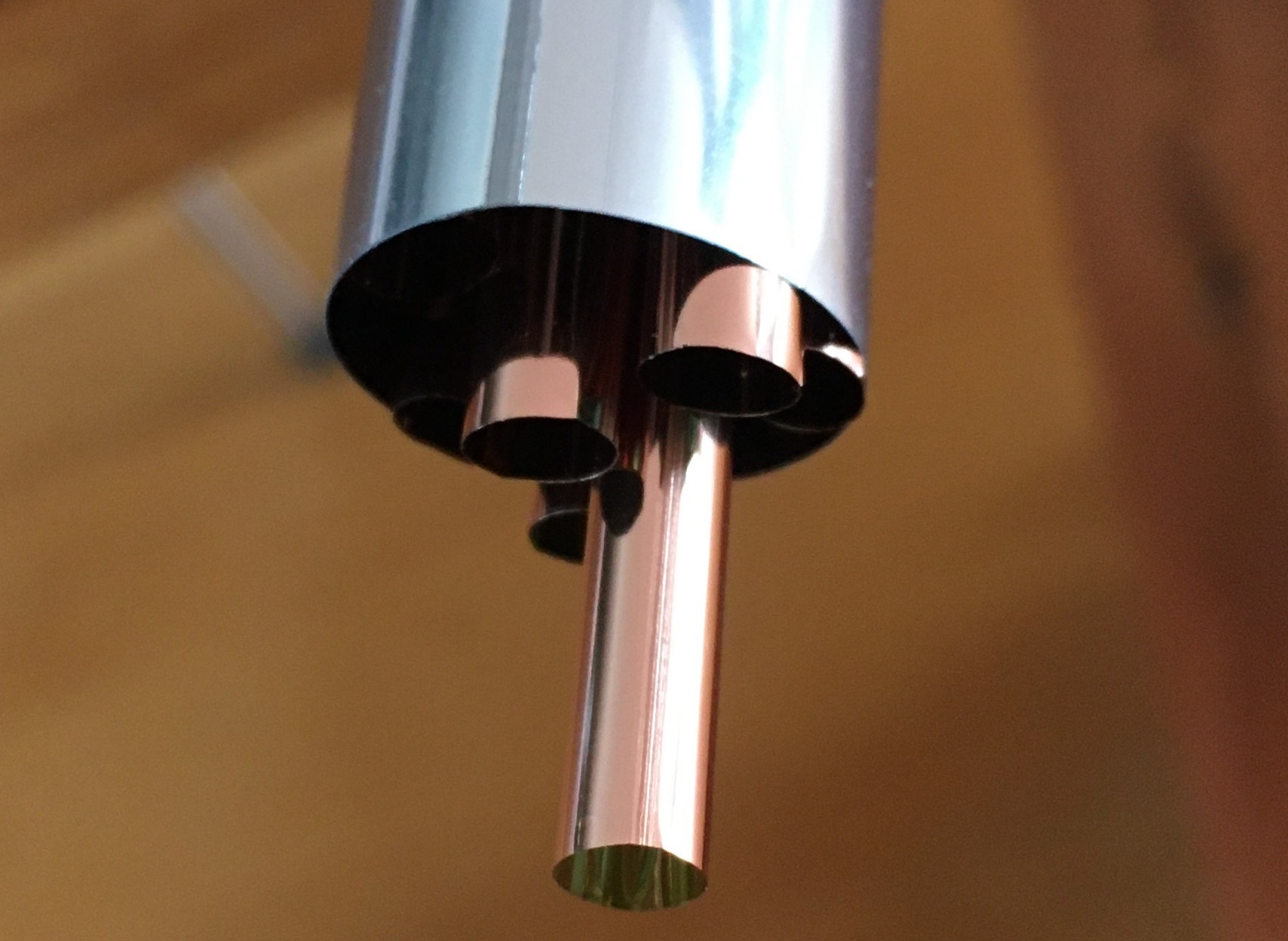}~
\includegraphics[width=0.45\textwidth]{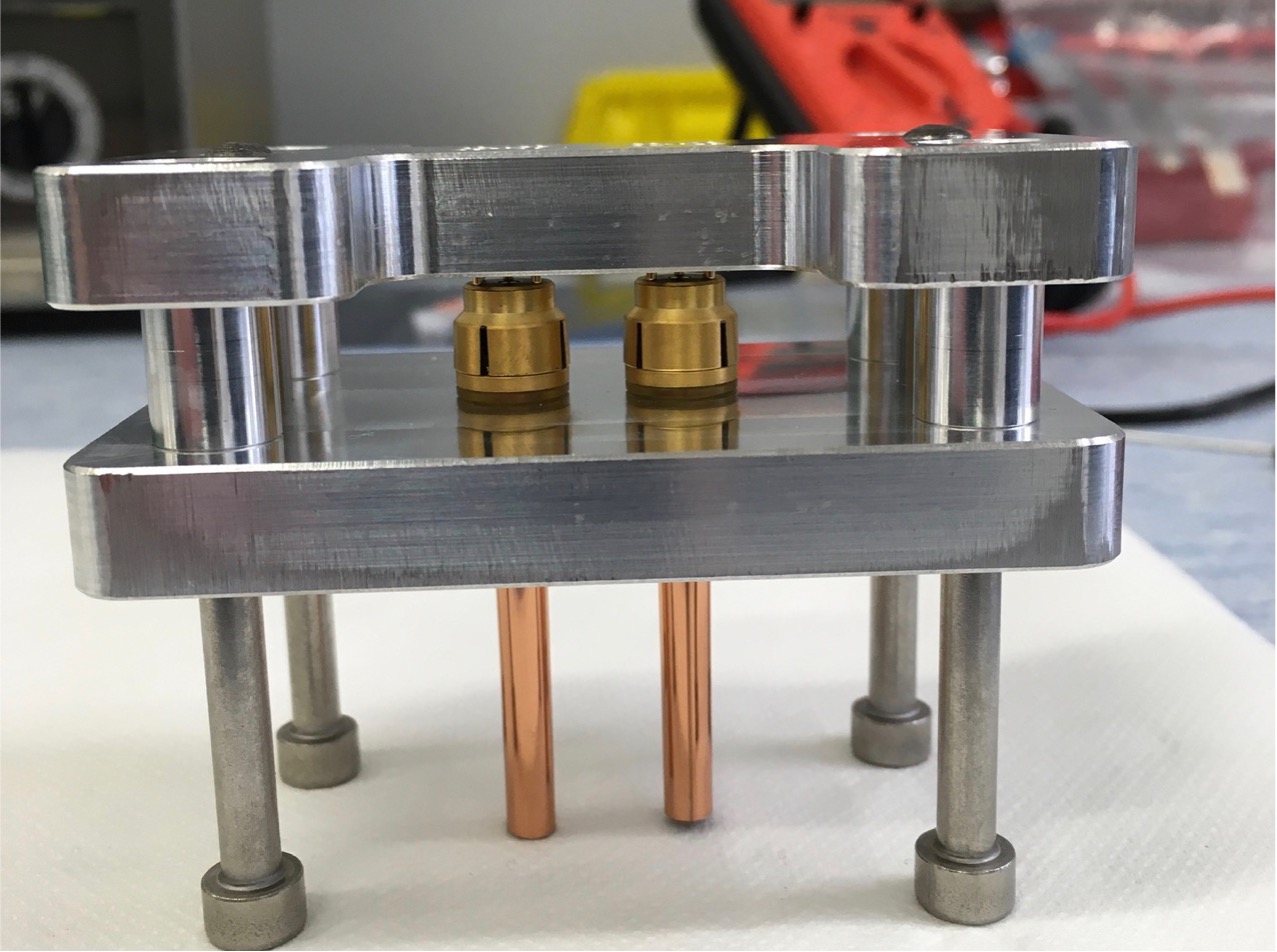}
\caption{Left: pre-production of straws with a diameter of 4.82~mm and a wall thickness of 19~$\mu$m. Right: test of signal connectivity and high-voltage stability of individual components. }    \label{fig:spectrometer:prototest}
\end{figure}

\begin{figure}[p]
\centering
\includegraphics[width=0.7\textwidth]{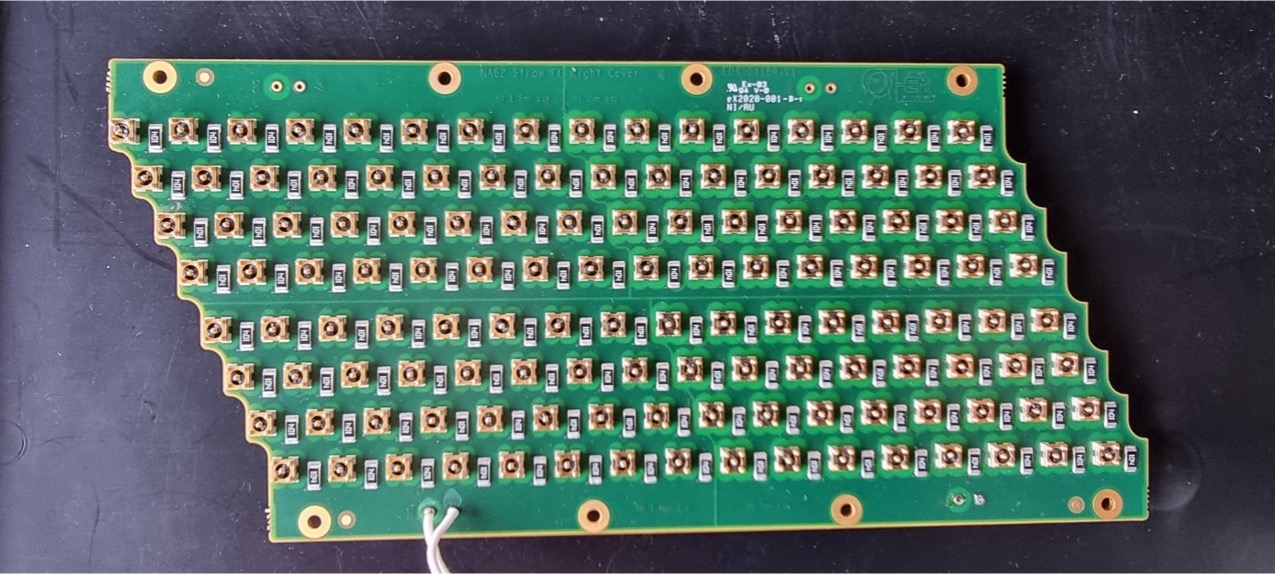}
\caption{Prototype of a HV board for the new straw spectrometer.}
\label{fig:spectrometer:hv_board}
\end{figure}

Investigations of possible technological solutions for the straw connectivity, design of a new high-voltage board, and a pre-production of straw tubes with a diameter of 4.82~mm have already been started (Figs.~\ref{fig:spectrometer:connectivity}, \ref{fig:spectrometer:prototest}, \ref{fig:spectrometer:hv_board}).

%%%%%%%%%%%%%%%%

\begin{figure}[p]
\centering
\includegraphics[width=0.37\textwidth]{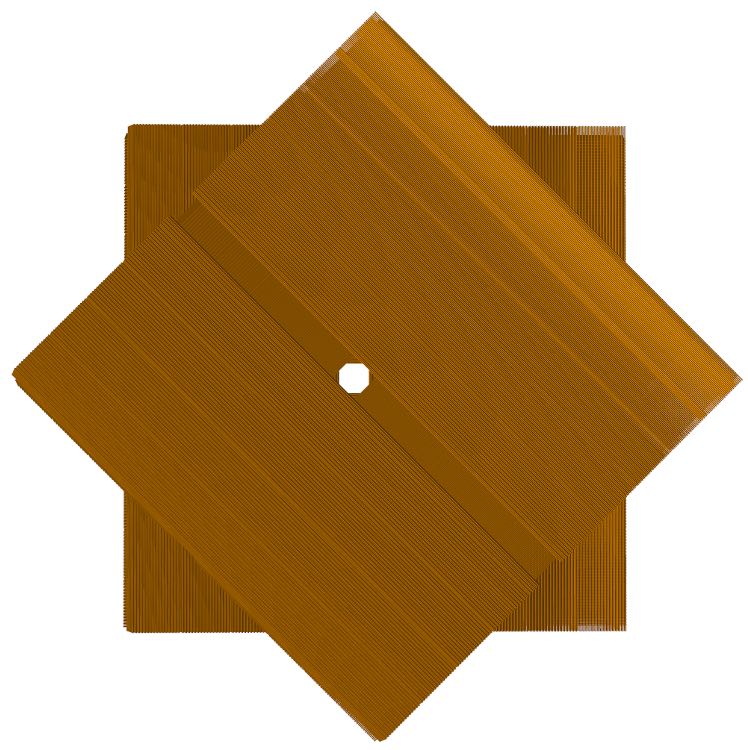}%
\includegraphics[width=0.37\textwidth]{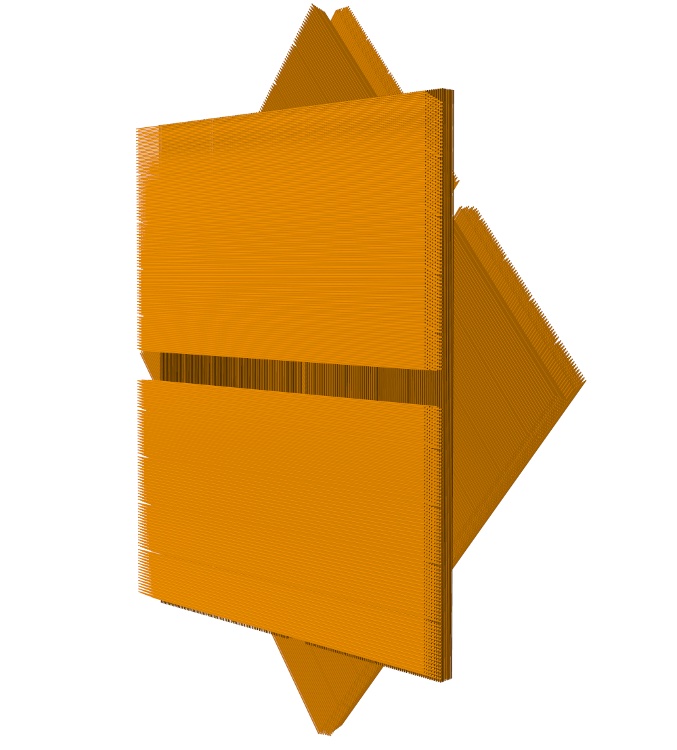}
\vspace{-2mm}
\caption{
\label{fig:straw_visualization2}
Geant4 visualisation of a new straw chamber: (left) front view; (right) tilted back view.}
\end{figure}

\begin{figure}[p]
\centering
\includegraphics[width=0.5\textwidth, page=1, trim={0.0cm 0.0cm 1.8cm 0.0cm}, clip]{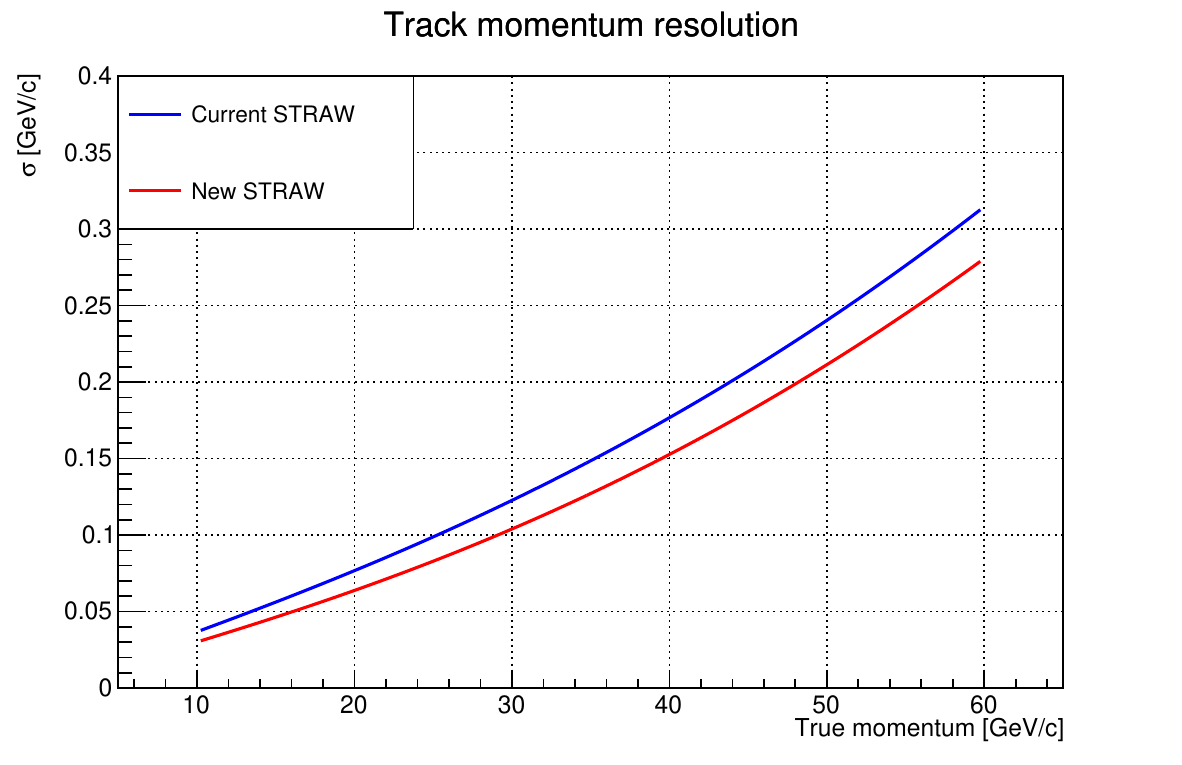}%
\includegraphics[width=0.5\textwidth, page=3, trim={0.0cm 0.0cm 1.8cm 0.0cm}, clip]{straw_resolutions.pdf}
\vspace{-8mm}
\caption{
\label{fig:straw_resolution}
Preliminary comparison of resolutions of track momentum (left) and track $\theta_X$ angle (right) between the existing NA62 spectrometer (blue) and the new spectrometer with $12~\mu$m mylar thickness (red). A similar improvement is observed in the reconstructed $\theta_Y$ angle.}
\end{figure}

\subsection{Detectors downstream of the decay volume}

\subsubsection{Ring-imaging Cherenkov counter (RICH)}

The NA62 RICH, which uses neon at atmospheric pressure as the radiator, is well suited for operation within the HIKE programme in terms of the mechanical structure (vessel, mirror support, end-caps). Major changes only concern the Cherenkov light sensors and the two flanges hosting them. Improvement of the geometrical acceptance for negative particles is also being considered.

\paragraph{New photodetectors}

The present NA62 RICH is equipped with Hamamatsu R7400-U03 phototubes with a time resolution of 240~ps for single photons, quantum efficiency (QE) of $\sim$20$\%$ and with a distance between the centres of adjacent sensors of 18~mm \cite{Anzivino_2020}. This distance, constrained by the sensor size, gives the main contribution to the resolution on the single hit position ($\sim$4.7~mm) and consequently to the overall resolution on the ring radius ($\sim$1.5~mm), the main parameter driving the performance on the particle identification (PID) of the RICH detector \cite{Anzivino_2018}. The single hit time resolution together with the number of hits associated to each ring determines the overall time resolution of the RICH for positive tracks of 80~ps on average.

The main requirement for RICH operation at HIKE Phase~1 is a ring time resolution of 20--30~ps, in order to reduce the coincidence window with the upstream detectors (beam tracker and KTAG). The RICH is the only downstream detector that could reach such resolution for the kaon decay products. Besides, a reduction of the sensor size would improve the particle identification performance.
Silicon photomultiplier (SiPMs)
meet these requirements: SiPMs with 100~ps time resolution and QE of 40\% or above are already available from the main manufacturers.

The other fundamental parameter to improve is the number of Cherenkov photons collected for each ring, depending on the geometrical acceptance of the RICH, on the number of photon originally emitted and on the QE of the sensor.

\begin{figure}
\centering
\includegraphics[width=0.5\textwidth]{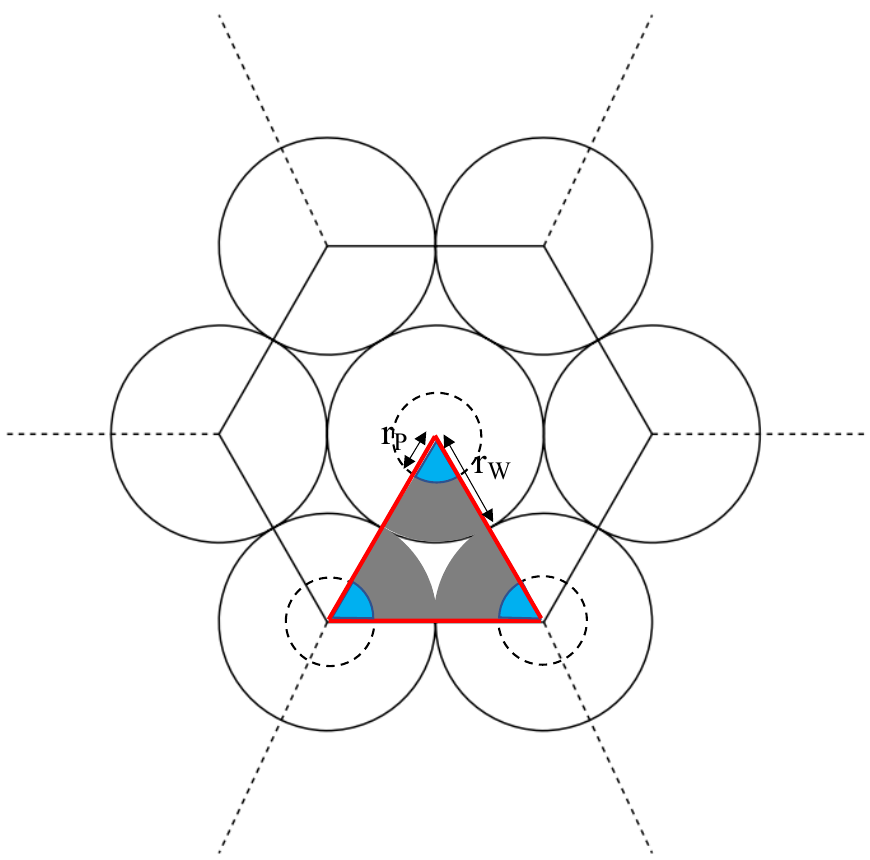}
\caption{The NA62 RICH elementary cell derived from the positions of the photomultipliers in the two flanges.}
\label{fig:elementary-triangle}
\end{figure}

Let's first consider the geometrical filling factor for the present RICH flanges housing photomultipliers and the future ones instrumented with SiPM. 
For the NA62-RICH, the red triangle in Fig.~\ref{fig:elementary-triangle} is the elementary cell unit that, once replicated, can describe the NA62-RICH flanges. 
The radius of the photocathode is $r_P=3.5$~mm and the radius of the Winston cone is $r_W=9.0$~mm.
In that triangle, three regions can be identified:
\begin{itemize}
\item Between the Winston cones, the white area inside the triangle in Fig.~\ref{fig:elementary-triangle}, corresponding to $9.3\%$ of the elementary cell unit, that is completely passive (not instrumented).
\item The grey region, where the Cherenkov photons have to be reflected by the Winston cone (mylar surface, average reflectivity equal to $85\%$) at least once in order to reach the photocathode; the fraction between this surface and the area of the triangle is $77.0\%$.
\item The blue region where the Cherenkov photons hit directly the photocathode of the R7400 (the fraction of triangle area is $13.7\%$).
\end{itemize}
The relative ``geometrical'' inefficiency introduced by the flange layout is therefore $9.3\%+(1-0.85)\times 77.0\%=20.9\%$.
In the SiPM case the geometrical inefficiency can be strongly reduced: for a $9\times9~\rm{mm}^2$ sensors with a spacing between
the sensor of 0.2~mm, the inefficiency would be $4.5\%$. Such geometrical inefficiency sums up with the inefficiency due to the SiPM micro-structures, that in many SiPM already available on the market is at the level of $20\%$ and will be reduced in the coming years.
From the above considerations, the contribution of ``geometrical inefficiency'' (at the micro and macro scales) in the calculation of the expected number of hits for the HIKE upgrade of the RICH can be considered equal to the current NA62-RICH.

Considering a similar average yield of Cherenkov photons at HIKE with respect to the NA62 RICH (radiator and vessel length will not be changed), the only parameter that can increase the number of collected photons is the QE of the new sensors.
Taking into account a factor~2 improvement in the QE ($40\%$ for SiPMs against $20\%$ of the NA62-RICH PMs), the number of hits per ring and the track time resolution with the new configuration have been evaluated. The time resolution for pion momentum of 15 and 45~GeV/$c$
(the limits of the RICH working region) for the NA62 RICH, and those expected for the future HIKE-RICH instrumented with SiPMs, are listed in Table~\ref{tab:RICH1}. The latter meets the HIKE requirements.

%%%%%%%%%%%%%%%%%%%%%%

\newpage

\begin{table}[tb]
\begin{center}
\caption{Comparison of the time resolutions of the NA62-RICH and the HIKE-RICH.}
\vspace{-2mm}
\begin{tabular}{l|c|c}
\hline
& NA62 RICH & HIKE RICH \\ 
\hline
Sensor type & PMT & SiPM \\
Sensor time resolution & 240~ps & 100~ps \\
Sensor quantum efficiency & 20\% & 40\% \\
Number of hit for $\pi^+$ at 15~GeV/$c$ & 7 & 14 \\
Number of hit for $\pi^+$ at 45~GeV/$c$ & 12 & 24 \\
Time resolution for $\pi^+$ at 15~GeV/$c$ & 90~ps & 27~ps \\
Time resolution for $\pi^+$ at 45~GeV/$c$ & 70~ps & 20~ps \\
\hline
\end{tabular}
\label{tab:RICH1}
\end{center}
\vspace{-4mm}
\end{table}

\begin{figure}[tb]
\centering
\includegraphics[width=0.5\textwidth]{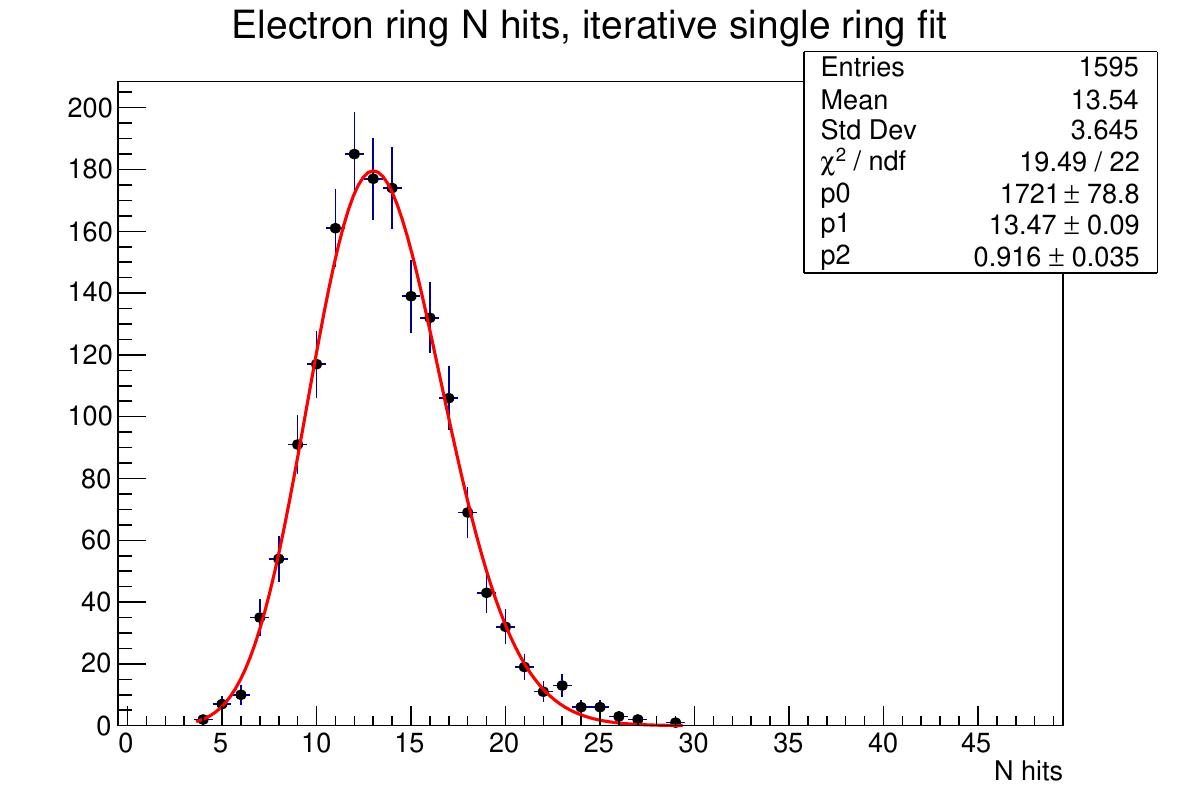}%
\includegraphics[width=0.5\textwidth]{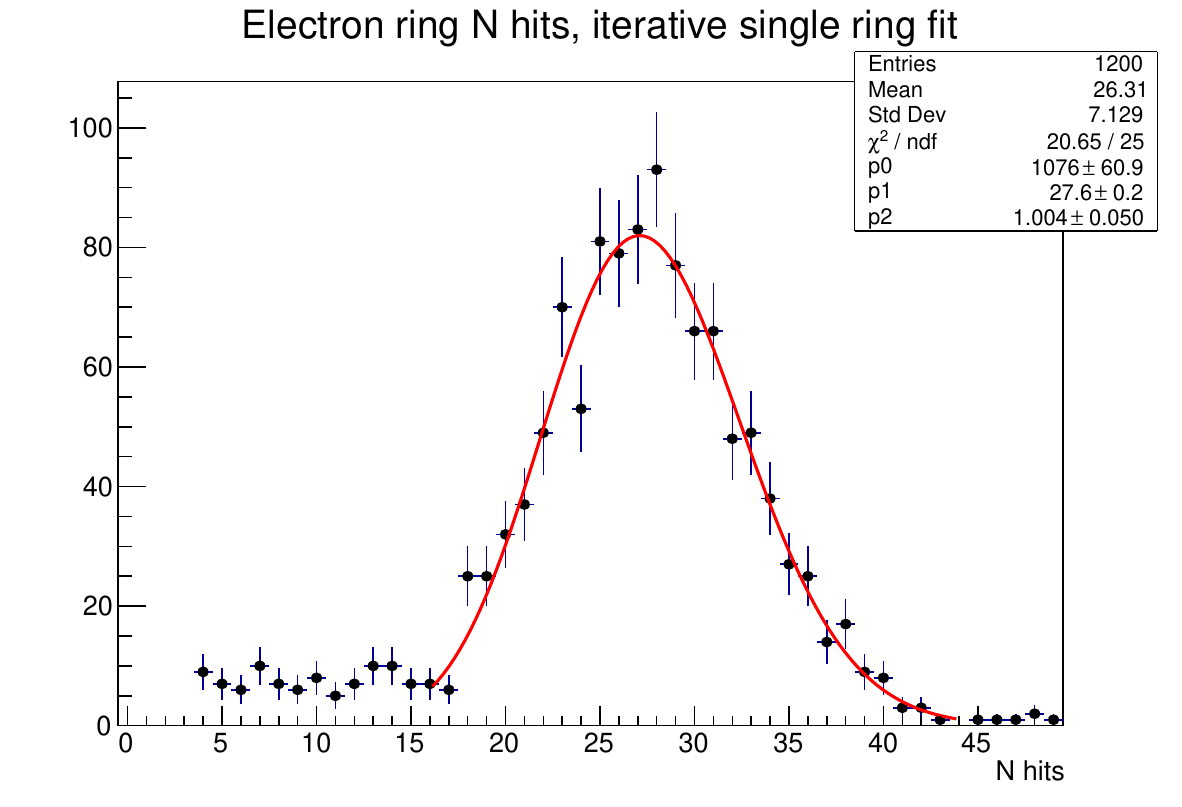}
\vspace{-5mm}
\caption{Number of hits in the rings produced positrons from reconstructed $K^+\to\pi^0e^+\nu$ decays for the official simulation of the NA62-RICH (left) and for the new proposed layout based on $9\times9~{\rm mm}^2$ SiPM (right). In the right plot low counts in number of hits appear from rings only partially inside the RICH geometrical acceptance due to the higher QE of the SiPM.}
\label{fig:RICH_upgrade_nhit}
\end{figure}

\begin{figure}[h]
\centering
\includegraphics[width=0.9\textwidth]{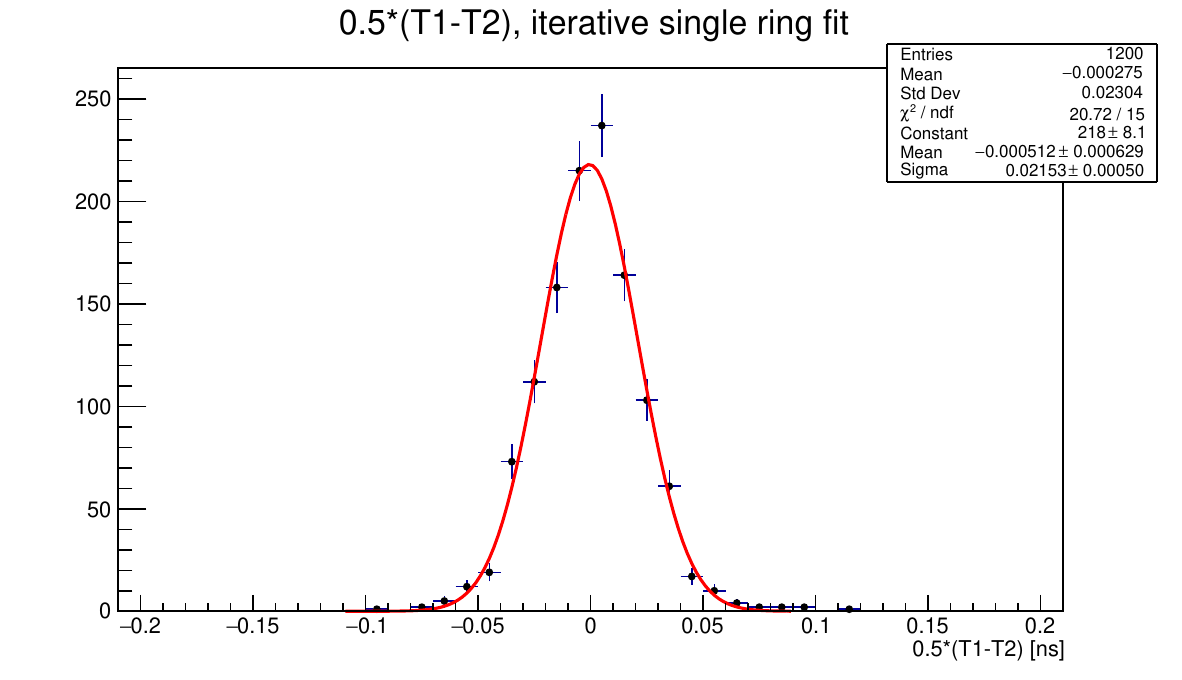}
\caption{Time resolution of the new proposed layout based on 9x9 mm$^2$ SiPM for the ring produced by the positron in the $K^+\to\pi^0e^+\nu$ decay. The resolution is calculated from the time difference between two equally populated sub-samples obtained splitting, for each event, the sample of hits belonging to the positron ring.}
\label{fig:RICH_upgrade_tres}
\end{figure}

The above considerations are confirmed by a Monte Carlo study of the RICH, where in the NA62 simulation the PMs are replaced by $9\times9~{\rm mm}^{2}$ SiPMs with a 100~ps time resolution and QE of 40\%.
Fig.~\ref{fig:RICH_upgrade_nhit} shows the number of reconstructed hits for the ring produced by the positron coming from $K^+\to\pi^0 e^+\nu$ decay for the standard NA62-RICH (left) and for the new simulation (right): the number of signals more than doubles in the new proposed configuration. The smaller sensor size reduces the probability of two Cherenkov photons hitting the same sensor, further increasing the number of reconstructed signals. 
The higher number of signals and the improved sensor time resolution result in a much improved time measurement of the positron ring at the level of 20~ps (Fig.~\ref{fig:RICH_upgrade_tres}).  

\begin{figure}[ht]
\centering
\includegraphics[width=0.8\linewidth]{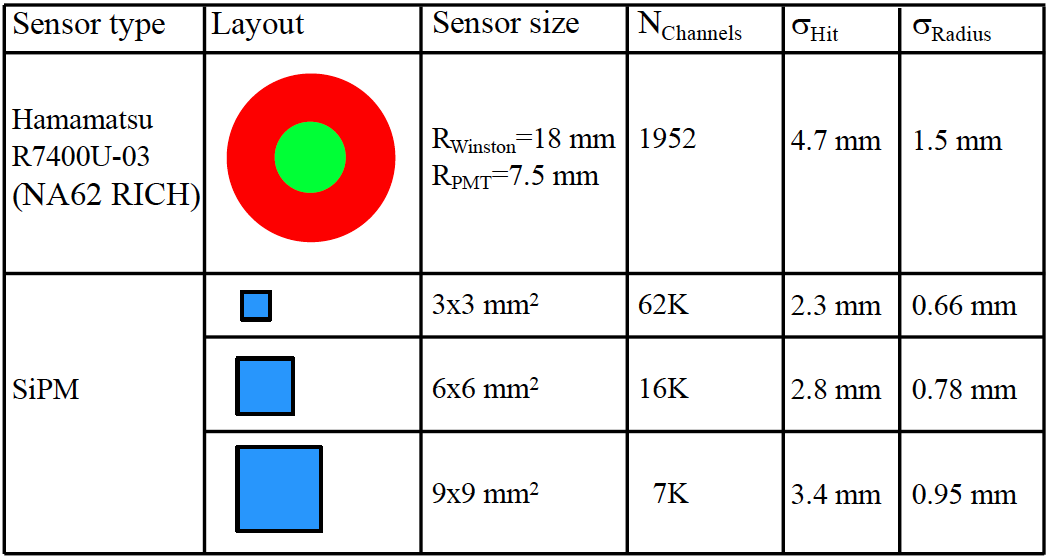}
\caption{Layout of the present RICH NA62 sensors compared to SiPM candidates (in scale) and the corresponding performance in terms of the ring radius resolution.}
\label{fig:RICH1}
\end{figure}

Replacement of the light sensors, necessary to improve the time resolution, also represents an opportunity to improve the RICH performance in terms of particle identification. A smaller sensor size and an improved QE will allow to establish a more optimal RICH working point, improving both the muon rejection and pion identification efficiency in the $K^+\to\pi^+\nu\bar\nu$ analysis. Fig.~\ref{fig:RICH1} illustrates the improvement in the ring radius resolution $\sigma_{\rm Radius}$ achieved by sensor size reduction and QE improvement. For a SiPM size smaller than $3\times3$~mm$^2$, the contributions to the single hit resolution coming from misalignment of the reflecting mirrors (0.6~mm) and neon (radiator) dispersion (2.1~mm) become dominant. 

The SiPM option of $9\times 9$~mm$^2$ size satisfies the HIKE requirements and allows to keep the number of channels at a reasonable level. In considering SiPMs as photodetector candidates for the HIKE RICH, mitigation of the following drawbacks is being evaluated:
\begin{itemize}
\item[-] Dark count: for a 800~ps coincidence window, and an annulus area of $7\times 10^{4}~{\rm mm}^2$ considered for evaluation of the possible particle identification hypotheses, a dark count rate of several kHz/mm$^2$ would produce a non-negligible number of spurious hits. To lower the contamination to the level of few percent, the SiPM should be operated at low temperature. A possible layout of the flanges housing the SiPM and the cooling system is discussed in Section~\ref{RICH-mechanics}.
\item[-] Cross-talk, strongly dependent on the SiPM type, can be reduced by cooling the SiPM, and will be a consideration for the choice of the final sensor. Cross-talk leads to an extra contribution (for a small fraction of events) to the hit space resolution, and consequently to the ring radius resolution.
\item[-] Ageing/radiation hardness: the RICH flanges are located at a distance of 1.5~m from the beam pipe and are not traversed by the bulk of the particle flux. Nevertheless the radiation level in the high intensity environment should be investigated. Cooling of the SiPMs would reduce the effects of radiation damage.
\end{itemize}
%To add: 
Alternative photodetectors are the MCP-PMTs considered for the HIKE KTAG detector and described in Section~\ref{sec:ktag}.

\begin{figure}[tb]
\centering
\includegraphics[width=0.45\linewidth]{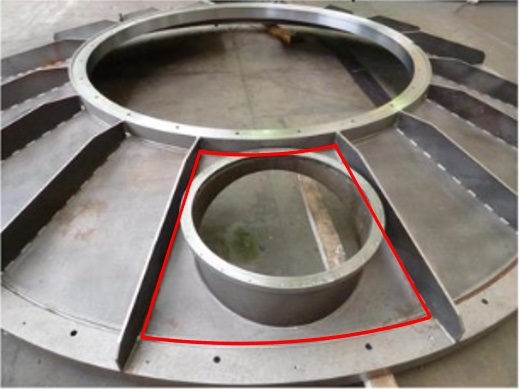}
\caption{The upstream RICH end-cap after its construction in 2013. The circular hole hosts the sensor flange in the NA62 RICH. The red contour delimits the region that can be instrumented with new photo-sensors.}
\label{fig:RICH2}
\end{figure}

\paragraph{New mechanics}
\label{RICH-mechanics}

The NA62 RICH vessel, end-caps and the mirror support panel will not be changed in the HIKE configuration, apart for minor modifications. 
Concerning the mirrors, a replacement or a re-aluminisation should be
considered to recover from the deterioration of the coating layer already observed in 2014 during their installation.
The new reflecting surface must match, in terms of reflectivity as a function of light wavelength, the quantum efficiency of the photodetectors, so both the reflective and coating layers will be chosen accordingly.

The change of sensors will request modification of the two
flanges hosting them. This represents an opportunity to increase the RICH acceptance for negatively-charged particles (note that the NA62 RICH is optimised
for the $K^+\to\pi^+\nu\bar\nu$ decay). Simulations show that increasing the instrumented area from 5700 cm$^2$ to 7600~cm$^2$ would lead to a good acceptance for negative tracks. The end-cap region that can be instrumented without compromising the vacuum-proof mechanics is shown in Fig.~\ref{fig:RICH2}. In the redesign of the new flanges, a photosensor cooling system will be introduced, which must avoid
inducing a temperature gradient in the RICH radiator gas. An adequate system to guarantee thermal insulation between the sensor flanges and the vessel will be implemented.  

\subsubsection{Timing detector}

The timing detector will allow to have a redundant and complementary measurement of the transit time of charged particles in the downstream detector region with respect to the RICH. The information of the timing plane can be used for efficiency studies, trigger purposes and to veto accidental activity. The timing detector can also be used to have more than one time measurement for pions and muons with momentum below the Cherenkov thresholds in the RICH of 12 and 8~GeV/$c$, respectively. 
In the $K^+\to\pi^+\nu\nu$ analysis of HIKE, the timing detector is used mainly for the multiplicity veto, i.e., the veto on extra tracks and shower fragments, while the RICH provides the downstream track time used for matching the upstream track, as in NA62. As the principal role of the timing detector for the most demanding analysis is as a veto, the principal timing concern is to minimise the random veto rate, and the required time resolution is determined by the expected hit rate. Like the other veto detectors for HIKE, the resolution required is of the order of 100~ps.
%The requested time resolution is about 100~ps.

\begin{figure}[tb]
\centering
\includegraphics[width=0.8\linewidth]{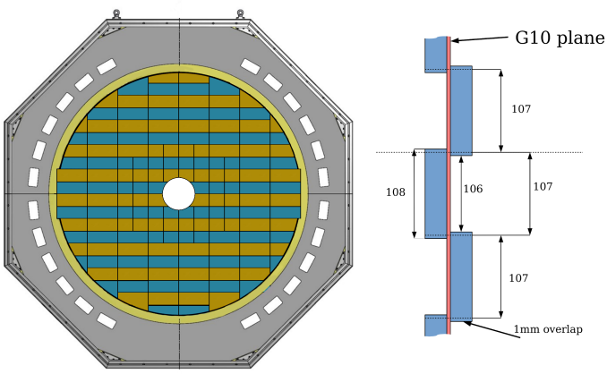}
\caption{Layout of the present NA62 CHOD detector in the $xy$ and $xz$ planes.}
\label{fig:CHOD1}
\end{figure}

For the construction of this detector HIKE will profit from the expertise gained by NA62 with the NA48-CHOD and NA62-CHOD detectors. The NA48-CHOD, consisting of two planes of scintillator
slabs of approximately $100\times6$~cm$^2$ size, cannot be used in the HIKE environment; the high particle rate (and the high probability of more than one track hitting the same counter) will not allow to 
correct the measured time for the light propagation time in the scintillator to the photo-sensors depending on the track impact point.
On the contrary, the NA62-CHOD layout (Fig.~\ref{fig:CHOD1}),
consisting of tiles of approximately $10\times 13$~cm$^2$ (in the central region) will be suitable for HIKE. The expected maximum rate per tile of 2.8~MHz (inferred from experience with NA62 data taking) is affordable and can be lowered further by reducing the dimension of the tiles near the beam pipe, where the particle rate is higher. 

The NA62-CHOD time resolution (500~ps) is limited by the light transportation outside the detector acceptance via wavelength shifting fibers. In the HIKE setup, the photosensors will be attached
directly to the tiles to avoid the degradation. In this layout the sensors will be irradiated by the downstream particle flux. The photosensor type will be chosen profiting from the studies foreseen for other detectors (KTAG, RICH, HCAL, MUV).  To further improve the time resolution, an additional detector plane can be considered.

%cost estimate from Piccini
% all tiles 10x10cm = 400 tiles. 2 sensors per tile = 800 channels
%2 planes = 1600 channels
% TDC

\subsubsection{Main electromagnetic calorimeter}
\label{sec:calorimeter}
 
A main electromagnetic calorimeter (MEC) at forward angles is crucial for any kaon programme. In particular, for HIKE Phase 1, this detector serves as the principal photon veto for the measurement of $K^+\to\pi^+\nu\bar\nu$ and is essential for the reconstruction of the final states for decays like $K^+\to\ell^+\nu$, $K^+\to\pi^+\ell^+\ell^-$, and $K^+\to\pi^+\gamma\gamma$. For Phase~2, the electromagnetic calorimeter reconstructs the $\pi^0$ in the 
$K_L\to\pi^0\ell^+\ell^-$ decays,
serves for particle identification, and helps to reject events with extra photons. 
Many of the same issues arise in the design of the electromagnetic calorimeter for the $K^+$ and $K_L$ phases. 
We therefore seek a design for a fast calorimeter with excellent photon detection efficiency and energy resolution to be used in all phases of the HIKE programme, including the possible future $K_L\to\pi^0\nu\bar\nu$ measurement. 

\paragraph*{Performance requirements}

The principal performance requirements for the electromagnetic calorimeter are excellent energy resolution and intrinsic detection efficiency for high-energy photons, good two-cluster separation for photons, and excellent time and double-pulse resolution.

It is natural to inquire as to whether the liquid-krypton calorimeter (LKr)~\cite{Fanti:2007vi}
currently used in NA62 can be reused for HIKE.
The energy, position, and time resolution of the LKr calorimeter were measured in NA48 to be
\begin{align}
\frac{\sigma_E}{E} &=0.0042\oplus\frac{0.032}{\sqrt{E {\rm (GeV)}}}\oplus\frac{0.09}{E {\rm (GeV)}}, \\
\sigma_{x, y} & = 0.06~{\rm cm} \oplus \frac{0.42~{\rm cm}}{\sqrt{E {\rm (GeV)}}}, \\
\sigma_t & = \frac{2.5~{\rm ns}}{\sqrt{E {\rm(GeV)}}}.
\end{align}
The efficiency
and energy resolution of the
LKr calorimeter appear to be satisfactory for all phases of HIKE.
Studies of $K^+\to\pi^+\pi^0$ decays with NA48 data and tests conducted in 2006 with tagged photons from an electron beam confirmed that the LKr has an inefficiency of less than $10^{-5}$ for photons with energies above 10~GeV, providing the needed rejection for forward photons~\cite{NA62+07:M760}.
This result was fully confirmed by NA62. Notwithstanding the presence of a much larger amount of material upstream of the LKr calorimeter in NA62 than in NA48, a study of single-photon efficiency underpinning the NA62 measurement of ${\cal B}(K^+\to\pi^+\nu\bar{\nu})$ and used to obtain a limit on ${\cal B}(\pi^0\to X_{\rm inv})$
found an inefficiency of about $10^{-5}$ at 20~GeV, slightly decreasing at higher energies~\cite{NA62:2020pwi}.
The LKr time resolution, however, is a significant issue. 
For both HIKE phases, a faster calorimeter will be necessary in order to avoid unacceptable losses to random veto; in addition, the signals must be limited to 20~ns duration FWHM to resolve overlapping clusters.
For the $K_L$ phase, 
the calorimeter must also provide the neutral event time measurement, and must have a time resolution of 100~ps or better for the reconstruction of $\pi^0$ mesons with energies of a few~GeV,
in order to tolerate a coincidence rate of up to 100~MHz while holding random-veto losses to at most 25\% with an adequate safety margin.
Additionally, the size of the LKr calorimeter inner bore would limit the beam solid angle and hence the kaon flux during the $K_L$ phase. 

A new calorimeter that would meet these requirements is described below. However, the existing LKr calorimeter is a versatile instrument with a track record of nearly 30~years of success. Before it is decommissioned, it is essential to thoroughly validate the solution that will replace it. Indeed, there is an ongoing effort in HIKE to evaluate possible upgrades which would allow the continued use of the LKr calorimeter during the commissioning and early phases of the $K^+$ programme if needed.

\paragraph{The new HIKE electromagnetic calorimeter} 
\label{sec:klever_mec}

The baseline design for the HIKE MEC is a shashlyk calorimeter patterned on the PANDA FS calorimeter~\cite{PANDA:2017sxm}, in turn
based on the calorimeter designed for the KOPIO experiment~\cite{Atoian:2007up}. This design featured modules $110\times110$~mm$^2$ in cross section made of 
alternating layers of 0.275-mm-thick lead absorber and 1.5-mm-thick injection-moulded polystyrene scintillator.
This composition has a radiation length of 3.80 cm and a sampling fraction of 39\%. 
The scintillator layers were optically divided into four $55\times55$~mm$^2$ segments; the scintillation light was collected by WLS fibres traversing the stack longitudinally and read out at the back by avalanche photodiodes (APDs). 

With a $3\times3$-module prototype of this design, KOPIO measured energy and time resolutions of
\begin{align}
\frac{\sigma_E}{E} &= \frac{(2.74\pm0.05)\%}{\sqrt{E~{\rm(GeV)}}} \oplus (1.96\pm0.10)\%, \\
\sigma_t &= \frac{(14\pm2)~{\rm ps}}{E~{\rm (GeV)}} \oplus \frac{(72\pm14)~{\rm ps}}{\sqrt{E~\rm(GeV)}},
\end{align}
over the range 100--500~MeV~\cite{Atoian:2007up}.
The relatively large value of the constant term was ascribed to longitudinal leakage, given the shallow $15.9X_0$ depth of prototype modules. A full-depth module for HIKE would be at least $25X_0$ or about 90~cm thick.
PANDA extended the measurement of the energy resolution to the range 1--20~GeV for a similar prototype with a depth of $20X_0$, confirming the value of the stochastic term and obtaining a smaller value for the constant term, 1.3\%~\cite{Morozov:2009zza,PANDA:2017sxm}.
These results demonstrate that the fine-sampling shashlyk design is capable of providing the same energy resolution as the LKr while meeting the time resolution requirements for HIKE. Note that the KOPIO and PANDA protoypes made use of Kuraray Y11 WLS fibres, which have an emission decay time of about 7~ns. The timing performance might benefit further from the use of faster fibres, such as Saint Gobain BCF-92 or 
Kuraray YS1 and YS2, all of which have decay times of about 3~ns. 

The choice to use APDs rather than photomultiplier tubes (PMTs) in PANDA was dictated by the lower quantum efficiency of the available PMTs at green wavelengths, as well as to eliminate the need for a costly high-voltage system with cumbersome cabling (the APDs featured on-board power supplies).   
For HIKE, these advantages could be procured by updating the design to use silicon photomultipliers (SiPMs) instead of APDs. 
While SiPMs are certainly a more cost-effective solution, new generation metal-package PMTs, such as the square, $18\times18~{\rm mm}^2$ R7600U-20 from Hamamatsu, offer green-extended spectral response, very fast timing (a 3.2~ns FWHM pulse), and gains of $10^6$ at 800~V, and are thus excellent candidates for evaluation as alternatives to SiPMs, with presumable advantages in terms of radiation hardness and possibly dynamic range.   

The final choice of transverse module size and readout granularity has yet to be determined. 
HIKE simulations assume that electromagnetic clusters are resolved if more than 6~cm apart. The effective Moli\`ere radius of the PANDA/KOPIO design is~59~mm\footnote{This value was misstated in the Letter of Intent.}, so readout cells as large as $55\times55$~mm$^2$ could be reasonable. Optimization studies are in progress, and smaller cells could be used if needed. 
\begin{figure}
    \centering
    \includegraphics[width=0.52\textwidth]{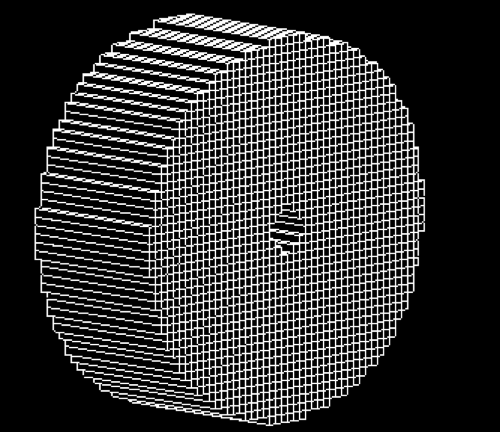}
    \caption{Wireframe drawing of the MEC as implemented in Geant4 for the NA62 Monte Carlo.}
    \label{fig:mec}
\end{figure}

The new calorimeter is optimized for the neutral beam and
will have an inner bore of at least 12~cm in radius to allow the passage of a beam with an opening angle of 0.4~mrad. The bore could be widened to as much as 15~cm to allow the penumbra of beam photons and neutral hadrons to pass through and be intercepted by the small-angle vetoes. For the $K^+$ phase, a smaller inner aperture would be required; in NA62, this angular region is covered by the IRC (\Sec{sec:irc}), and a new IRC based on the same shashlyk technology would be used together with the MEC during the $K^+$ phase. The sensitive area has an outer radius of 125~cm. Fig.~\ref{fig:mec} shows a wireframe drawing of the calorimeter as implemented in the NA62 Geant4-based Monte Carlo for HIKE.

\begin{figure}
\centering
\includegraphics[width=0.7\textwidth]{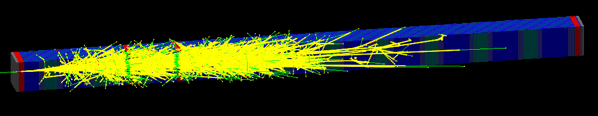}
\caption{Geant4 simulation of a 5~GeV photon showering in a MEC module.}
\label{fig:shashlyk_shower}
\end{figure}
\begin{figure}
\centering
\includegraphics[width=0.6\textwidth]{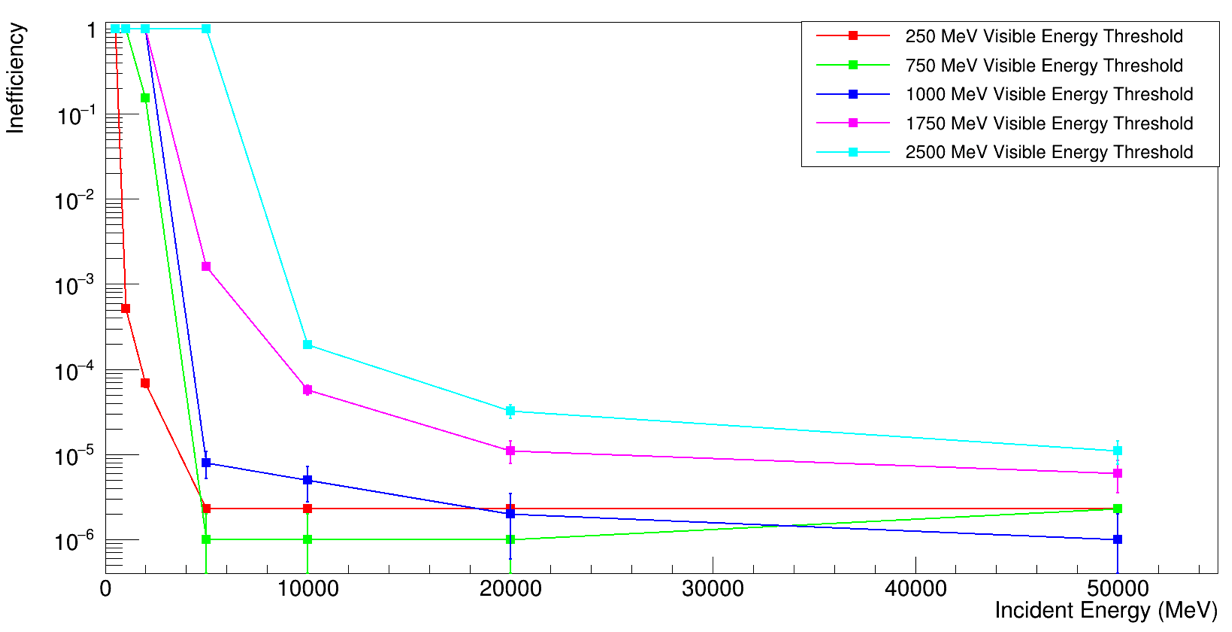}
\caption{Detection efficiency as a function of incident particle energy, for various thresholds on visible energy, from the Geant4 simulation.}
\label{fig:shashlyk_efficiency}
\end{figure}

The inefficiency for photon detection assumed for the MEC is 1 for energies below 100~MeV, falling exponentially to $10^{-3}$ at 1~GeV, $10^{-4}$ at 5~GeV, and then to $10^{-5}$ at 15~GeV, in accordance with experience with the LKr as discussed above. 
The model of the shashlyk calorimeter developed for the HIKE Monte Carlo does not include the light and signal readout, which are still being defined, but does contain a detailed representation of the module structure, allowing studies of the efficiency as a function of visible energy deposition. Fig.~\ref{fig:shashlyk_shower} shows the simulated interaction of a 5~GeV photon in a module of the calorimeter. Fig.~\ref{fig:shashlyk_efficiency} shows results obtained from the simulation for the photon detection inefficiency as a function of incident energy for various thresholds on visible energy. The simulation, which confirms the visible energy fraction of 39\%, demonstrates that the fine-sampling shashlyk design satisfies the efficiency requirements, 
as expected. 

In addition to the basic criteria on energy resolution, efficiency, time resolution, and two-cluster separation, ideally, the MEC would provide information useful for particle identification. 
The fine transverse segmentation of the MEC will play an important role: a simple cut on cluster RMS with the existing LKr can suppress up to 95\% of pion interactions in NA62 data. Fast digitisation of the signals from the MEC is expected to provide additional $\gamma/n$ discrimination. For both the $K^+$ and $K_L$ phases, the MEC will be backed up with hadronic veto calorimeters as discussed in \Sec{sec:hcal}.

\begin{figure}
    \centering
    \includegraphics[width=0.8\textwidth]{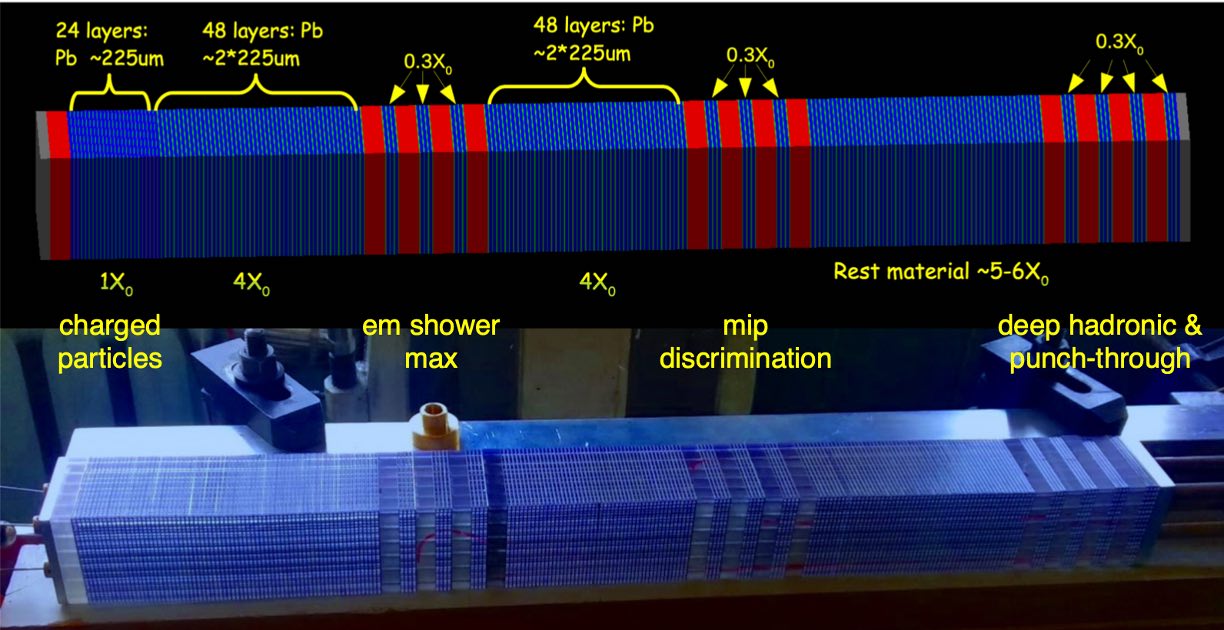}\\ \vspace*{5mm}
    \includegraphics[width=0.8\textwidth]{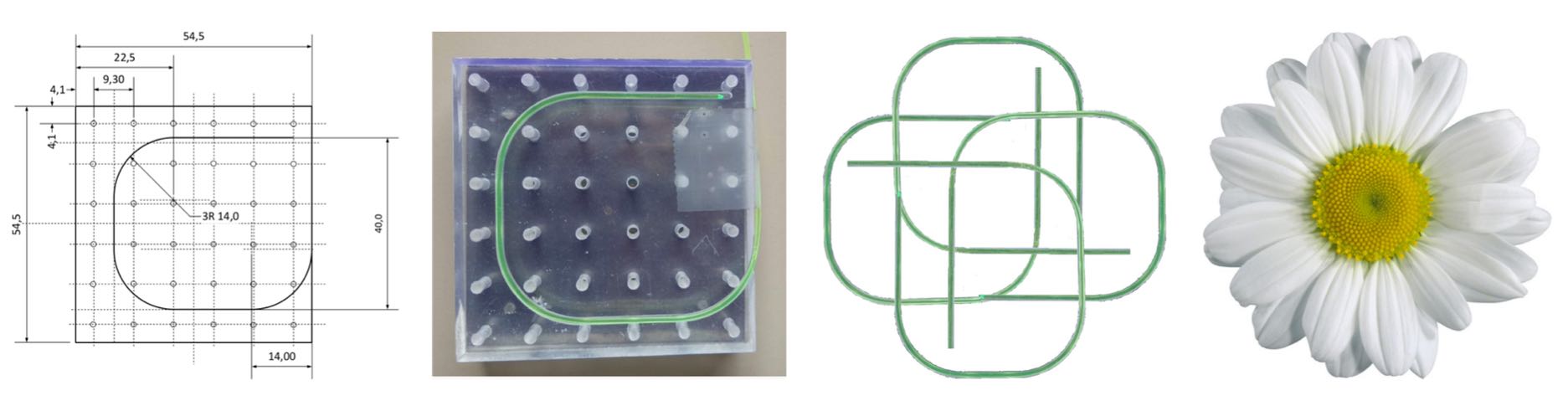}\\
    \caption{Top: Geant4 model of small prototype for {\it romashka} calorimeter, featuring spy tiles placed at key points in the shashlyk stack, together with a photograph of the module constructed at Protvino. Bottom: Fiber routing scheme for independent readout of spy tiles, giving rise to the name {\it romashka} (chamomile).}
    \label{fig:klever_spy_tiles}
\end{figure}
We are also experimenting with concepts to obtain information on the longitudinal shower development from the shashlyk design. One possible concept makes use of ``spy tiles'', 10-mm thick scintillator bricks
incorporated into the shashlyk stack but optically isolated from it and
read out by separate WLS fibres. The spy tiles are located at key points
in the longitudinal shower development: near the front of the stack,
near shower maximum, and in the shower tail, as illustrated in 
Fig.~\ref{fig:klever_spy_tiles}. This provides longitudinal sampling of the shower
development, resulting in additional information for $\gamma/n$ separation.
The prototype shown in  Fig.~\ref{fig:klever_spy_tiles} was constructed at Protvino in early 2018. Its basic functionality was tested in the OKA beamline in April 2018, and more comprehensive tests were carried out in September 2019 at DESY in collaboration with LHCb. Simulations suggest (and preliminary test beam data validate, to a certain extent) that the {\it romashka} design with spy tiles can give at least an order of magnitude of additional neutron rejection relative to what can be obtained from the transverse segmentation of the calorimeter alone, providing an overall suppression of up to 99.9\% for neutron interactions. 
The small prototype has a cross sectional area of only $55\times55$~mm$^2$ (one readout cell) and a depth of $14 X_0$ (60\% of the design depth), and both transverse and longitudinal leakage significantly complicated attempts to measure the time resolution. For electrons with $1 < E < 5$~GeV, the measured time resolution was about 200~ps and virtually constant; we expect that this can be significantly improved. The time resolution for hits on the spy tiles (independently of information from the shashlyk stack) was on the order of 500--600~ps, which may be difficult to improve. This is not expected to be a problem, however: the main shashlyk signal establishes the event time and the association of the PID information from the shashlyk tiles is based on the segmentation, with occupancies per cell of at most a few tens of kHz on the innermost layers.  

Although promising, the {\it romashka} design is not the only solution under investigation for obtaining information on the longitudinal shower development. Other concepts being explored, besides variations on the spy tile readout such as on-board SiPMs, include alternatives such as two-sided front/back readout and explicit segmentation into two or more modules in depth. 

The radiation resistance of the scintillator is a potential concern, especially for the $K_L$ phase. 
Precise dose rate calculations have yet to be performed, but an estimate suggests a dose of 4~kGy/yr to the scintillator for the innermost layers. This estimate would suggest that radiation damage, while a concern, is likely manageable. Radiation robustness may be a factor in the final choice of scintillator.

\begin{figure}
\centering
\includegraphics[width=0.4\textwidth]{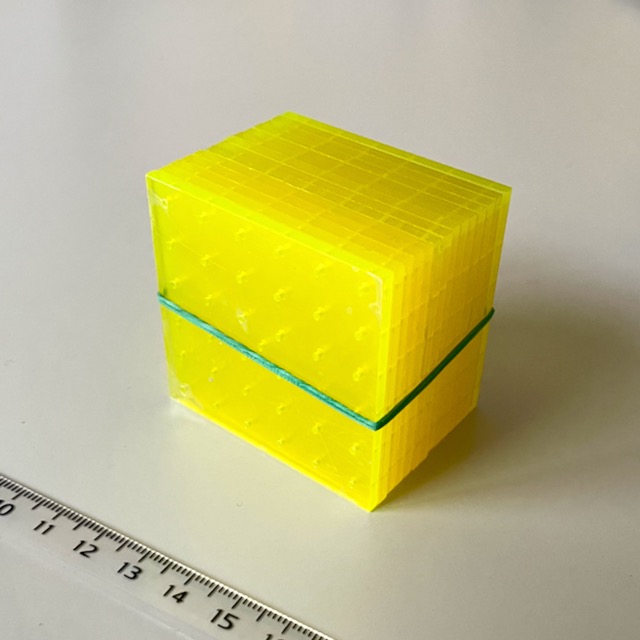}
\hspace{0.075\textwidth}
\includegraphics[width=0.4\textwidth]{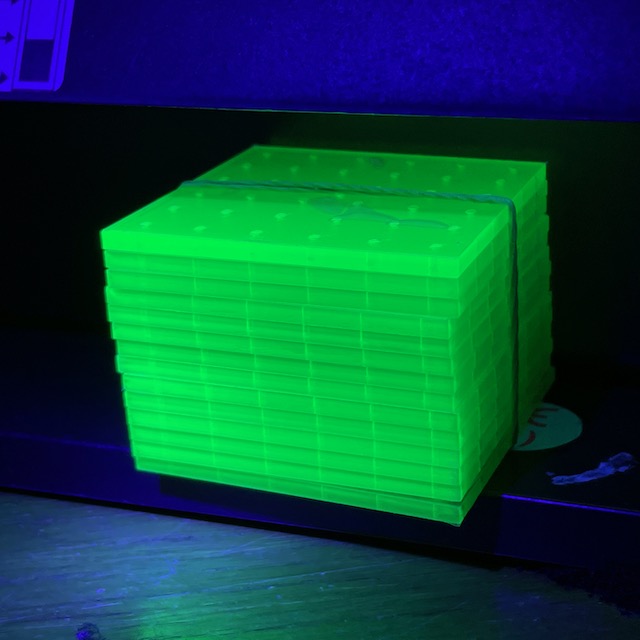}
\caption{Shashlyk tiles made of perovskite nanocomposite scintillator tested as an alternative to conventional scintillator, in ambient light (left) and under ultraviolet light (right).}
\label{fig:nc}
\end{figure}
Although current information suggests that optimized formulations of conventional polystyrene scintillator are sufficiently luminous, fast, and radiation resistant, in synergy with the AIDAinnova project NanoCal, we are evaluating the advantages that can be obtained with less conventional choices for the light emitter (e.g., perovskite~\cite{Gandini:2020aaa, Dec:2022aaa} or chalcogenide~\cite{Liu:2017aaa} quantum dots) or matrix material (e.g., polysiloxane~\cite{Acerbi:2020itd}). In particular, we have participated in a recent head-to-head experimental comparison of small shashlyk prototypes made from conventional scintillator, specifically, the extrusion-moulded polystyrene scintillator formulated at IHEP Protvino for KOPIO~\cite{Atoian:2007up}, with 1.5\%~PTP and 0.04\%~POPOP, and nanocomposite scintillators consisting of 0.2\%~cesium lead bromide (CsPbBr$_3$) or chloride (CsPbCl$_3$) nanocrystals in PMMA\footnote{Glass To Power SpA, Rovereto TV, Italy.}, with and without secondary fluors. These nanocomposites emit at around 520~nm (\Fig{fig:nc}) and are expected to be a very fast and bright alternative to conventional scintillators; the comparatively long wavelength emission and use of PMMA as a matrix material is expected to confer good radiation hardness. However, test beam results from 2022 and 2023 have so far shown significantly more light output from the conventional scintillator than from the nanocomposites. 

We remain interested in the use of novel scintillation materials to improve timing and increase radiation hardness, and continue to collaborate in the testing of successively optimised nanocomposite scintillator formulations,
as well as of heterogenous calorimeter structures and longitudinal segmentation schemes. Meanwhile, however,
we are pushing forward with the baseline design using conventional materials. We are designing a full-depth prototype of the baseline solution with conventional scintillator and a uniform shashlyk stack for beam testing in 2024, with particular attention to the sourcing of scintillating materials and other components for full-scale production.

Fermilab has been producing high-quality, low-cost extruded polystyrene scintillator for many years, but has only recently begin to produce scintillators via injection moulding. 
We have obtained information on the cost and production time of injection-moulded shashlyk tiles from Fermilab. The production of 1 million tiles (enough for the entire MEC) could be produced in about a year at a cost that is entirely compatible with the pricing assumptions made during the LoI phase. We are ready to move forward with a partnership to produce an initial lot of tiles for evaluation purposes. Injection-moulded scintillator can also be sourced from the Institute for Scintillation Materials (ISMA)\footnote{\url{isma.kharkov.ua}} at the National Academy of Sciences of Ukraine, in Kharkiv. The injection-moulding facility at ISMA was established to produce modules for the NICA calorimeter, which also makes use of the PANDA/KOPIO design. Notably, ISMA can also manufacture other components for shashlyk calorimetry, such as perforated lead foils.

We will use the opportunity provided by the test of the full-depth prototype in 2024 to study solutions for digitising the MEC signals.
For the beam tests of conventional and nanocomposite scintillators described above, the shashlyk prototypes were read out with Hamamatsu 13360-6050 SiPMs ($6\times6$~mm$^2$, 50~$\mu$m pixel size) and fast amplifier with a gain of 4, providing signals of duration 50~ns FWHM. This value could easily be reduced to 20~ns or less with moderate additional shaping. To obtain the best possible time resolution and double-pulse separation, we presume that it will be necessary to fully digitise the waveforms at high frequency (0.5--1~GS/s). We have therefore developed a four-channel prototype digitising readout based on Analog Devices 9680-1000 ADC evaluation boards, with 1~GS/s sampling at 14~bits, read out with a custom interface board with a Xilinx Kintex UltraScale+ FPGA. This solution will provide the sampling and dynamic range needed to for full evaluation of the performance requirements for the MEC readout system, as well as a starting point for its design. 

\paragraph{Upgrade possibilities for the LKr calorimeter}
\label{sec:LKr_calo}
As noted above, the LKr energy resolution meets the HIKE requirements, while the time resolution must be substantially improved. This would require a major upgrade of the LKr readout electronics.%, as discussed below.

The readout of the NA62 liquid krypton calorimeter is based on the measurement of the initial current induced by the charge generated by the showers, to have a fast response and to separate pile-up pulses. The electronics chain for each cell is composed of three elements:
\begin{itemize}
\item a preamplifier, sitting in the cold liquid,  to collect the charge with an integration time of about 150~ns;
\item a so-called transceiver module, mounted immediately after the cryostat feed-throughs, whose function is to restore the fast signal, removing the pole of the integration in the preamplifier, and to prepare a differential signal to drive long cables to the next stage;
\item a shaper, to have a narrow signal (70~ns), followed by an FADC to digitise the signal at a rate of 40~MHz.
\end{itemize}

The four-fold increase in beam intensity will produce more pile-up events, whose detection could be difficult with the existing chain. To cope with this, the baseline for a new readout structure will be the reduction of the shaping time to the minimum possible of about 28~ns (the transceiver output has a risetime of about 20~ns which adds to the time constants of the shaper) with a reduction of the amplitude of about 40\%, and subsequent digitization at 160~MHz, which is above the Nyquist limit, but which could help in identifying superposition of pulses looking at the width of the pulse.

The increase of intensity will also worsen local non-linear responses due to the effect of the space charge.
This effect, at low densities, reduces the electric field at the anode and increases it at the cathode. There is then a threshold value after which the field at the anode is zero and above this value an increasing fraction of the cell is unresponsive. An increase in intensity will not only worsen this behaviour for the hottest cells, but it will also extend the area affected by this problem. 
However, this effect does not cause a degradation of the performances for veto operation: energy clusters in the affected area are still reconstructed (and so they can be used for vetoing purposes) but their energy are underestimated by ${\cal O}(10\%)$. The only case in which this effect could increase the photon detection inefficiency is when the cluster energy is below 1~GeV, as the energy of the most-energetic cell in the cluster could be underestimated and go below the 250~MeV threshold used for cluster reconstruction. However, this type of events is not particularly dangerous: i) since the maximum allowed pion momentum is around 45~GeV, there are still around 30~GeV from additional particles in the rest of the detector that can be exploited to veto the event; ii) it is very unlikely that clusters with energy below 1~GeV reach the main calorimeter, as they are
usually emitted at larger angles.

With the current NA62 beam intensity we see already that in the hottest cells we are near or just above the critical value, clearly seen both from the decrease in the response across the burst and from a modulation of the response across the $x$ coordinate of the cells.

Space charge effects can be mitigated by an increase of the high voltage: this effect is directly proportional to the squared root of the energy density created by ions and inversely proportional to the value of the high voltage across the cell. 
Therefore, to maintain the same performance as now for a factor 4 increased intensity, the high-voltage needs to be increased by a factor of two.
Going from the actual 3.5~kV to an easily achievable 5~kV will worsen the value of the critical parameter by a factor of $\sqrt{2}$; the feasibility of increasing it up to 7~kV in order to retrieve the same performance as now is under investigation.

An additional improvement for all analyses that do not use the calorimeter as a veto will be to increase the size of the actual Intermediate Ring Calorimeter (IRC, \Sec{sec:irc}), to cover the area most affected by the space charge.

It is clear that the effort needed to upgrade the LKr to fulfill the HIKE specifications is such that a solution based on a new detector is preferable.
In addition the cost of maintaining the cryogenics infrastructure up-to-date would need to be evaluated.
However, the LKr calorimeter remains a valuable option for the early commissioning and data taking phases.

\subsubsection{Intermediate-ring calorimeter (IRC)}
\label{sec:irc}

In NA62, the IRC is a small, ring-shaped calorimeter with inner (outer) radii of 60~(145)~mm that sits just upstream of the LKr to provide photon veto coverage for the angular interval corresponding to the inactive region near the inner surface of the LKr cryostat. The inner and outer radii are not concentric; the inner bore is offset in the horizontal direction by 10~mm, so that it is centered on the beam axis. The NA62 IRC is a shashlyk calorimeter with layers consisting of 1.5~mm of lead and 1.5~mm of scintillator, segmented into quadrants and read out with PMTs.

In HIKE, as discussed in \Sec{sec:calorimeter}, the IRC will be used to extend the coverage of the calorimeter to small radii, and performance specifications similar to those for the calorimeter will be necessary. 
If the LKr used is during the $K^+$ phase, the IRC will cover the region of highest rates on the LKr, where space-charge effects are important. 
The new MEC, not needing a cryostat, will have minimum dead space at its inner radius. However, it could be convenient to have a bore of 15~cm or more to allow passage of the neutral beam halo during the $K_L$ phase of the experiment. In that case, during the $K^+$ phase, the IRC would be used in the same way with the MEC as it would with the LKr. 
For the $K_L$ phase, the IRC will then be moved downstream of the calorimeter and used to veto photons at the small radii occupied by the penumbra of the neutral beam. 

Ideally, the same instrument would be used for both phases, but due to the horizontal displacement of the $K^+$ beam at the location of the IRC, changes to the geometry may be necessary. 
If two different IRCs are needed, they can certainly be built with identical technology, namely, as shashlyk calorimeters with geometry similar to that for the NA62 IRC and sampling and readout granularity as for the HIKE MEC, providing more light for better time resolution and higher readout granularity for better rate resistance.

\subsubsection{Hadron calorimeter (HCAL)}
\label{sec:hcal}

In NA62 the hadron calorimeter is one of the main detectors for $\pi/\mu$ identification and separation.
It consists of two modules, called MUV1 and MUV2, and is followed by a fast muon veto plane (MUV3)~\cite{NA62:2017rwk}.
The MUV1-3 system achieves an average muon mis-identification probability of ${\cal O}$(\num{e-6}) over the momentum range from 10 to about \SI{50}{GeV/$c$}, while preserving 85\% of the pions.
Despite the NA62 performances would fully satisfy the HIKE requirements for $\pi/\mu$ separation, the currently employed detector layout (based on scintillator strips) would not be capable of sustaining the 4--6 times higher intensity environment.
The increase in the total particle rate on the calorimeter introduces dead time and ambiguities in the reconstruction of energy deposits, due to pile up activity in the long scintillator strips.
First signs of such limitation have already been observed in NA62 during period of data taking at higher intensity.

Therefore, a new hadron calorimeter (HCAL) with a cellular design will be constructed for HIKE to both reduce the rate on each channel and improve the time resolution.
Additionally, the HIKE HCAL will measure the energy deposited in each scintillating plane individually, thus providing a more efficient rejection of catastrophically-interacting muons (those releasing most to all of the energy in the calorimeter by bremsstrahlung or nuclear interactions) with respect to the NA62 setup.

\paragraph{HCAL baseline design}
\label{para:hcaldesign}

\begin{figure}
    \centering
    \raisebox{1pt}{%
    \includegraphics[width=0.31\linewidth]{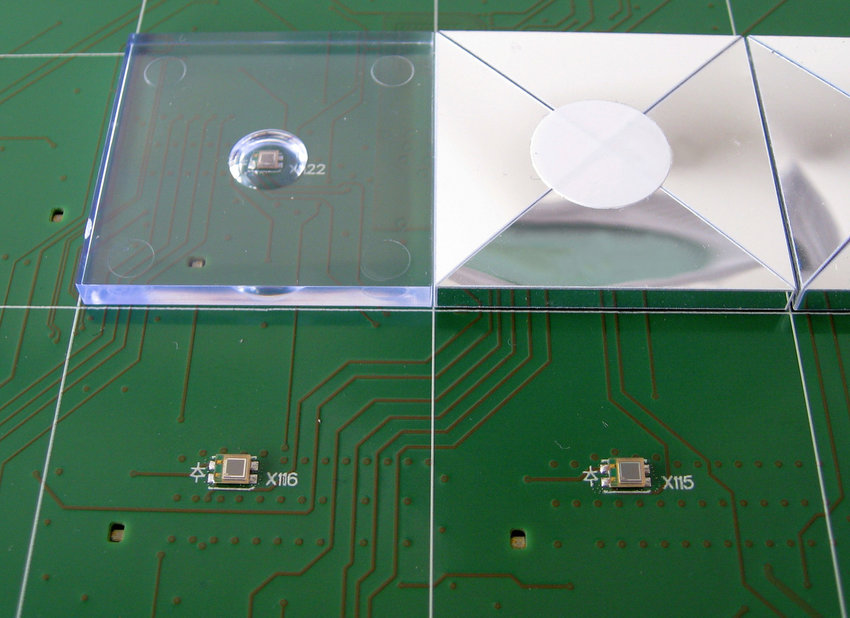}}
    \hfill
    \includegraphics[width=0.65\linewidth]{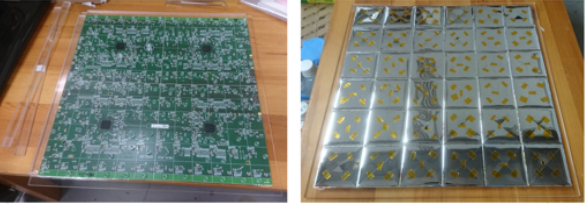}
    \caption{Left: \qtyproduct[product-units=power]{3x3}{\cm} scintillator tiles of the CALICE AHCAL, wrapped and unwrapped, mounted on a common readout board with SiPMs~\cite{Sefkow:2018rhp}.
    Centre and right: back and front sides of a readout board with \qtyproduct[product-units=power]{6x6}{\cm} scintillator tiles~\cite{Tsuji:2019zuj}.}
    \label{fig:CaliceTiles}
\end{figure}

The HCAL design is oriented  on the highly granular analogue hadron calorimeter 
(AHCAL), developed by the CALICE collaboration for a detector at a future $e^+e^-$ collider~\cite{Sefkow:2018rhp}.
The CALICE AHCAL is an iron-scintillator sandwich calorimeter with the active layers being built from 
scintillating tiles, which are read out via SiPMs, surface-mounted on an underlying PCB (Fig.~\ref{fig:CaliceTiles}).
At a future collider, high granularity is needed to improve jet energy resolution with particle-flow methods. 
While jet reconstruction is not necessary in HIKE, a similar tile design is nevertheless 
required for optimum $\pi/\mu$ separation in the HIKE high-rate environment. 

\begin{figure}
\centering
\includegraphics[width=0.48\linewidth]{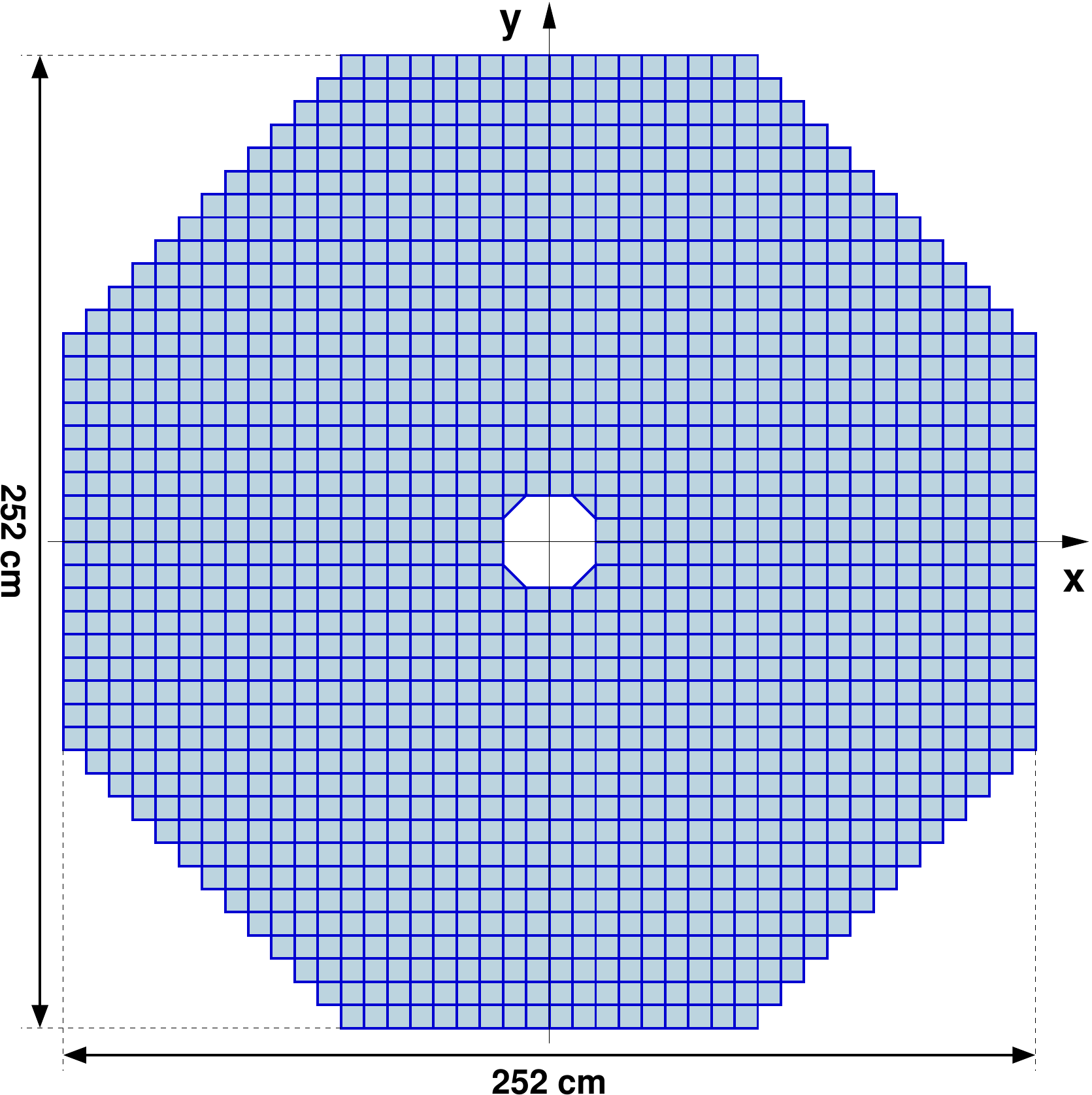}
\vspace{-1mm}
\caption{Transverse HCAL layout with 1440 readout cells of \qtyproduct[product-units=power]{6x6}{\cm} cross section.
The octagon has an inner radius of 126~cm, and leaves space for a beam pipe with 12~cm radius.}
\label{fig:HCAL_Layout}
\end{figure}

Using an iron absorber with Moli\`ere radius of \SI{1.72}{\cm}, a cell size of 
\qtyproduct[product-units=power]{6x6}{\cm}, corresponding to the \SI{6}{\cm} strip width in the NA62 MUV1 detector, 
allows a sufficient distinction between hadronic pion showers
and electromagnetic showers from catastrophically-interacting muons.
A corresponding HCAL layout with octagonal shape, of \SI{126}{\cm} inner radius, is shown in Fig.~\ref{fig:HCAL_Layout}.
It consists of a grid of \numproduct{42x42}~cells which, after the subtraction of the beam pipe region and the corners to obtain the octagon shape, results in \num{1440} cells in each layer.
We plan for in total 40 iron absorber layers of \SI{3}{\cm} thickness, interleaved with layers of scintillating tiles. 
With a tile thickness of \SI{6}{\mm} and sufficient space for a PCB with surface-mounted components, 
the HCAL has a longitudinal length of about \SI{1.80}{\m} and comprises \num{7.2}~nuclear interaction lengths. 

\begin{figure}
\centering
\raisebox{12mm}{%
\includegraphics[width=0.45\linewidth]{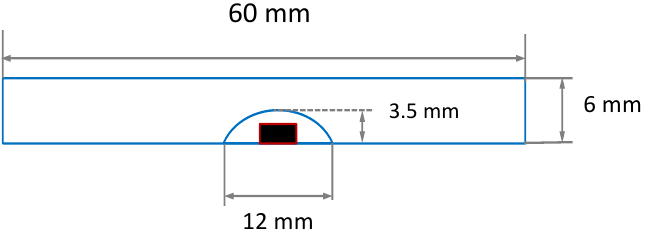}}
\hfill
\includegraphics[width=0.5\linewidth]{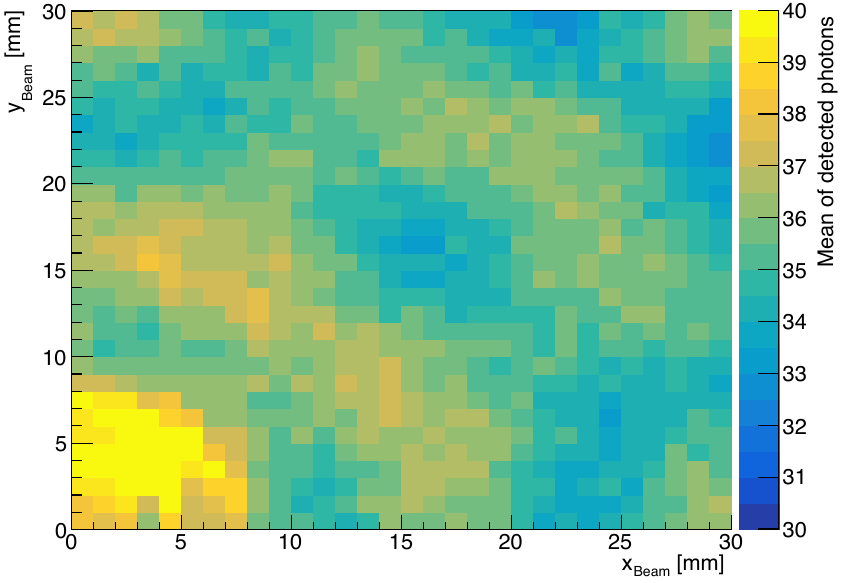}
\vspace{-2mm}
\caption{Left: dimensions of an HCAL scintillator tile. Both the size of the cavity and the tile thickness have been optimised by simulations. Right: average yield of photo electrons as a function of the position of a through-going minimum-ionising particle, as determined by simulation. 
Shown is one quadrant of an HCAL tile.}
\label{fig:HCALTile}
\end{figure}

The scintillating tiles are made of a polystyrene-based plastic scintillator, which is available from several companies and institutes. 
The layout of one tile is shown in Fig.~\ref{fig:HCALTile}~(left). 
The cavity, housing a SiPM with \qtyproduct[product-units=power]{6x6}{\mm} sensitive area, 
has the shape of a spherical cap with a width of \SI{12}{\mm} and a depth of \SI{3.5}{\mm}. 
The tile thickness of \SI{6}{\mm} has been optimised for both light yield and uniformity. 
Fig.~\ref{fig:HCALTile}~(right) shows the result of a simulation of through-going muons at different positions of the tile. 
The simulation takes into account all known effects as attenuation length, wrapping with reflective foil, SiPM quantum efficiency, etc.
The mean number of detected photo electrons is about 36 with a uniformity of better than $\pm\SI{10}{\%}$ over practically the whole tile surface.   
Such a uniformity is fully sufficient for the separation of hadronic and electromagnetic showers.

The SiPMs need a relatively large dynamic range, i.e. a large number of pixels,
for a linear energy measurement for both minimum-ionising particles 
and electromagnetic and hadronic showers. 
Candidates for such SiPMs are e.g. the existing Hamamatsu types S13360 or S14160 with
more than \num{14000} pixels and a sensitive area of \qtyproduct[product-units=power]{6x6}{\mm}. 

Each SiPM is connected to a common readout board (see Fig.~\ref{fig:CaliceTiles} for the CALICE AHCAL layout) of
\qtyproduct[product-units=power]{36x36}{\cm} size, which carries $\numproduct{6x6}=\num{36}$
scintillating tiles, wrapped in reflective foil and glued to the PCB.
The front-end electronics together with a digitizing ASIC occupy the 
back side of the readout board. The digitization therefore takes place directly 
next to the photo detectors, ensuring an excellent time resolution.
We plan to use the PicoTDC, currently developed at CERN, which can time-stamp signals with \SI{12}{ps} precision~\cite{PicoTDCDataSheet} 
and measures charges via time-over-threshold.
Another suitable chip is the HGCROC~\cite{Bouyjou:2022wii}, 
which has been developed for the high-rate environment of the CMS HGCAL.

Assuming the general design shown in Fig.~\ref{fig:HCAL_Layout} with 1440 cells 
in each layer and in total 40 scintillating layers, 
the number of readout boards is 1800 (not all fully equipped) and the total number of channels is 57600.
The possibility of a reduced channel count by e.g. using larger tiles in the 
outer region or less active layers is being studied at the moment.

\paragraph{Expected performance and further R\&D}
\label{para:hcalperformance}

The intrinsic time resolution of a single scintillating tile is expected to be of $\mathcal{O}(\SI{750}{ps})$ for a single MIP, 
derived from measurements for the CALICE AHCAL~\cite{Emberger:2022epg}\footnote{Within CALICE a time resolution of $\pm\,\SI{466}{ps}$ 
was measured for \qtyproduct[product-units=power]{3x3x0.3}{\cm} tiles with about half the light yield.}.  
Together with the readout electronics, the time resolution of a MIP should be of $\mathcal{O}(\SI{900}{ps})$ or better, following the same CALICE AHCAL measurement.
Given that a minimum-ionising muon of sufficient energy hits 40 tiles, the time resolution of a MIP is expected to be at most $\pm\,\SI{200}{ps}$.
In the short-term future, the time resolution will be verified and optimised with single-tile prototypes.

The planned HCAL tile size of \SI{6}{cm} is the same as the NA62 MUV1 strip width (and half of the MUV2 strip width). 
Furthermore, the expected light yield per MIP is larger than in the NA62 strip-like layout and, in addition, the HIKE HCAL is fully segmented in longitudinal direction. 
Therefore, the $\pi$/$\mu$ separation cannot be worse than in NA62, which already would be sufficient for the HIKE purposes. 
Ongoing and future simulations of the full HCAL detector, including clustering and employment of classification methods as Boosted Decision Trees and neural nets, 
will yield the real separation power of the HIKE HCAL. 
Following those studies, we aim to find an optimum between the separation power and the cost of the detector, which is mainly driven by the number of readout channels.
Possible options for optimisation are e.g. larger tile sizes or less active layers. 
However, since these studies are still on-going, the cost estimate 
refers to the baseline design above.
Further R\&D has still to go into the readout electronics. In particular, a decision on the digitizing ASIC will be taken in the near-term future.

\subsubsection{Muon veto detector (MUV)}

The muon veto detector (MUV) vetoes muons 
and contributes to the muon identification and suppression 
at a event-building-farm or the analysis level. It is placed following the 
HCAL behind a \SI{80}{\cm} thick iron wall 
(corresponding to about 5 additional nuclear interaction lengths).
The MUV needs a time resolution of the order of \qtyrange{100}{150}{ps} to keep the loss of signal due to random activity at the level of a few percent, 
as it is presently achieved by NA62 with a time resolution of about \SI{500}{ps}.

The MUV will be built from scintillating tiles with direct photodetector readout. 
Two feasible options exist. The first scenario would be layers similar to the HCAL layers with scintillating tiles read out by SiPMs. 
To obtain the required time resolution, the tiles would have \SI{5}{cm} thickness to ensure high light yields and possibly smaller transversal dimensions. 
The second MUV option is a copy of the NA62 MUV3 detector~\cite{NA62:2017rwk}, 
also consisting of scintillating tiles with a thickness of \SI{5}{cm} or more, but read out 
by PMTs from the back side. In contrast to NA62, where the tile front faces measure 
\qtyproduct[product-units=power]{22x22}{\cm}, the tile size would need to be reduced
to both reduce the light path and to be able to cope with the higher rate in HIKE.

\subsubsection{Small-angle electromagnetic calorimeter (SAC)}

For the measurements of $K^+\to\pi^+\nu\bar\nu$, 
as well as other measurements of rare decays requiring hermetic photon vetoes, the coverage of the veto system must extend down to zero in the polar angle to intercept photons that would otherwise escape the detector through the downstream beam pipe.
Vetoing photons at small angle is easier for the $K^+$ beam than for the $K_L$ beam, because the $K^+$ beam can be diverted and dumped outside the acceptance of the small-angle calorimeter. For the $K_L$ beam, instead, the SAC sits directly in the neutral beam; it must reject photons from $K_L$ decays that would otherwise escape via the downstream beam exit while remaining relatively insensitive to the very high flux of neutral hadrons in the beam, so that the experiment is not blinded by the random vetoes from these hadrons. The design and construction of the SAC for the $K_L$ beam is thus a unique challenge, particularly for the measurement of $K_L\to\pi^0\nu\bar\nu$ further in the future. 
Note that for much of HIKE Phase~2 (specifically, for $K_L\to\pi^0\ell^+\ell^-$ measurements) a production angle of 2.4~mrad is preferred. For running at the smallest production angles, the SAC might not be installed because of the extremely high total neutral beam rate.
The R\&D for the SAC is well underway.

The following indicative efficiency requirements can be identified on the basis of  $K\to\pi\nu\bar{\nu}$ sensitivity studies for both $K^+$ and $K_L$:
\begin{itemize}
\item For $E < 5~{\rm GeV}$, the SAC can be blind. For the both the $K^+$ and $K_L$ beams, background processes that are not otherwise efficiently vetoed do not have photons on the SAC with $E < 5~{\rm GeV}$.
\item For photons with $5~{\rm GeV} < E < 30~{\rm GeV}$, the SAC inefficiency must be less than 1\%.
\item Only for photons with $E > 30~{\rm GeV}$ must the inefficiency be very low ($<10^{-4}$).
\end{itemize}
Although these SAC efficiency requirements are not intrinsically challenging, from the simulations of the neutral beam 
with a production angle of 8~mrad (the maximum angle foreseen),
there are about 130 MHz of $K_L$ mesons, 440~MHz of neutrons, and 40~MHz of high-energy ($E>5$~GeV) beam photons incident on the SAC, and the required efficiencies must be attained while maintaining insensitivity to the nearly 600~MHz
of neutral hadrons in the beam. 
In order to keep the false veto rate to an acceptable level, the hadronic component must be reduced to at most a few tens of MHz, so that the total 
accidental rate is dominated by the beam photons and in any case significantly less than the 100~MHz target.
These requirements lead to the following considerations:

\begin{itemize}
\item The SAC must be as transparent as possible to the interactions of neutral hadrons. In practice, this means that the nuclear interaction length $\lambda_{\rm int}$ of the SAC must be much greater than its radiation length $X_0$.
\item The SAC must have good transverse segmentation to provide $\gamma/n$
discrimination. 
\item It would be desirable for the SAC to provide additional information useful for offline $\gamma/n$ discrimination, for example, from longitudinal segmentation, from pulse-shape analysis, or both.
\item The SAC must have a time resolution of 100~ps or less.
\item The SAC must have double-pulse resolution capability at the level of a few ns.
\end{itemize}
In addition, in five years of operation, the SAC will be exposed to a neutral hadron fluence of about $10^{14}$~$n$~1~MeV~eq/cm$^{2}$, as well as a dose of up to several MGy from photons.

\begin{figure}
\centering
\includegraphics[width=0.5\textwidth]{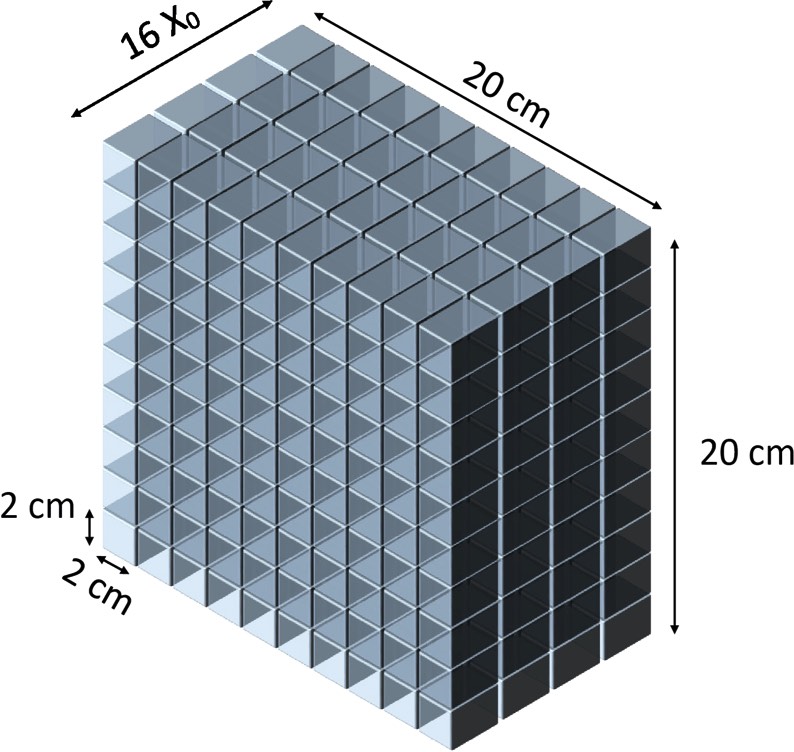}
\caption{Dimensional sketch of a SAC based on dense, high-$Z$ crystals with both transverse and longitudinal segmentation.}
\label{fig:klever_sac}
\end{figure}

One possible design that is well-matched to the above requirements would be to use a highly segmented, homogeneous calorimeter with dense, high-$Z$ crystals providing very fast light output. As an example, the small-angle calorimeter for the PADME experiment used an array of 25~lead fluoride (PbF$_2$) crystals of $30\times30\times140~{\rm mm}^3$ dimensions. PbF$_2$ is a Cherenkov radiator and provides very fast signals.
For single crystals read out with PMTs, a time resolution of 81~ps and double-pulse separation of 1.8~ns were obtained for 100--400~MeV electrons~\cite{Frankenthal:2018yvf}. 

At the doses expected, radiation-induced loss of transparency to Cherenkov light could be significant for PbF$_2$, as suggested by existing studies with ionising doses of up to ${\cal O}(10~{\rm kGy})$~\cite{Cemmi:2021uum,Kozma:2002km,Anderson:1989uj}. However, these studies also found significant annealing and dose-rate effects in PbF$_2$, as well as the effectiveness of bleaching with UV light. If the effects of radiation damage to PbF$_2$ prove to be a significant problem, a good, radiation-hard alternative could be optimised lead tungstate (PbWO$_4$, PWO) 
\cite{PANDA:2011hqx,Auffray:2016xtu,Follin:2021kgn}. In particular, ultrafast PWO (PWO-UF) with a decay time constant of 640 ps, good light yield, and high radiation tolerance has recently been developed~\cite{Korzhik:2022xln}.
In collaboration with the authors of \cite{Cemmi:2021uum}, we have performed transmission measurements with a 40-mm path length on PbF$_2$ and PWO-UF crystals exposed to gamma rays from a $^{60}$Co source at the Calliope facility.\footnote{ENEA Casaccia, \url{https://www.casaccia.enea.it}}
After irradiation to 361 kGy, the PbF$_2$ samples\footnote{SICCAS, Shanghai, China} demonstrated a $\sim20$\% loss of transmission at 400~nm, with the onset of transparency at 320~nm, up from 250~nm before irradiation. Substantial recovery of the transmission of the PbF$_2$ crystals was observed after several days of annealing in natural light. By contrast, even after irradiation to 2106 kGy, the PWO-UF sample\footnote{Crytur, Prague, Czech Republic} showed no increase in the wavelength of onset of transparency (350~nm) and a loss of transparency of only about 5\% at longer wavelengths.

For the SAC, a design with longitudinal segmentation is under study, as shown in Fig.~\ref{fig:klever_sac}. This design would feature four layers of $20\times20\times40$~mm$^3$ PbF$_2$ or PWO-UF crystals (each $\sim$4$X_0$ in depth). To minimise leakage, the gaps between layers would be kept as small as possible. Compact PMTs such as Hamamatsu's R14755U-100 could fit into a gap of as little as 12~mm. 
This tube has ideal timing properties for the SAC, including 400~ps rise and fall times and a transit-time spread of 200~ps.
Readout with SiPMs would facilitate a compact SAC design even further, but may require advances in SiPM radiation resistance and timing performance.

\begin{figure}
\centering
\includegraphics[width=0.5\textwidth]{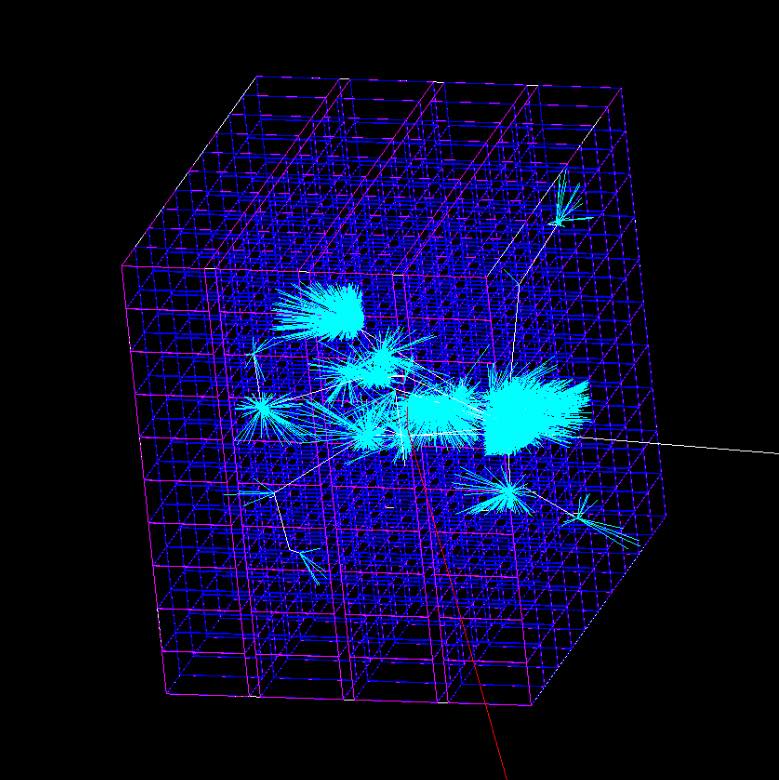}
\caption{Geant4 simulation of the SAC showing a 1-GeV photon incident from the right. White tracks show photons, cyan tracks show Cherenkov photons, and the red track shows a negatively charged particle.}
\label{fig:sac_interaction}
\end{figure}

\begin{figure}
\centering    
\vspace{-2mm}
\includegraphics[width=0.47\textwidth]{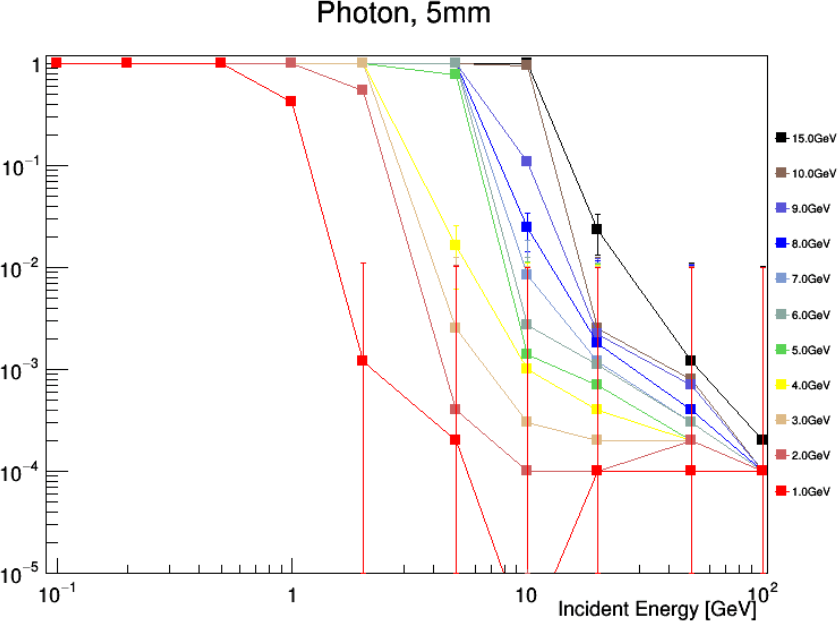}\\
\includegraphics[width=0.47\textwidth]{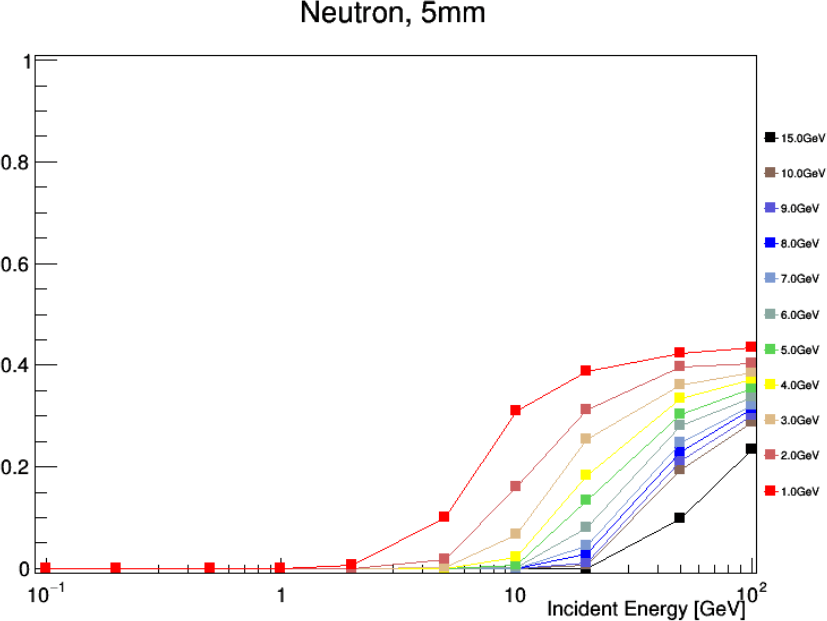}%
\hspace{0.02\textwidth}%
\includegraphics[width=0.47\textwidth]{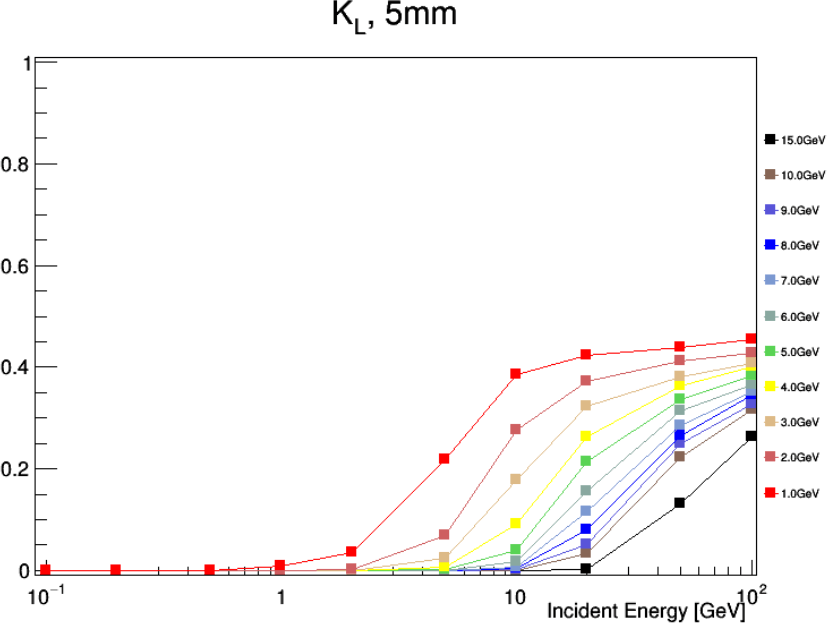}
\vspace{-2mm}
\caption{SAC inefficiency for photons (top) and  efficiency for neutrons and $K_L$ mesons (bottom left and right) vs.\ incident particle energy, for various thresholds on observed incident particle energy as measured from the number of Cherenkov photons produced.}
\vspace{-2mm}
\label{fig:sac_eff}
\end{figure}

A Geant4 simulation of the SAC has been developed for the HIKE Monte Carlo. This simulation can be used to study the performance with different crystals and geometries (the readout response is not yet simulated).
Fig.~\ref{fig:sac_interaction} shows the simulation of an interaction of a 1-GeV photon in the SAC with PbF$_2$
crystals and a 5-mm gap width between layers for readout with SiPMs. The Cherenkov photon yield is 85,000 per GeV of incident photon energy. There is a $\sim$10\% loss of shower containment at the highest photon energies, which is not seen to depend on the size of the gaps between layers over the interval of 0--20~mm. This is considered to be an acceptable trade-off for the purpose of maintaining the detector relatively transparent to hadrons. The response 
to the main components of the neutral beam has been studied. Fig.~\ref{fig:sac_eff} shows the inefficiency of the SAC for photon detection and the efficiency of the SAC for neutron and $K_L$ detection as functions of incident particle energy, for various thresholds set on the measured incident particle energy from the number of Cherenkov photons produced. With a threshold set at the number of Cherenkov photons produced by a photon of 4--5 GeV, the SAC efficiency requirements are satisfied, while about 30\% of the neutral hadrons leave signals above threshold, contributing to the random veto.   

\begin{figure}
\centering
\includegraphics[height=0.2\textheight]{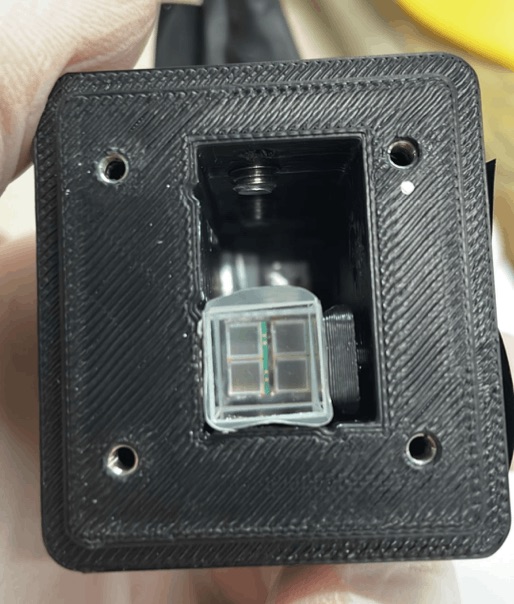}\hspace{0.1\textwidth}
\includegraphics[height=0.2\textheight]{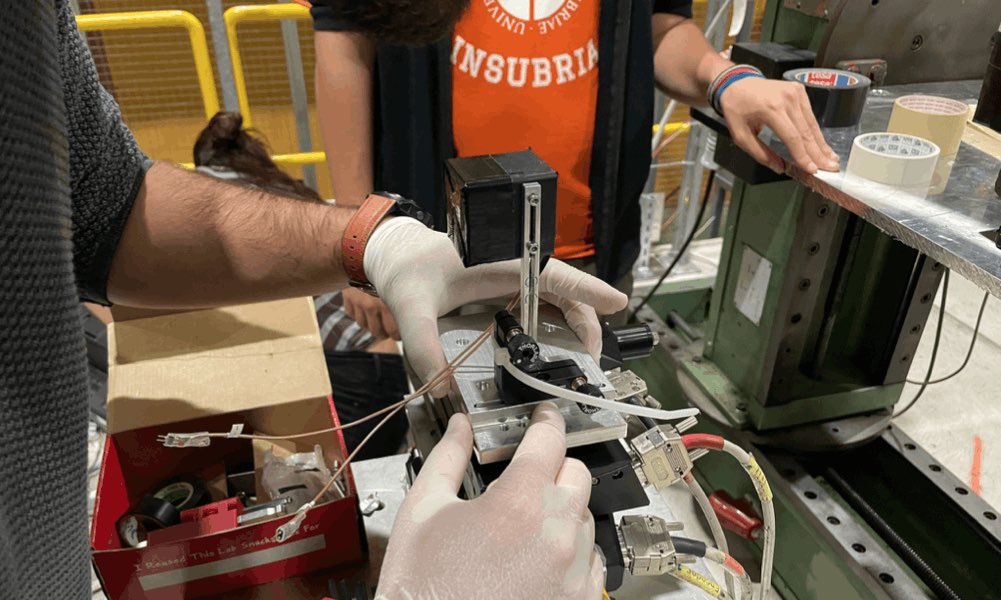}
\caption{Left: Single crystal read out with $2\times2$ matrix of four $4\times4$~mm$^2$ SiPMs. Right: Installation of crystal on H2 beam line during fall 2022 test beam.}
\label{fig:crilin_tb}
\end{figure}

R\&D work on the SAC is currently being carried out in synergy with other collaborations, facilitated in part by participation in the AIDAinnova research network. 
CRILIN, an electromagnetic calorimeter under development for the International Muon Collider Collaboration, is an independently proposed, highly granular, longitudinally segmented, fast crystal calorimeter with SiPM readout and performance requirements similar to those for the SAC~\cite{Ceravolo:2022rag}. Much development work is being carried out in collaboration with CRILIN, with particular emphasis on the SiPMs, front-end electronics, and signal readout, as well as on solutions for detector mechanics and SiPM cooling. The first test beam measurements with individual PbF$_2$ and PWO crystals were performed in summer 2021 at the Frascati BTF and the SPS North Area, followed by additional tests in fall 2022 in which some of the first commercially available samples of PWO-UF were also tested.
These tests were focused on understanding the best possible time resolution that can be obtained, studying the systematics of light collection in the small crystals, and validating the CRILIN choices of SiPMs and the design of the readout amplifier. In autumn 2022, single $10\times10\times40$~mm$^3$ crystals of PbF$_2$ (4.3$X_0$) and PWO-UF (4.5$X_0$) were exposed to high-energy (60--120 GeV) electron beams at the SPS H2 beamline. 
Each crystal was viewed by a matrix of four Hamamatsu 14160-4010 SiPMs ($4\times4$~mm$^2$, 10~$\mu$m pixel size), which were read out in pairs, providing independent readout channels for the left and right sides of the crystal (Fig.~\ref{fig:crilin_tb}).
The SiPM was chosen to represent a plausible choice in the current state of the art for high-speed response, short pulse width, and good radiation resistance.
The signals from the SiPMs were amplified
with the CRILIN electronics and digitized at 5 GS/s. 

\begin{figure}
\centering
\includegraphics[width=0.96\textwidth]{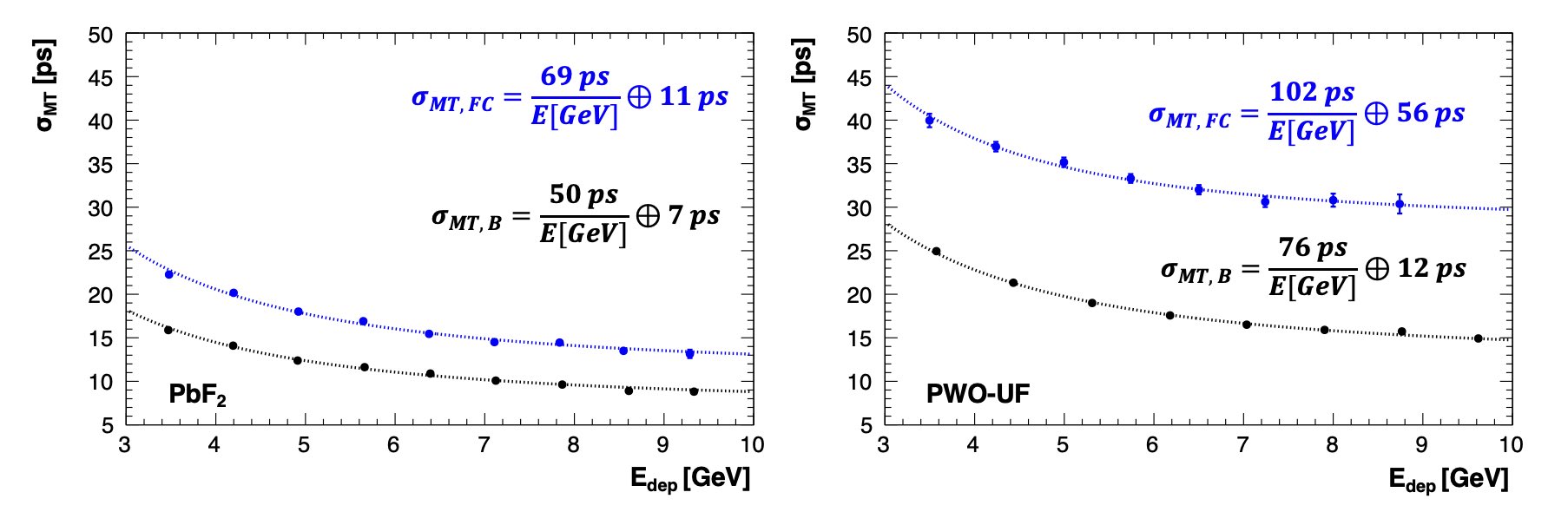}
\vspace{-3mm}
\caption{Time resolution obtained with single $10\times10\times40$~mm$^2$ crystals of PbF$_2$ (left) and PWO-UF (right) as a function of deposited energy, from beam tests with high energy electrons. The blue (black) curves are for data taken in ``front'' (``back'') configurations.}
\label{fig:sac_time_res}
\end{figure}

Analysis of the test beam data has recently been completed and submitted for publication~\cite{Cantone:2023fac}. Fig.~\ref{fig:sac_time_res} shows the time resolution obtained for single crystals as a function of deposited energy, in the ``front'' (SiPMs on downstream end of crystals, viewing upstream) and "back" (SiPMs on upstream end of crystals, viewing downstream) configurations. For both crystal types, the time resolution is significantly better in the ``back'' configuration: due to the highly directional nature of the Cherenkov light, the light arriving at the downstream end of the crystal is highly localized, giving rise to a dependence of the light arrival time on the position of particle incidence that is thoroughly randomized when the light is reflected upstream. 
Due to the scintillation component, the light yield for PWO-UF ($\sim$0.6~p.e./MeV) is approximately twice that for PbF$_2$ ($\sim$0.3~p.e./MeV), but due to the purely Cherenkov emission for PbF$_2$, its time resolution is slightly better.
The time resolution obtainable from the combination of either crystal, PbF$_2$ or PWO-UF, with the chosen SiPM and CRILIN electronics is excellent: about 12~ps for PbF$_2$ and 19~ps for PWO-UF in ``back'' configuration for an energy deposit of 5~GeV (corresponding to the SAC threshold). This is significantly better than required for the SAC and offers ample prospect for greatly reducing losses to random veto beyond the initial expectations.
The CRILIN electronics performed very well; HIKE should evaluate the possibility of faster shaping to obtain better double-pulse discrimination.

\begin{figure}
\centering
\includegraphics[height=0.175\textheight]{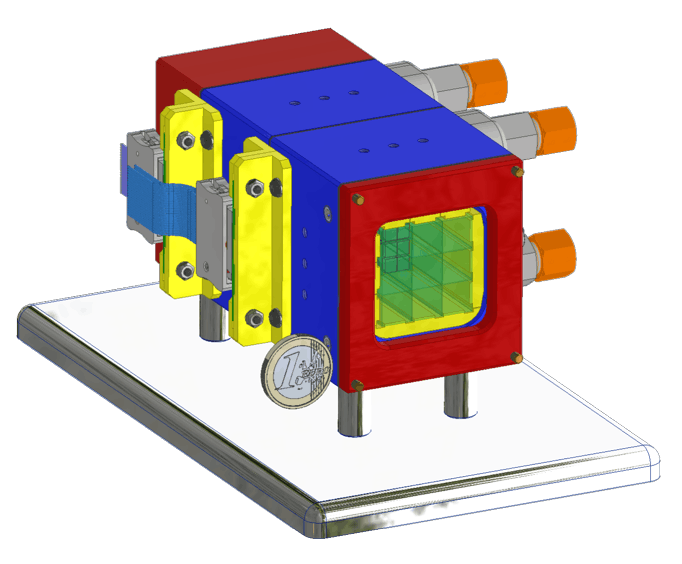}
\includegraphics[height=0.175\textheight]{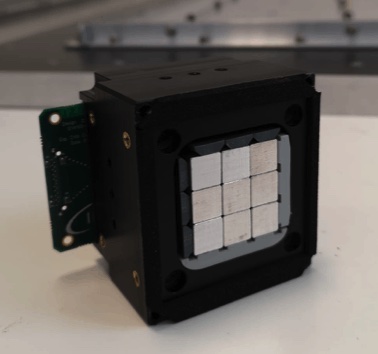}
\includegraphics[height=0.175\textheight]{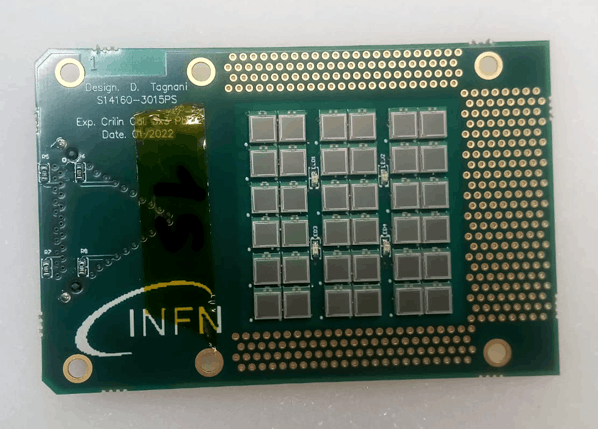}
\caption{Two-layer $3\times3$ crystal CRILIN prototype: mechanical drawing (left), assembled front module (center), and PCB with SiPMs for one layer (right).}
\label{fig:crilin}
\end{figure}
Tests of a 
two-layer, $3\times3$ crystal array are currently in progress as the next step of CRILIN development.
The prototype (Fig.~\ref{fig:crilin}) features a scheme for SiPM cooling. The light is read out in the same manner as for the single crystals tested in 2021--2022, with two channels per crystal, for a total of 36~channels. Modular electronics for power distribution, SiPM signal processing, and control have also been developed. This prototype can be populated with both PbF$_2$ and PWO-UF crystals.
In July 2023, it was tested with 450~MeV electrons at the Frascati BTF, and, at the time of writing (August 2023), is currently being tested with 20--120~GeV electrons in the H2 beamline at CERN.
These tests will validate the segmentation scheme, measure the time resolution with reconstructed clusters, study the effects of design optimisations such as the choice of crystal surface treatments and wrapping materials, and evaluate the performance of the engineering solutions adopted.

As noted above, the full SAC design with 4 layers of 4-cm crystals suffers slightly from shower leakage, while the interaction probability for hadrons in the beam is about 30\%. A possibility to make the SAC more hermetic for photon showers and less sensitive to hadrons is to exploit the effects of the coherent interactions of high-energy photons in oriented crystals to induce prompt electromagnetic showering. 
This is the same technique to be used for optimisation of the photon converter in the neutral beam (\Sec{sec:beam}).
Because of the relative ease in producing high-$Z$ optical crystals of high quality, there are good prospects for using this technique to construct a compact
calorimeter with a very small radiation length, referred to the primary interaction~\cite{Bandiera:2019tmg}. A decrease by a factor of 5 in the effective radiation length for a 4-mm thick PWO crystal has been observed for 120~GeV electrons incident to within 1~mrad of the crystal axis~\cite{Bandiera:2018ymh}. Considering that the SAC acceptance for photons from $K_L\to\pi^0\pi^0$ decays in the fiducial volume extends at most to $\pm2$~mrad, it should be possible to orient the crystals to enhance the probability for photon conversion. This would allow the thickness of the SAC layers to be reduced, decreasing the rate of neutral hadron interactions while maintaining shower containment and photon detection efficiency.

The potential gains from aligning 
the crystals,
as well as the procedures and mechanics needed, are under study.
In summer 2021, in a joint test with the CRILIN and STORM collaborations in H2, a tagged photon beam was used to measure the enhancement of shower processes in thicker PbF$_2$ and PWO crystals as a function of angular alignment, correlating variables such as the energy dispersed in the crystal and multiplicity of charged particles produced in the shower with the production of Cherenkov and scintillation light in the crystals. STORM collected similar data with electron beams in summer 2021 and 2022. The data are still under analysis, but as a preliminary generalisation, the effective radiation length for crystals of 2--4 cm thickness is reduced by $\sim$30\% by alignment, with an angular acceptance of about 1~mrad~\cite{Monti-Guarnieri:2022ezh}.

\begin{figure}
\centering
\includegraphics[width=0.7\textwidth]{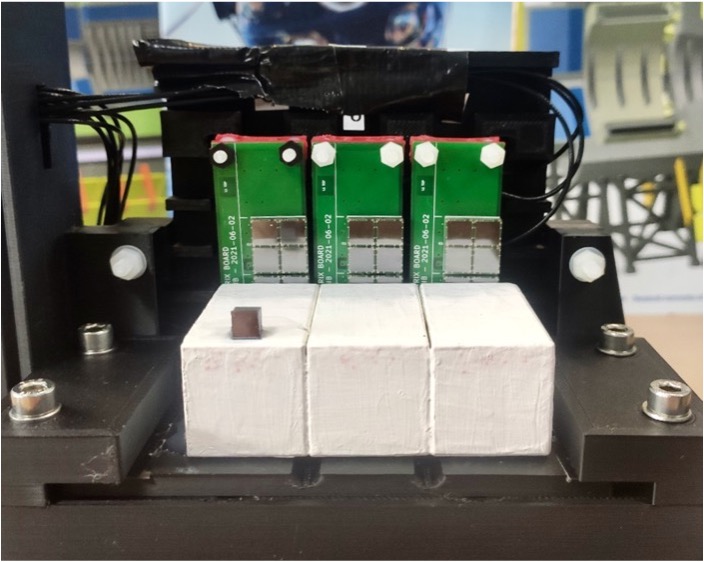}
\vspace{-1mm}
\caption{Array of three PWO-UF crystals with aligned axes tested in August 2023.}
\label{fig:oreo_pwo_row}
\end{figure}
Together with the OREO collaboration, which is continuing the work of STORM, we are constructing a PWO-UF calorimeter prototype
in which the axes of the crystals on the front layer are aligned. In addition to allowing studies of the variation of the energy deposition, radiation length and Moli\`ere radius with beam energy and crystal alignment, the OREO prototype will serve as a platform for developing the procedures and mechanics needed to build a calorimeter with aligned crystals. Much progress has been made over the past year in the areas of measuring the crystallographic quality (mosaicity, axis uniformity, and absence of internal stresses) and determining the axis orientation of PWO-UF crystals both via high-resolution X-ray diffraction and photoelastic analysis with laser conoscopy, as well as in defining a proof-of-concept calorimeter geometry with a Geant4 model including the shower development in oriented crystals, and above all, developing the techniques and equipment needed for the alignment of a matrix of scintillator crystals~\cite{Bandiera:2023xxx}.
The alignment system is based on motorized optomechanical components and features laser interferometric control of the crystal orientation during the entire procedure; it was used in July 2023 to align the axis of three $25\times25\times40$~mm$^3$ crystals to within 500~$\mu$rad for gluing. 
The alignment of this three-crystal matrix, shown in Fig.~\ref{fig:oreo_pwo_row}, has recently been tested in the T9 beamline at the CERN PS with electron and hadron beams. Preliminary results show an increase in the scintillation yield for all three crystals, suggesting that close alignment between the crystallographic axes has been achieved. More sensitive tests in the SPS H2 beamline with high-energy electrons are planned imminently.

The next step is to develop a final design for the HIKE SAC that combines elements of the CRILIN and OREO prototypes, with the final, optimized choice of crystal dimensions and CRILIN-like stackable layer mechanics that permit the crystals to be aligned. Because of the need for strong shaping of the SiPM pulses in order to provide double-pulse resolution at the level of a few~ns, and because of the uncertainties regarding the development of SiPMs of sufficient radiation hardness, this design will likely feature readout with compact PMTs.

\subsubsection{Hadron sampling calorimeter (HASC)}

The original NA62 setup
was improved by the addition of a hadron sampling calorimeter (HASC-1) adjacent to the beampipe downstream of the muon detector. The primary purpose of this detector is to reduce the $K\to\pi^+\pi^+\pi^-$ background to the $K^+\to\pi^+\nu\bar\nu$ decay, by vetoing the topology in which the $\pi^-$ undergoes hadronic interaction in the first STRAW chamber, while an energetic $\pi^+$ travels in the beampipe undetected by the IRC and emerges downstream. Analysis of the NA62 Run~1 dataset has revealed that HASC-1 is also efficient as a photon veto, providing a 30\% reduction of the $K^+\to\pi^+\pi^0$ background to the $K^+\to\pi^+\nu\bar\nu$ decay. Consequently, a second calorimeter (HASC-2) was installed in 2021 at the HASC-1 longitudinal position, symmetrically with respect to the beam axis, which improves further the $K^+\to\pi^+\pi^0$ rejection.

Each of the HASC-1 and HASC-2 stations consists of 9~identical modules. 
Each module is a sandwich of 120~lead/scintillator alternating tiles, with a total volume of $10\times 10\times 120 ~\mathrm{cm}^3$. The sampling ratio is 4:1, the scintillator tiles having a dimension of $100 \times 100 \times 4~{\rm mm}^3$ with a lead thickness is 16~mm. Each module is organised in 10~longitudinal readout sections, each scintillator tile of every section being optically coupled with a wavelength shifting (WLS) optical fiber of 1~mm$^2$ cross-section. At the rear side of each module there are 10~optical connectors, originally designed to be coupled with 3$\times$3~mm$^2$ green-sensitive micro-pixel avalanche photodiodes (MAPD) (currently the S12572-015C Hamamatsu SiPM sensors are used).

The FE electronics and SiPMs installed on the HASC-2 station are cooled down to 21ºC with a custom-made system  
consisting of three Peltier thermoelectric coolers / module and a water-air heat exchanger used to blow cold air in the modules end-cap cases. The temperature is maintained constant by a temperature controller – MCU based – which regulates the Peltier supply voltage via a PID routine. For HIKE, this scheme will be extended to HASC-1,
improving the quality of SiPM signals and reducing ageing due to radiation to an acceptable level.

Signal rates observed in the HASC during the NA62 data taking in 2022 at nominal beam intensity are summarised in Table~\ref{tab:hasc_rates}. The rates fall rapidly with the distance from the beampipe. The modules next to the beampipe contain $35\%$ and $25\%$ of the total activity in the HASC-1 and HASC-2 stations, respectively.

\begin{table}[tb]
\centering
\caption{Signal rates observed in the HASC during the NA62 data taking in 2022.}
\vspace{-2mm}
\begin{tabular}{l|c}
\hline
Detector part & Rates in NA62 setup at nominal intensity\\
\hline
HASC-1 & 1~MHz\\
HASC-1, most active module & 0.35~MHz\\
HASC-1, most active channel & 0.05~MHz\\
HASC-2 & 0.21~MHz  \\
HASC-2, most active module & 0.05~MHz\\
HASC-2, most active channel & 6~kHz\\
\hline
Total (HASC-1 and HASC-2) & 1.21~MHz \\
\hline
\end{tabular}
\vspace{-2mm}
\label{tab:hasc_rates}
\end{table}

The NA62 HASC provides near-optimal geometric coverage for the relevant topologies of the $K^+\to\pi^+\pi^+\pi^-$ and $K^+\to\pi^+\pi^0$ backgrounds to the $K^+\to\pi^+\nu\bar\nu$ decay, and is suitable also for HIKE Phase~1. Assuming the HIKE beam profile to be similar to the NA62 one, we expect at most a rate of 200~kHz in the most active channel for the HIKE HASC. The main issue in the operation of the current HASC at the HIKE beam intensity is the random veto. It is determined chiefly by the time resolution of the detector, presently at the level of 370~ps FWHM for HASC-2 and 430~ps for HASC-1, in agreement with Hamamatsu specifications~\cite{Hamamatsu:S1272-015C}. For HIKE, the resolution should be improved to $\mathcal{O}(100~\mathrm{ps})$. The Hamamatsu S13362-3050DG SiPMs achieving this value are available on the market~\cite{Hamamatsu:S13362-3050DG}, and represent a promising candidate for the upgrade. Table~\ref{tab:HASC_SiPM_performances} shows a comparison between the current HASC SiPMs and the candidate for the upgrade. Another appealing feature of the new SiPMs is the embedded two-stage thermoelectric cooling which greatly simplifies the current cooling scheme.

\begin{table}[tb]
\centering
\caption{\label{tab:HASC_SiPM_performances} Summary of SIPM characteristics for the model currently employed in the NA62 HASC and the one proposed for the HIKE upgrade.}
\vspace{-2mm}
\begin{tabular}{l|c|c}
\hline
SIPM model & S12572-015C (current) & S13362-3050DG (upgrade)\\
\hline
Photon detection efficiency     & 25\% & 40\% \\
Typical dark count & 1000~kcps & 25 kcps \\
Gain & $2.3\times 10^5$ & $1.7\times 10^6$ \\
Time resolution FWHM & 400~ps & 110~ps~\cite{Kravchenko_2017}\\
\hline
\end{tabular}
\vspace{-3mm}
\end{table}

An alternative option for the HASC upgrade, which is currently under study, involves PMTs. A particularly interesting example is Hamamatsu H14220A, which could be used to read out individual scintillator plates (instead of groups of six, as done presently), eliminating pileup due to detecting a convoluted signal hence offering lower rise time ($\tau_\mathrm{r}$) and fall time ($\tau_{\mathrm{f}}$). The main drawback of reading single scintillators is the lower number of collected photons ($N_{\mathrm{ph}}$), compared with SiPM readout, which could have a negative impact on the time resolution which is known to be proportional with $\sqrt{\tau_\mathrm{r}\tau_\mathrm{f}/N_{\mathrm{ph}}}$~\cite{Gundacker_2013}.

\section{Data acquisition and high-level trigger}
\label{sec:readout}

Particle rate in HIKE Phase~1 is about 3~GHz in the beam tracker and 200~MHz for $K^+$ identification. Single-particle rates in the downstream detectors are about 50~MHz, including both $K^+$ decay products and the muon halo of the beam. The neutral beam of HIKE Phase~2 is 50\% more intense that the Phase~1 charged beam. The small-angle calorimeter (SAC) in the neutral beam will encounter rates of about 100~MHz, and the other detectors encounter rates of about 20~MHz. 
The high particle rates will give rise to a harsh radiation environment in the HIKE experimental cavern.
To limit the impact of radiation on the DAQ system, we aim to minimise the amount of electronics located in the cavern, particularly electronics related to data processing and storage, and use radiation-tolerant optical links and error-correcting data formats such as lpGBT.

The foreseen DAQ architecture will need first of all a timing distribution system and related optical fibers. 
Given the number of detectors involved, we plan about 50 Readout Units, complemented by the backend computing infrastructure (servers and GPU/FPGAs) and associated network equipment. 
%(for details of the HIKE DAQ, see Section~\ref{sec:readout}).

\subsection{Readout boards}
\label{sec:ro-boards}

%The readout boards serve two purposes: to host devices that timestamp digital signals produced by the detectors, and to encode and transmit these data away. 
The readout boards digitise and timestamp the signals from the detector electronics, and encode and transmit data to the DAQ.
The readout boards are located in the high-radiation environment close to the detector, so radiation-hard/radiation-tolerant technologies will be used. 
The time of signals from the detectors will be recorded either with time-digital converters (TDC), analogue-to-digital converters (ADC), or dedicated ASIC devices. The 64-channel PicoTDC is under development at CERN and is anticipated to be ready before HIKE Phase~1. The PicoTDC can timestamp signals at 12~ps precision~\cite{PicoTDCDataSheet} and outputs data in the lpGBT format, allowing simple readout boards to be used that only convert electric signals from the PicoTDC into optical signals that are sent to the rest of the DAQ system.
%RFSoC (Radio Frequency System on %Chip) are FPGA devices on which %sixteen 14-bit ADCs with 1\,GHz %sampling frequency can be %implemented, and the GBT format could %be implemented here also, again %simplifying the readout boards. The 
Possible commercial solutions for fast FADC-based readout boards, to be adapted to the requirements of the DAQ in synergy with the manufacturer, are under investigation. The HIKE Phase~1 beam tracker data will be handled by dedicated ASICs.
% Grand total channels: 102600 channels TDC, 
% 19500 LKr, 8000 MEC
The number of detector channels read in HIKE is estimated to be about~100k TDC and 20k ADC channels.
%The number of detector channels read by TDC or ADC in HIKE Phase~1 is estimated to be about~50000. Approximately half of the detector channels will be read by TDC devices, the other half by ADCs. Therefore an estimated 400 TDC devices with 64 channels each, and 1600 ADC devices with~16 channels each, will be required for HIKE Phase~1. For HIKE Phase~2, 
%other LAVs do NOT fit in Phase 2: the 5000 channels of the LAV
%$\mathcal{O}(2000)$ channels of the MEC, and 400 channels of the SAC will be read using ADCs, requiring about 800 ADC devices with 16 channels each. Spy tiles in the MEC would contribute an additional 8000 channels that can be read using TDCs, in which case about 125 TDC devices with 64 channels each will be needed.

%For data transfer, as well as clock, timing, commands distribution, and monitoring, the FELIX board is a promising option.

\subsection{Streaming readout}
\label{sec:streaming-readout}

Moving the bulk of the DAQ hardware away from the high-radiation environment is achievable by implementing a streaming readout system, where detector signals are collected by radiation-tolerant readout boards and then transferred immediately, without an external trigger signal and without complicated processing or buffering. The transferred data are received and buffered in the RAM of a set of servers, and eventually transmitted to a computing farm for selection and filtering.

A possible architecture for the HIKE DAQ, fulfilling the requirements of all the phases of the experiment, will consist of a series of readout units implemented with powerful servers hosting one or more FELIX boards, developed by ATLAS and exploited in several HEP use cases.
The version of the FELIX PCIe card being finalised for LS3 has 24 optical links with a 9.6~Gb/s line rate and supports several data formats. When operating with the radiation-tolerant lpGBT format, the effective data bandwidth is 8.96~Gb/s per link.
NA62 is conducting tests for a triggerless readout based on the FELIX
board, which is the building block of a purely software trigger system. A FELIX-based triggerless readout for the NA62 veto counter is in place already and has been working since 2022. This configuration has allowed us to test the Felix firmware modifications needed for the NA62 TDCs and the integration of the Felix software with the NA62 readout farm.
NA62 plans to extend the use of this system to the CHANTI detector in 2024, allowing for a further step towards the triggerless scenario. The CHANTI detector will have more data links than the veto counter, so it will be possible to stress the acquisition with higher rates.
The above configurations have been tested with an asynchronous trigger initiated by the NA62 L1 processing: this is an important step to validate the readout before implementing a triggerless (synchronous) system.
The FELIX system also takes care of distributing clock, timing and commands to the front-ends, as well as receiving monitoring data for detector control purposes. In addition, FELIX firmware could be implemented to further process the data, for example to perform zero-suppression algorithms.
Another option for interfacing the detector data with the computing farm would be the evolution of the PCI40 board developed for the LHCb and ALICE experiments.

As modern servers can be equipped with 1~TB of RAM, the data recorded during a whole SPS spill (4.8~s) may be buffered, greatly relaxing any latency requirement on the data processing as the inter-spill period (at least 9.6~s) can be exploited. 

The readout units will be connected on one side to the front-end electronics of the different detectors, and on the other to a computing farm via a switched network: event selection and event building software will run on the servers of the computing farm.
A boost in the speed of selection and event building could come from the installation in these servers of GPUs and/or FPGA boards which can be used to perform for example specific pattern recognition tasks.
While the readout servers will need to be at the experiment's premises (less than 500~m from the readout boards), the computing farm may be located either close to the readout units, or in the Prevessin Computing Centre.

Additional computing equipment will be used to control, configure and monitor the hardware, using modern software tools, able to detect and autonomously solve possible problems, in order to minimize data losses and down-time.

The main challenges of a streaming readout system are the transfer and processing of large quantities of data, mainly generated by FADCs. Reading out the full complement of data from calorimetry is challenging and costly (although not impossible). These high data volumes could  be considerably reduced by only reading channels that contain non-zero energy, i.e. by implementing a zero-suppression (ZS) algorithm in the readout boards. 
Therefore, profiting from the advances in electronics technology and carefully following the developments done for LHC, R\&D will be carried out to reduce the data volumes at the FADC readout boards level. Adding more intelligence in the front-end can be risky in a high-radiation environment. Nevertheless, in HIKE the most exposed detectors upstream will have TDC-type front-end, with most of the intelligence in the readout. The more downstream calorimetry-type detectors will be in a moderate radiation environment for which the implementation of zero-suppression using FPGAs in readout boards should be feasible.

With the advance of the detector readout design, we plan to evaluate in parallel whether there will be the need of a buffering mechanism of mildly zero-suppressed data implemented at the FADC-based readout boards. These boards would send data only upon generation of a software trigger, based either on the information from other detectors or only on the information of that detector, and would have enough local memory to store data for a time greater than the trigger latency.

%An alternative architecture could also be evaluated whereby buffering of mildly zero-suppressed data is implemented at the FADC-based readout boards and those send data only upon generation of a software trigger based on the information from other detectors. In this case, the FADC-based boards should have enough local memory to store data for a time greater than the software trigger latency.

%\subsection{Cost estimate}

%The proposed architecture will need first of all a timing distribution system and the installation of the required optical fibers. We expect this will cost about 100 KCHF. Given the number of detectors involved, we plan about 50 Readout Units at 15 KCHF each, for a total of about 750 KCHF. The network equipment will cost about 50 KCHF. The cost of the backend computing infrastructure, with servers and GPU/FPGA, will be about 400 KCHF. The grand total is therefore about 1.3 MCHF. 
\section{Online and offline computing }
\label{sec:computing}

The baseline computing model is a scaled version of the current NA62 system. A flexible HIKE software platform is currently under development, on the basis of the NA62 software. HIKE will profit from the use of dedicated machines in the Pr\'evessin Computing Centre for many, if not all, of the described functions below, provided enough band capability is present between the experiment site and the computing centre.

%%%%%%%%%%%%

\subsection{Online monitoring, calibration and data quality}

The HIKE online monitoring system will be based on an updated model of the current one used in NA62.
At present in NA62, a subset of the data are sent from farm nodes running the event-building processes to dedicated four monitoring computers; 
each event is then processed by the standard reconstruction software.
The reconstructed events are then forwarded to a last machine where the different streams are recombined for displaying by dedicated processes. A set of low-level monitoring histograms produced by the reconstruction processes are also combined and displayed, for the purpose of quick data quality checks. 

For HIKE, a multi-level analysis pipeline, where each level performs more complex and time consuming tasks, should be implemented to shorten the latency between the acquisition of the data and the first feedback. Low-level (data corruption) and high-level (data quality) algorithms will perform error detection and alert the shift crew and the run control software. As often as possible automatic recovery procedures will be implemented to minimise data losses. A set of low-level calibration constants will also be computed during this first reconstruction and used both for monitoring purposes and to be fed back into the acquisition system. Depending on the details of the trigger scheme, the computing farm could also perform some of data filtering tasks. % as well.

Reconstructed events will be processed with data quality algorithms, also used during full offline data processing (see next section), yielding output information collected over a whole run (about 1500~bursts) to be reviewed by the detector experts.

%%%%%%%%%%%%%%%%%%

Calibration and data quality monitoring strategy will be built following the NA62 experience and existing infrastructure. Calibration will be performed in several stages. First, geometrical detector alignments will be initially obtained from special ``muon'' runs, where TAXes are closed and the spectrometer magnet is switched off to obtain straight penetrating tracks. Later, these geometrical alignments will be refined and monitored using kaon decays in standard data-taking conditions. A similar procedure to that of NA62 for relative timing alignment between different detectors (``T0s'') will be developed, to achieve an average precision at the picosecond level, with a width corresponding to the individual timing resolutions of each detector. This will be done on a per-burst basis. During processing information on detector-specific calibrations will also be obtained, on a per-burst basis, like energy conversions, dead/noisy channels etc, which will be fed back to the analysis software.
Raw timing alignments within individual detectors, like slewing corrections, will be obtained from data and assumed to be constant over a certain data taking period, although they are permanently monitored. 

Data quality will be additionally monitored during full offline processing 
by software specific to each detector, but also including more common properties like efficiencies for different (online or offline) trigger algorithms. 
Graphical and numerical outputs will be produced for every detector, together with a database containing the monitored (and calibrated) quantities and alarms raised if certain thresholds are passed for one burst.

%%%%%%%%%%%%%%

\subsection{Data processing and distributed computing}

Based on experience, the NA62 processing model, with some improvements and changes, can be adapted for HIKE. In NA62, raw data processing and user analysis is performed only on CERN computing clusters, while Monte Carlo production is performed using the international GRID, including CERN Grid resources.

Processing of raw data for HIKE will be performed at CERN and using GRID resources (with a split still to be defined depending on the details of available computing and data sizes at the time of data taking), due to the increased amount of data. The splitting of the processed data according to trigger and/or physics analysis interests (``filtering'') must be refined to obtain stream sizes that are smaller and easier to handle. Stronger selection criteria will be applied, and the amount of information written for every selected event (slimming) will be reduced.

Large-scale generation of Monte Carlo simulations will be performed exclusively on the Grid. The existing NA62 Grid framework is a one-of-its-kind system, created specifically for NA62 and providing an easy-to-use and fully automated production system \cite{NA62:Grid}. With an uptime of 99.99\% over the last ten years, it is a proven system that we envisage to extend to HIKE. Besides its simulation and reconstruction capabilities, our Grid framework can be extended to handle user analysis, in case this becomes a necessity for HIKE.

\subsection{Data reduction model}

Due to the expected increase in the amount of data (due to the increased intensity) and in the size of data (due to the increase of signal channels), the same strategy as in NA62, based on filtering and slimming, will be followed and strengthened for HIKE.
For filtering, overlap between different filtering streams must be minimised, while several filter outputs may be combined at analysis level.
For slimming, after reconstruction no low-level hit information will be stored, discarding as much as possible any information that can be:
\begin{enumerate}
\item Recomputed from stored information.
\item Likely not needed at analysis level.
\end{enumerate}
In specific cases of detector studies, or when an analysis requires access to low-level information, a selected list of events/bursts will be reconstructed again. The low-level information  
will be kept for those small dedicated samples for the duration of the study.

Generally, most analyses make use of standard high-level objects (downstream tracks, vertices) without the need to access most of the information stored in the underlying objects from which they were built (clusters, tracks, candidates). Those high level objects could therefore contain only the subset of information commonly used (time and position at associated sub-detectors) and stored in high-level analysis files which would be the basis for most user analyses. Again, some dedicated studies occasionally will require access to the full reconstructed objects, but these are generally aiming at deeper understanding of detector effects and to develop procedures to deal with them. In most cases the final procedure will not rely on the full reconstruction information, but only on that already available in the high-level objects. In addition, such procedures can generally be standardised and applied to several analyses without further study. It will therefore be sufficient to reconstruct a subset of data with full information for the duration of the study.

\section{Timescale, infrastructure and cost estimate}

\subsection{Infrastructure}

HIKE will be housed in TCC8 and ECN3 where NA62 is located at present. An evaluation of the necessary modifications to the experimental area has been carried out in collaboration with the ECN3 Task Force~\cite{Brugger:2022}, the North Area Consolidation Project (NA-CONS)~\cite{Kadi:2018, Kadi:2019, Kadi:2021} and the Conventional Beams Working Group~\cite{Gatignon:2650989}, where topics such as requirements for beam infrastructure, vacuum, cooling and ventilation, electrical distribution, handling and transport of detectors, access to the cavern, IT infrastructure, gas distribution, cryogenic systems, civil engineering and radiation protection have been addressed. The full list is provided in Ref.~\cite{Charalambous:2022}. In order to ensure compatibility with the North Area consolidation and as a central access point to all CERN service groups, discussion for the implementation of all modifications is steered through the NA-CONS Technical Coordination Committee (TCC) meetings, while integration aspects are handled in the Integration Committee for Experimental Areas (ICEA) as part of NA-CONS. 
%The best match between the HIKE requirements and the %infrastructure studies will be further investigated for the %Proposal.
No important issues have been identified.
The integration of Phase 1 is complete while Phase 2 integration is still in progress but no important issues have been identified.
%and will be completed by Autumn 2023.
A more detailed integration plan will be folded into the Technical Design Report. 
% \begin{itemize}
%   \item Vacuum:
% \end{itemize} 
% \begin{itemize}
%   \item Cooling and ventilation:
% \end{itemize}
% \begin{itemize}
%   \item Radio Protection (RP)  
% \end{itemize}
% \begin{itemize}
%   \item Electrical distribution :
%   \end{itemize}
% \begin{itemize}
%   \item Civil engineering:
%   \end{itemize}
The layout of the HIKE detector setup does not require any civil engineering work before at least LS4.
%The minimal changes and updates in the infrastructure will %have a positive impact on the foreseen costs. 
Furthermore, the requirements for the electrical distribution and the detector cooling will be similar to that of NA62. The vacuum system with seven cryopumps went through thorough maintenance in 2021 and will be suitable to provide a good vacuum at the $10^{-6}$~mbar level for HIKE.
% \textbf{Note Follow-up with Johannes (and Markus) in week 43: Can we avoid TCC8/ECN3 separation of air volumes when staying with 2x$10^{13}$/spill . Are we ok for HIKE? }
Power consumption will stay the same or reduced depending on a possible upgrade to the powering equipment of the dipole magnet, since the dipole is the main power consumer by far (3.1~MW in continuous mode).
CO$_2$ consumption for the STRAW detector is small and will not increase. 

%\subsection{LKr maintenance (until LS4?) (Danielsson, Bryman)}
To keep the NA62 LKr calorimeter operational as a backup for HIKE Phase~1, continuous maintenance of the cryogenics is needed, and
%, besides the installation of a new readout back end, 
the legacy components of the readout chain should be reviewed.
%, all the actually bad behaving channels should be fixed, the spare supply should be reviewed and possibly implemented. 
Actions in this direction will be started already in next Year-End-Technical-Stop (YETS) and procedures for regular review intervention will be defined.

From the detector point of view, the change from Phase 1 to Phase 2, to move or remove sub-detectors, can fit in a YETS (order of six months). Switching between kaon and beam-dump modes takes about 15~minutes, and can be made during a scheduled pause in the data collection.

%A period of beam commissioning should also be considered.

%Studies for the upgrade of Target and TAX to allow for the intensity increase are included in the PBC Conventional Beam Working Group document~\cite{Gatignon:2650989}. %The HIKE Collaboration will bear the cost of the experimental setup which has a bulk figure of XXX.

\subsection{Timescale and schedule}
\label{sec:schedule}

%The possibility to measure ${\rm BR}(K_L\to\pi^0\nu\bar{\nu})$ in Phase~3 demonstrates that the HIKE will continue to be able produce extremely interesting measurements even over a longer term;
%It remains our intention to pursue Phase 3, as expressed in the LoI; 
%however, because of the long time scale and significant additional resources needed for its realization, in addition to the need for further studies to clarify certain feasibility aspects, including those that you have commented upon, we have decided to make approval of Phase 3 the subject of a separate proposal. 

A standard year of operation is taken to be 200 days and 3000 spills per day. 
%A Full Year Equivalent (FYE) is defined as 200 days and 3000 %spills per day. 
%HIKE requires spill length of 4.8 sec. 
%Years of operation are taken to be FYEs. 
Considering the uncertainties on the start and end of LS slots, the concept of years between long shutdowns (LS) will not be used in this document. 
%Instead, we use FYEs 
%define a full year equivalent (FYE) as a data taking period of 200 days and 3000 spills per day. 
%and we count FYEs from the time T0 when first beam will be available. 
Instead, we count standard years from the time T0 when first beam will be available.
We note, however, that the programme has enough built-in flexibility to allow for the accommodation of LS
when the overall CERN SPS schedule will be settled.
%NA62 collects about $1.8 \times 10^{18}$ POT per FYE, corresponding to about $5\times 10^{12}$ Kaon decays in the fiducial volume per FYE.
%HIKE Phase 1 will collect $2 \times 10^{13}$ Kaon decays in the fiducial volume per FYE, corresponding to $7 \times 10^{18}$ POT per FYE.
%Hence HIKE corresponds to a factor 4 increase in POT and Kaon decays with respect to NA62.
%, and the number of kaon decays per POT is the same for both experiments (about $3 \times 10^{-6}$ kaon decays per POT). 
%HIKE, with new or upgraded detectors and readouts to profit the most from the increased beam intensity, will improve the acceptance of those kaon decays and keep the random veto under control at much higher intensity. Further studies are on-going for the proposal.

The number of kaons decaying in the fiducial volume per proton-on-target (POT) is different for each phase of the experiment, as the production angle, secondary beam acceptance and decay volume definition are different for each phase. The beam intensity is expressed in terms of protons per pulse (ppp) over a spill length of 4.8~sec.

For Phase~1 (Section~\ref{sec:phase1}), there are $2.5\times10^{-6}$ $K^+$ decays in the fiducial volume per POT. %With 3000 good spills per day, 200 days per year, 
In a standard year and with a beam intensity of $1.3\times10^{13}$~ppp (four times NA62 nominal intensity equal to $33\times10^{11}$~ppp), this gives $2\times10^{13}$ $K^+$ decays per year, i.e. to a factor four increase in POT and $K^+$ decays with respect to NA62.

For Phase~2 (Section~\ref{sec:phase2}), the neutral beam is produced at 2.4~mrad with a solid-angle bite of $\pi(0.4~{\rm mrad})^2 = 0.503$~$\mu$sr, leading to $5.4\times10^{-5}$ $K_L$ at the final collimator plane per POT. With an acceptance of 5.9\%, this corresponds to $3.2\times10^{-6}$ $K_L$ decays in the fiducial volume per POT, or $3.8\times10^{13}$ $K_L$ decays per year at a primary intensity of $2\times10^{13}$~ppp (six times NA62).

The sharing between kaon and beam-dump modes is a matter of scientific scheduling, that by 2031 will take into account the physics priorities at that time and the results from current experiments (NA62, LHCb, Belle~II and FIPs experiments). At HIKE, kaon physics is the highest priority while the sensitivity to FIPs is an extension to the flavour programme that adds value and scheduling flexibility. Three beam sharing scenarios  summarised in Table~\ref{tab:my_label} are considered. In all scenarios, we assume a (conservative) beam intensity of $2\times 10^{13}$~ppp in the dump mode.
%although an increase in intensity to $2.4\times 10^{13}$~ppp is possible and would be welcome by HIKE since the experiment is not limited by the instantaneous beam intensity in the dump mode (Section~\ref{sec:FIPS}).
%and this would allow to collect data in dump mode in less %integrated time. 
%We also note that the compatibility of the dump mode with the $K_L$ beam line design is being checked by the beam group (that would have implications in the scenario B described below).
%Three scenarios are presented for simplicity
%, and the only-kaon scenario is for comparison.
%The scenarios are 
Scenario A is the kaon-only scenario, presented for comparison.
%is strongly towards kaon physics, with occasional yearly periods in dump mode as NA62, reaching an integrated $10^{19}$
Scenario B reaches $10^{19}$~POT in dump-mode by the end of Phase 1. This would allow HIKE to reach the deliverables of Phase~1 ($K^+$, 4+1 standard years) and Phase~2 ($K_L$, 5+1 standard years), including $10^{19}$~POT in dump-mode (about 1 standard year), in 12 standard years. 
We note that one year is added to the minimum time for each phase to account for contingency.
%The second scenario (B) considered is a yearly share between kaon and dump of about 2/3 and 1/3 respectively.
%, reaching an integrated POT of $\sim 5 \times 10^{19}$ POT  in dump mode in 4 integrated FYEs. 
%In this scenario, HIKE would reach the deliverables of Phase~1 ($K^+$, 4 FYEs + 1 contingency) and Phase 2 ($K_L$ with tracking, 5 FYEs + 1 contingency), including $\sim 5\times 10^{19}$ POT in dump-mode (4~FYEs) by the end of Phase~2, in about 15 FYEs.
Scenario (C) involves a 50\%--50\% yearly share between kaon and dump modes in the first 8 standard years only, reaching an integrated $5\times 10^{19}$~POT in dump mode in 4~integrated standard years.
In this scenario, HIKE would reach the deliverables of Phase~1 ($K^+$, 4+1 standard years) and Phase~2 ($K_L$, 5+1 standard years), including $5\times 10^{19}$~POT in dump mode (4 standard years) by the end of Phase~2, in about 15 standard years. HIKE would welcome the increase of about 25\% in intensity in dump-mode that is deemed feasible at this point of the TAX and beamline design (Section~\ref{sec:beam}) as this would allows us to collect the same integrated $5\times 10^{19}$~POT in dump mode with a time sharing of 60--40\% in the first 8 standard years.
We also note that HIKE in dump mode is not limited by the beam intensity (integrated or instantaneous) since the detector rates are very low compared to the kaon mode (see Section~\ref{sec:fips} for more details).

The scenario C is equivalent to the scenario presented by the SHADOWS collaboration, and is assumed to produce the HIKE and SHADOWS combined sensitivity curves for FIPs searches shown in Section~\ref{sec:fips_sensitivity}. The possibility of switching rapidly between kaon and beam-dump modes adds flexibility to the programme and opportunities for optimisation and best exploitation of the available beam time, also when fitting into the overall SPS schedule.

%We suggest that scenario C is used for combining the FIPs sensitivity with HIKE and SHADOWS.
%Scenario C is the scenario currently compatible with SHADOW's requests.

%The third scenario (C) considered is a yearly share between kaon and dump of about 50\% respectively in the first 8 FYEs, reaching an integrated POT of $\sim 5 \times 10^{19}$ POT in 4 integrated FYEs. In this scenario, HIKE would reach the deliverables of Phase 1 ($K^+$, 4 FYEs + 1 contingency) and Phase 2 ($K_L$ with tracking, 5 FYEs + 1 contingency), including $\sim 5\times 10^{19}$ POT in dump-mode (4 FYEs), in about 15 FYEs.

%kaon plus dump, scenario C as example of presence of both that uses the beam at best, flexible and best use of beam schedule. both have priorities but able to adapt them and optimise them to push more physics in. Nobody knows which physics urgencies will be in 2031. Search for BSM is CERN priority.

\begin{table}[t]
\caption{Scenarios of sharing between kaon and beam dump physics, with the time fraction in kaon mode, and years needed to complete HIKE phases.}
\vspace{-2mm}
\centering
\begin{tabular}{c|c|c|c|c}
\hline
Scenario & Fraction of time & Integrated POT & Years to & Years to \\
& in kaon mode & in dump mode & Phase~1 goals & Phase~1+2 goals \\
& over HIKE lifetime &        &               &     \\
\hline
A  & 100\% & -- & 5 & 11 \\
(for comparison) &  &  & & \\
%A & 92\% & $10^{19}$ & 6 & 12 \\
\hline
%B & 73\% & $5\times 10^{19}$& 7 & 15 \\
B & 92\% & $10^{19}$ & 6 & 12 \\
\hline
% C & 73\% 50\% in first 8 years) & & 9 & 15 \\
C & 50\% in first 8 years,  & $5\times 10^{19}$& 9 & 15 \\
& 100\% afterwards &  &   &   \\
%\hline
%\hline
%Kaons only & 100\% &  - & 5 & 11 \\
\hline
\end{tabular}
\label{tab:my_label}
\end{table}

\begin{figure}[p]
\begin{center}
\includegraphics[width=\textwidth]{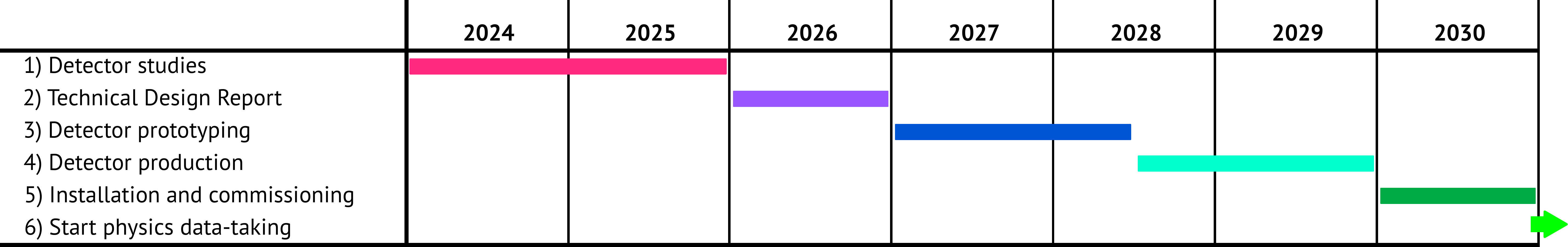}
\vspace{-6mm}
\caption{Tentative HIKE timeline from present to the first beam.}
\label{fig:chart}
\end{center}
\end{figure}

In summary, assuming scenario C, HIKE Phases~1 and 2 will cover a total of 15 standard years, including 4 standard years in dump mode and 11 standard years in kaon mode. The timeline from present to first beam is presented in Fig.~\ref{fig:chart}.

%\begin{itemize}
%\item 2024--2025: detector studies;
%\item 2026: Technical Design Report;
%\item 2027--2028: prototyping; 
%\item 2028--2029: production; 
%\item 2030: installation and possible commissioning with lower-intensity beam;
%\item 2031: start of data acquisition with a high-intensity beam.
%\end{itemize}

%%%%%%%%%%%%%%%%%%%%

\subsection{Cost estimate}

The human, technical and financial resources needed for HIKE are being evaluated. The cost of the main HIKE detectors (see Section~\ref{sec:detectors} for details) is presented in Table~\ref{tab:costs}.
%HIKE Phase~2 will see the addition of the Main Electromagnetic Calorimeter, for an extra 5~MCHF. 
%We expect this will cost about 100 KCHF. Given the number of detectors involved, we plan about 50 Readout Units at 15 KCHF each, for a total of about 750 KCHF. The network equipment will cost about 50 KCHF. The cost of the backend computing infrastructure, with servers and GPU/FPGA, will be about 400 KCHF. The grand total is therefore about 1.3 MCHF. 
%The total for Phase~1 is 22.3~MCHF.
%HIKE Phase~2 will see the addition of the Main Electromagnetic Calorimeter, for an extra 5~MCHF. 
%The remaining costs for Phase~3 (KLEVER) will be mainly the increase in the number of Large Angle Vetoes from 12 to 25 units (for an extra 9~MCHF). 
The numbers are only intended to give an idea of the financial extent of the project. A refined estimate will be prepared for the Technical Design Report.
Operation costs will be similar to today (indexed by inflation). The present host lab services are free of charge, and we assume they will remain so. Collaboration institutes will be charged a fair share of the Maintenance and Operation Costs, according to common practice for CERN experiments. A summary of the detector configuration for HIKE Phases~1 and 2 is presented in Table~\ref{tab:hike-detectors}.

\begin{table}[p]
\centering
\caption{Preliminary group interests and detector costs for HIKE.}
\vspace{-2mm}
\begin{tabular}{l|l|c}
\hline
Detector & Group & Cost (MCHF)  \\ 
\hline
Kaon ID (KTAG) & UK & 0.5 \\
%& Cherenkov - new photo-detectors  \\
\hline
Beam tracker & Italy, CERN, UK, & 3 \\
%& Silicon detector \\
       & Belgium, Canada, France &      \\
\hline
Charged particle veto (CHANTI) & Switzerland & 0.4 \\
%& 6 stations, with SiPMs \\
\hline
Veto counter (VC) & Switzerland & 0.3 \\
%3 stations (SciFi technology) \\
\hline
ANTI-0 & Germany & 0.4 \\
%& 1 plane, scint. tiles and SiPMs\\
\hline
Large Angle Vetos (LAV) & UK & 8 \\
%12 modules, Shashlyk technology and SiPMs\\
\hline
STRAW & CERN, Kazakhstan, & 3.5 \\
& Slovakia, Czech Republic &    \\
\hline
%LKr & CERN &  0.3 \\ 
%or & & \\
Main calorimeter & Italy & 5 \\
%& LKr or Shashlyk technology and SiPMs \\
\hline
Small Angle Calorimeter (SAC) & Italy & 2 \\
%& High-Z crystals \\
%Main Elect. Calorimenter (MEC) & 5 & Shashlyk technology and SiPMs \\
\hline
Pion ID (RICH) & Italy, Mexico & 0.8 \\
%& Cherenkov - new photo-detectors  \\
\hline
Timing detector & Belgium & 0.4 \\
%& 2 planes, scint. tiles and SiPMs \\
\hline
HCAL & Germany & 1.5 \\
%& Shashlyk technology and SiPMs \\
\hline
Muon plane & Germany & 0.2 \\
%& 1 plane, scint.tiles and SiPMs \\
\hline
HASC & Romania & 0.2 \\
%& Using SiPMs \\
\hline
DAQ, computing & CERN, Italy, Spain, Mexico, US & 1.3 \\
\hline
Total & & 27.5 \\
\hline
\end{tabular}
\label{tab:costs}
\end{table}

%%%%%%%%%%%%%%%%%

\begin{table}[p]
\centering
\caption{Detector configurations for HIKE Phases~1 and 2. The evolution is indicated versus the current NA62 configuration for Phase~1, and versus the upgraded Phase~1 configuration for Phase~2.}
\vspace{-2mm}
\begin{tabular}{lccl}
\hline
Detector & Phase~1 & Phase~2 & Comment \\
\hline
Cherenkov tagger & upgraded & removed & faster photo-detectors\\
Beam tracker & replaced & removed & 3D-trenched or monolithic silicon sensor \\
Upstream veto detectors & replaced & kept & SciFi \\
Large-angle vetos & replaced & kept & lead/scintillator tiles\\
Downstream spectrometer & replaced & kept & STRAW (ultra-thin straws)\\
Pion identification (RICH) & upgraded & removed & faster photo-detectors\\
Main EM calorimeter & replaced & kept & fine-sampling shashlyk \\
Timing detector & upgraded & kept & higher granularity\\
Hadronic calorimeter & replaced & kept & high-granularity sampling\\
Muon detector & upgraded & kept & higher granularity\\
Small-angle calorimeters & replaced & kept & oriented high-Z crystals\\
HASC & upgraded & kept & larger coverage\\
DAQ, computing & upgraded & kept & fully-software trigger \\
\hline
\end{tabular}
\label{tab:hike-detectors}
\end{table}

%\begin{sidewaysfigure}
%%\begin{table}
%\begin{tabular}{lccll}
%\hline
%Detector & Phase~1 & Phase~2 & Comment & Preliminary group interests \\
%\hline
%Cherenkov tagger & upgraded & removed & faster photo-detectors & UK\\
%Beam tracker & replaced & removed & 3D-trenched or monolithic silicon sensor & CERN, Italy, UK, Belgium, Canada, France \\
%Upstream veto detectors & replaced & kept & SciFi & Switzerland \\
%Large-angle vetos & replaced & kept & lead/scintillator tiles & UK \\
%Downstream spectrometer & replaced & kept & STRAW (ultra-thin straws) & CERN, Kazakhstan, Slovakia, Czech Republic \\
%Pion identification (RICH) & upgraded & removed & faster photo-detectors & Italy, Mexico \\
%Main EM calorimeter & replaced & kept & fine-sampling shashlyk & Italy \\
%Timing detector & upgraded & kept & higher granularity & Belgium \\
%Hadronic calorimeter & replaced & kept & high-granularity sampling & Germany \\
%Muon detector & upgraded & kept & higher granularity & Germany \\
%Small-angle calorimeters & replaced & kept & oriented high-Z crystals & Italy \\
%HASC & upgraded & kept & larger coverage & Romania\\
%DAQ, computing & upgraded & kept & fully-software trigger & CERN, Italy, Spain, Mexico, US \\
%\hline
%\end{tabular}
%\caption{Detector configurations for HIKE Phases~1 and 2. The evolution is indicated versus the current NA62 configuration for Phase~1, and versus the upgraded Phase~1 configuration for Phase~2.}
%\label{tab:hike-detectors}
%%\end{table}
%\end{sidewaysfigure}

\section*{Acknowledgements}

We are grateful to Andrzej Buras, Andreas Crivellin, Felix Kahlhoefer,
Marc Knecht, and David Marzocca for their help and advice on the theoretical aspects of this proposal. We would like to thank the various CERN groups who helped with and gave input to this proposal, especially in the beam section, and the Physics Beyond Colliders members for their support.

\printbibliography[heading=bibintoc] %biblatex

\end{document}